%% file: main.tex
\documentclass[10pt,twocolumn,letterpaper]{article}

\usepackage[pagenumbers]{cvpr}

\definecolor{cvprblue}{rgb}{0.21,0.49,0.74}
\usepackage[pagebackref,breaklinks,colorlinks,allcolors=cvprblue]{hyperref}
\usepackage{colortbl}
\usepackage{makecell}

\def\paperID{1933} 
\def\confName{CVPR}
\def\confYear{2025}

\title{EvEnhancer: Empowering Effectiveness, Efficiency and Generalizability for Continuous Space-Time Video Super-Resolution with Events}

\author{
    Shuoyan Wei$^{1,2}$\ \ \ 
    Feng Li$^{3,}$\footnotemark[1]\ \ \ \ 
    Shengeng Tang$^{3}$\ \ \ 
    Yao Zhao$^{1,2}$\ \ \ 
    Huihui Bai$^{4,1,2,}$\footnotemark[1] \\
    $^{1}$Institute of Information Science, Beijing Jiaotong University\\
    $^{2}$Beijing Key Laboratory of Advanced Information Science and Network Technology\\
    $^{3}$School of Computer Science and Information Engineering, Hefei University of Technology\\
    $^{4}$Tangshan Research Institute of Beijing Jiaotong University\\
{\tt\small\{shuoyan.wei, yzhao, hhbai\}@bjtu.edu.cn, \{fengli, tangsg\}@hfut.edu.cn}}

\begin{document}
\maketitle
\footnotetext[1]{Corresponding authors}
\input{sec/0_abstract}    
\input{sec/1_intro}

\input{sec/2_rel}

\input{sec/3_meth}

\input{sec/4_exp}
\input{sec/5_con}

\section*{Acknowledgment} 
This work was partly supported by the National Natural Science Foundation of China (62302141, 62331003, 62120106009), the Joint Funds of the National Natural Science Foundation of China (U23A20314), the Fundamental Research Funds for the Central Universities (JZ2024HGTB0255), Beijing Natural Science Foundation (L223022), and the Natural Science Foundation of Hebei Province (F2024105029). 

{
    \small
    \bibliographystyle{ieeenat_fullname}
    \bibliography{main}
}
\input{fig/exp/exp_suppl_event}
\input{fig/exp/exp_suppl_event2}
\input{fig/exp/exp_suppl_adobe240}
\input{fig/exp/exp_suppl_gopro_t12s6}
\input{fig/exp/exp_suppl_gopro_t6sx}
\end{document}

%% file: sec/0_abstract.tex
\begin{abstract}
Continuous space-time video super-resolution (C-STVSR) endeavors to upscale videos simultaneously at arbitrary spatial and temporal scales, which has recently garnered increasing interest. However, prevailing methods struggle to yield satisfactory videos at out-of-distribution spatial and temporal scales. On the other hand, event streams characterized by high temporal resolution and high dynamic range, exhibit compelling promise in vision tasks. This paper presents EvEnhancer, an innovative approach that marries the unique advantages of event streams to elevate effectiveness, efficiency, and generalizability for C-STVSR. Our approach hinges on two pivotal components: 1) Event-adapted synthesis capitalizes on the spatiotemporal correlations between frames and events to discern and learn long-term motion trajectories, enabling the adaptive interpolation and fusion of informative spatiotemporal features; 2) Local implicit video transformer integrates local implicit video neural function with cross-scale spatiotemporal attention to learn continuous video representations utilized to generate plausible videos at arbitrary resolutions and frame rates. Experiments show that EvEnhancer achieves superiority on synthetic and real-world datasets and preferable generalizability on out-of-distribution scales against state-of-the-art methods. Code is available at \url{https://github.com/W-Shuoyan/EvEnhancer}.

\end{abstract}

%% file: sec/1_intro.tex
\section{Introduction}
\label{sec:intro}

\input{fig/performance/performance}
Video super-resolution (VSR) is a fundamental low-level vision task, aiming to recover high-resolution (HR) videos from low-resolution (LR) inputs. Nevertheless, video resources are mostly archived with reduced spatial resolution and constrained frame rates (\textit{i.e.}, temporal resolution) practically, necessitating the conversion to HR and high-frame-rate (HFR) ones for various applications~\cite{zhang2019two, wang2020dual,wang2023compression, yoo2023video,xu2024ibvc,li2024enhanced} and user experiences~\cite{hou2022perceptual,rahimi2023spatio,wu2024perception}.

To address this problem, a naive approach is to cascade separate VSR~\cite{wang2019edvr, li2019fast, haris2019recurrent, li2020learning, chan2021basicvsr, chan2022basicvsr++} and video frame interpolation (VFI)~\cite{niklaus2017video1, reda2019unsupervised, xu2019quadratic, lee2020adacof, chi2020all} or in turn, which overlooks the correlations between two sub-tasks, leading to suboptimal exploitation of spatiotemporal information in videos. Space-time video super-resolution (STVSR)~\cite{haris2020space, xiang2020zooming, kim2020fisr, hu2022spatial, wang2022bi, wang2023stdan, hu2023cycmunet+} integrates VSR and VFI into a unified framework that super-resolves the videos with intermediate frame interpolation via one-stage process. Both the above approaches are limited to fixed discrete space and time magnifications, and cannot provide arbitrary controls. Some researchers study continuous VFI~\cite{jiang2018super, bao2019depth, huang2022real, zhang2023extracting} or continuous VSR~\cite{lu2023learning, li2024savsr, huang2024arbitrary, shang2025arbitrary} methods, where the former is capable of variable frame interpolation while the latter enables arbitrary spatial upsampling including non-integer scales. However, these methods operate in isolation, failing to simultaneously ensure spatial and temporal flexibility.

To achieve this, research on continuous STVSR (C-STVSR) \cite{chen2022videoinr, chen2023motif, lu2024hr} algorithms has emerged, which typically learn video implicit neural representation (INR) to decode 3D spatiotemporal coordinates into RGB values, thus generating HR while HFR videos arbitrarily. These methods mainly decouple the learning for video INR into spatial INR and temporal INR to predict continuous spatial and temporal feature domains separately. This fails to fully capitalize on the spatiotemporal dependencies in the continuous domain, thus hindering the reconstruction performance, particularly at scales falling outside the training distribution. Besides, existing methods \cite{chen2022videoinr, chen2023motif} approximate intermediate motion fields between consecutive frames captured from conventional frame-based cameras, which is insufficient to infer accurate inter-frame temporal features when motions are vast or non-linear. Recently, event cameras famous as bio-inspired asynchronous sensors with high temporal resolution and high dynamic range, providing a promising solution to fulfill the inter-frame motion fields in video reconstruction~\cite{yu2021training,han2021evintsr,jing2021turning,lu2023learning,lin2020learning,tulyakov2021time,he2022timereplayer}. Nonetheless, the potential of events in C-STVSR remains largely unexplored.

In light of the above discussion, we propose an innovative event-driven approach EvEnhancer, which adeptly integrates the spatiotemporal synergies between frames and events to learn continuous video representations for effective and generalized C-STVSR. First, we present an event-adapted synthesis module (EASM) that executes event-modulated alignment and bidirectional recurrent compensation to explore the event information for long-term holistic motion trajectory modeling. It incorporates motion cues between input frames with event-based modulation forward and backward to acquire latent inter-frame features. We then recurrently propagate the event stream across time and fuse it with aligned frame features in both directions to maximize the gathering of temporal information.

In contrast to the decoupling strategies~\cite{chen2022videoinr, chen2023motif, lu2024hr} for video INR, we propose a local implicit video transformer (LIVT) that integrates local implicit video neural function with spatiotemporal attention. LIVT calculates cross-scale 3D attention based on the queried space-time coordinates in the continuous domain and the features from EASM over a local sampling grid. In this way, we can directly and efficiently obtain an accurate unified video INR that predicts the RGB values of queried coordinates, enabling to super-resolve videos in arbitrary frames and resolutions. As shown in Figure~\ref{fig:prf}, we conduct experiments with synthetic GoPro~\cite{nah2017deep} and Adobe240~\cite{su2017deep} datasets and the real-world BS-ERGB~\cite{tulyakov2021time} dataset. Experiments demonstrate that EvEnhancer achieves superior effectiveness and efficiency in super-resolving videos in arbitrary spatial and temporal scales within training distribution (In-dist.) against existing state-of-the-art methods. We also validate its preferable generalizability at out-of-distribution (OOD) scales. 

The main contributions are as follows:
\begin{itemize}
    \item We propose EvEnhancer that subtly marries the unique advantages of event streams with video frames for C-STVSR. Extensive experiments validate its superior performance over recent state-of-the-art methods.  
    \item We propose the event-adapted synthesis module (EASM) that enables long-term holistic motion trajectory modeling to acquire informative spatiotemporal features.
    \item We propose the local implicit video transformer (LIVT) that integrates local implicit video neural function with cross-scale spatiotemporal attention to learn accurate unified continuous video INR efficiently.
\end{itemize}

%% file: fig/performance/performance.tex
\begin{figure}[t]
  \centering
\includegraphics[width=1.0\linewidth]{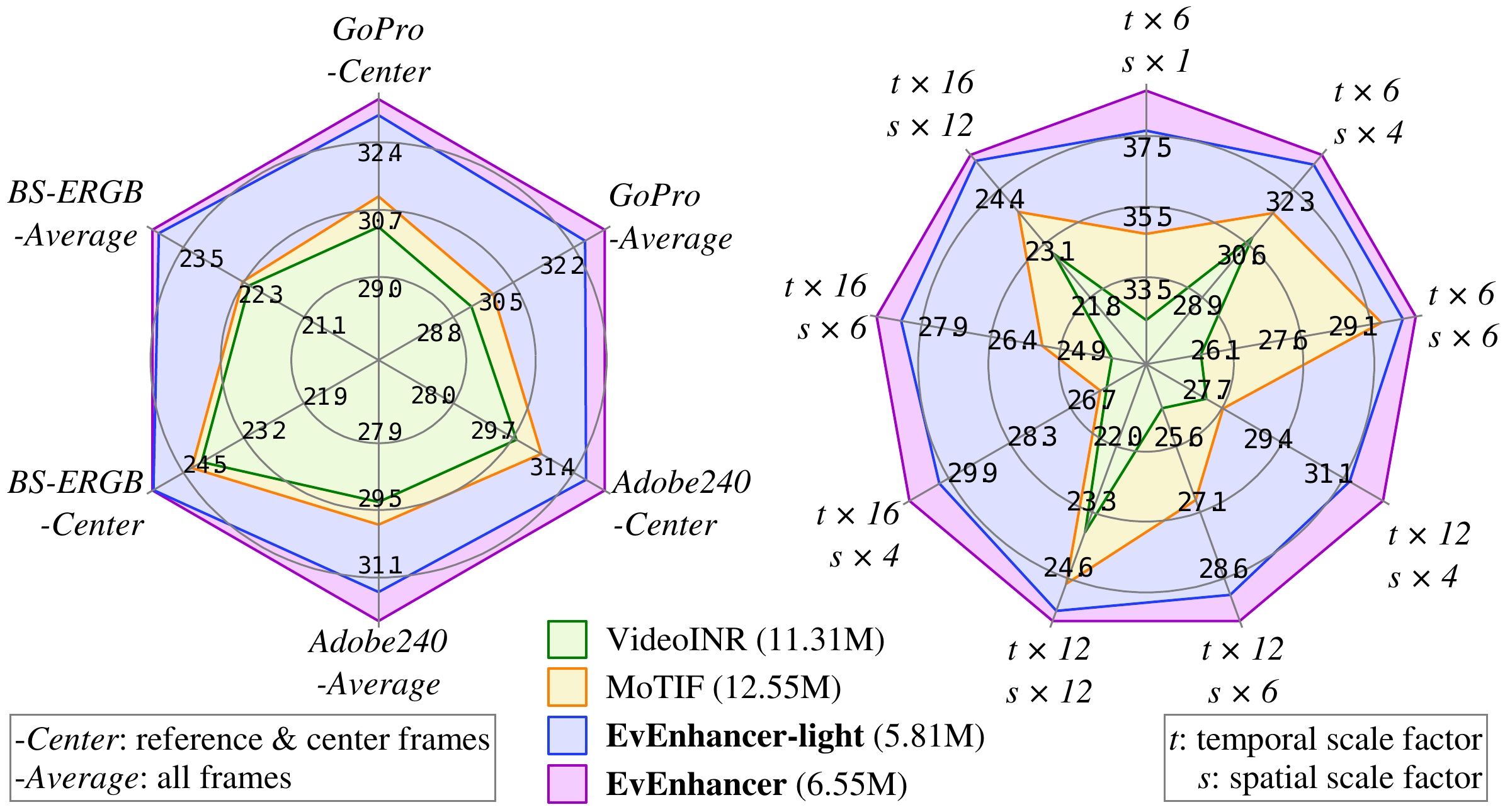}
   \footnotesize
  \begin{tabular}{c@{\hspace{0.5cm}}c}
  (a) Different datasets (In-dist.) &
  (b) Different upsampling scales
  \\
  \end{tabular}
  \begin{tabular}{@{}c@{\hspace{0.1cm}}|@{\hspace{0.1cm}}c@{\hspace{0.1cm}}c@{}}  \includegraphics[width=0.32\linewidth]{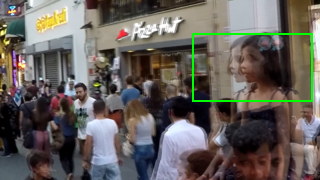}&
  \includegraphics[width=0.32\linewidth]{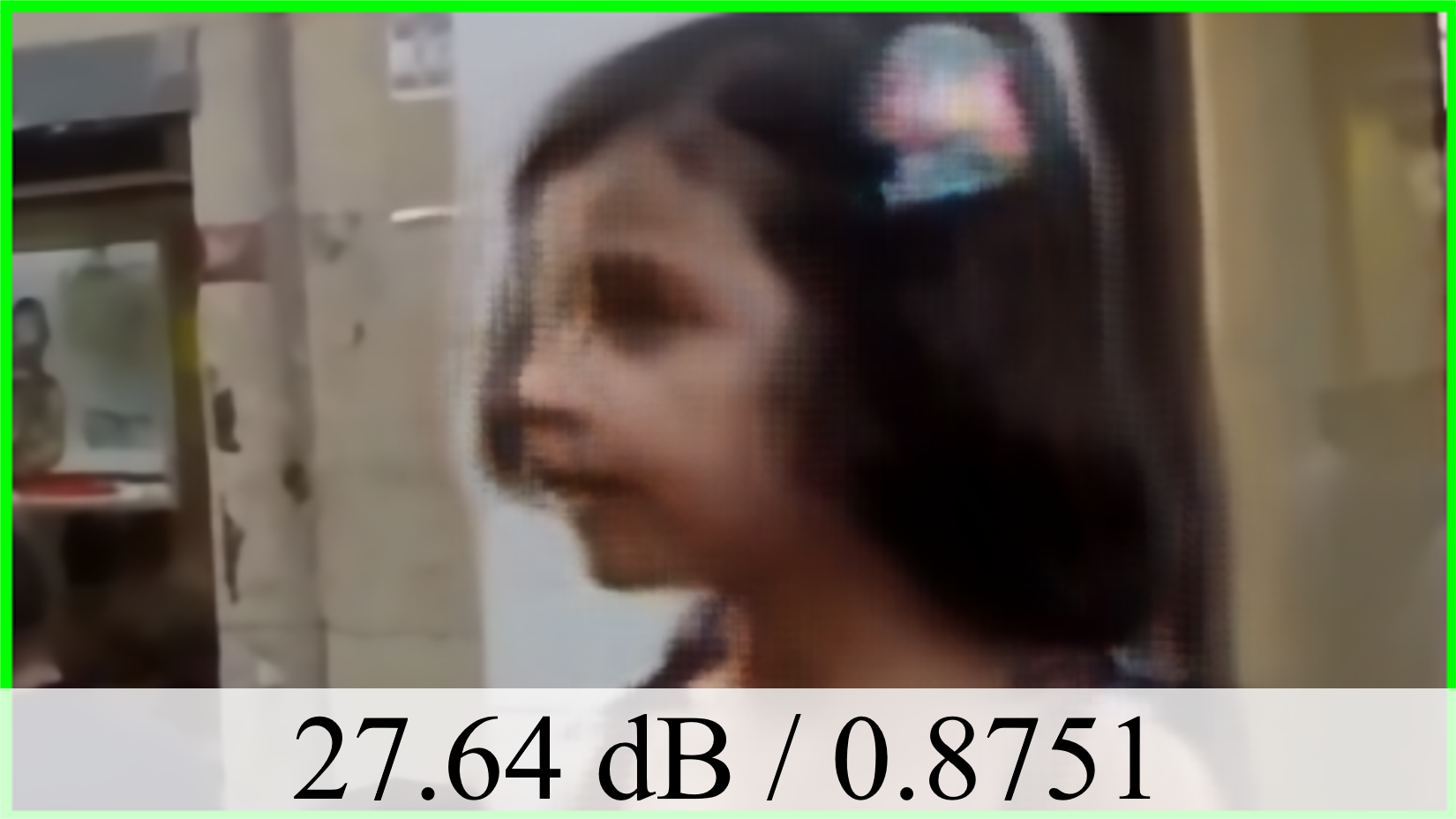}&
  \includegraphics[width=0.32\linewidth]{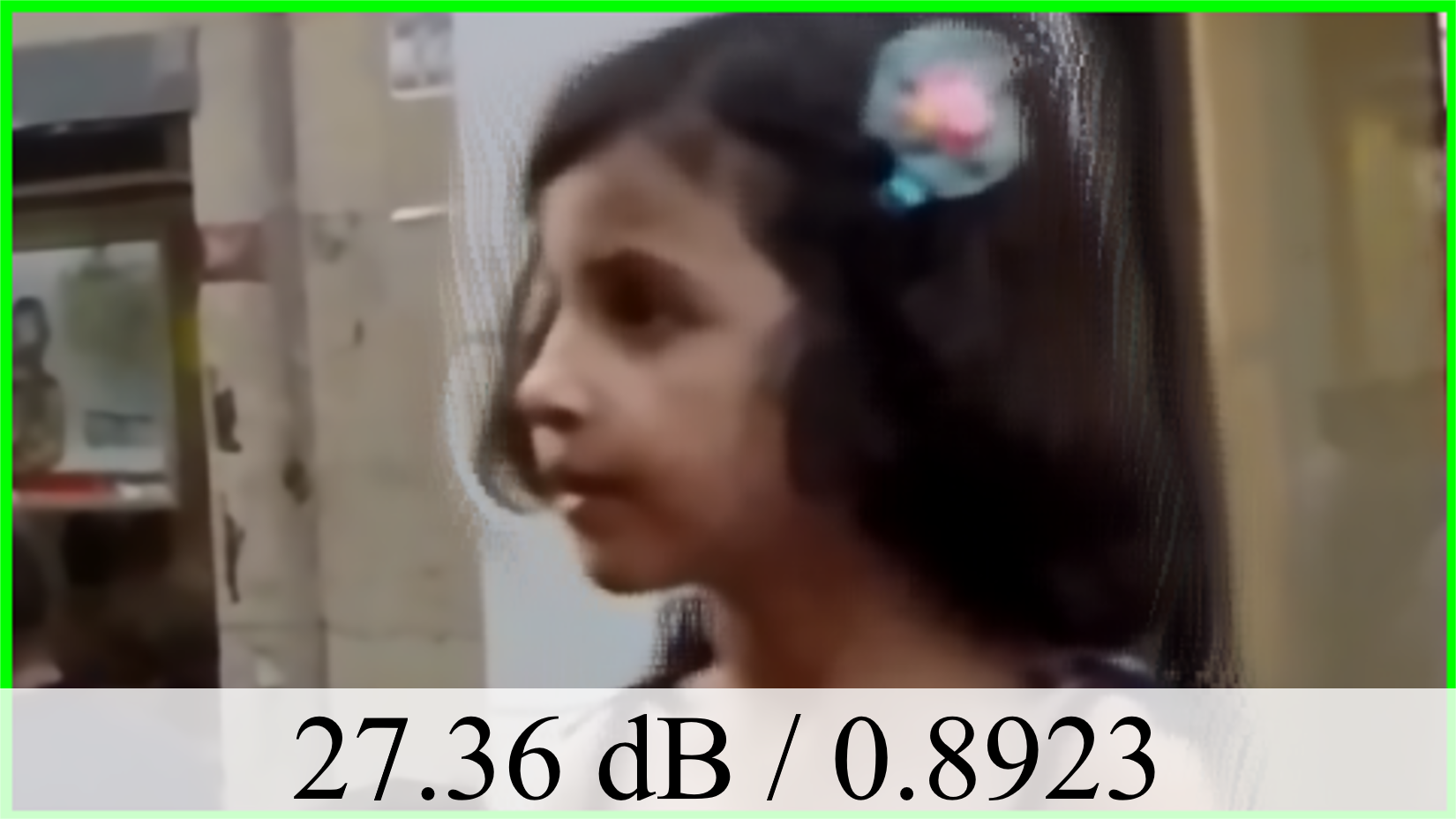} \\
  \scriptsize Overlapping LR Frames &
  \scriptsize VideoINR\cite{chen2022videoinr}&
  \scriptsize MoTIF\cite{chen2023motif} \\
  \includegraphics[width=0.32\linewidth]{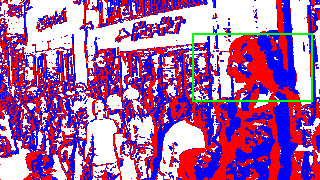}&
  \includegraphics[width=0.32\linewidth]{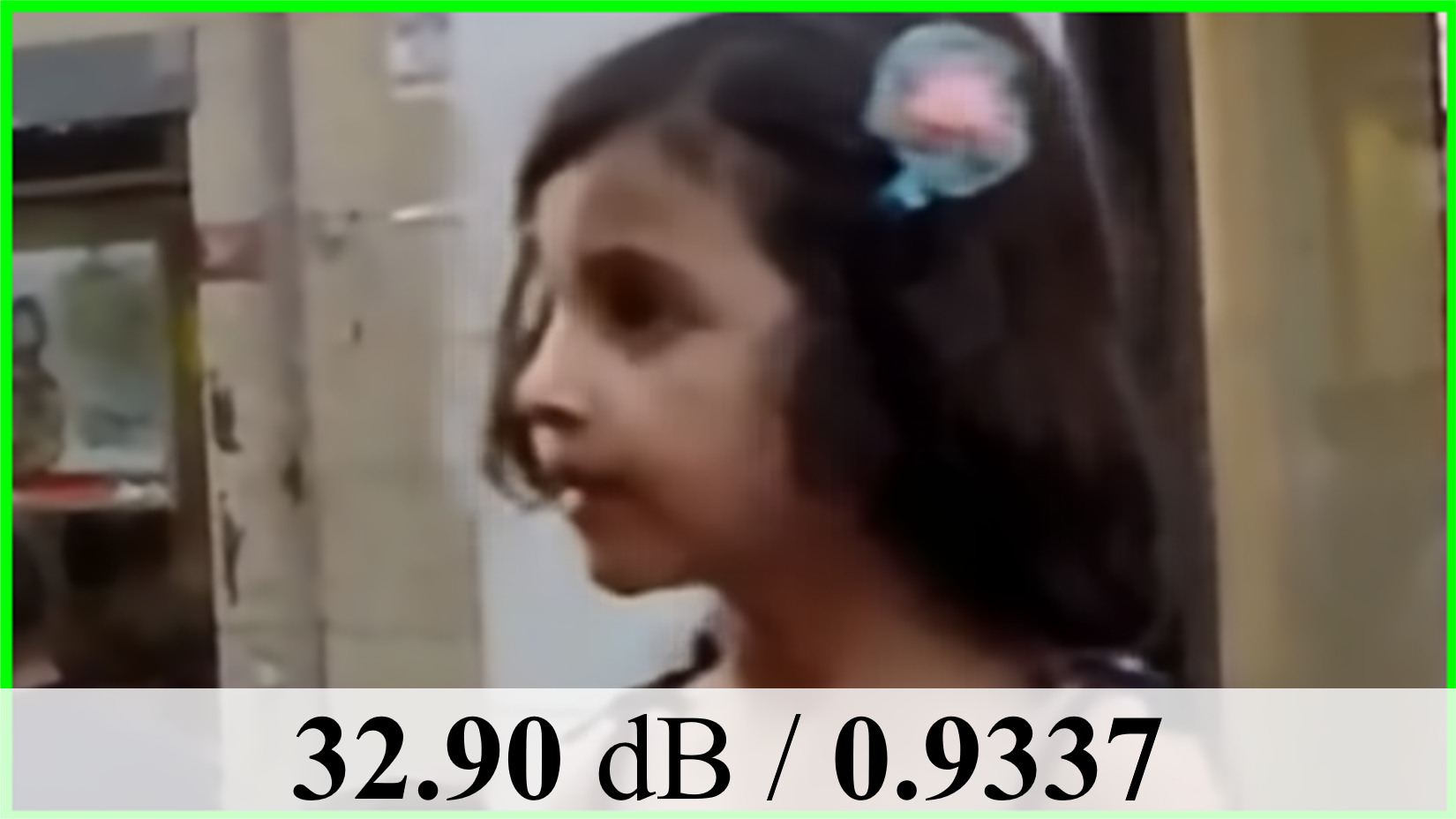}&
  \includegraphics[width=0.32\linewidth]{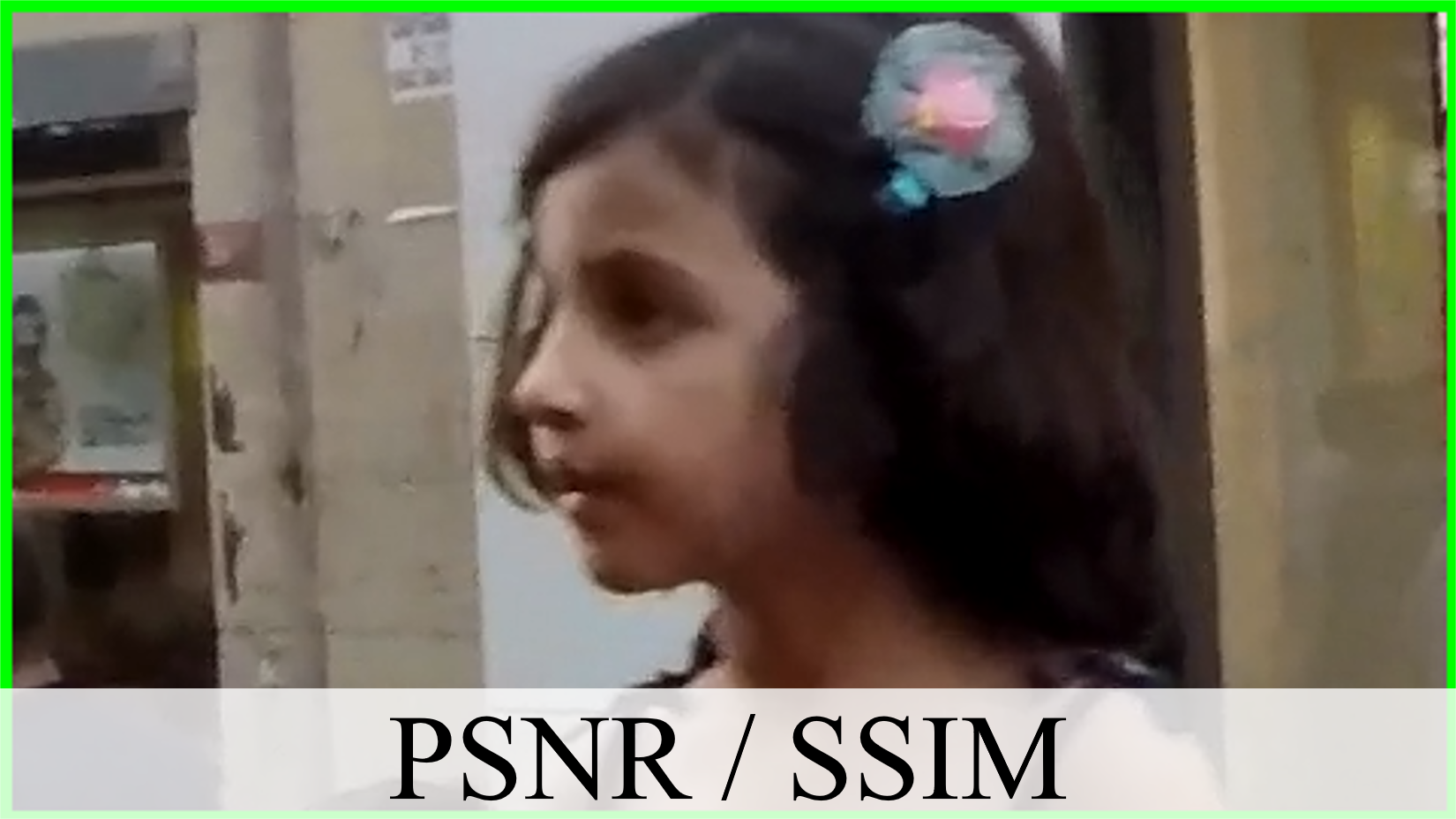}\\
  \scriptsize LR Events &
  \scriptsize \textbf{EvEnhancer (Ours)} &
  \scriptsize GT
  \\
  \multicolumn{3}{c}{(c) Visualization comparison}
  \end{tabular}
  \caption{Performance comparison of different C-STVSR methods. (a) PSNR (dB) comparison for in-distribution (In-dist.) spatiotemporal upsampling scale (temporal scale \(t=8\), spatial scale \(s=4\)) on the different datasets including GoPro \cite{nah2017deep}, Adobe240 \cite{su2017deep} and BS-ERGB \cite{tulyakov2022time}. (b) PSNR (dB) comparison for different spatiotemporal upsampling scales on GoPro. (c) Visualization comparison for In-dist. scale on GoPro.}
  \label{fig:prf}
\end{figure}

%% file: sec/2_rel.tex
\section{Related Work}
\label{sec:rel}

\noindent{\bf Video Super-Resolution (VSR).} Conventional VSR aims to reconstruct HR videos from LR counterparts, which heavily relies on motion estimation and compensation between frames. Earlier methods address this problem by learning frame-level optical flows~\cite{li2020learning,wang2019learning}. Later methods apply deformable convolution \cite{wang2019edvr,tian2020tdan,chan2021basicvsr} or 3D convolution~\cite{huang2017video,isobe2020video} to conduct implicit feature-level motion compensation, exhibiting improved effectiveness and efficiency. Some methods draw inspiration from attention mechanism~\cite{wang2018non,isobe2020revisiting}, which incorporate spatial-temporal attention \cite{yi2019progressive,li2020mucan,yu2022memory,zhou2022revisiting}or video transformer~\cite{liu2022learning, liang2022recurrent, qiu2023learning, zhou2024video} to model inter-frame correlations. However, these methods can only upsample at fixed scales.

\noindent{\bf Video Frame Interpolation (VFI).} VFI involves interpolating potential frames between consecutive video frames to enhance the temporal resolution. Flow-based methods~\cite{xu2019quadratic, park2021asymmetric, kalluri2023flavr} leverage optical flow networks~\cite{dosovitskiy2015flownet,ilg2017flownet,sun2018pwc} to estimate the bidirectional flows between frames. Kernel-based methods \cite{niklaus2017video1, niklaus2017video2, cheng2020video} utilize local convolution to map inter-frame motion, which synthesizes target pixels by convolving over input frames. In addition, some recent works focus on splatting-based methods~\cite{niklaus2020softmax, niklaus2023splatting,hu2023video} or transformer-based methods~\cite{lu2022video, shi2022video, zhang2023extracting, liu2024sparse}, showing significant advances. These methods almost learn pixel-level motions for interpolation, which are effective in linear motion but struggle with complex non-linear motion or occlusions.

\begin{figure*}[t]
    \centering 
    \includegraphics[width=\textwidth]{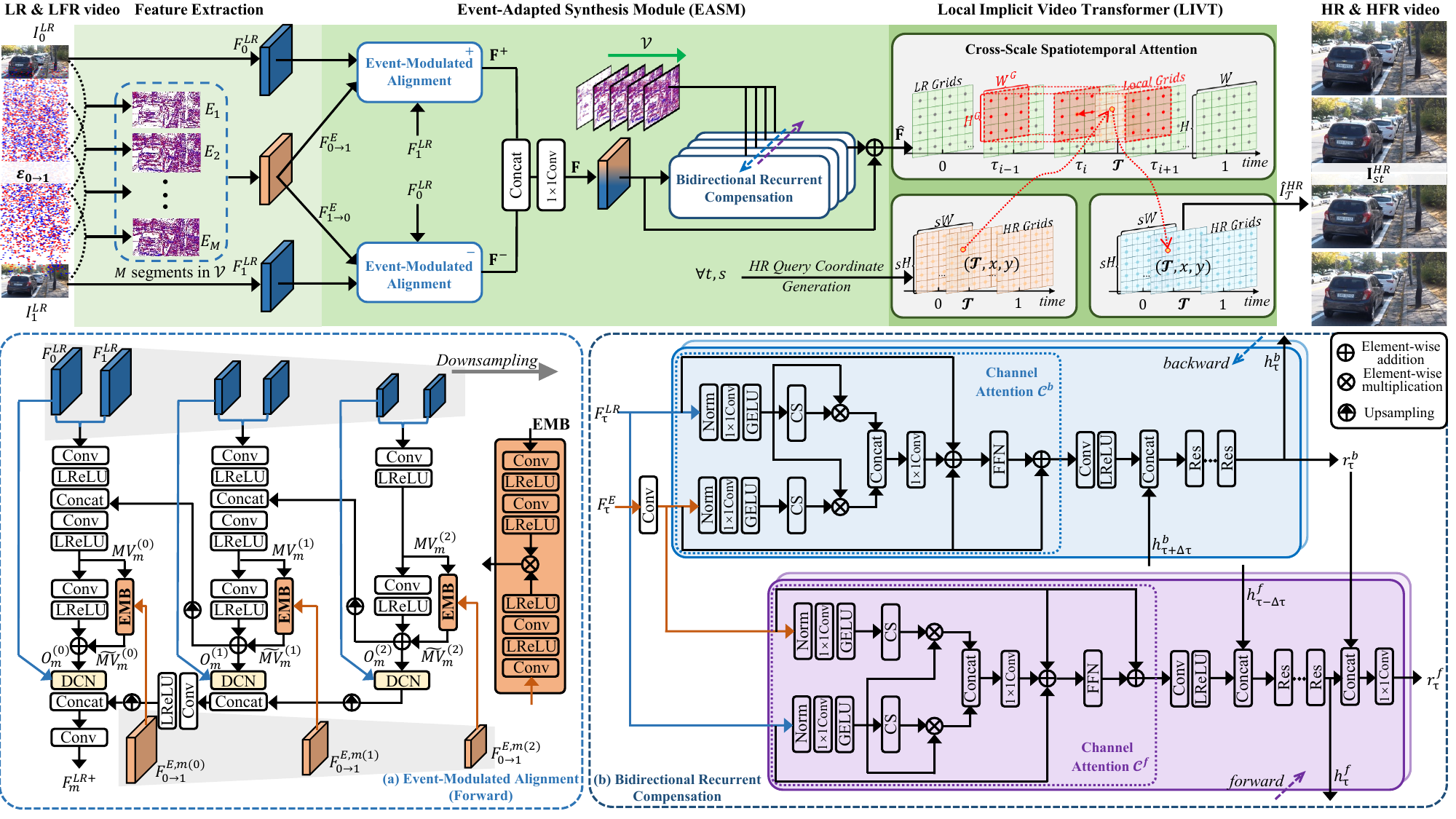}
    \caption{The overall architecture of our EvEnhancer, an event-driven C-STVSR method which consists of an event-adapted synthesis module (EASM), and a local implicit video transformer (LIVT), where EASM contains two steps: (a) event-modulated alignment (EMA), and (b) bidirectional recurrent compensation (BRC). ``CS'': channel squeeze, ``FFN'': feed-forward neural network, ``Res'': residual blocks.}
    \label{fig:network}
\end{figure*}

\noindent{\bf Space-Time Video Super-Resolution (STVSR).} STVSR can be seen as a combination of VSR and VFI, which reconstruct HR videos spatially and temporally. In literature, most existing methods including traditional~\cite{shechtman2005space,shahar2011space} and deep learning methods~\cite{xiang2020zooming, geng2022rstt, huang2024scale} are prone to achieving the STVSR with fixed spatial and temporal scales, denoted by F-STVSR. To mitigate the temporal scale constraints, TMNet \cite{xu2021temporal} proposes controlled frame interpolation with temporal modulation. SAFA~\cite{huang2024scale} introduces a scale adaptive feature aggregation network that iteratively estimates motion with trainable block-wise scale selection. Though these methods achieve temporal flexibility to some extent, they still fail in arbitrary spatial scales.

\noindent{\bf Event-Based VSR \& VFI.} Event cameras are characterized by high temporal resolution, high dynamic range, and low latency, promoting many VFI~\cite{tulyakov2021time,tulyakov2022time,wu2022video, sun2023event, kim2023event} and VSR~\cite{han2021evintsr,jing2021turning,lu2023learning} works. E-VSR~\cite{jing2021turning} introduces the pioneering event-based VSR method that utilizes event-based asynchronous interpolation to synthesize asynchronous neighboring frames. EGVSR~\cite{lu2023learning} unleashes the potential of events in arbitrary spatial VSR, which learns implicit neural representations from both RGB frames and events. EvTexture \cite{kai2024evtexture} regards the voxelized events as high-frequency signals to improve the textures in RGB frames, yielding promising VSR reconstruction. As for event-based VFI, TimeLens \cite{tulyakov2021time} and TimeLens++ \cite{tulyakov2022time} innovatively integrate deformation and synthesis processes. CBMNet \cite{kim2023event} utilizes cross-modality information to directly estimate inter-frame motion fields. REFID \cite{sun2023event}, as a flexible method, exhibits robustness even with blurred reference frames. Notably, these methods achieve upsampling only in the spatial or temporal domain independently, and few studies address a unified approach for both, \textit{i.e.} event-based STVSR.
 
\noindent{\bf Implicit Neural Representation (INR) in STVSR.} As a method of continuous representation of signals, INR is a common means of continuous super-resolution, including image \cite{chen2021learning, yang2021implicit, lee2022local, chen2023cascaded, he2024latent} and video \cite{chen2022videoinr, lu2023learning, chen2023motif, lu2024hr} tasks.
Inspired by LIIF \cite{chen2021learning}, VideoINR \cite{chen2022videoinr} pioneers continuous STVSR (C-STVSR) by learning separate spatial and temporal INRs. Building on this, MoTIF \cite{chen2023motif} introduces forward spatiotemporal INR to enhance motion field estimation, but it retains the paradigm of decoupling video INR across multiple domains, which lacks robustness across varying scales. 
Similarly, HR-INR~\cite{lu2024hr}, a recent event-based C-STVSR method, also adheres to the same decoupling paradigm as VideoINR~\cite{chen2022videoinr}. Besides, HR-INR conducts holistic event-frame extraction for long-range motion dependencies and requires repeatedly extracting input frames to realize frame temporal super-resolution, introducing redundancy.
Instead, we break the inherent paradigm. We devise the EASM that executes event-modulated alignment and bidirectional recurrent compensation for long-term motion trajectory modeling. Based on this, our proposed LIVT achieves a unified spatiotemporal INR through the high temporal resolution grids generated from events, thus achieving better generalization with lower weights.

%% file: sec/3_meth.tex
\section{Proposed Method}
\label{sec:method}
Given LR and low-frame-rate (LFR) videos and corresponding event streams as inputs, we present an event-driven C-STVSR approach to reconstruct HR and HFR videos at arbitrary spatial and temporal scales. Before elaborating on the components of our methodology, we first provide the details for the event representation in this work.

\subsection{Event Representation}
\label{sec:event}
Event cameras asynchronously capture the logarithmic intensity change $\Delta \log(I_{x,y,\tau})$ of each pixel at coordinate $(x,y)$ within the RGB image $I_{x,y,\tau}$, where $\tau$ denotes the timestamp of an event stream $\mathcal{E}=(x,y,\tau,p)$. $p$ represents the polarity of $\mathcal{E}$ triggered at $(x,y)$, with 1 and -1 representing the change directions respectively. Compared to conventional RGB images, a raw event stream is constituted by large amounts of sparse points, which is capable of 
continuously capturing intensity change with high temporal resolution. However, processing such an unstructured sparse format in deep networks is non-trivial. Following existing event-based method~\cite{sun2023event}, we represent the event stream into a 3D spatiotemporal voxel grid $\mathcal{V}\in\mathbb{R}^{H\times W\times (M+1)}$ linearly, where $H\times W$ denotes the spatial size, and then discretize the temporal dimension into $M$ event segments $[E_1,E_2,...,E_M]$. Each segment records the intensity information within a time window of $\mathcal{E}$, which is often set to be a small constant (see Figure~\ref{fig:network}).

\subsection{Overview}
\label{sec:3-1}
Let $\mathbf{I}^{LR}=I_{0}^{LR}, I_{1}^{LR}\in\mathbb{R}^{H\times W\times 3}$ denote two LR RGB frames and $\mathcal{E}_{0 \rightarrow 1}$ denotes their corresponding event stream. We propose EvEnhancer that addresses the task of super-resolving $\mathbf{I}^{LR}$ to $(t+1)$ HR video frames with any spatial scale $s\geq1$ and temporal scale $t\geq1$. The overall framework is illustrated in Figure~\ref{fig:network}, which consists of an event-adapted synthesis module (EASM) and a local implicit video transformer (LIVT).

Generally, we first extract initial features on frames and event segments, resulting in RGB frame features $F^{LR}_0$, $F^{LR}_1$, and event features $\mathbf{F}^E_{0\rightarrow1}=\{F^{E,m}_{0\rightarrow1}\}^M_{m=1}$ for $M$ segments. Considering that the latent inter-frame features between RGB frames can be captured in both forward and backward directions, we also reverse the time and polarity of $\mathcal{V}$ to derive backward event features $\mathbf{F}^E_{1\rightarrow0}$. Then, EASM models long-term holistic motion trajectories with the guidance of events to generate a spatiotemporal feature sequence $\widehat{\mathbf{F}}$ with high temporal resolution. Finally, LIVT integrates cross-scale spatiotemporal attention with the local implicit function to learn continuous video INR that decodes arbitrary space-time coordinates into RGB values, thus yielding HR video frames $\mathbf{I}^{HR}_{st}$ with HFR.

\subsection{Event-Adapted Synthesis Module}
\label{sec:3-2}
As depicted in Figure~\ref{fig:network}, EASM contains two steps: 1) Event-modulated alignment incorporates motion cues between input frames with event modulation to acquire latent inter-frame features. Based on the event features and reverse ones, we deploy such alignment in parallel to model the motions forward (``+'') and backward (``-''). 2) Bidirectional recurrent compensation propagates the event stream across time and fuses it with acquired features in both directions to maximize the gathering of temporal information.

\noindent\textbf{Event-Modulated Alignment (EMA)}. Given input frame features $F^{LR}_0$ and $F^{LR}_1$, our goal here is to infer the features of intermediate frames according to $M$ event segment features. Motivated by the effectiveness of the PCD module in video restoration~\cite{wang2019edvr,xiang2020zooming}, we construct a customized pyramid and event-modulated structure that progressively learns the motion offsets and feature alignment. In the following, we take the forward alignment as an example for clarity.

Specifically, for the forward alignment, as shown in Figure~\ref{fig:network}(a), we implement a 3-level pyramid by downsampling the frame and event features ($F^{LR}_0$, $F^{LR}_1$, $F^{E,m}_{0\rightarrow1}$) to 3 different spatial scales \(\times 1\), \(\times \frac{1}{2}\), \(\times \frac{1}{4}\) via stride convolutions. At each level $l$, we first calculate an initial motion vector by a convolutional block (``Conv+LeakyReLU (LReLU)'') on concatenated frame features, which is then fused with the learned offset $O_{m}^{(l-1)}$ from the $(l-1)$-th level to obtain a motion vector ${MV}_{m}^{(l)}$ near the $m$-th timestamp. Then, we design an event modulation block (EMB) to generate modulated $\widetilde{MV}_{m}^{(l)}$. The modulation is controlled by the event feature $F^{E,m}_{0\rightarrow1}$. Compared to previous PCD-based methods~\cite{wang2019edvr,xu2021temporal,xiang2020zooming}, our event-based approach can furnish more deterministic motion cues to facilitate frame synthesis, formulated as
\begin{equation}
    \widetilde{MV}_{m}^{(l)} = \mathcal{M}^{(l)}({MV}_{m}^{(l)}, F^{E,m(l)}_{0\rightarrow1}),
\end{equation}
where $F^{E,m(l)}_{0\rightarrow1}$ denotes the downsampled event feature at level $l$. Then, we can learn the offset ${O}_{m}^{(l)}$ from ${MV}_{m}^{(l)}$ and $\widetilde{MV}_{m}^{(l)}$. Following~\cite{wang2019edvr}, we use the deformable convolutional network (DCN)~\cite{zhu2019deformable} for feature alignment and interpolation. The deformable alignments at different levels are cascaded and aggregated coarse-to-fine for accuracy improvement via the pyramid workflow. 

Notably, as there are $M$ segment features $\{F^{E,m}_{0\rightarrow1}\}^M_{m=1}$, by the event-modulated alignment, we can produce $M$ intermediate frame features $\mathbf{F}^+=[F^{LR+}_1, F^{LR+}_2,...,F^{LR+}_M]$. Similarly, for the backward alignment, we get $\mathbf{F}^-=[F^{LR-}_1, F^{LR-}_2,...,F^{LR-}_M]$. The two are integrated by channel concatenation with a convolutional block, forming the final features $\mathbf{F}=\{F^{LR}_{\tau}\}$ with the input features $F^{LR}_0$, $F^{LR}_1$, where the timestamp $\tau=0, \frac{1}{M+1},....,\frac{M}{M+1},1$ is aligned with the input frames and $M$ event segments.

\noindent{\bf Bidirectional Recurrent Compensation (BRC).} Given that the information encapsulated within each event segment is confined to a very narrow time window, after the modulated alignment, we propose bidirectional recurrent compensation that iteratively fuses the event and frame features to model the global temporal information inherent in two modalities, as shown in Figure~\ref{fig:network}(b). In each recurrent iteration, the compensation block receives the event feature $F^E_{\tau}$ from the current timestamp $\tau$, the forward hidden state from $h^f_{\tau-\Delta \tau}$ the previous timestamp $\tau-\Delta \tau$, and the backward hidden state $h^b_{\tau+\Delta \tau}$ from the next timestamp $\tau+\Delta \tau$, and the aligned frame feature $F^{LR}_{\tau}\in\mathbf{F}$. Here, $\tau\pm\Delta \tau$ denotes the former timestamp and the latter one. Hence, the recurrent fusion process at timestamp $\tau$ is formulated as
\begin{equation}
    \begin{aligned}
    r_{\tau}^{b}, h_{\tau}^{b} &= \mathcal{F}^b(\mathcal{C}^b(F^{LR}_{\tau}, F^E_{\tau}), h_{\tau+\Delta \tau}^{b}), \\
    r_{\tau}^{f}, h_{\tau}^{f} &= \mathcal{F}^{f}(\mathcal{C}^f(F^{LR}_{\tau}, F^E_{\tau}), h_{\tau-\Delta \tau}^{f}, r_{\tau}^{b}),\\
    \end{aligned}
    \label{eq:BER}
\end{equation}
where $r_{\tau}^{b}$ and $h_{\tau}^{b}$ denote the output and hidden states respectively in the backward direction, and $r_{\tau}^{f}$ and $h_{\tau}^{f}$ correspond to the forward. $\mathcal{C}^f$ and $\mathcal{C}^b$ calculate the channel attention on event and frame features during the forward and backward recurrent iterations respectively. By recurrently propagating the event information across time, we can comprehensively exploit the high temporal resolution properties from events to refine the frame features and maximize the gathering of temporal information. 
Finally, a bidirectional feature fusion is run by the element-wise addition between the input feature fed into BRC and its output bidirectional features to retain valid information.

\begin{figure}[t]
    \centering 
    \includegraphics[width=0.475\textwidth]{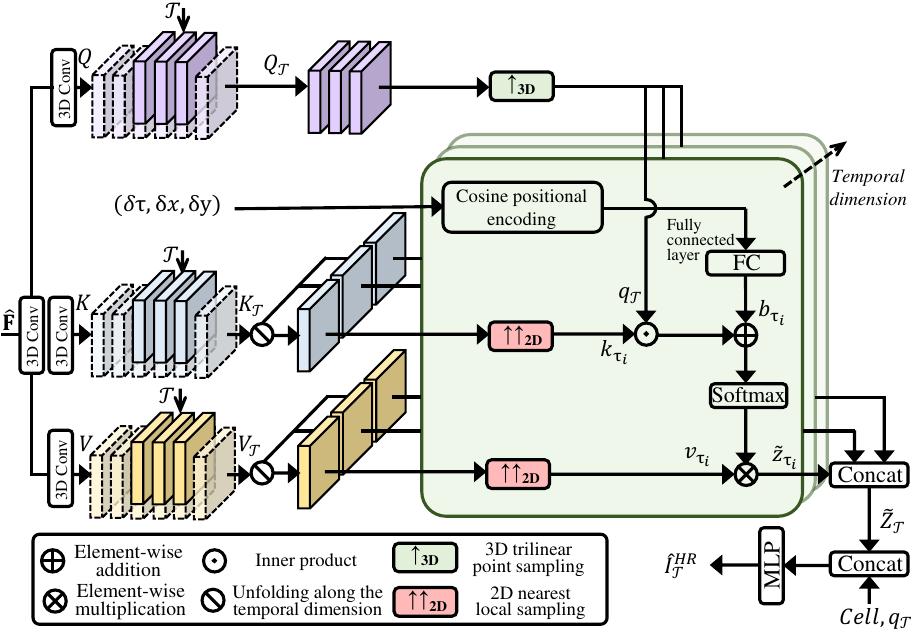}
    \caption{The structure of our local implicit video transformer (LIVT), which integrates the 3D local spatiotemporal attention with implicit neural function to learn continuous video INR to reconstruct HR and HFR video frames.}
    \label{fig:STCLA}
\end{figure}

\subsection{Local Implicit Video Transformer}
\label{sec:3-3}
After the EASM, we obtain the discrete temporal feature sequence $\widehat{\mathbf{F}}$. The other key in this method is to learn continuous video INR that the feature at the queried space-time coordinates $(\mathcal{T},x,y)$ is transferred to any target spatial scale $s$ and temporal scale $t$ in Figure~\ref{fig:network}. Previous C-STVSR methods~\cite{chen2022videoinr, chen2023motif, lu2024hr} learn spatial and temporal INR separately to realize video INR, which fails to fully capitalize on the spatiotemporal dependencies in the continuous domain. In this work, we propose LIVT which calculates cross-scale 3D attention based on the queried space-time coordinates in the continuous domain over a local sampling grid to learn an accurate unified video INR directly.

Assuming the temporal length and spatial size of the local grid in LR space are $T^{G}$ and $H^{G}\times W^{G}$, respectively. For the full temporal range $[0,1]$, to generate a frame at the timestamp $\mathcal{T}\in[0,1]$, we only need to select the $T^G$ nearest elements to $\mathcal{T}$ along the time axis.
We denote the set of candidate timestamps $\tau_i$ as $\mathbf{S}_{\tau} \subseteq [0, \frac{1}{M+1},....,\frac{M}{M+1},1]$, and the size of the $\lvert \mathbf{S}_{\tau} \rvert=T^G$, which is selected following the rule as
\begin{equation}
    \begin{aligned}
     \mathbf{S}_{\tau} = \underset{\mathbf{S}_{\tau}}{\arg \min}\sum_{\tau_{i} \in \mathbf{S}_{\tau}} \lvert \tau_{i} - \mathcal{T} \rvert.
    \end{aligned}
    \label{eq:select}
\end{equation}

As shown in Figure~\ref{fig:STCLA}, with the LR feature sequence $\widehat{\mathbf{F}}$ from EASM, we encode them into three embedding spaces through 3D convolution: key $\{K\}$, query $\{Q\}$, and value $\{V\}$. Notably, we do not calculate the spatiotemporal attention across the overall time range. Based on Eq.~\ref{eq:select}, we conduct temporal selection on $K$, $Q$, and $V$, which allows the learning for 3D local attention in a sub-temporal space near the target timestamp $\mathcal{T}$. On top of this, we employ 3D trilinear sampling on $Q$ to get the query \(q_\mathcal{T} \in \mathbb{R}^{s^{2}HW \times 1 \times C}\) for a certain frame, where $C$ is the channel number. However, since both $k_\mathcal{T}$ and $v_\mathcal{T}$ have size \(s^{2}HW \times T^{G}H^{G}W^{G} \times C\), it will suffer from expensive computational costs if we directly calculate the attention. To solve this problem, we propose to unfold the sequence along the temporal dimension and calculate cross-scale attention separately on $T^{G}$ embeddings. Thus, we sample the key $k_{\tau_i}$ and value $v_{\tau_i}$ from $K$ and $V$ respectively using 2D nearest local sampling at each time slice $\tau_i$. Moreover, we encode and reshape the spatiotemporal relative coordinates $(\delta \tau, \delta x, \delta y)\in(-1,1)$ from each query point to all pixel points within its local grid, obtaining the spatiotemporal bias \(b_{\tau_{i}}\in\mathbb{R}^{s^2HW\times H^G W^G\times3}\). The cosine positional encoding used here is an extension of~\cite{chen2023cascaded}, formulated by 
\begin{equation}
    \begin{aligned}
    g(\delta \mathbf{C})=[&\sin(2^0\delta \mathbf{C}),\cos(2^0\delta \mathbf{C}),...,\\
    &\sin(2^{L-1}\delta \mathbf{C}),\cos(2^{L-1}\delta \mathbf{C})],
    \end{aligned}
    \label{eq:pos_enc}
\end{equation}
where $\delta \mathbf{C}=\{(\delta \tau, \delta x, \delta y)\}$ is the spatiotemporal relative coordinates and $L$ is a hyper-parameter set to 10. 

We compute the inner product between $q_\mathcal{T}$ and $k_{\tau_{i}}$ and add it with the \(b_{\tau_{i}}\) to obtain the local attention map at the timestamp $\mathcal{T}$ after a Softmax operation. After the matrix multiplication with $v_{\tau_{i}}$, we can derive the output embedding \(\widetilde{z}_{\tau_{i}}\). This procedure is described as
\begin{equation}
    \begin{aligned}
    \widetilde{z}_{\tau_{i}} = Softmax(\frac{q_{\mathcal{T}}k_{\tau_{i}}^ \mathrm{T}}{\sqrt{C}}+b_{\tau_{i}}) \otimes v_{\tau_{i}}.
    \end{aligned}
    \label{eq:attention}
\end{equation}
We concatenate the output embeddings \(\widetilde{z}_{\tau_{i}}\) along the channel dimension to construct the complete continuous video INR \(\widetilde{Z}_\mathcal{T}\) to yield the features corresponding to continuous space-time coordinates. Following \cite{chen2021learning, chen2023cascaded, chen2022videoinr}, we decode the features into RGB values using a 5-layer MLP with the previous query \(q_\mathcal{T}\) and cell decoding to reconstruct the target RGB frame \(\hat{I}_\mathcal{T}^{HR}\).

%% file: sec/4_exp.tex
\input{tab/tab_id}
\input{tab/tab_ood}

\section{Experiments}
\label{sec:exp}

\subsection{Experimental Setup}
\label{sec:4-1}
\noindent{\bf Dataset.} 
In our experiments, we train all the models on the Adobe240 dataset~\cite{su2017deep} which contains 133 video sequences, where 100 for training, 16 for validation, and 17 for testing.
We evaluate the performance of our models on both synthetic Adobe240~\cite{su2017deep} and GoPro~\cite{nah2017deep} datasets and real-world BS-ERGB \cite{tulyakov2022time} and ALPIX-VSR~\cite{lu2023learning} datasets. 
For the synthetic events, we use the event simulation method vid2e \cite{gehrig2020video} to generate simulated event data.
Given a spatial scale \(s\) and a temporal scale \(t\), we select consecutive (\(t+1\)) frames as a clip and select the first and the last frames for \(s:1\) bicubic downsampling as the LR and LFR inputs fed into the network to generate HR and HFR frame sequences, and the original (\(t+1\)) frames are ground-truth (GT).
Following~\cite{kai2024evtexture}, the event voxels are bicubic-downsampled as the LR event inputs, thus ensuring the spatial size is aligned with the LR frames.

\noindent{\bf Implementation Details.} 
Firstly, for the LR frames, we use a $5\times 5$ convolutional layer with LReLU and 5 residual blocks to extract RGB features. As for voxelized event segments, we use the $3\times 3$ convolution and another 5 residual blocks. In EMA, all the convolutional layers are with the kernel size of $3\times 3$ activated by LReLU. We use the DCN~\cite{zhu2019deformable} for feature alignment. In BRC, besides the first $5\times 5$ convolution for event feature extraction and $1\times 1$ convolutions in the channel attention mechanisms and feed-forward network~\cite{sun2023event}, all the convolutional layers are also with the kernel size of $3\times 3$. In LIVT, we use a $3\times 3\times 3$ convolutional layer to extract the spatiotemporal features. Then, another three 3D convolutions with the same kernel size are used to get the key $K$, query $Q$, and value $V$. Finally, we use a 5-layer MLP with  $[256,256,256,256, 3]$ dimensions and GELU activations to produce the HR and HFR frames. In this work, we implement two model variants EvEnhancer and EvEnhancer-light. For the former, we set the number of event segments $M=7$ and all the convolutional layers have 64 channels. For EvEnhancer-light, we reduce the event segment $M$ to 5 and the channel number of the convolutional layers in LIVT to 16. 

\noindent{\bf Training Settings.} 
Referring~VideoINR \cite{chen2022videoinr}, the training for our model comprises two stages. In the first stage, we set \(s=4\) and \(t=8\), training for 450K iterations on Adobe240~\cite{su2017deep}. In the second stage, we uniformly sample \(s\) in \([1, 4]\) and fine-tune for another 150K iterations on the same dataset. We use the Adam optimizer \cite{kingma2014adam} with \(\beta_{1}=0.9\) and \(\beta_{2}=0.999\), and apply cosine annealing to decay the learning rate from \(1 \times 10^{-4}\) to \(1 \times 10^{-7}\) over 150K iterations. For data augmentation, we randomly crop \(32 \times 32\) spatial regions from downsampled images and event segments with random rotations and horizontal flips. The model is optimized with Charbonnier loss \cite{lai2017deep} using \(\epsilon^2=1 \times 10^{-6}\). 

\noindent{\bf Evaluation.} 
To measure our models, we calculate the PSNR (dB) and SSIM on the Y channel. 

\subsection{Comparison with State-of-the-Art Methods}
\label{sec:4-2}
We comprehensively consider frame-based and event-based one-stage or two-stage state-of-the-art methods for comparison: 
1) two-stage event-based STVSR methods consisting of event-based VFI (TimeLens \cite{tulyakov2021time}, REFID \cite{sun2023event}, CBMNet-L \cite{kim2023event}) and event-based VSR (EGVSR \cite{lu2023learning}, EvTexture \cite{kai2024evtexture}); 
2) one-stage frame-based F-STVSR methods (Zooming Slow-Mo~\cite{xiang2020zooming}, TMNet \cite{xu2021temporal}, SAFA \cite{huang2024scale});
3) one-stage frame-based C-STVSR methods (VideoINR \cite{chen2022videoinr}, MoTIF \cite{chen2023motif});
4) one-stage event-based C-STVSR methods (HR-INR \cite{lu2024hr}).
For fair comparisons, we use the models of most VFI methods and all STVSR methods trained or fine-tuned on the same Adobe240 dataset \cite{su2017deep} as ours. Notably, due to the limitation that the HR-INR source code is not yet available, the results of HR-INR are derived from its original paper.

\input{tab/tab_real}
\input{tab/tab_alpix}
\input{fig/exp/exp}

\noindent\textbf{{Quantitative Comparison}}. Table~\ref{tab:id} reports the results for the F-STVSR task, where the spatial and temporal scales are the same as the training distribution (In-dist.). ``-\emph{Center}'' represents the average performance of the reconstructed left, right, and center frames. ``-\emph{Average}'' represents the average performance of all reconstructed frames. Notably, VideoINR-\emph{fixed} is trained specifically for single-frame interpolation. We can see that both our EvEnhancer and EvEnhancer-light significantly outperform existing methods in all settings with much fewer model parameters, which achieves more than 1 dB PSNR improvements. Then, we also evaluate our models on the C-STVSR task, where spatial and temporal scales are out of the training distribution. The results are illustrated in~Table~\ref{tab:ood}. It can be observed that both our models exceed almost all existing methods at various spatial and temporal scales, especially at the challenging large temporal scale, achieving superior effectiveness and generalization ability.
In particular, our EvEnhancer-light and EvEnhancer use lighter-weight models and perform better than HR-INR~\cite{lu2024hr} at most temporal and spatial scales, especially at OOD scales.
We further validate the effectiveness of our method on the real-world BS-ERGB dataset under the setting of fixed spatial scale \(s=4\) but varying temporal scales \(t\). The comparisons in Table~\ref{tab:real} demonstrate that our method yields the best quantitative performance under all conditions. 
Compared to the cascaded event-based VFI and VSR methods, our EvEnhancer simultaneously learns temporal interpolation and spatial VSR, thus achieving superior performance. Due to the high temporal resolution of events, equipped by EASM, our EvEnhancer enables holistic motion trajectory modeling, thus outperforming frame-based methods. Benefiting from the cross-scale attention in LIVT, it also performs preferably in generalization.
To validate the effectiveness of our EvEnhancer for real-world LR events, we conduct experiments on the real-world ALPIX-VSR dataset~\cite{lu2023learning} (image resolution: 3264$\times$2448, event resolution: 1632$\times$1224) for 2$\times$ VSR. As shown in Table~\ref{tab:alpix}, our EvEnhancer significantly surpasses almost all existing methods without additional fine-tuning.

\input{tab/tab_efficiency}

\noindent\textbf{{Model Complexity}}. To validate the model efficiency of the proposed methods. Besides the model parameter comparison in Table~\ref{tab:id}, we compare the TFLOPs with existing C-STVSR methods \cite{chen2022videoinr,chen2023motif} in Table~\ref{tab:eff}, which are calculated on the upsampling of a clip with a resolution of 180 \(\times\) 320 by a spatial upsampling scale \(s=4\) at different temporal upsampling scales \(t\). VideoINR shows the fewest TFLOPs but more parameters and the worst performance (see Table~\ref{tab:id}). At the low temporal scale ($\leq 4$), MoTIF is more computationally efficient than ours. However, as the temporal scale increases, its computational costs increase and exceed both our models. The overall comparisons validate that EvEnhancer can achieve an optimal trade-off between effectiveness, generalizability, and efficiency.

\noindent\textbf{{Qualitative Comparison}}. With the qualitative results in Figure \ref{fig:exp}, compared to VideoINR~\cite{chen2022videoinr} and MoTIF~\cite{chen2023motif}, our models reconstruct the faithful HR frames with clearer textures. By computing the difference maps between generated frames and corresponding ground truth frames, we can see that our methods are more excellent at tackling non-linear motions and recovering more details. 
In Figure~\ref{fig:s_event}, we conduct comparisons on the real-world BS-ERGB dataset~\cite{tulyakov2022time}, where the spatiotemporal scales are OOD.
As we can see, our method can produce more preferable HR frames at any time. In Figure~\ref{fig:s_event2},~\ref{fig:s_adobe}, and~\ref{fig:s_gopro_t12s6}, there are more results for In-dist. and OOD scales on the BS-ERGB~\cite{tulyakov2022time}, Adobe240~\cite{su2017deep}, and GoPro~\cite{nah2017deep} datasets, where the results demonstrate the superior effectiveness of our models. In Figure~\ref{fig:s_gopro_t12sx}, we fix the temporal scale between two input LR frames as 6 and perform arbitrary spatial VSR. As seen in the reconstruction performance of the center frame at timestamp $\mathcal{T}=0.5$, both our EvEnhancer-light and EvEnhancer can recover more textures.

\begin{figure}[t]
    \centering 
    \includegraphics[width=0.475\textwidth]{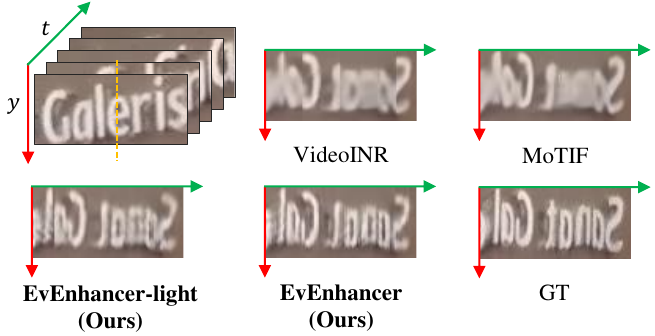}
    \caption{Comparison of temporal profile on the GoPro dataset~\cite{nah2017deep} (\(t=12,s=6\)). We select a column (orange dotted lines) and observe the changes across time.}
    \label{fig:tempora1}
\end{figure}

\begin{figure}[t]
    \centering 
    \includegraphics[width=0.45\textwidth]{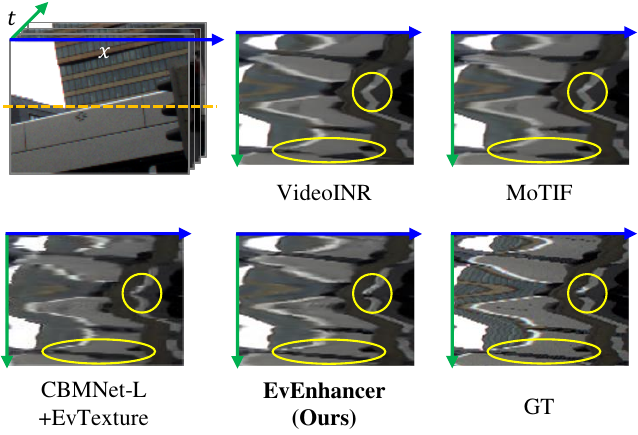}
    \caption{Comparison of temporal profile on the BS-ERGB dataset~\cite{tulyakov2022time} (\(t=4,s=4\)). We select a row (orange dotted lines) and observe the changes across time.}
    \label{fig:tempora2}
\end{figure}

\noindent\textbf{{Temporal Consistency}}. In Figure~\ref{fig:tempora1}, we visualize the temporal profiles of VideoINR~\cite{chen2022videoinr}, MoTIF~\cite{chen2023motif}, and our EvEnhancer on the GoPro dataset~\cite{nah2017deep}. We can observe that, the results of VideoINR and MoTIF contain obvious noise, blurs, and heavy flickering artifacts, which indicates their poor temporal consistencies. In contrast, the profiles of EvEnhancer-light can guarantee better consistency but still contain discontinuity and artifacts. Our full model EvEnhancer shows more pleasant and smoother temporal profiles. Also, we visualize the temporal profiles on the real-world BS-ERGB dataset~\cite{tulyakov2022time} in Figure~\ref{fig:tempora2}, including event-based VFI + VSR methods (CBMNet-L~\cite{kim2023event} + EvTexture~\cite{kai2024evtexture}). Our EvEnhancer maintains the best consistency.

\subsection{Ablation Studies}
\label{sec:4-3}
We conduct ablation studies by keeping the same training strategy on the Adobe240 dataset \cite{su2017deep}, and testing on the GoPro dataset \cite{nah2017deep} at both In-dist. ($t = 8$, $s = 4$) and OOD ($t = 12$, $s = 6$) spatiotemporal scales to validate the effectiveness of our methodology design. 

\input{tab/tab_Ab_Enc}
\input{tab/tab_EMA}
\input{tab/tab_BRC}

\noindent{\bf Event-Adapted Synthesis Module.} 
Table \ref{tab:Ab_Enc} presents the ablations of EASM design, which involves the event-modulated alignment (EMA), bidirectional recurrent compensation (BRC), and bidirectional feature fusion (BF). The model with only EMA achieves the worst performance. When we introduce BRC in the model, we can see the improvement in performance. Besides, with the integration of BRC, BF, and EMA, the model achieves the best at both In-dist. and OOD scales. We can also observe that the model with only a single forward alignment can suffer from performance degradation. The results demonstrate the effects of EASM design.
Table \ref{tab:EMA} shows that 1$\times$ scale alignment plays a more critical role. However, the multi-scale manner (\(1\times\), \(\frac{1}{2}\times\), \(\frac{1}{4}\times\)) captures richer motion cues compared to the single-scale setting, leading to improved performance.
Table \ref{tab:BRC} presents the further ablations of BRC, which involves bidirectional (forward and backward) recurrence in BRC. We also investigate the influence of channel attention mechanisms in BRC. As we can see, the model using only the forward or backward compensation shows the worst performance. When we incorporate attention in each direction, the performance increases. Moreover, by implementing bidirectional compensation with attention both forward and backward, the model performs the best, which achieves significant improvements.

\input{tab/tab_Ab_Dec}
\input{tab/tab_positional}

\noindent{\bf Local Implicit Video Transformer.} Table~\ref{tab:Ab_Dec} investigates the effect of LIVT design. We first implement the model with a 2D decoupling strategy as videoINR~\cite{chen2022videoinr} but introduce the local attention, concatenate cell decoding, and query, and we can see that the model achieves better performance than the model without any components. When we replace the 2D decoupling with our unified 3D video INR, the model produces results with higher PSNR and SSIM, especially at OOD scales. In addition, the concatenation of the cell~\cite{chen2021learning} and previous query~\cite{chen2022videoinr} brings a larger gain at OOD scales than In-dist., which validates their effects. 
Furthermore, we conduct ablation studies on the attention mechanism and positional encoding in LIVT. In LIVT, the query is obtained from the large-scale features at the HR grid after trilinear upsampling, while the key and value are obtained from the small-scale at the local LR grids nearest to the query. Therefore, we call it ``cross-scale'', which can capture spatiotemporal dependencies across LR and HR scales. Our cross-scale attention can be seen as a 3D cross-scale derivation of neighborhood attention~\cite{hassani2023neighborhood}. As illustrated in Table~\ref{tab:positional}, it exhibits suboptimal performance if we use this neighborhood attention directly.
Besides, we encode and reshape the spatiotemporal relative coordinates $(\delta \tau, \delta x, \delta y)\in(-1,1)$ from each query point to all pixel points within its local grid via the cosine positional encoding~(Eq.~\ref{eq:pos_enc}). Here, we investigate its impact by comparing it with another learnable positional encoding scheme~\cite{mildenhall2021nerf}. Table~\ref{tab:positional} shows that the model with cosine positional encoding is superior to the learnable one.

\input{tab/tab_Ab_Param}

\noindent{\bf Hyper-Parameter Setting.} Here, we investigate the impact of event segments \(M\) in EASM, local grid size in LIVT, and the channel of learning video INR. As shown in Table~\ref{tab:Ab_Param}, selecting an insufficient number of event segments, INR channels, or local grid size can lead to a degradation in reconstruction quality. Conversely, an over-large number of each can significantly decrease the model efficiency. Considering the balance between performance and complexity, we implement the EvEnhancer-light and EvEnhancer using the settings in the last two rows as our baseline models to compare with other methods in this work.

%% file: tab/tab_id.tex
\begin{table*}
  \caption{Quantitative comparisons for in-distribution (In-dist.) spatiotemporal upsampling scales (\(t=8, s=4\)). 
  $\dagger$ indicates the model trained on the same Adobe240 dataset~\cite{su2017deep} as ours,
  and $\ddag$ indicates the model fine-tuned on the Adobe240 dataset. \textbf{Bold} and \underline{underline} indicate the best and the second-best performance, respectively. Metrics: PSNR (dB) / SSIM.}
  \fontsize{8}{12}\selectfont
  \centering
  \begin{tabular}{cc|c|ccccccc}
    \hline
        VFI Method&
        VSR Method&
        Events&
        GoPro-\emph{Center} &
        GoPro-\emph{Average} &
        Adobe240-\emph{Center} &
        Adobe240-\emph{Average} &
        Params (M)\\
    \hline
        TimeLens~\cite{tulyakov2022time} &
        EGVSR~\cite{lu2023learning} &
        \checkmark &
        28.41 / 0.8307 &
        27.42 / 0.8077 &
        26.64 / 0.7675 &
        25.13 / 0.7298 &
        72.20+2.45 \\
        TimeLens~\cite{tulyakov2022time} &
        EvTexture~\cite{kai2024evtexture} &
        \checkmark &
        30.50 / 0.8784 &
        28.51 / 0.8478 &
        28.80 / 0.8337 &
        26.12 / 0.7789 &
        72.20+8.90 \\
    \hline
        CBMNet-L~\cite{kim2023event} &
        EGVSR~\cite{lu2023learning} &
        \checkmark &
        26.14 / 0.7674 &
        23.55 / 0.7018 &
        25.15 / 0.7311 &
        22.89 / 0.6735 &
        22.23+2.45 \\
        CBMNet-L~\cite{kim2023event} &
        EvTexture~\cite{kai2024evtexture} &
        \checkmark &
        27.58 / 0.7966 &
        23.92 / 0.7163 &
        26.74 / 0.7777 &
        23.38 / 0.6986 &
        22.23+8.90 \\
    \hline
    REFID~\cite{sun2023event}$^\dagger$ &
        EGVSR~\cite{lu2023learning} &
        \checkmark &
        28.74 / 0.8364 &
        27.97 / 0.8171 &
        27.28 / 0.7844 &
        26.55 / 0.7637 &
        15.91+2.45 \\
        REFID~\cite{sun2023event}$^\dagger$ &
        EvTexture~\cite{kai2024evtexture} &
        \checkmark &
        30.86 / 0.8784 &
        29.14 / 0.8489 &
        29.42 / 0.8424 &
        27.69 / 0.8059 &
        15.91+8.90 \\
    \hline
        \multicolumn{2}{c|}
        {Zooming Slow-Mo~\cite{huang2024scale}$^\dagger$} &
        &
        30.69 / 0.8847 &
        - &
        30.26 / 0.8821 &
        - &
        11.10 \\
        \multicolumn{2}{c|}{TMNet~\cite{xu2021temporal}$^\ddag$} &
        &
        30.14 / 0.8692 &
        28.83 / 0.8514 &
        29.41 / 0.8524 &
        28.30 / 0.8354 &
        12.26 \\
        \multicolumn{2}{c|}
        {SAFA~\cite{huang2024scale}$^\dagger$} &
        &
        31.28 / 0.8894 &
        30.22 / 0.8761 &
        30.97 / 0.8878 &
        30.13 / 0.8782 &
        4.94 \\
    \hline
        \multicolumn{2}{c|}
        {VideoINR-\emph{fixed}~\cite{chen2022videoinr}$^\dagger$} &
        &
        30.73 / 0.8850 &
        - &
        30.21 / 0.8805 &
        - &
        11.31 \\
        \multicolumn{2}{c|}{VideoINR~\cite{chen2022videoinr}$^\dagger$} &
        &
        30.26 / 0.8792 &
        29.41 / 0.8669 &
        29.92 / 0.8746 &
        29.27 / 0.8651 &
        11.31 \\
        \multicolumn{2}{c|}{MoTIF~\cite{chen2023motif}$^\dagger$} &
        &
        31.04 / 0.8877 &
        30.04 / 0.8773 &
        30.63 / 0.8839 &
        29.82 / 0.8750 &
        12.55 \\
    \hline
        \multicolumn{2}{c|}{HR-INR~\cite{lu2024hr}$^\dagger$} &
        \checkmark &
        31.97 / \textbf{0.9298} &
        32.13 / \textbf{0.9371} &
        31.26 / \textbf{0.9246} &
        31.11 / \textbf{0.9216} &
        8.27 \\
    \hline
        \multicolumn{2}{c|}{\textbf{EvEnhancer-light}} &
        \checkmark &
        \underline{33.11} / 0.9242 &
        \underline{32.73} / 0.9203 &
        \underline{31.90} / 0.9033 &
        \underline{31.47} / 0.8988 &
        5.81 \\
        \multicolumn{2}{c|}{\textbf{EvEnhancer}} &
        \checkmark &
        \textbf{33.52} / \underline{0.9295} &
        \textbf{33.30} / \underline{0.9279} &
        \textbf{32.43} / \underline{0.9129} &
        \textbf{32.18} / \underline{0.9116} &
        6.55 \\
    \hline
  \end{tabular}
  \label{tab:id}
\end{table*}

%% file: tab/tab_ood.tex
\begin{table*}
  \caption{Quantitative comparisons for out-of-distribution (OOD) spatiotemporal upsampling scales on the GoPro dataset \cite{nah2017deep}.}
  \fontsize{8}{11}\selectfont
  \centering
  \begin{tabular}{cc||ccccc|cccc}
    \hline
        \makecell{Temporal\\Scale}& 
        \makecell{Spatial\\Scale}& 
        \makecell{TMNet~\cite{xu2021temporal}$^\ddag$} &
        \makecell{SAFA~\cite{huang2024scale}$^\dagger$} &
        \makecell{VideoINR~\cite{chen2022videoinr}$^\dagger$} &
        \makecell{MoTIF~\cite{chen2023motif}$^\dagger$} &
        \makecell{HR-INR~\cite{lu2024hr}$^\dagger$} &
        \makecell{\textbf{EvEnhancer}\\\textbf{-light}} &
        \makecell{\textbf{EvEnhancer}} \\
    \hline
        $t=6$ & 
        $s=4$ &
        30.49 / 0.8861 &
        31.68 / 0.9068 &
        30.78 / 0.8954 &
        31.56 / 0.9064 &
        - &
        \underline{33.10} / \underline{0.9266} &
        \textbf{33.41} / \textbf{0.9300}
        \\
        $t=6$ & 
        $s=6$ &
        - &
        - &
        25.56 / 0.7671 &
        29.36 / 0.8505 &
        - &
        \underline{29.84} / \underline{0.8603} &
        \textbf{30.12} / \textbf{0.8675}    
        \\
        $t=6$ & 
        $s=12$ &
        - &
        - &
        24.02 / 0.6900 &
        \textbf{25.81} / \textbf{0.7330} &
        - &
        25.41 / 0.7261 &
        \underline{25.50} / \underline{0.7323}     
        \\
    \hline
        $t=12$ & 
        $s=4$ &
        26.38 / 0.7931 &
        27.95 / 0.8249 &
        27.32 / 0.8141 &
        27.77 / 0.8230 &
        28.87 / 0.8854 &
        \underline{31.17} / \underline{0.8969} &
        \textbf{32.07} / \textbf{0.9116}
        \\
        $t=12$ & 
        $s=6$ &
        - &
        - &
        24.68 / 0.7358 &
        26.78 / 0.7908 &
        27.14 / 0.8173 &
        \underline{28.94} / \underline{0.8425} &
        \textbf{29.54} / \textbf{0.8579}
        \\
        $t=12$ & 
        $s=12$ &
        - &
        - &
        23.70 / 0.6830 &
        24.72 / 0.7108 &
        - &
        \underline{25.25} / \underline{0.7248} &
        \textbf{25.45} / \textbf{0.7345}        
        \\
    \hline
        $t=16$ & 
        $s=4$ &
        24.72 / 0.7526 &
        26.38 / 0.7832 &
        25.81 / 0.7739 &
        25.98 / 0.7758 &
        27.29 / 0.8556 &
        \underline{30.15} / \underline{0.8762} &
        \textbf{30.93} / \textbf{0.8918}         
        \\
        $t=16$ & 
        $s=6$ &
        - &
        - &
        23.86 / 0.7123 &
        25.34 / 0.7527 &
        26.09 / 0.7954 &
        \underline{28.34} / \underline{0.8283} &
        \textbf{28.86} / \textbf{0.8434}
        \\
        $t=16$ & 
        $s=12$ &
        - &
        - &
        22.88 / 0.6659 &
        23.88 / 0.6923 &
        - &
        \underline{25.10} / \underline{0.7217} &
        \textbf{25.25} / \textbf{0.7310}
        \\
    \hline
        $t=6$ & 
        $s=1$ &
        - &
        - &
        32.34 / 0.9545 &
        34.77 / 0.9696 &
        \underline{38.53} / \textbf{0.9735} &
        37.68 / 0.9667 &
        \textbf{38.80} / \underline{0.9714}         
        \\
    \hline
  \end{tabular}
  \label{tab:ood}
  
\end{table*}

%% file: tab/tab_real.tex
\begin{table}[ht]
    \centering
    \caption{Quantitative comparisons for different temporal upsampling scales \(t\) (fixed \(s=4\)) on the BS-ERGB dataset \cite{tulyakov2022time} .}
    \setlength{\tabcolsep}{3.0pt}
    \fontsize{6.6}{9}\selectfont
    \centering
    \begin{tabular}{c|c|ccc}
    \hline
        Method &
        Events &
        \(t=4\) &
        \(t=6\) &
        \(t=8\) \\
    \hline
        TimeLens\cite{tulyakov2022time}+EGVSR\cite{lu2023learning} &
        \checkmark &
        22.98/0.6762 &
        22.01/0.6592 &
        21.33/0.6467 \\
        TimeLens\cite{tulyakov2022time}+EvTexture\cite{kai2024evtexture} &
        \checkmark &
        23.58/0.7088 &
        22.48/0.6887 &
        21.74/0.6746\\
    \hline
        CBMNet-L\cite{kim2023event}+EGVSR\cite{lu2023learning} &
        \checkmark &
        22.74/0.6856 &
        21.93/0.6705 &
        21.29/0.6657\\
         CBMNet-L\cite{kim2023event}+EvTexture\cite{kai2024evtexture} &
        \checkmark &
        23.40/0.7199 &
        22.47/0.7020 &
        21.78/0.6962\\
    \hline
        REFID\cite{sun2023event}$^\dagger$+EGVSR\cite{lu2023learning} &
        \checkmark &
        - &
        - &
        22.76/0.6478 \\
        REFID\cite{sun2023event}$^\dagger$+EvTexture\cite{kai2024evtexture} &
        \checkmark &
        - &
        - &
        22.77/0.6527 \\
    \hline
        SAFA\cite{huang2024scale}$^\dagger$ &
        &
        24.32/0.7314 &
        23.23/0.7120 &
        22.41/0.6954\\
        VideoINR\cite{chen2022videoinr}$^\dagger$ &
        &
        24.06/0.7290 &
        23.00/0.7098 &
        22.23/0.6945 \\
        MoTIF\cite{chen2023motif}$^\dagger$ &
        &
        24.21/0.7296 &
        23.15/0.7099 &
        22.37/0.6939 \\
    \hline
        \textbf{EvEnhancer-light} &
        \checkmark &
        \underline{25.32}/\underline{0.7311} &
        \underline{24.54}/\underline{0.7184} &
        \underline{24.16}/\underline{0.7100} \\
        \textbf{EvEnhancer} &
        \checkmark &
        \textbf{25.44}/\textbf{0.7338} &
        \textbf{24.70}/\textbf{0.7215} &
        \textbf{24.29}/\textbf{0.7138}\\
    \hline
  \end{tabular}
  \label{tab:real}
\end{table}

%% file: tab/tab_alpix.tex
\begin{table}[ht]
    \centering
    \caption{2$\times$ VSR performance on the ALPIX-VSR dataset~\cite{lu2023learning}.}
    \setlength{\tabcolsep}{3.0pt}
    \fontsize{7.2}{9}\selectfont
    \centering
    \begin{tabular}{@{\hskip 0.01\linewidth}c@{\hskip 0.01\linewidth}|@{\hskip 0.01\linewidth}cccc@{\hskip 0.01\linewidth}}
    \hline
    Method
    & VideoINR~\cite{chen2022videoinr}
    & MoTIF~\cite{chen2023motif}
    & EGVSR~\cite{lu2023learning}
    & \textbf{EvEnhancer}
    \\
    \hline
    PSNR / SSIM
    & 32.53 / 0.9383
    & 38.61 / \underline{0.9636}
    & \underline{39.31} / 0.9635
    & \textbf{40.84} / \textbf{0.9786}
    \\
    \hline
    \end{tabular}
    \label{tab:alpix}
\end{table}

%% file: fig/exp/exp.tex
\setlength{\tabcolsep}{0.1pt}
\renewcommand{\arraystretch}{0.1}

\begin{figure*}[t]
    \centering\
    \begin{subfigure}{0.11\linewidth}
        \begin{tabular}{c}
            \includegraphics[width=1\linewidth]{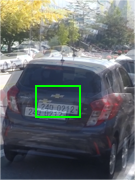}
            \\ \\ \\
            \footnotesize
            LR Frames
            \\ \\ \\ \\ \\ \\ \\
            \\ \\ \\ \\ \\ \\ \\
            \includegraphics[width=1\linewidth]{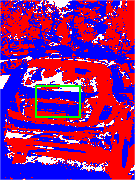}
            \\ \\ \\
            \footnotesize
            LR Events
            \\ \\ \\ \\ \\ \\ \\ \\ \\
            \\ \\ \\ \\ \\ \\ \\ \\ \\
        \end{tabular}
    \end{subfigure}
    \hfill
    \begin{subfigure}{0.88\linewidth}
         \begin{tabular}{cc}
            \scriptsize
            \rotatebox[origin=c]{90}{\makecell{VideoINR\\ \cite{chen2022videoinr}}}  &
            \begin{tabular}{c@{\hskip 0.003\linewidth}c@{\hskip 0.003\linewidth}c@{\hskip 0.003\linewidth}c@{\hskip 0.003\linewidth}c@{\hskip 0.003\linewidth}c@{\hskip 0.003\linewidth}c@{\hskip 0.003\linewidth}c}
                \includegraphics[width=0.115\linewidth]{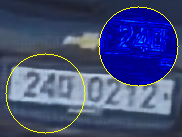}&
                \includegraphics[width=0.115\linewidth]{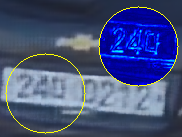}&
                \includegraphics[width=0.115\linewidth]{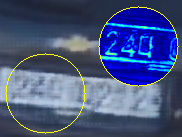}&
                \includegraphics[width=0.115\linewidth]{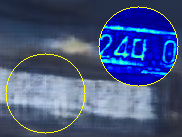}&
                \includegraphics[width=0.115\linewidth]{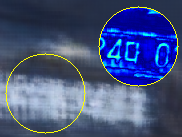}&
                \includegraphics[width=0.115\linewidth]{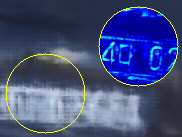}&
                \includegraphics[width=0.115\linewidth]{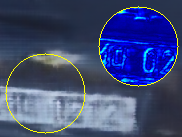}&
                \includegraphics[width=0.115\linewidth]{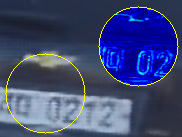}
            \end{tabular}
            \\ \\ \\
            \scriptsize
            \rotatebox[origin=c]{90}{\makecell{MoTIF\\ \cite{chen2023motif}}} &
            \begin{tabular}{c@{\hskip 0.003\linewidth}c@{\hskip 0.003\linewidth}c@{\hskip 0.003\linewidth}c@{\hskip 0.003\linewidth}c@{\hskip 0.003\linewidth}c@{\hskip 0.003\linewidth}c@{\hskip 0.003\linewidth}c}
                \includegraphics[width=0.115\linewidth]{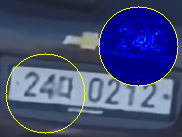}&
                \includegraphics[width=0.115\linewidth]{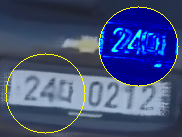}&
                \includegraphics[width=0.115\linewidth]{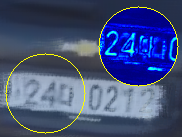}&
                \includegraphics[width=0.115\linewidth]{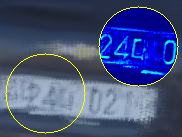}&
                \includegraphics[width=0.115\linewidth]{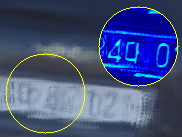}&
                \includegraphics[width=0.115\linewidth]{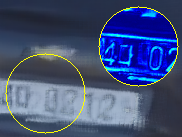}&
                \includegraphics[width=0.115\linewidth]{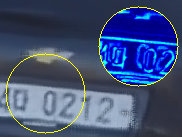}&
                \includegraphics[width=0.115\linewidth]{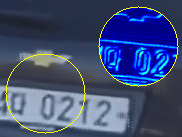}
            \end{tabular}
            \\ \\ \\
            \scriptsize
            \rotatebox[origin=c]{90}{\makecell{\textbf{EvEnhancer}\\\textbf{(Ours)}} } &
            \begin{tabular}{c@{\hskip 0.003\linewidth}c@{\hskip 0.003\linewidth}c@{\hskip 0.003\linewidth}c@{\hskip 0.003\linewidth}c@{\hskip 0.003\linewidth}c@{\hskip 0.003\linewidth}c@{\hskip 0.003\linewidth}c}
                \includegraphics[width=0.115\linewidth]{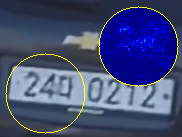}&
                \includegraphics[width=0.115\linewidth]{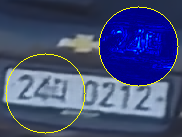}&
                \includegraphics[width=0.115\linewidth]{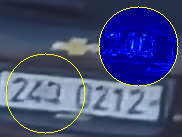}&
                \includegraphics[width=0.115\linewidth]{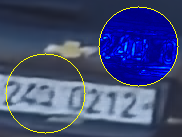}&
                \includegraphics[width=0.115\linewidth]{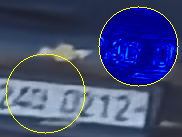}&
                \includegraphics[width=0.115\linewidth]{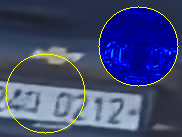}&
                \includegraphics[width=0.115\linewidth]{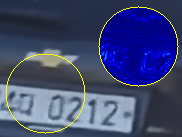}&
                \includegraphics[width=0.115\linewidth]{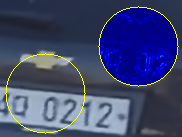}
            \end{tabular}
            \\ \\
            \scriptsize
            \rotatebox[origin=c]{90}{\makecell{\textbf{EvEnhancer}\\\textbf{-light(Ours)}} } &
            \begin{tabular}{c@{\hskip 0.003\linewidth}c@{\hskip 0.003\linewidth}c@{\hskip 0.003\linewidth}c@{\hskip 0.003\linewidth}c@{\hskip 0.003\linewidth}c@{\hskip 0.003\linewidth}c@{\hskip 0.003\linewidth}c}
                \includegraphics[width=0.115\linewidth]{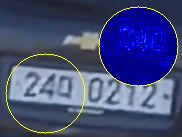}&
                \includegraphics[width=0.115\linewidth]{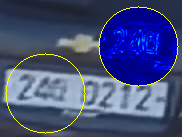}&
                \includegraphics[width=0.115\linewidth]{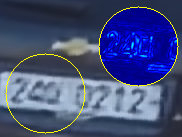}&
                \includegraphics[width=0.115\linewidth]{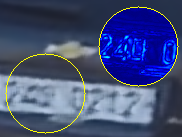}&
                \includegraphics[width=0.115\linewidth]{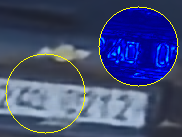}&
                \includegraphics[width=0.115\linewidth]{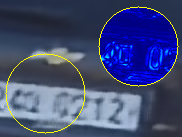}&
                \includegraphics[width=0.115\linewidth]{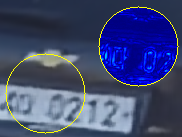}&
                \includegraphics[width=0.115\linewidth]{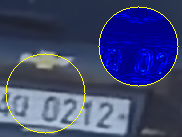}
            \end{tabular}
            \\ \\
            \begin{tabular}{c}
            \small
            \rotatebox[origin=c]{90}{GT}
            \\ \\ \\ \\ \\ \\ \\ \\ \\ \\
            \end{tabular}
            &
            \begin{tabular}{c@{\hskip 0.003\linewidth}c@{\hskip 0.003\linewidth}c@{\hskip 0.003\linewidth}c@{\hskip 0.003\linewidth}c@{\hskip 0.003\linewidth}c@{\hskip 0.003\linewidth}c@{\hskip 0.003\linewidth}c}
                \includegraphics[width=0.115\linewidth]{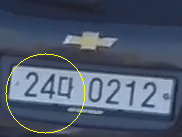}&
                \includegraphics[width=0.115\linewidth]{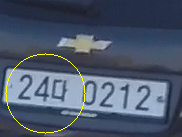}&
                \includegraphics[width=0.115\linewidth]{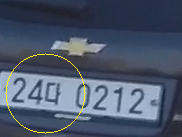}&
                \includegraphics[width=0.115\linewidth]{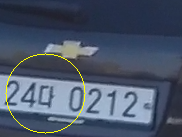}&
                \includegraphics[width=0.115\linewidth]{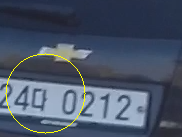}&
                \includegraphics[width=0.115\linewidth]{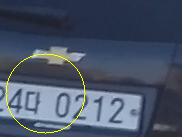}&
                \includegraphics[width=0.115\linewidth]{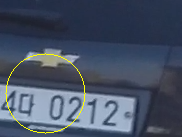}&
                \includegraphics[width=0.115\linewidth]{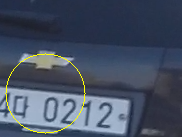}
            \\
            \footnotesize
            \(\mathcal{T}=0\)&
            \footnotesize
            \(\mathcal{T}=0.125\)&
            \footnotesize
            \(\mathcal{T}=0.25\)&
            \footnotesize
            \(\mathcal{T}=0.375\)&
            \footnotesize
            \(\mathcal{T}=0.5\)&
            \footnotesize
            \(\mathcal{T}=0.625\)&
            \footnotesize
            \(\mathcal{T}=0.75\)&
            \footnotesize
           \( \mathcal{T}=0.875\)
            \end{tabular}
            \\ \\ \\
         &
            \multicolumn{1}{r}{        \includegraphics[width=0.5\linewidth]{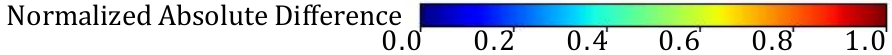}}
        \end{tabular}
    \end{subfigure}
    \caption{
       Qualitative comparison for In-dist. scale (\(t=8,s=4\)) on the GoPro dataset \cite{nah2017deep}. We compare the normalized absolute difference maps (yellow boxes) for the same regions in each frame as in the GT frames.
    }
    \label{fig:exp}
\end{figure*}
\renewcommand{\arraystretch}{1.}

%% file: tab/tab_efficiency.tex
\begin{table}[t]
    \centering
    \caption{TFLOPs comparisons of C-STVSR models at different upsampling temporal scales \(t\) (fixed \(s=4\)).  * indicates the repeated iterations during inference on an NVIDIA GeForce RTX 3090 GPU due to their large computations.}
    \setlength{\tabcolsep}{3.0pt}
    \fontsize{8}{10}\selectfont
    \centering
    \begin{tabular}{c|cccc}
    \hline
    Temporal
    & VideoINR~\cite{chen2022videoinr} 
    & MoTIF~\cite{chen2023motif} 
    & \textbf{EvEnhancer}
    & \textbf{EvEnhancer}  \\
    Scale
    & (11.31M)
    & (12.55M)
    & \textbf{-light} (5.81M)
    & (6.55M) \\
    \hline
    $t=2$
    & 2.011
    & 2.043
    & 1.876
    & 3.398 \\
    $t=4$
    & 2.395
    & ~~4.086*
    & 2.472
    & 4.647 \\
    $t=6$
    & 2.779
    & ~~6.129*
    & 3.067
    & 5.885 \\
    $t=8$
    & 3.163
    & ~~6.129*
    & 3.663
    & 7.129 \\
    $t=12$
    & 3.932
    & ~~10.22*
    & 4.854
    & 9.617\\
    $t=16$
    & 4.700
    & ~~12.26*
    & 6.044
    & 12.10 \\
    \hline
    \end{tabular}
    \label{tab:eff}
\end{table}

%% file: tab/tab_Ab_Enc.tex
\begin{table}[t]
    \centering
    \caption{Ablations on the designs of event-adapted synthesis module (EASM). ``fwd.'' and ``bwd.'' denote the forward and backward directions, respectively. Metrics: PSNR (dB) / SSIM.}
    \setlength{\tabcolsep}{3.0pt}
    \fontsize{8}{10}\selectfont
    \centering
    \begin{tabular}{ccc|cc}
    \hline
    EMA
    & BRC
    & BF
    & In-dist.
    & OOD\\
    \hline
    fwd. \& bwd.
    & 
    & 
    & 32.50 / 0.9166
    & 28.86 / 0.8407\\
    fwd. \& bwd.
    & \checkmark
    & 
    & 33.13 / 0.9252
    & 29.42 / 0.8539\\
    fwd.
    & \checkmark
    & \checkmark
    & 33.20 / 0.9262
    & 29.47 / 0.8556\\
    \rowcolor{gray!20}
    fwd. \& bwd.
    & \checkmark
    & \checkmark
    & \textbf{33.30} / \textbf{0.9279}
    & \textbf{29.54} / \textbf{0.8579}\\
    \hline
    \end{tabular}
    \label{tab:Ab_Enc}
\end{table}

%% file: tab/tab_EMA.tex
\begin{table}[t]
    \centering
    \caption{Ablations on the event-modulated alignment (EMA).  Metrics: PSNR (dB) / SSIM.}
    \setlength{\tabcolsep}{3.0pt}
    \fontsize{8}{10}\selectfont
    \centering
    \begin{tabular}{c@{\hskip 0.02\linewidth}|@{\hskip 0.02\linewidth}cc}
    \hline
    Alignment
    & In-dist.
    & OOD\\
    \hline
    1$\times$ scale
    & 33.21 / 0.9266 
    & 29.46 / 0.8560 \\
    1/2$\times$ scale
    & 32.98 / 0.9231 
    & 29.32 / 0.8520 \\
    1/4$\times$ scale
    & 32.92 / 0.9223
    & 29.26 / 0.8502 \\
    \rowcolor{gray!20}
    multi-scale
    & \textbf{33.30} / \textbf{0.9279}
    & \textbf{29.54} / \textbf{0.8579}\\
    \hline
    \end{tabular}
    \label{tab:EMA}
\end{table}

%% file: tab/tab_BRC.tex
\begin{table}[t]
    \centering
    \caption{Ablations on the bidirectional recurrent compensation (BRC). Metrics: PSNR (dB) / SSIM.}
    \setlength{\tabcolsep}{3.0pt}
    \fontsize{8}{10}\selectfont
    \centering
    \begin{tabular}{cc@{\hskip 0.02\linewidth}|@{\hskip 0.02\linewidth}cc}
    \hline
    Recurrent
    & Channel Attention
    & In-dist.
    & OOD\\
    \hline
    fwd.
    & 
    & 32.57 / 0.9176 
    & 28.96 / 0.8429 \\
    bwd.
    & 
    & 32.55 / 0.9171 
    & 28.94 / 0.8424 \\
    fwd.
    & \checkmark
    & 32.79 / 0.9211 
    & 29.12 / 0.8467 \\
    bwd.
    & \checkmark
    & 32.79 / 0.9207 
    & 29.11 / 0.8467 \\
    \rowcolor{gray!20}
    fwd. \& bwd.
    & \checkmark
    & \textbf{33.30} / \textbf{0.9279}
    & \textbf{29.54} / \textbf{0.8579}\\
    \hline
    \end{tabular}
    \label{tab:BRC}
\end{table}

%% file: tab/tab_Ab_Dec.tex
\begin{table}[t]
    \centering
    \caption{Ablations on the designs of local implicit video transformer (LIVT). ``LA": local attention, ``CE": concatenate cell decoding, ``PQ": concatenate the previous query. Metrics: PSNR (dB) / SSIM.}
    \setlength{\tabcolsep}{3.0pt}
    \fontsize{8}{10}\selectfont
    \centering
    \begin{tabular}{cccc|cc}
    \hline
    INR Mode
    & LA
    & CE
    & PQ
    & In-dist.
    & OOD\\
    \hline
    2D decoupling
    & \checkmark
    & \checkmark
    & \checkmark
    & 33.26 / 0.9272
    & 29.26 / 0.8525\\
    3D unification   
    &
    &
    &
    & 27.32 / 0.7975
    & 25.03 / 0.7032\\
    3D unification 
    & \checkmark
    &
    &
    & 33.27 / 0.9276
    & 29.47 / 0.8565\\
    3D unification 
    & \checkmark
    & \checkmark
    &
    & 33.30 / \textbf{0.9282}
    & 29.49 / 0.8570\\
    \rowcolor{gray!20}
    3D unification 
    & \checkmark
    & \checkmark
    & \checkmark
    & \textbf{33.30} / 0.9279
    & \textbf{29.54} / \textbf{0.8579}\\
    \hline
    \end{tabular}
    \label{tab:Ab_Dec}
\end{table}

%% file: tab/tab_positional.tex
\begin{table}[t]
    \centering
    \caption{Ablations on the attention mechanism and positional encoding in LIVT. Metrics: PSNR (dB) / SSIM.}
    \setlength{\tabcolsep}{3.0pt}
    \fontsize{8}{10}\selectfont
    \centering
    \begin{tabular}{cc@{\hskip 0.02\linewidth}|@{\hskip 0.02\linewidth}cc}
    \hline
    \makecell{Attention\\Mechanism}
    & \makecell{Positional\\Encoding}
    & \makecell{In-dist.}
    & \makecell{OOD}
    \\
    \hline
    Neighborhood~\cite{hassani2023neighborhood}
    & Cosine
    & 33.07 / 0.9248
    & 29.24 / 0.8492
    \\
    Cross-scale
    & Learnable~\cite{mildenhall2021nerf}
    & 31.92 / 0.9086 
    & 28.37 / 0.8294 \\
    \rowcolor{gray!20}
    Cross-scale
    & Cosine
    & \textbf{33.30} / \textbf{0.9279}
    & \textbf{29.54} / \textbf{0.8579}\\
    \hline
    \end{tabular}
    \label{tab:positional}
\end{table}

%% file: tab/tab_Ab_Param.tex
\begin{table}[t]
    \centering
    \caption{Ablations on hyperparameters including the number of event segments \(M\), the number of INR channels, and local grid size of \(T^{G} \times H^{G} \times W^{G}\). Metrics: PSNR (dB) / TFLOPs.}
    \setlength{\tabcolsep}{3.0pt}
    \fontsize{7}{9}\selectfont
    \centering
    \begin{tabular}{ccc|cc|c}
    \hline
    \(M\)
    & Local Grid
    & INR Channel
    & In-dist.
    & OOD
    & Params (M)
    \\
    \hline
    5
    & 3\(\times\)3\(\times\)3
    & 64
    & 32.32 / 6.742
    & 28.73 / 8.512
    & 6.548 \\
    9
    & 3\(\times\)3\(\times\)3
    & 64
    & 33.08 / 7.516
    & 29.46 / 8.852
    & 6.548 \\
    7
    & 1\(\times\)3\(\times\)3
    & 64
    & 33.15 / 4.609
    & 27.00 / 5.077
    & 6.253 \\
    7
    & 5\(\times\)3\(\times\)3
    & 64
    & 33.35 / 9.649
    & 29.60 / 12.29
    & 6.843 \\
    7
    & 3\(\times\)1\(\times\)1
    & 64
    & 32.99 / 3.769
    & 29.29 / 3.875
    & 6.155 \\
    7
    & 3\(\times\)5\(\times\)5
    & 64
    & 33.28 / 13.85
    & 29.52 / 18.30
    & 7.335 \\
    7
    & 3\(\times\)3\(\times\)3
    & 16
    & 33.01 / 4.003
    & 29.12 / 4.416
    & 5.810 \\
    \rowcolor{gray!20}
    5
    & 3\(\times\)3\(\times\)3
    & 16
    & 32.73 / 3.663
    & 28.94 / 4.266
    & 5.810 \\
    \rowcolor{gray!20}
    7
    & 3\(\times\)3\(\times\)3
    & 64
    & 33.30 / 7.129
    & 29.54 / 8.682
    & 6.548 \\
    \hline
    \end{tabular}
    \label{tab:Ab_Param}
\end{table}

%% file: sec/5_con.tex
\section{Conclusion}
\label{sec:conclusion}
In this paper, we introduce EvEnhancer, a novel approach that amalgamates the unique advantages of event streams to enhance the effectiveness, efficiency, and generalizability of C-STVSR. Our EvEnhancer is underpinned by two critical components: 1) the EASM capitalizes on the spatiotemporal correlations between frames and events to discern and learn long-term motion trajectories; and 2) the LIVT, which integrates a local implicit video neural function with cross-scale spatiotemporal attention to learn continuous video representations. Experiments indicate that EvEnhancer significantly surpasses current state-of-the-art methods. 

%% file: fig/exp/exp_suppl_event.tex
\setlength{\tabcolsep}{0.1pt}
\renewcommand{\arraystretch}{0.1}

\begin{figure*}[ht!]
    \centering
    \begin{tabular}{c@{\hskip 0.2\linewidth}c}
       \includegraphics[width=0.25\linewidth]{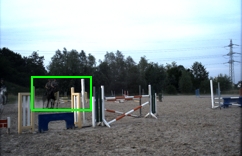}&
       \includegraphics[width=0.25\linewidth]{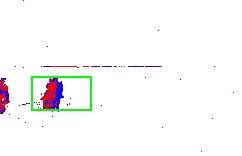} \\
       \footnotesize LR Frames &
       \footnotesize LR Events
       \\ \\ \\
    \end{tabular}
    \begin{tabular}{cc}
        \scriptsize
        \rotatebox[origin=c]{90}{\makecell{TimeLens\cite{tulyakov2021time}\\+EvTexture\cite{kai2024evtexture}}}  &
        \begin{tabular}{c@{\hskip 0.003\linewidth}c@{\hskip 0.003\linewidth}c@{\hskip 0.003\linewidth}c@{\hskip 0.003\linewidth}c@{\hskip 0.003\linewidth}c}
            \includegraphics[width=0.155\linewidth]{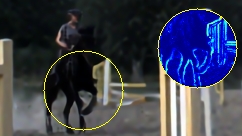}&
            \includegraphics[width=0.155\linewidth]{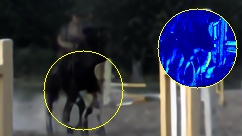}&
            \includegraphics[width=0.155\linewidth]{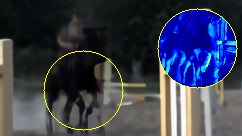}&
            \includegraphics[width=0.155\linewidth]{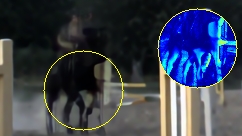}&
            \includegraphics[width=0.155\linewidth]{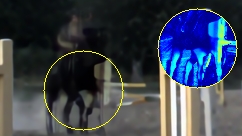}&
            \includegraphics[width=0.155\linewidth]{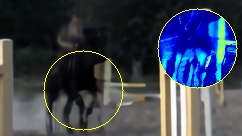}
        \end{tabular}
        \\ \\ \\
        \scriptsize
        \rotatebox[origin=c]{90}{\makecell{CBMNet-L\cite{kim2023event}\\+EvTexture\cite{kai2024evtexture}}}  &
        \begin{tabular}{c@{\hskip 0.003\linewidth}c@{\hskip 0.003\linewidth}c@{\hskip 0.003\linewidth}c@{\hskip 0.003\linewidth}c@{\hskip 0.003\linewidth}c}
            \includegraphics[width=0.155\linewidth]{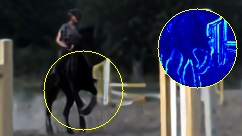}&
            \includegraphics[width=0.155\linewidth]{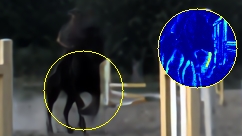}&
            \includegraphics[width=0.155\linewidth]{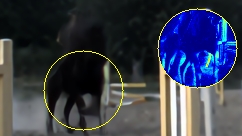}&
            \includegraphics[width=0.155\linewidth]{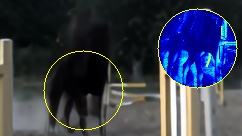}&
            \includegraphics[width=0.155\linewidth]{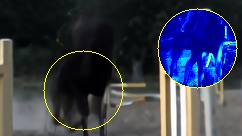}&
            \includegraphics[width=0.155\linewidth]{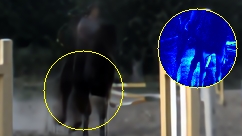}
        \end{tabular}
        \\ \\ \\
        \scriptsize
        \rotatebox[origin=c]{90}{\makecell{SAFA\\ \cite{huang2024scale}}}  &
        \begin{tabular}{c@{\hskip 0.003\linewidth}c@{\hskip 0.003\linewidth}c@{\hskip 0.003\linewidth}c@{\hskip 0.003\linewidth}c@{\hskip 0.003\linewidth}c}
            \includegraphics[width=0.155\linewidth]{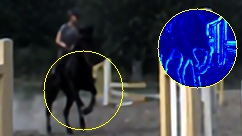}&
            \includegraphics[width=0.155\linewidth]{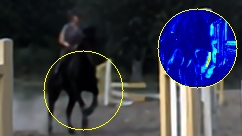}&
            \includegraphics[width=0.155\linewidth]{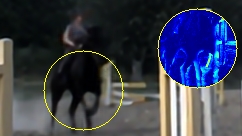}&
            \includegraphics[width=0.155\linewidth]{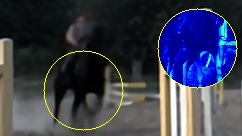}&
            \includegraphics[width=0.155\linewidth]{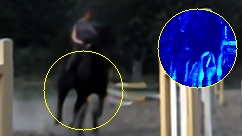}&
            \includegraphics[width=0.155\linewidth]{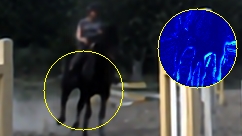}
        \end{tabular}
        \\ \\ \\
        \scriptsize
        \rotatebox[origin=c]{90}{\makecell{VideoINR\\ \cite{chen2022videoinr}}}  &
        \begin{tabular}{c@{\hskip 0.003\linewidth}c@{\hskip 0.003\linewidth}c@{\hskip 0.003\linewidth}c@{\hskip 0.003\linewidth}c@{\hskip 0.003\linewidth}c}
            \includegraphics[width=0.155\linewidth]{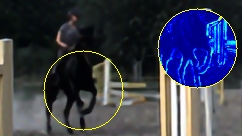}&
            \includegraphics[width=0.155\linewidth]{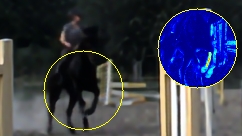}&
            \includegraphics[width=0.155\linewidth]{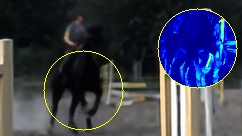}&
            \includegraphics[width=0.155\linewidth]{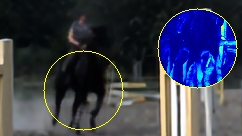}&
            \includegraphics[width=0.155\linewidth]{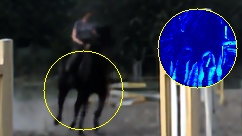}&
            \includegraphics[width=0.155\linewidth]{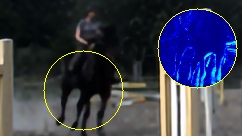}
        \end{tabular}
        \\ \\ \\ 
        \scriptsize
        \rotatebox[origin=c]{90}
        {\makecell{MoTIF\\ \cite{chen2023motif}}}  &
        \begin{tabular}{c@{\hskip 0.003\linewidth}c@{\hskip 0.003\linewidth}c@{\hskip 0.003\linewidth}c@{\hskip 0.003\linewidth}c@{\hskip 0.003\linewidth}c}
            \includegraphics[width=0.155\linewidth]{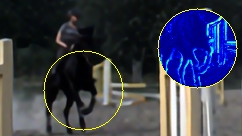}&
            \includegraphics[width=0.155\linewidth]{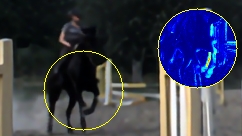}&
            \includegraphics[width=0.155\linewidth]{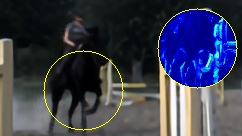}&
            \includegraphics[width=0.155\linewidth]{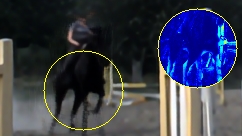}&
            \includegraphics[width=0.155\linewidth]{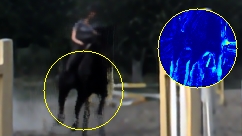}&
            \includegraphics[width=0.155\linewidth]{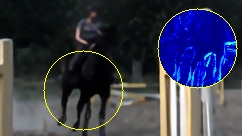}
        \end{tabular}
        \\ \\ \\
        \scriptsize
        \rotatebox[origin=c]{90}{\makecell{\textbf{EvEnhancer}\\\textbf{(Ours)}} } &
        \begin{tabular}{c@{\hskip 0.003\linewidth}c@{\hskip 0.003\linewidth}c@{\hskip 0.003\linewidth}c@{\hskip 0.003\linewidth}c@{\hskip 0.003\linewidth}c}
            \includegraphics[width=0.155\linewidth]{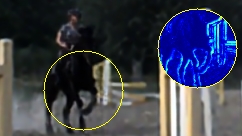}&
            \includegraphics[width=0.155\linewidth]{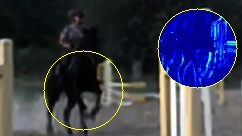}&
            \includegraphics[width=0.155\linewidth]{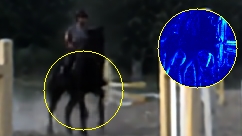}&
            \includegraphics[width=0.155\linewidth]{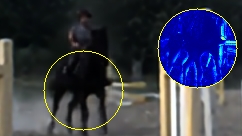}&
            \includegraphics[width=0.155\linewidth]{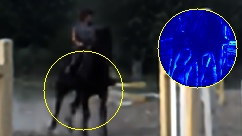}&
            \includegraphics[width=0.155\linewidth]{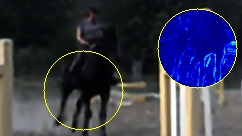}
        \end{tabular}
        \\ \\ \\
        \scriptsize
        \rotatebox[origin=c]{90}{\makecell{\textbf{EvEnhancer}\\\textbf{-light(Ours)}} } &
        \begin{tabular}{c@{\hskip 0.003\linewidth}c@{\hskip 0.003\linewidth}c@{\hskip 0.003\linewidth}c@{\hskip 0.003\linewidth}c@{\hskip 0.003\linewidth}c}
            \includegraphics[width=0.155\linewidth]{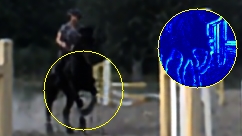}&
            \includegraphics[width=0.155\linewidth]{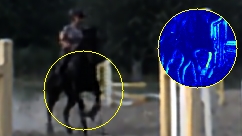}&
            \includegraphics[width=0.155\linewidth]{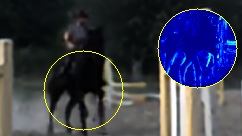}&
            \includegraphics[width=0.155\linewidth]{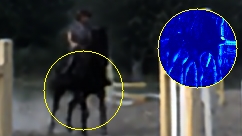}&
            \includegraphics[width=0.155\linewidth]{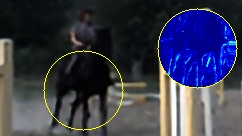}&
            \includegraphics[width=0.155\linewidth]{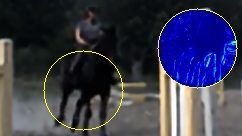}
        \end{tabular}
        \\ \\ \\
        \begin{tabular}{c}
        \small
        \rotatebox[origin=c]{90}{GT}
        \\ \\ \\ \\ \\ \\ \\ \\ \\ \\
        \end{tabular}
        &
        \begin{tabular}{c@{\hskip 0.003\linewidth}c@{\hskip 0.003\linewidth}c@{\hskip 0.003\linewidth}c@{\hskip 0.003\linewidth}c@{\hskip 0.003\linewidth}c}
            \includegraphics[width=0.155\linewidth]{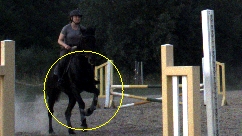}&
            \includegraphics[width=0.155\linewidth]{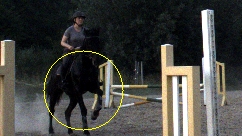}&
            \includegraphics[width=0.155\linewidth]{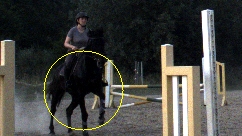}&
            \includegraphics[width=0.155\linewidth]{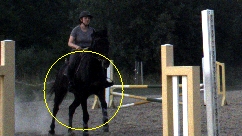}&
            \includegraphics[width=0.155\linewidth]{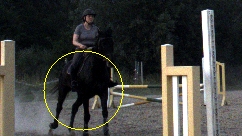}&
            \includegraphics[width=0.155\linewidth]{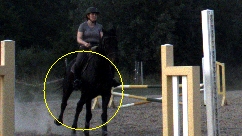}
        \\
        \\
        \footnotesize
        \(\mathcal{T}=0\)&
        \footnotesize
        \(\mathcal{T}=0.17\)&
        \footnotesize
        \(\mathcal{T}=0.33\)&
        \footnotesize
        \(\mathcal{T}=0.50\)&
        \footnotesize
        \(\mathcal{T}=0.67\)&
        \footnotesize
        \(\mathcal{T}=0.83\)
        \end{tabular}
        \\ \\
     &
        \multicolumn{1}{r}{        \includegraphics[width=0.5\linewidth]{fig/exp/legend.pdf}}
    \end{tabular}
    \caption{
       Qualitative comparison for OOD scale (\(t=6,s=4\)) on the BS-ERGB dataset \cite{tulyakov2022time}. We compare the normalized absolute difference maps (yellow boxes) for the same regions in each frame as in the GT frames.
    }
    \label{fig:s_event}
\end{figure*}
\renewcommand{\arraystretch}{1.}

%% file: fig/exp/exp_suppl_event2.tex
\setlength{\tabcolsep}{0.1pt}
\renewcommand{\arraystretch}{0.1}

\begin{figure*}[t]
    \centering\
         \begin{tabular}{cc}
            \rotatebox[origin=c]{90}{\makecell{\(\mathcal{T}=0\)\\}}  &
            \begin{tabular}{c@{\hskip 0.005\linewidth}c@{\hskip 0.005\linewidth}c@{\hskip 0.005\linewidth}c}
                \includegraphics[width=0.23\linewidth]{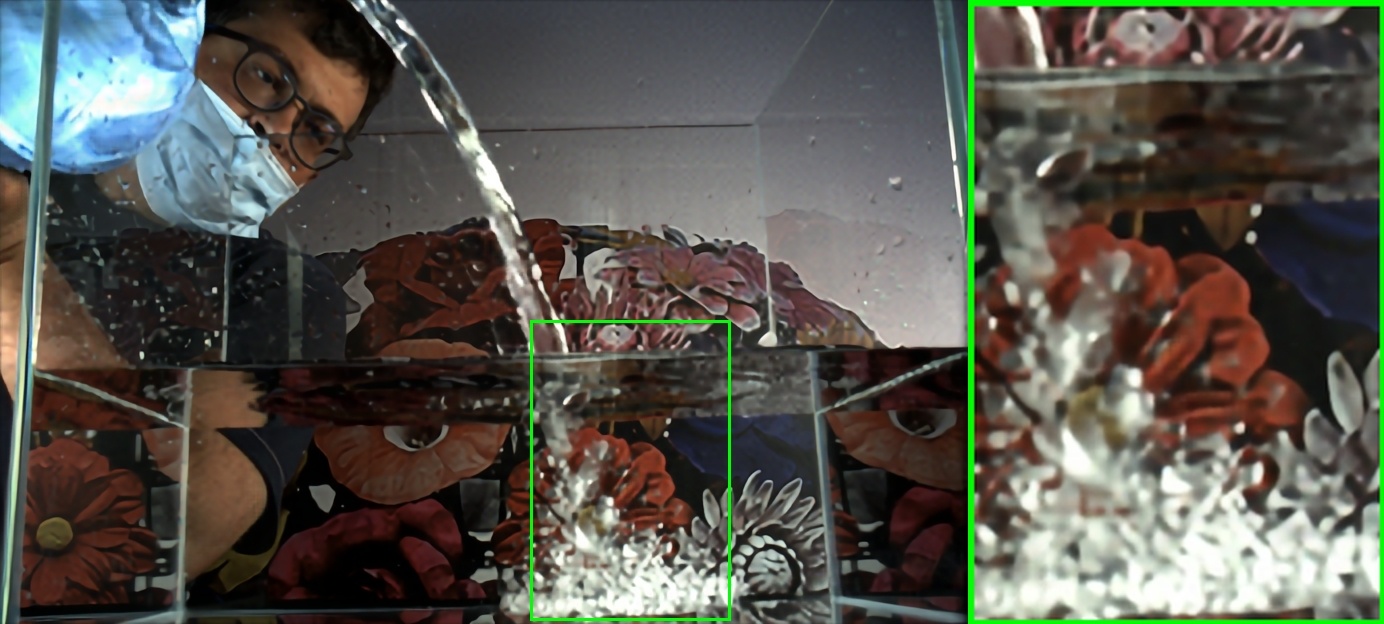}&
                \includegraphics[width=0.23\linewidth]{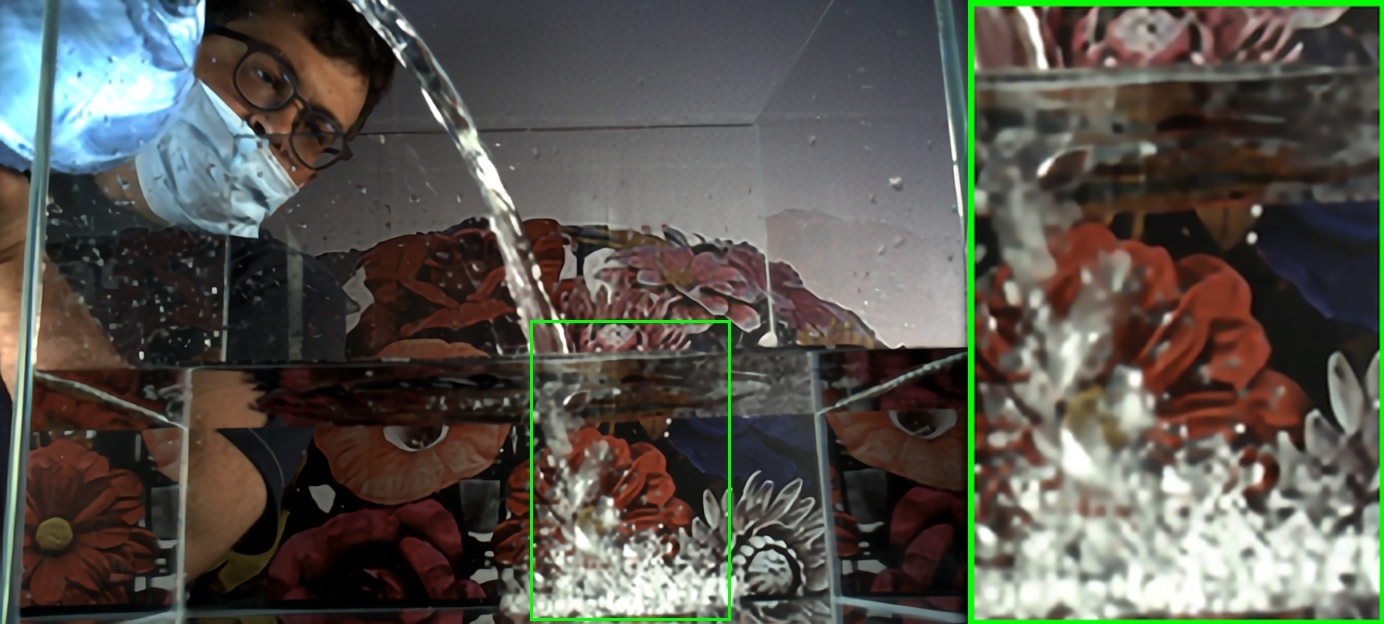}&
                \includegraphics[width=0.23\linewidth]{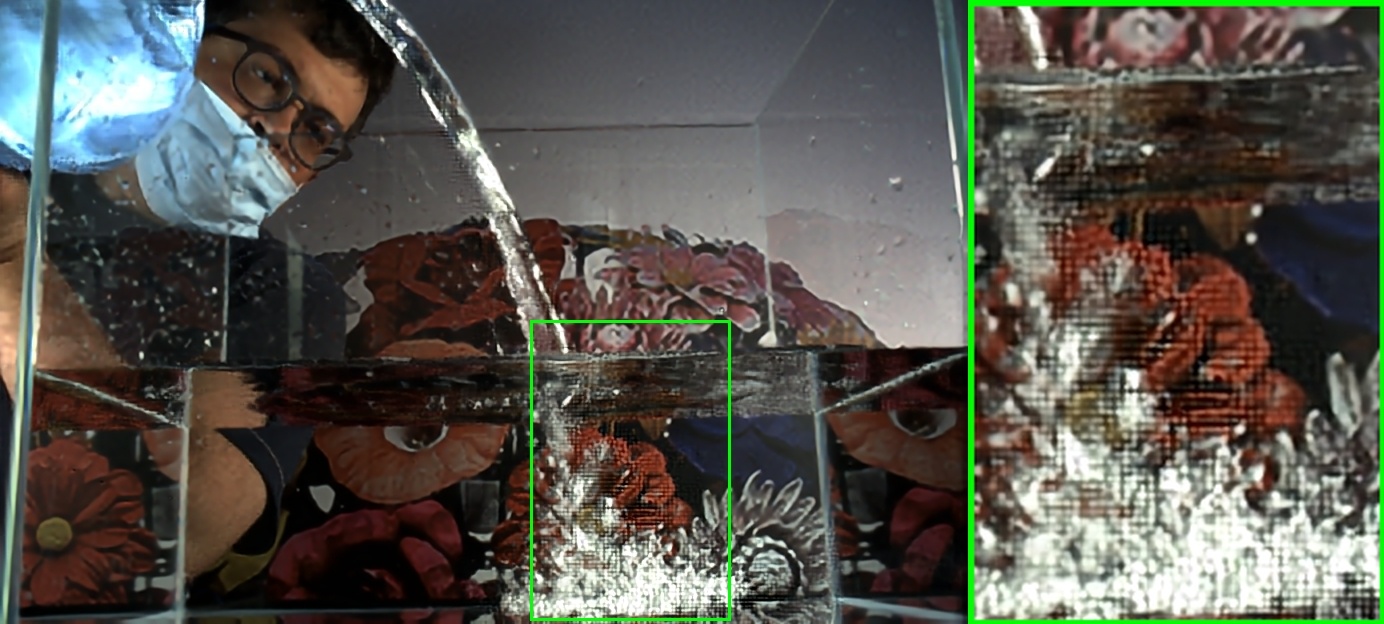}&
                \includegraphics[width=0.23\linewidth]{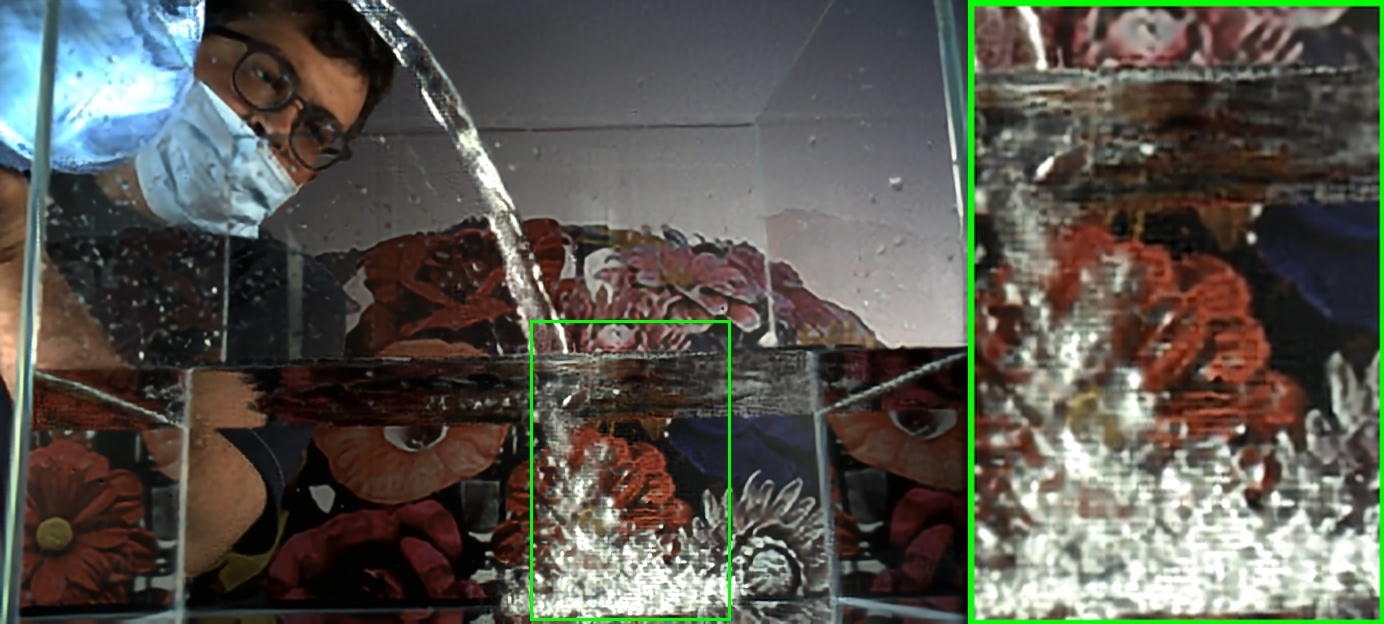}
            \end{tabular}
            \\ \\ \\
            \rotatebox[origin=c]{90}{\makecell{\(\mathcal{T}=0.125\)\\}}  &
            \begin{tabular}{c@{\hskip 0.005\linewidth}c@{\hskip 0.005\linewidth}c@{\hskip 0.005\linewidth}c}
                \includegraphics[width=0.23\linewidth]{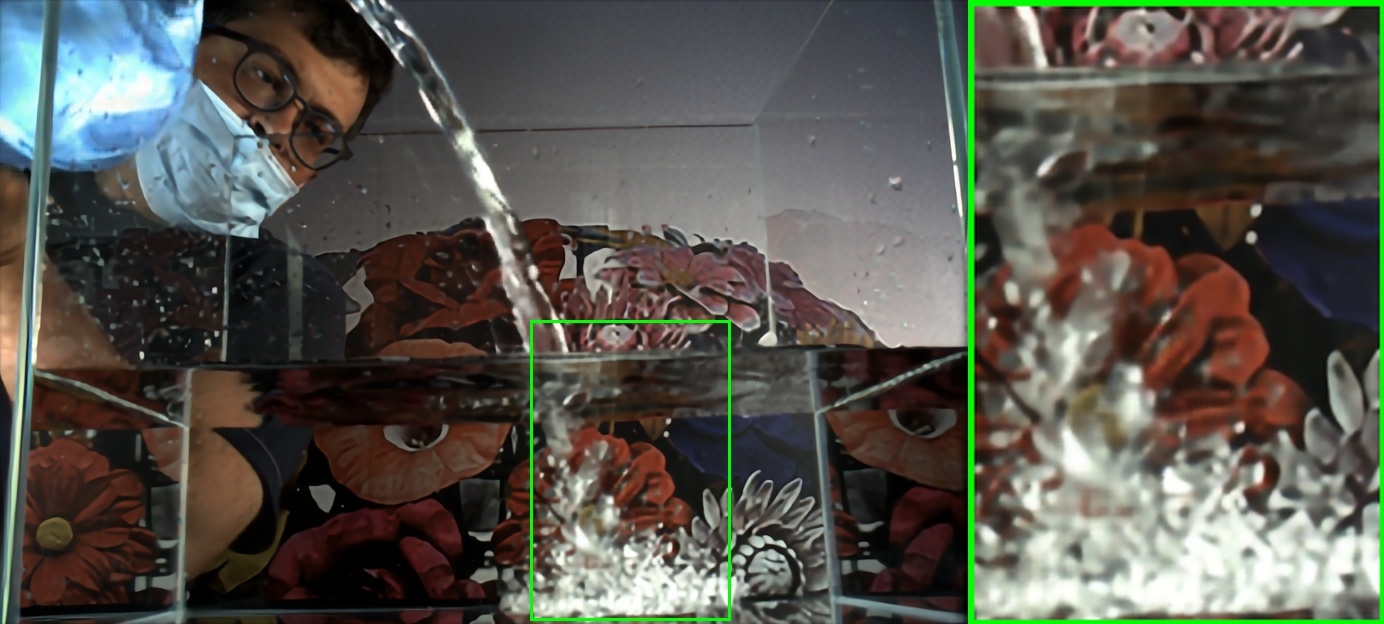}&
                \includegraphics[width=0.23\linewidth]{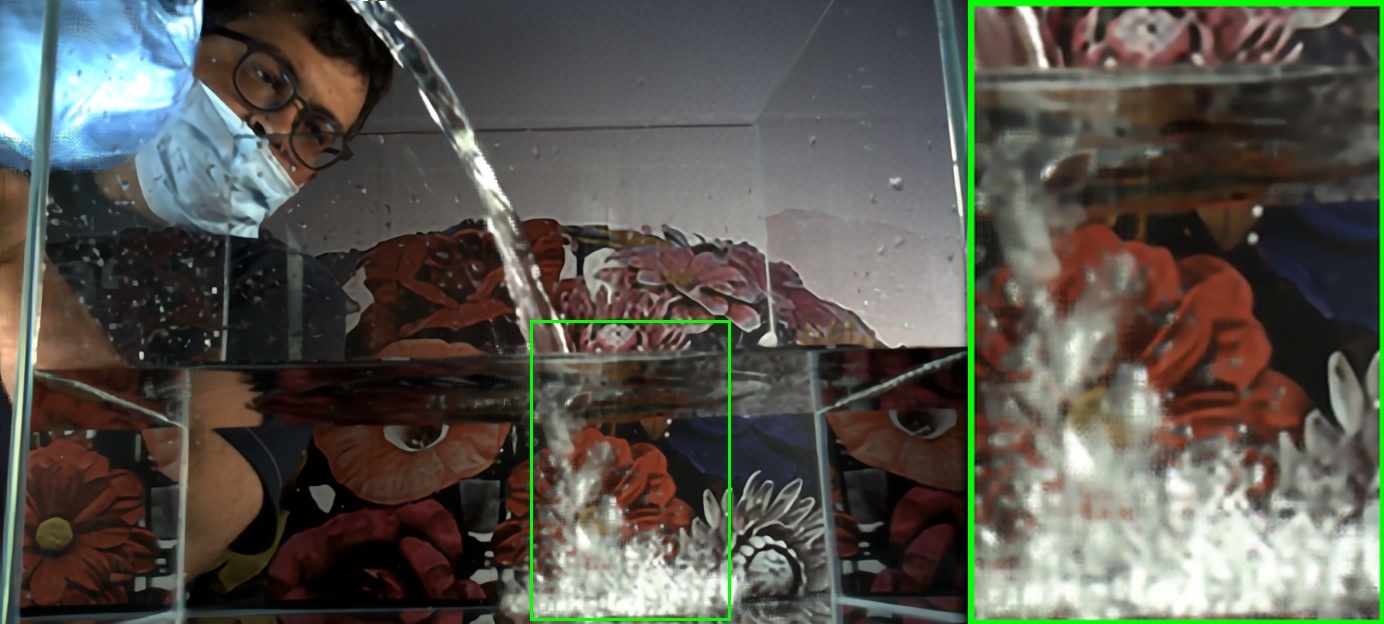}&
                \includegraphics[width=0.23\linewidth]{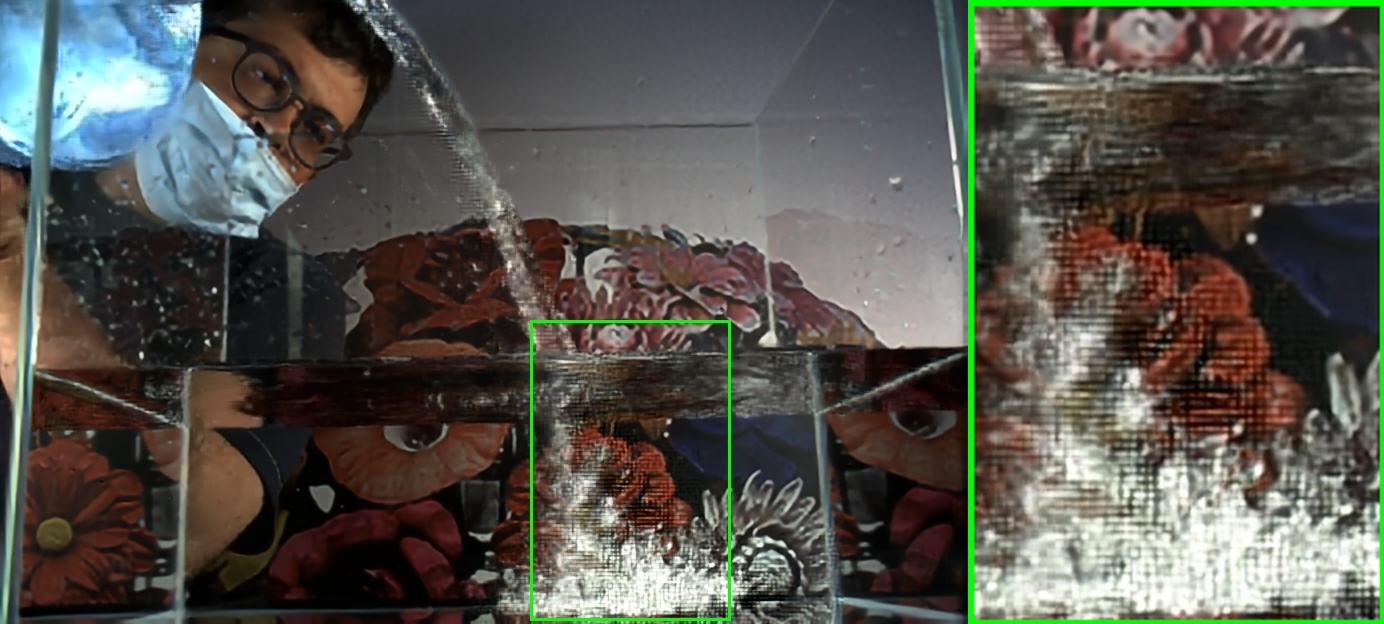}&
                \includegraphics[width=0.23\linewidth]{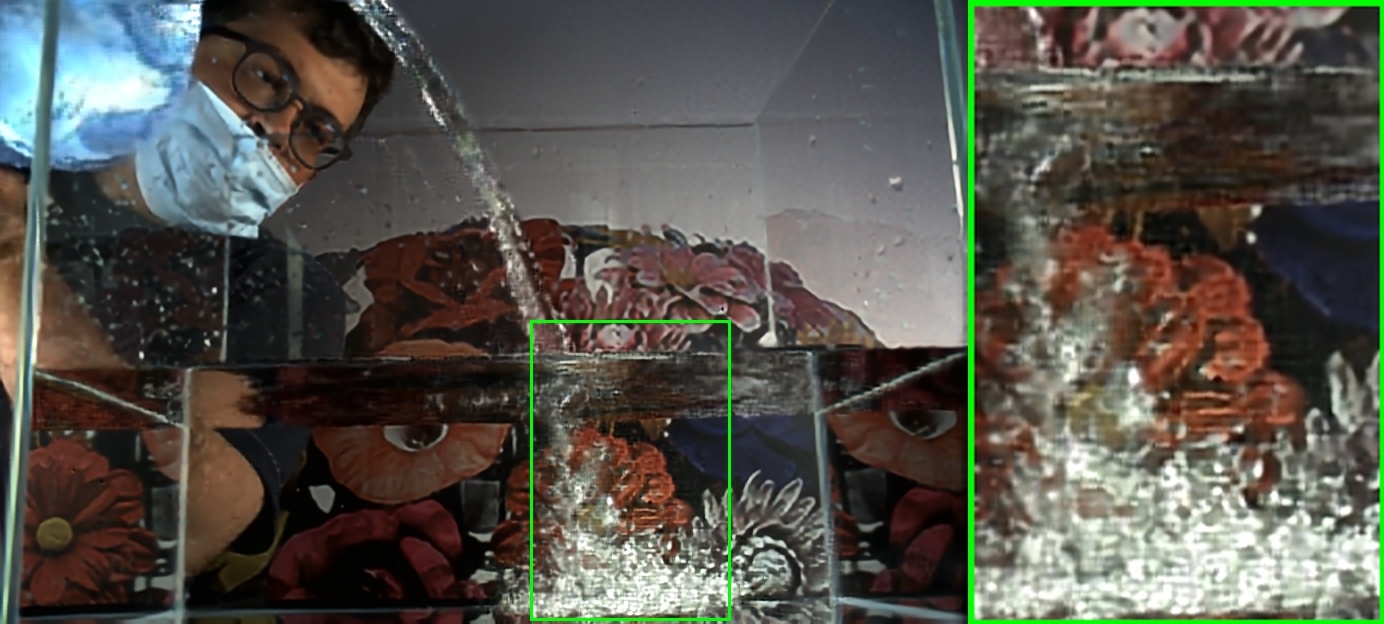}
            \end{tabular}
            \\ \\ \\
            \rotatebox[origin=c]{90}{\makecell{\(\mathcal{T}=0.25\)\\}}  &
            \begin{tabular}{c@{\hskip 0.005\linewidth}c@{\hskip 0.005\linewidth}c@{\hskip 0.005\linewidth}c}
                \includegraphics[width=0.23\linewidth]{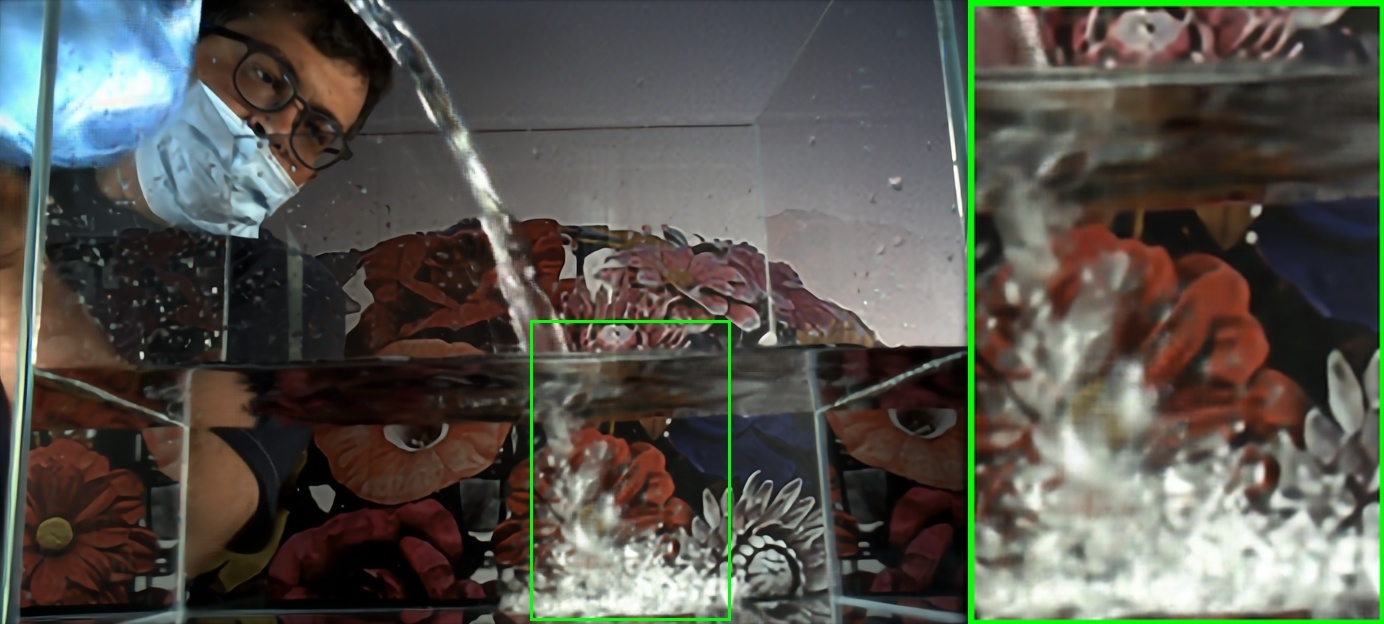}&
                \includegraphics[width=0.23\linewidth]{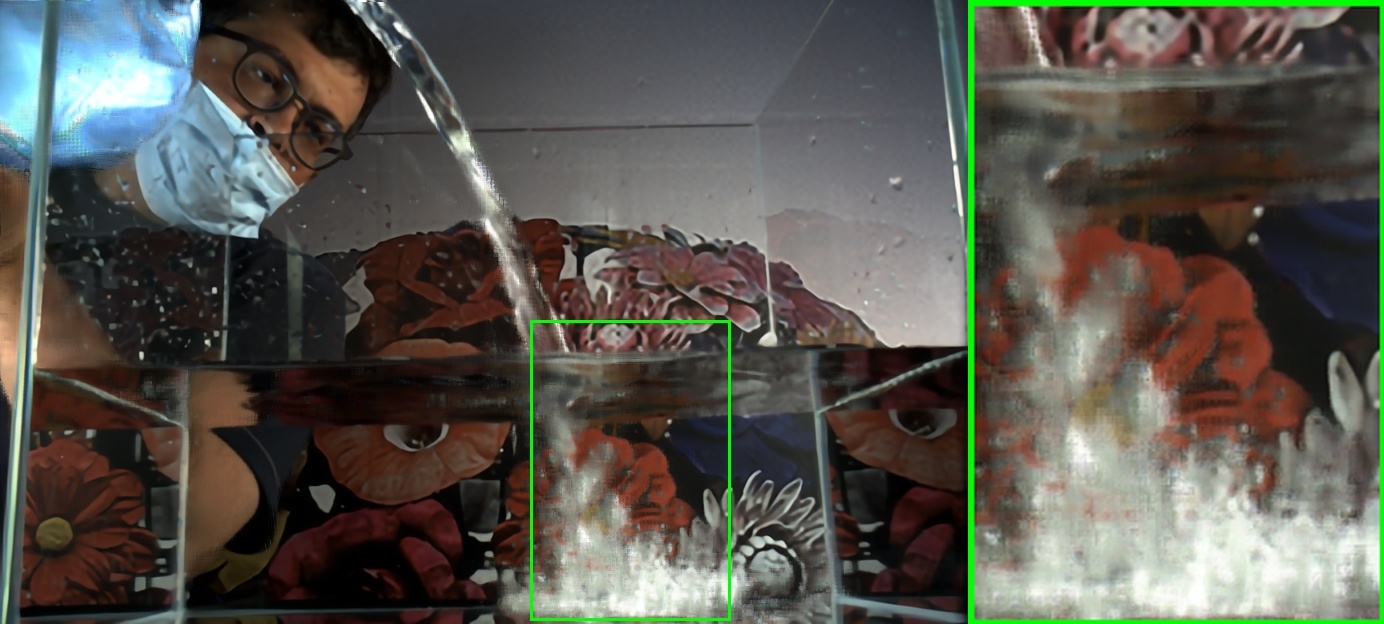}&
                \includegraphics[width=0.23\linewidth]{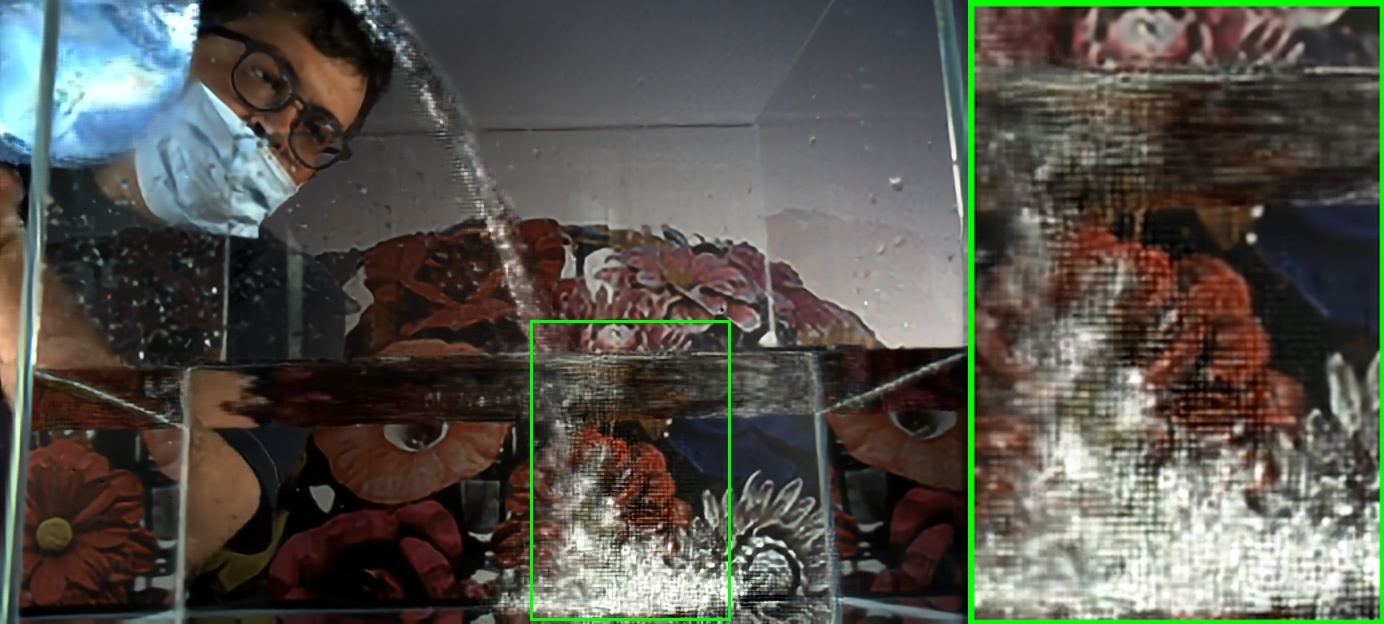}&
                \includegraphics[width=0.23\linewidth]{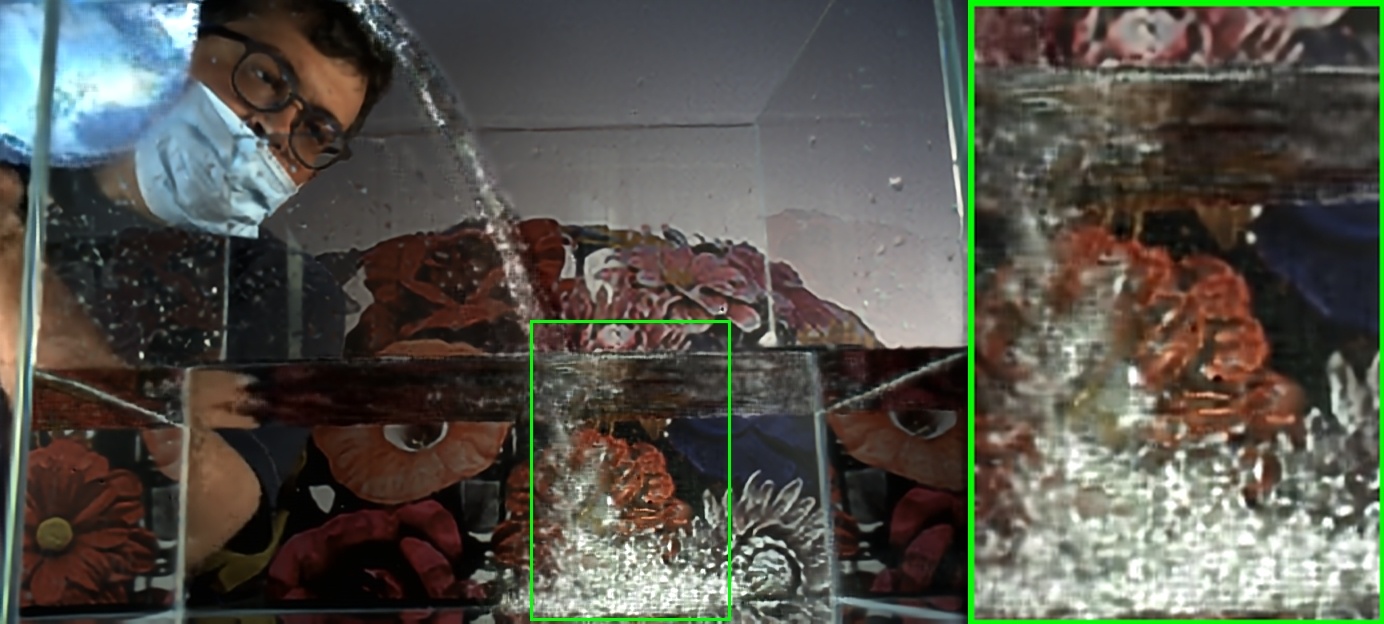}
            \end{tabular}
            \\ \\ \\
            \rotatebox[origin=c]{90}{\makecell{\(\mathcal{T}=0.375\)\\}}  &
            \begin{tabular}{c@{\hskip 0.005\linewidth}c@{\hskip 0.005\linewidth}c@{\hskip 0.005\linewidth}c}
                 \includegraphics[width=0.23\linewidth]{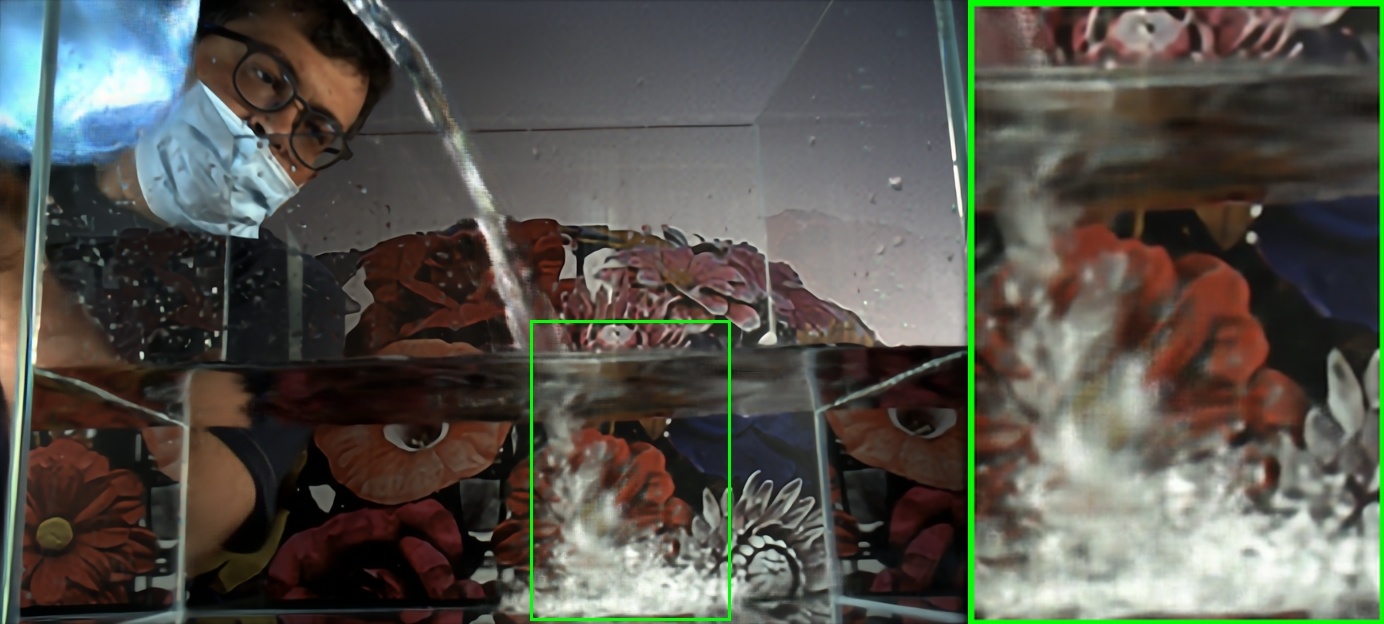}&
                \includegraphics[width=0.23\linewidth]{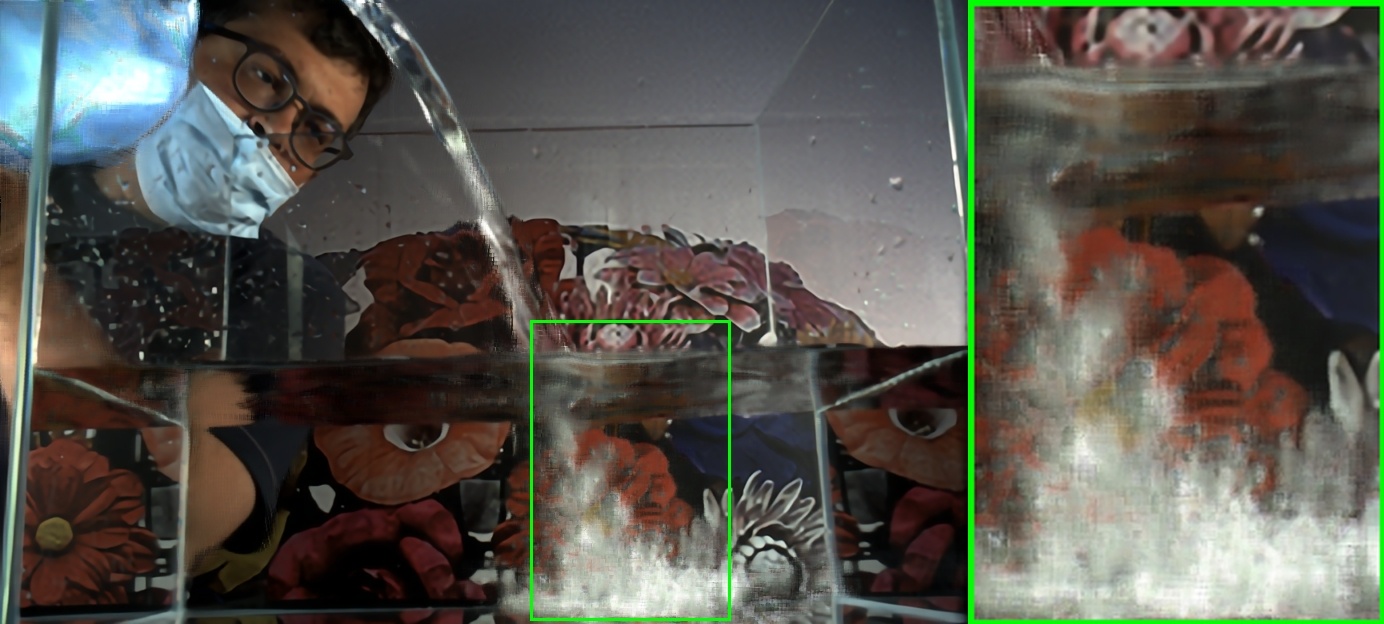}&
                \includegraphics[width=0.23\linewidth]{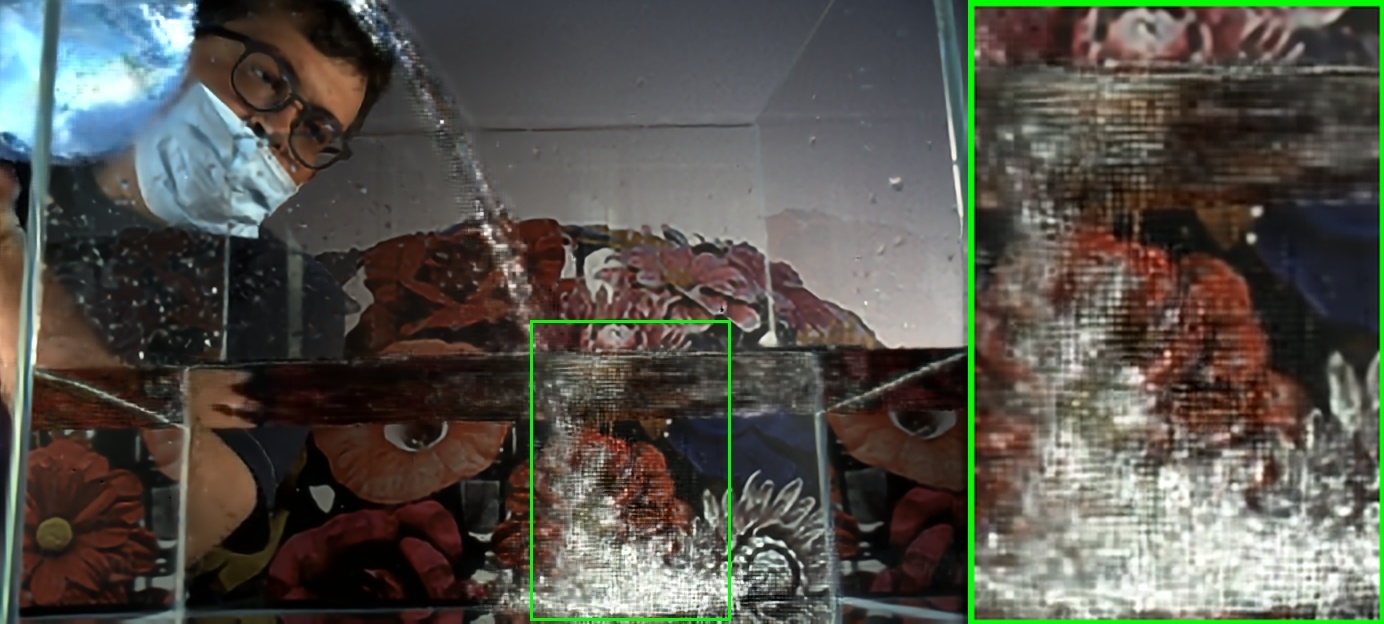}&
                \includegraphics[width=0.23\linewidth]{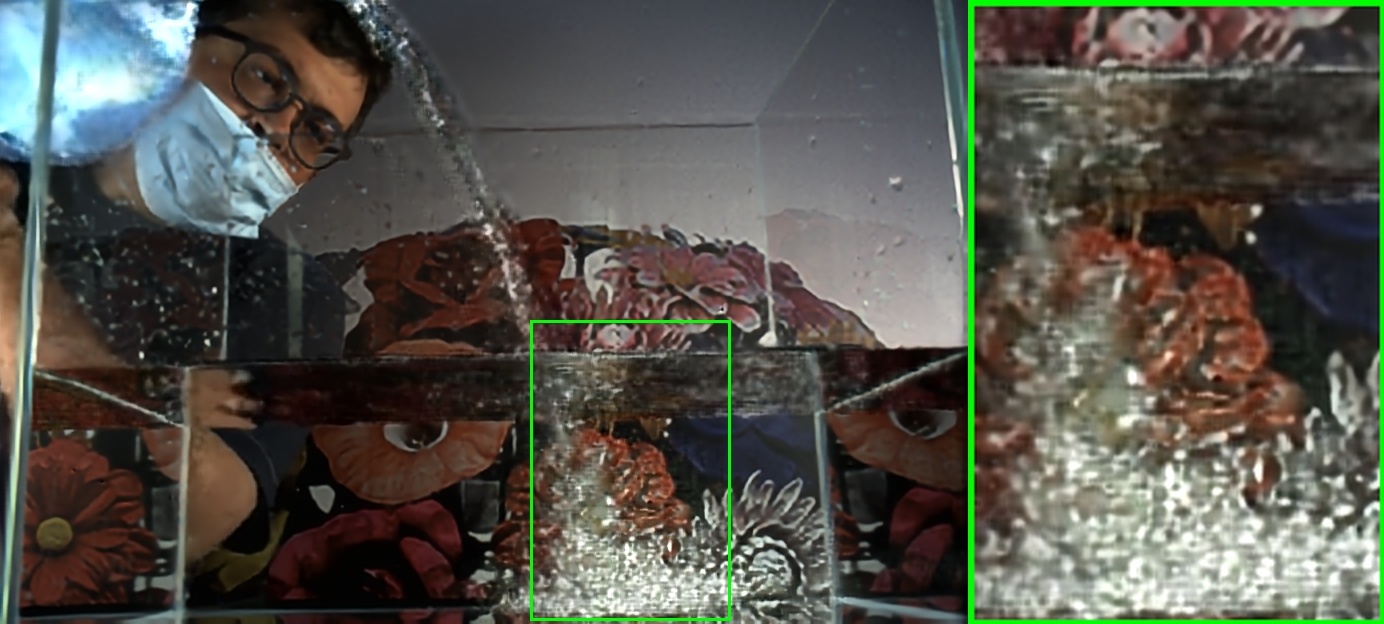}
            \end{tabular}
            \\ \\ \\
            \rotatebox[origin=c]{90}{\makecell{\(\mathcal{T}=0.5\)\\}}  &
            \begin{tabular}{c@{\hskip 0.005\linewidth}c@{\hskip 0.005\linewidth}c@{\hskip 0.005\linewidth}c}
                \includegraphics[width=0.23\linewidth]{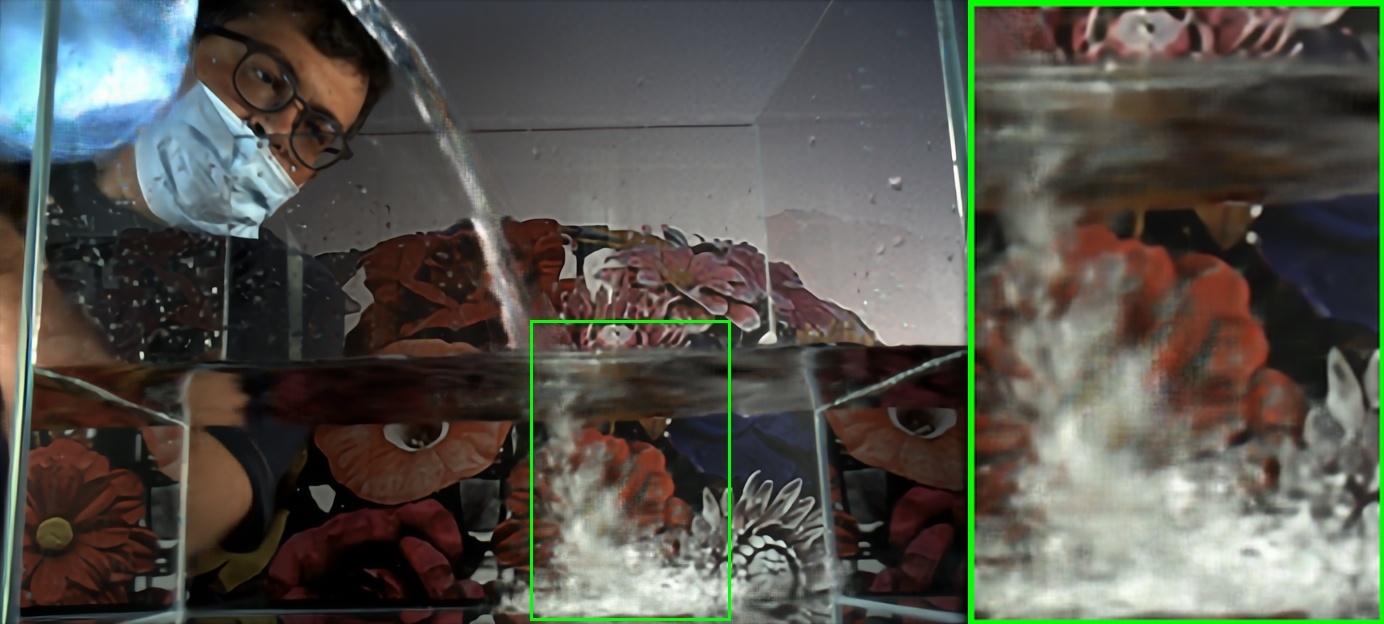}&
                \includegraphics[width=0.23\linewidth]{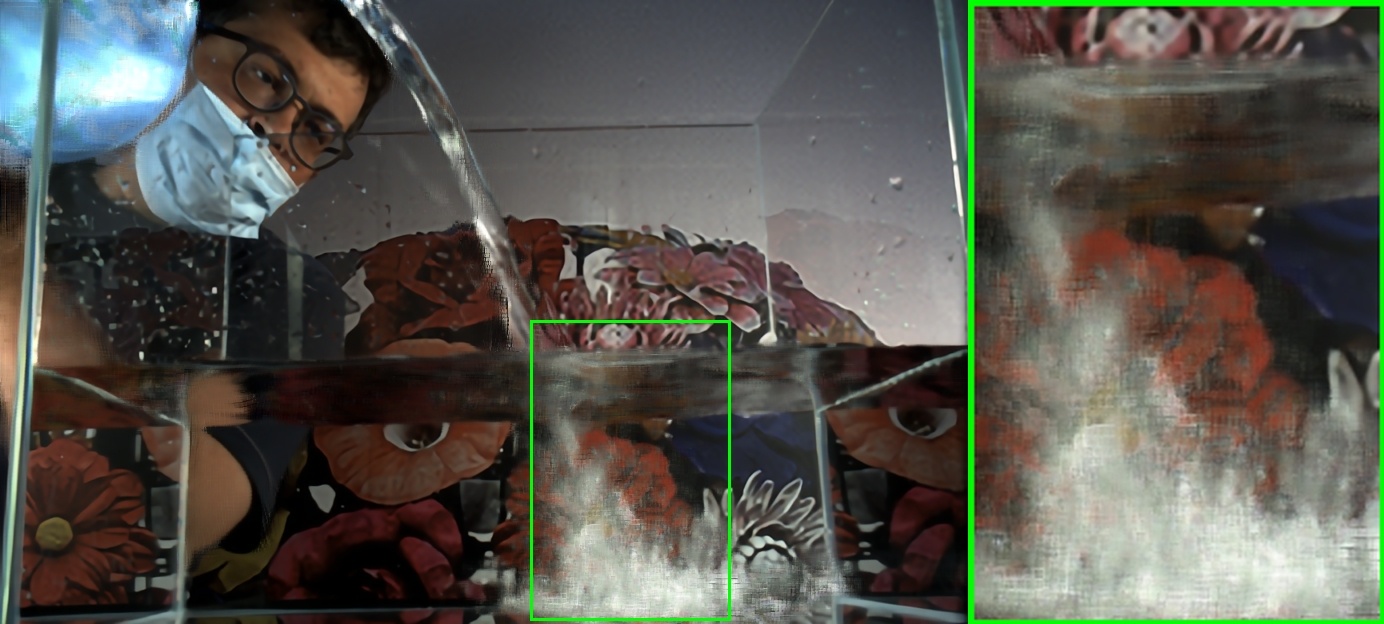}&
                \includegraphics[width=0.23\linewidth]{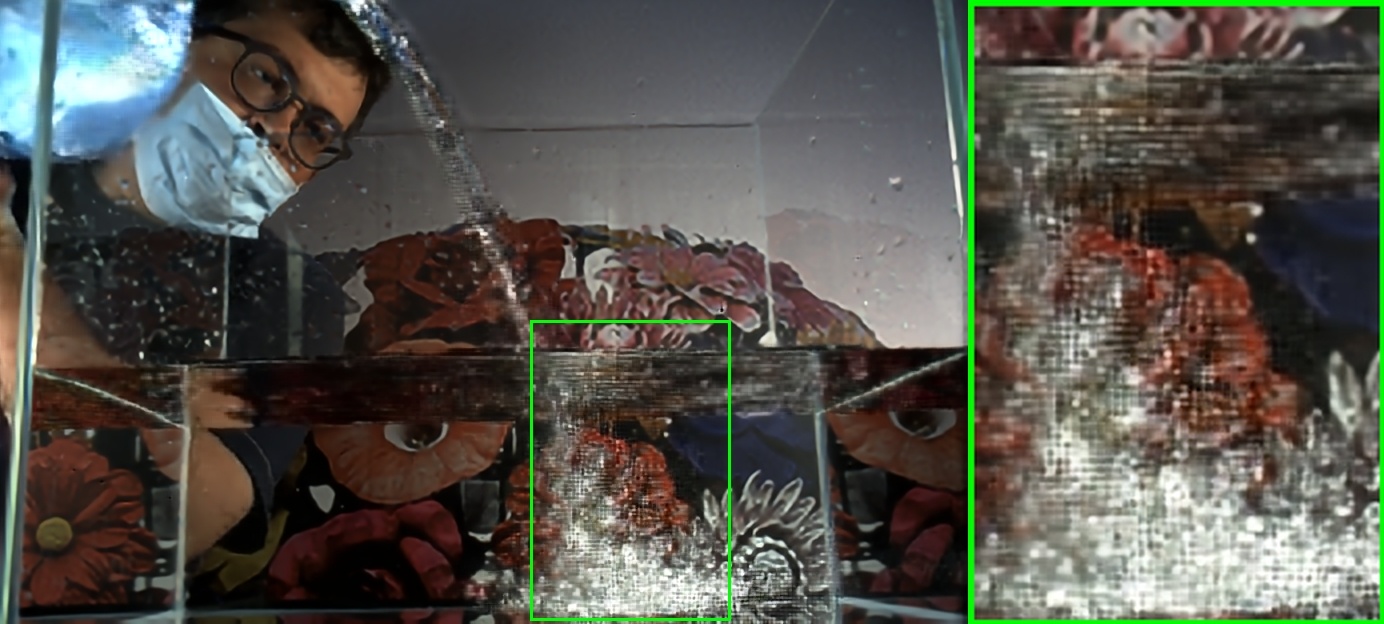}&
                \includegraphics[width=0.23\linewidth]{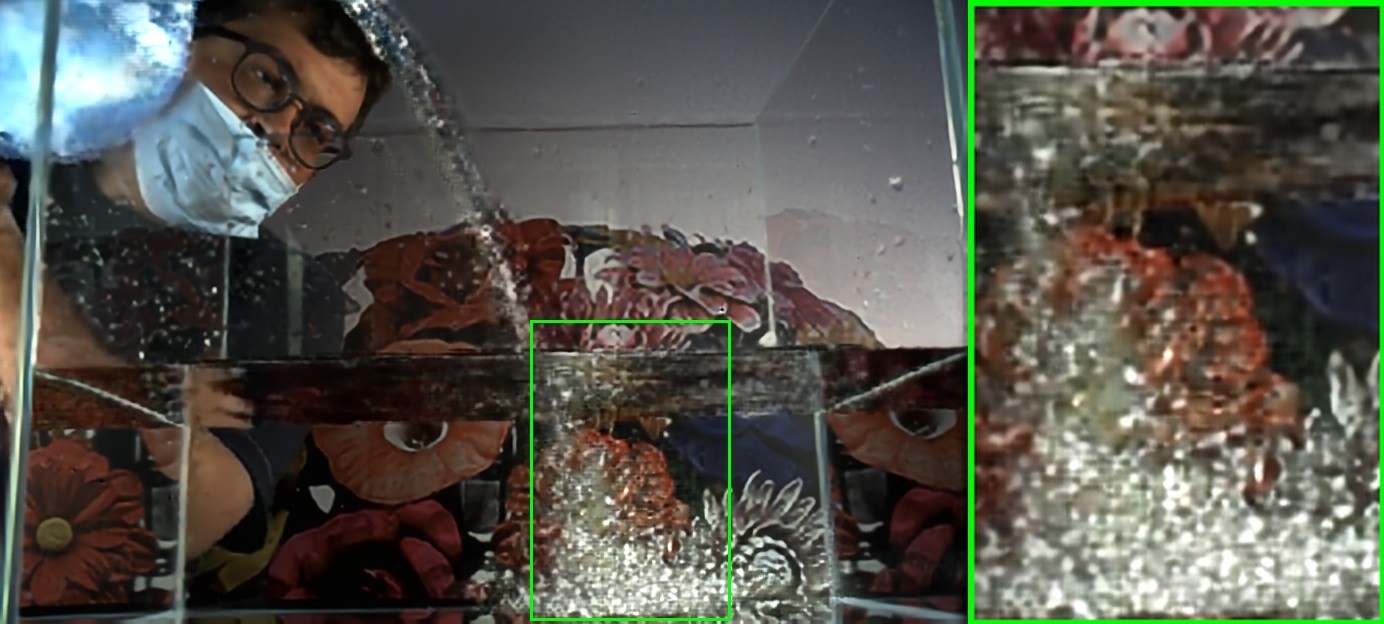}
            \end{tabular}
            \\ \\ \\
            \rotatebox[origin=c]{90}{\makecell{\(\mathcal{T}=0.625\)\\}}  &
            \begin{tabular}{c@{\hskip 0.005\linewidth}c@{\hskip 0.005\linewidth}c@{\hskip 0.005\linewidth}c}
                \includegraphics[width=0.23\linewidth]{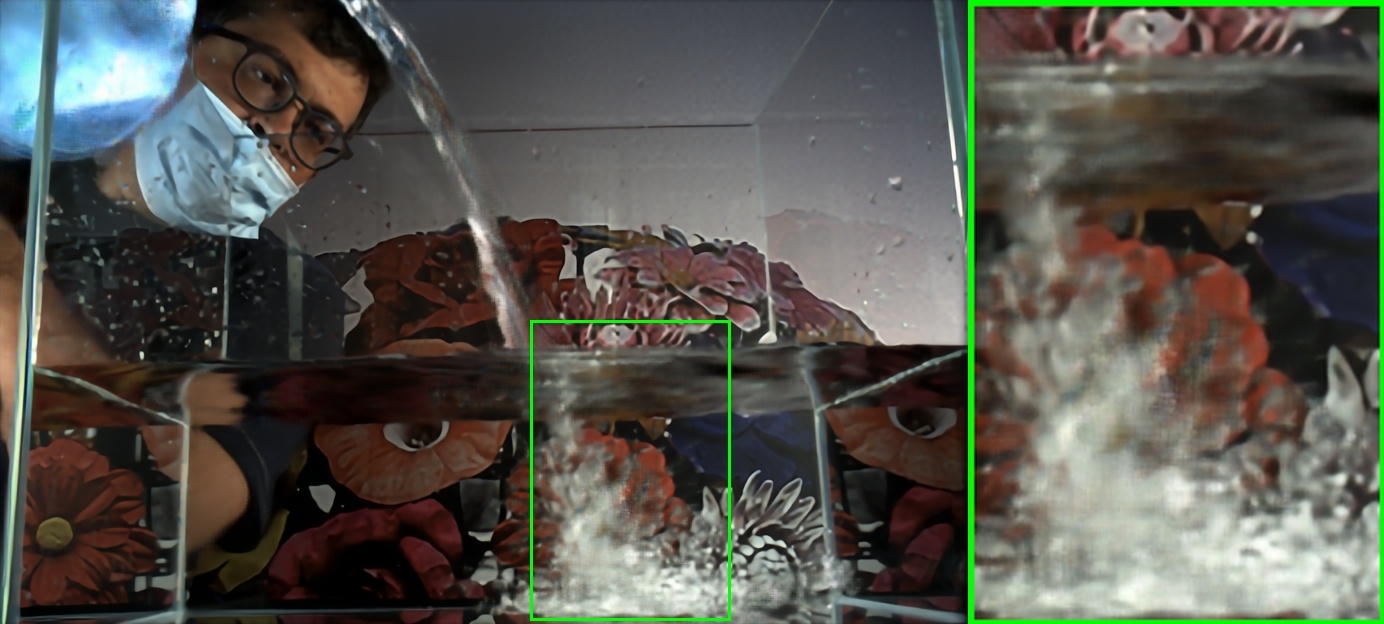}&
                \includegraphics[width=0.23\linewidth]{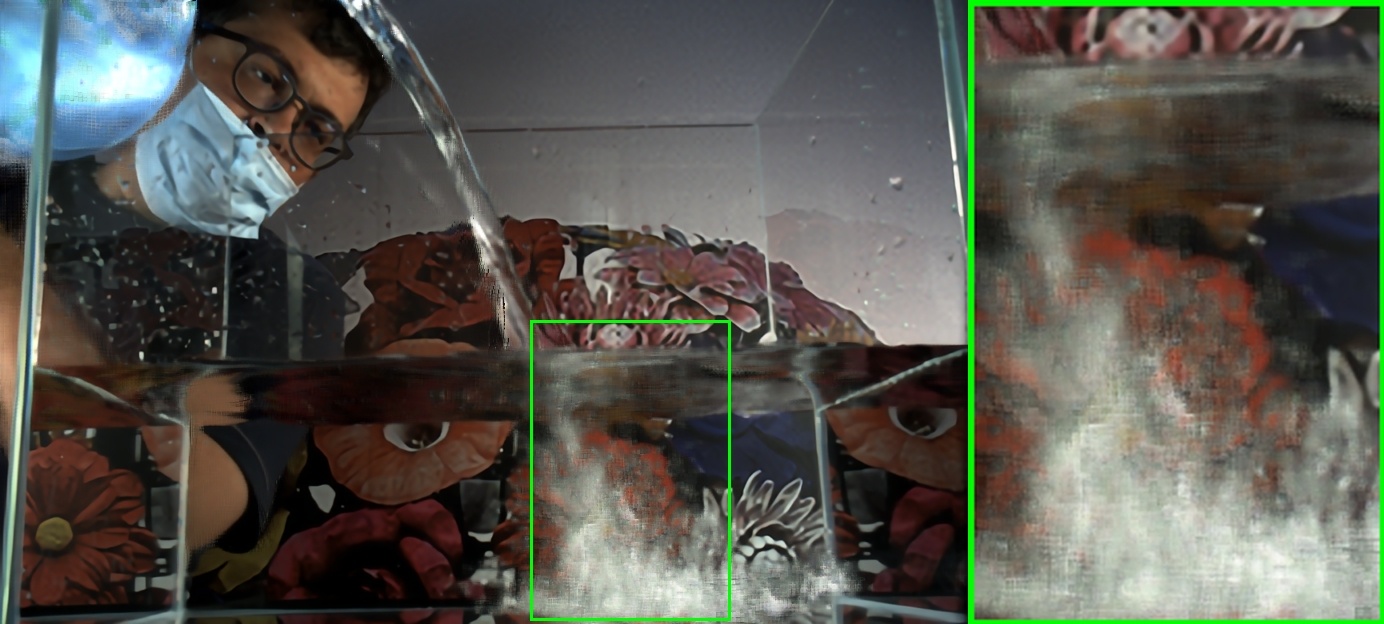}&
                \includegraphics[width=0.23\linewidth]{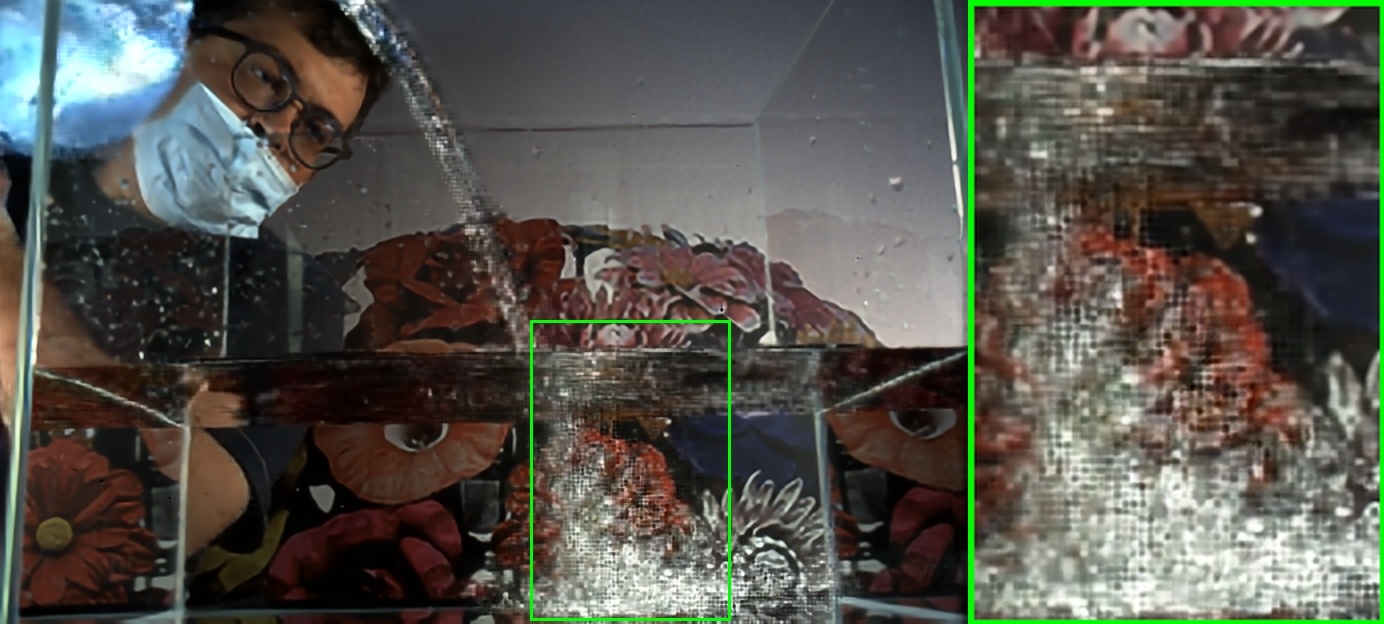}&
                \includegraphics[width=0.23\linewidth]{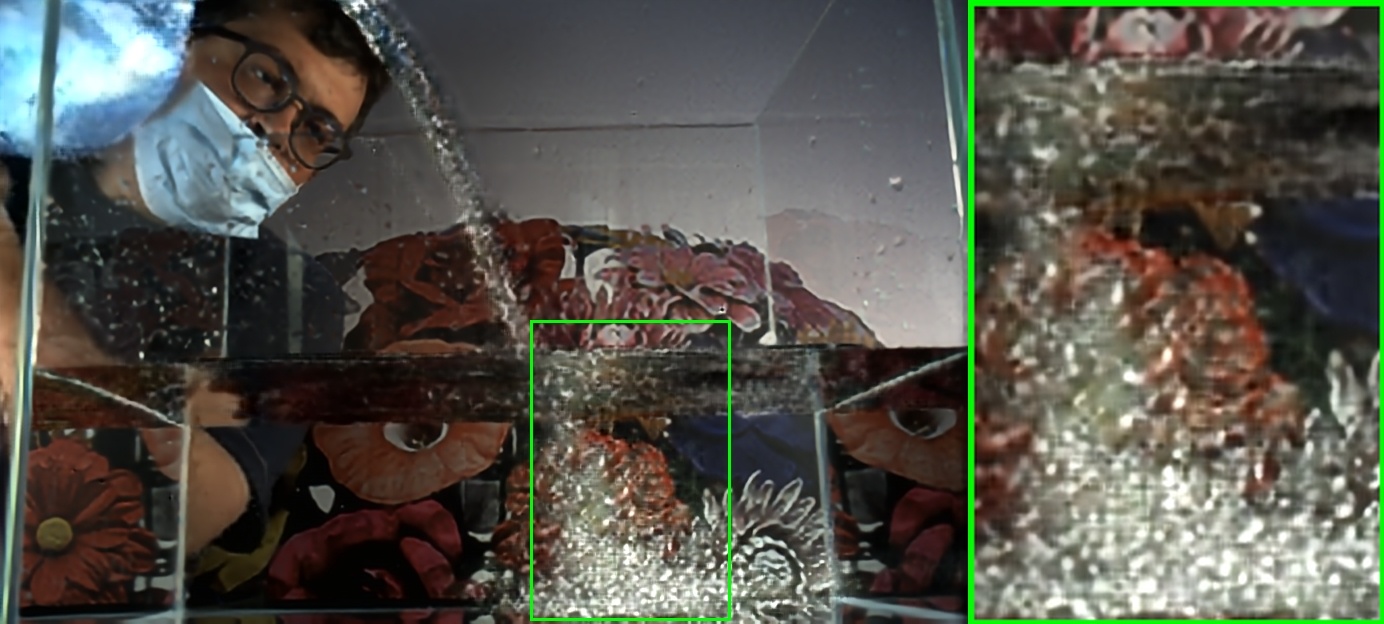}
            \end{tabular}
            \\ \\ \\
            \rotatebox[origin=c]{90}{\makecell{\(\mathcal{T}=0.75\)\\}}  &
            \begin{tabular}{c@{\hskip 0.005\linewidth}c@{\hskip 0.005\linewidth}c@{\hskip 0.005\linewidth}c}
                \includegraphics[width=0.23\linewidth]{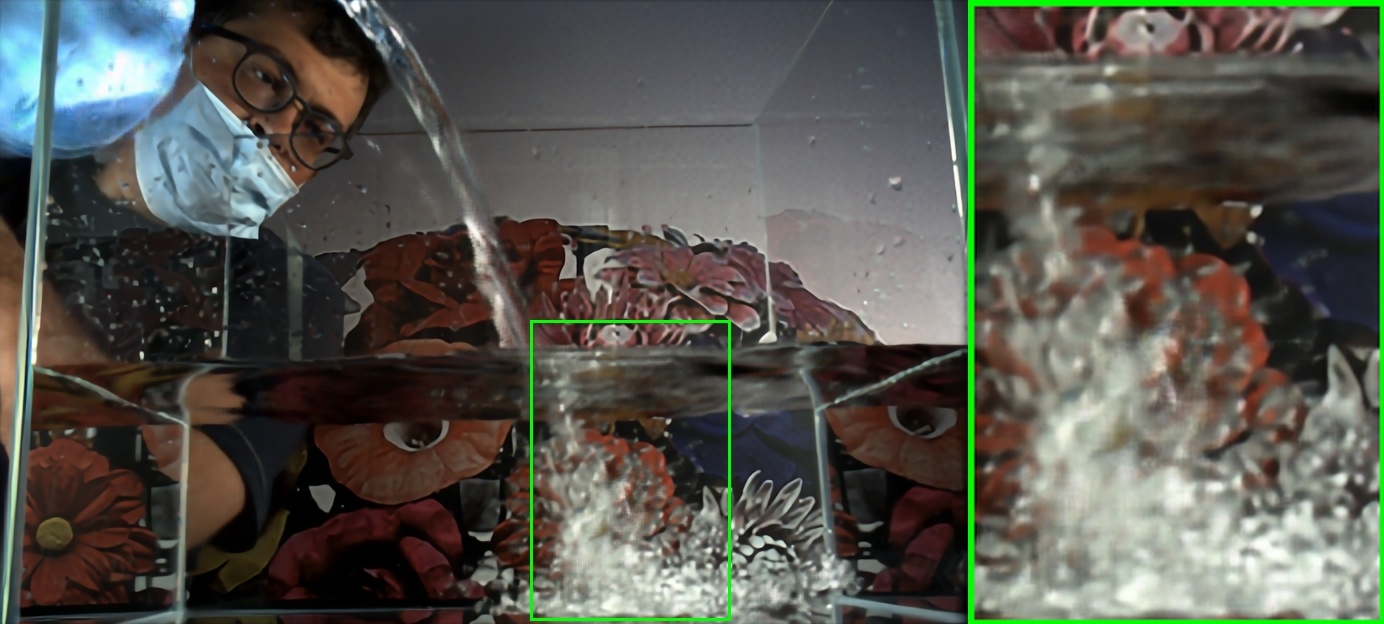}&
                \includegraphics[width=0.23\linewidth]{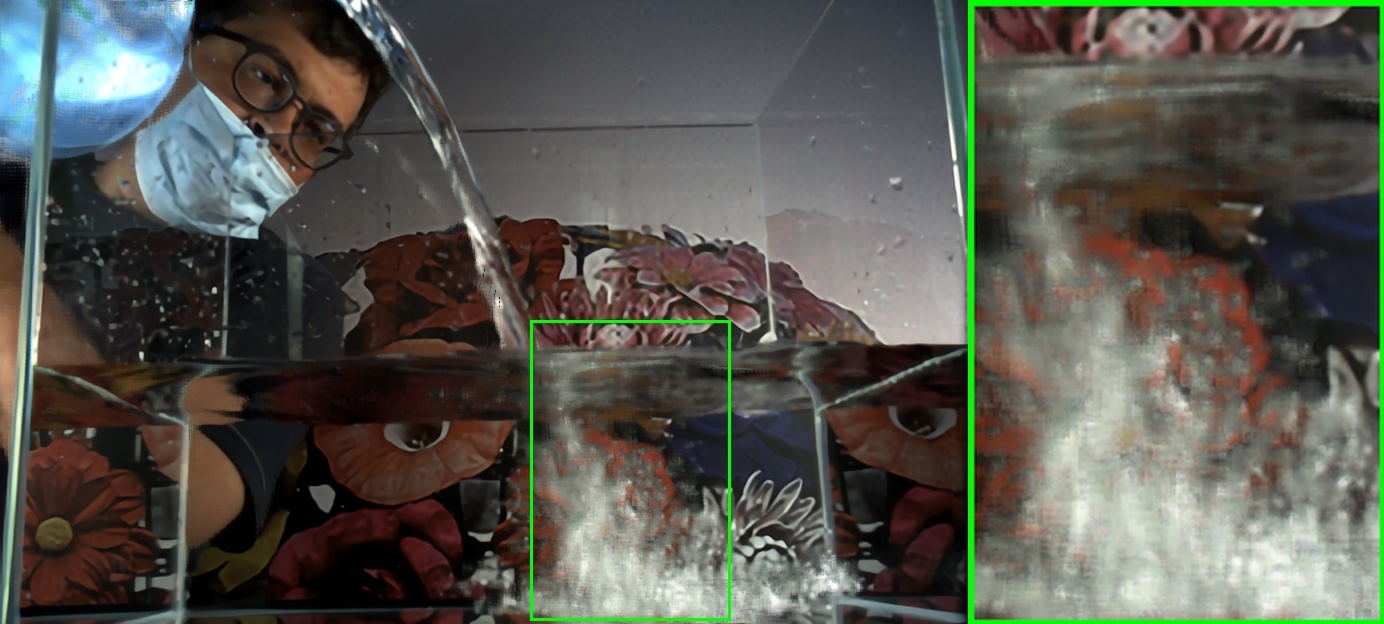}&
                \includegraphics[width=0.23\linewidth]{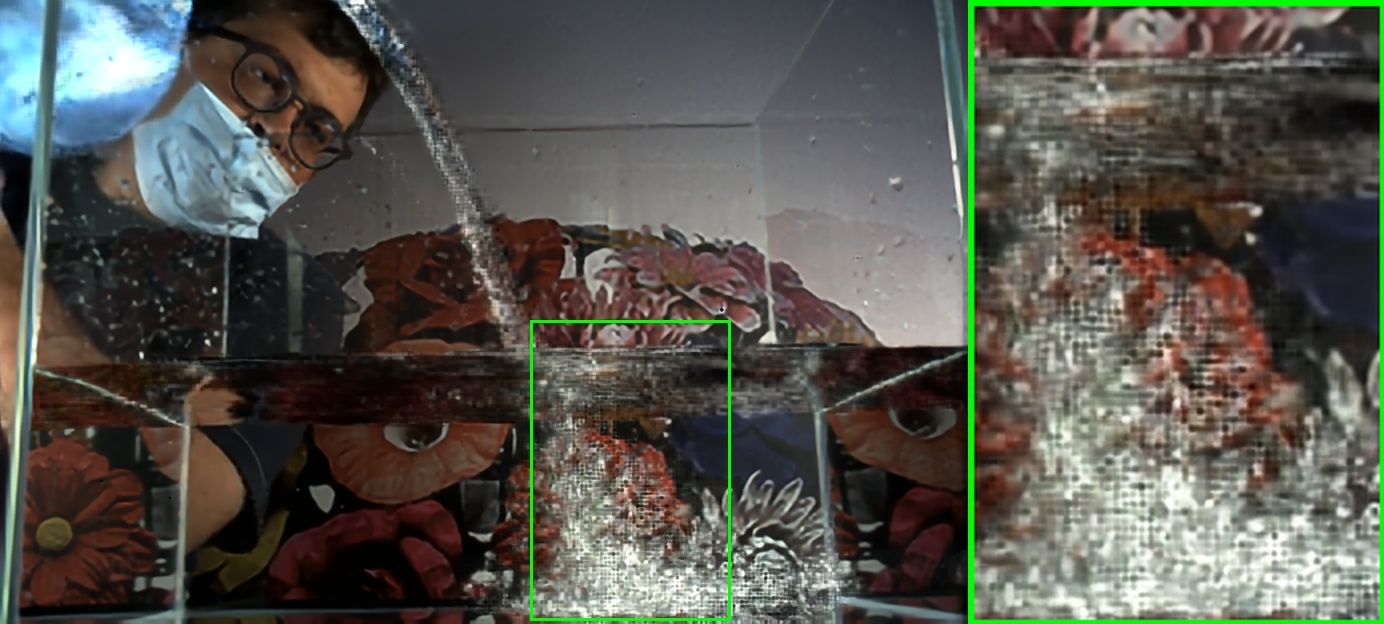}&
                \includegraphics[width=0.23\linewidth]{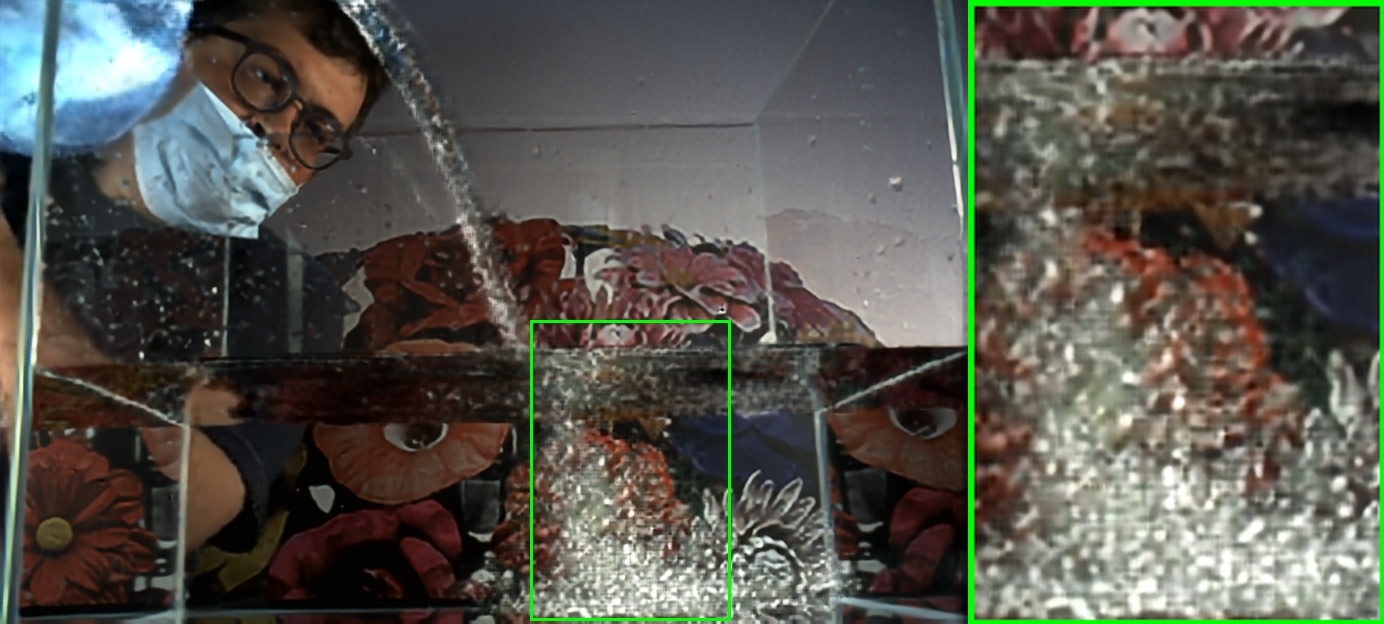}
            \end{tabular}
            \\ \\ \\
            \rotatebox[origin=c]{90}{\makecell{\(\mathcal{T}=0.875\)\\}}  &
            \begin{tabular}{c@{\hskip 0.005\linewidth}c@{\hskip 0.005\linewidth}c@{\hskip 0.005\linewidth}c}
                \includegraphics[width=0.23\linewidth]{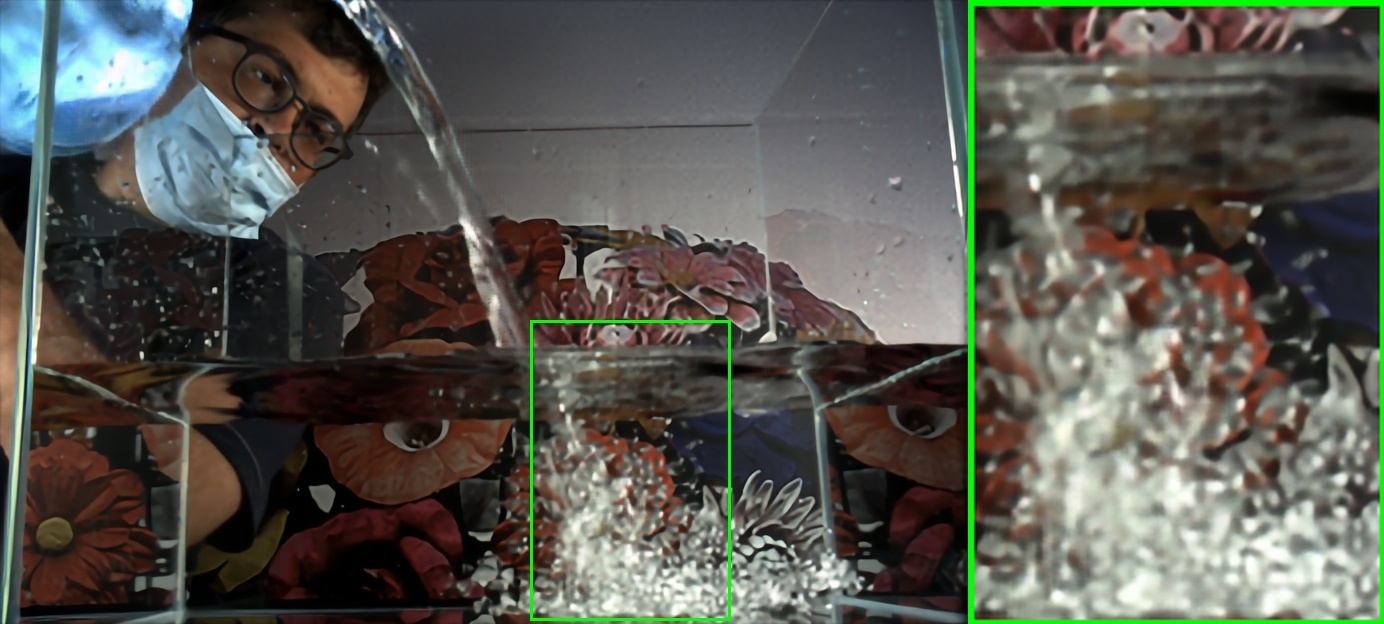}&
                \includegraphics[width=0.23\linewidth]{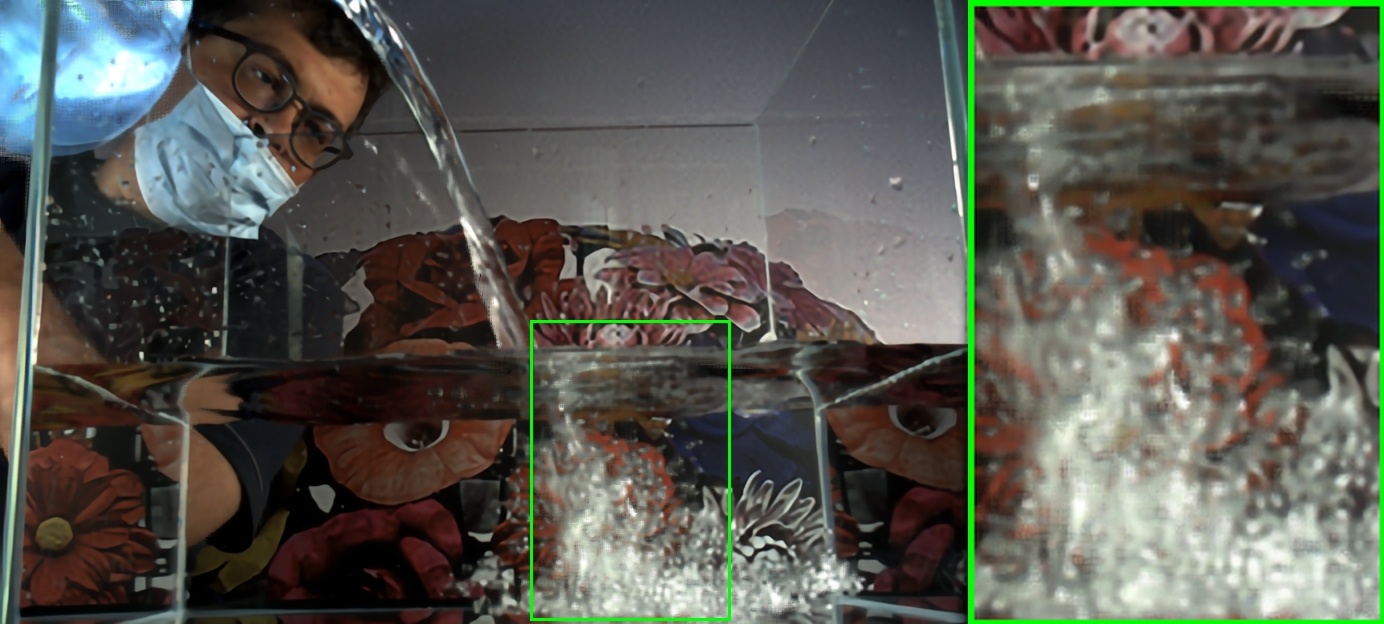}&
                \includegraphics[width=0.23\linewidth]{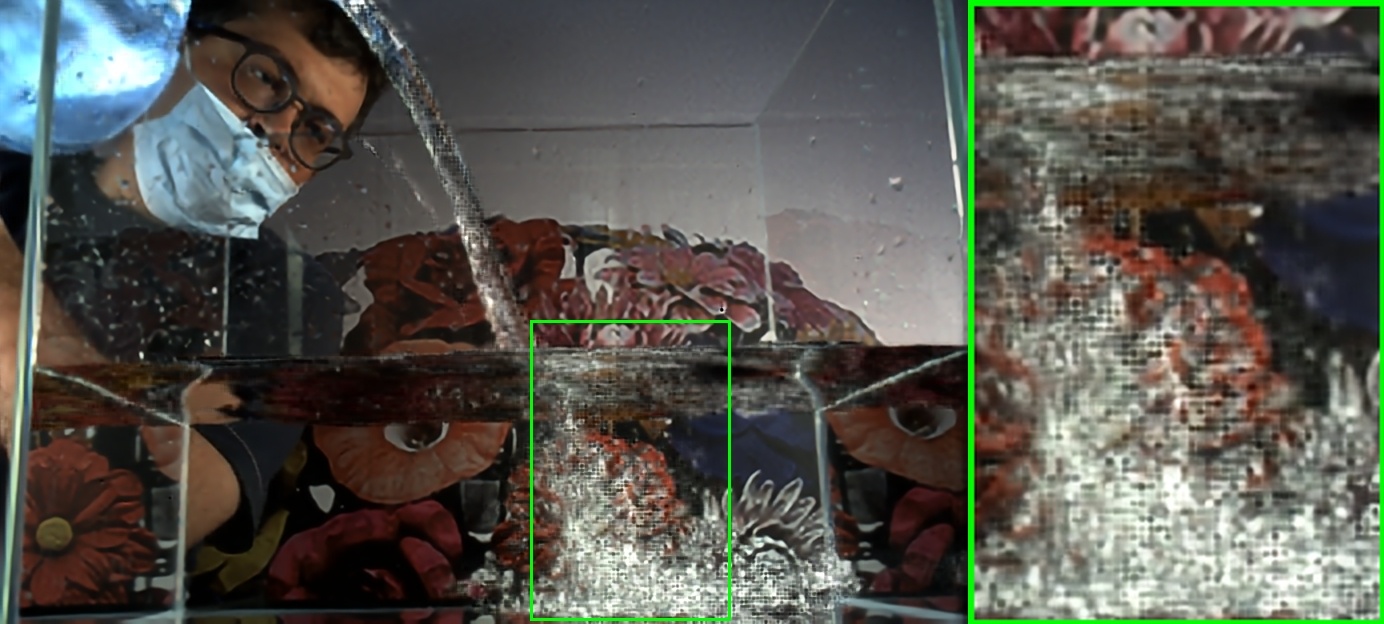}&
                \includegraphics[width=0.23\linewidth]{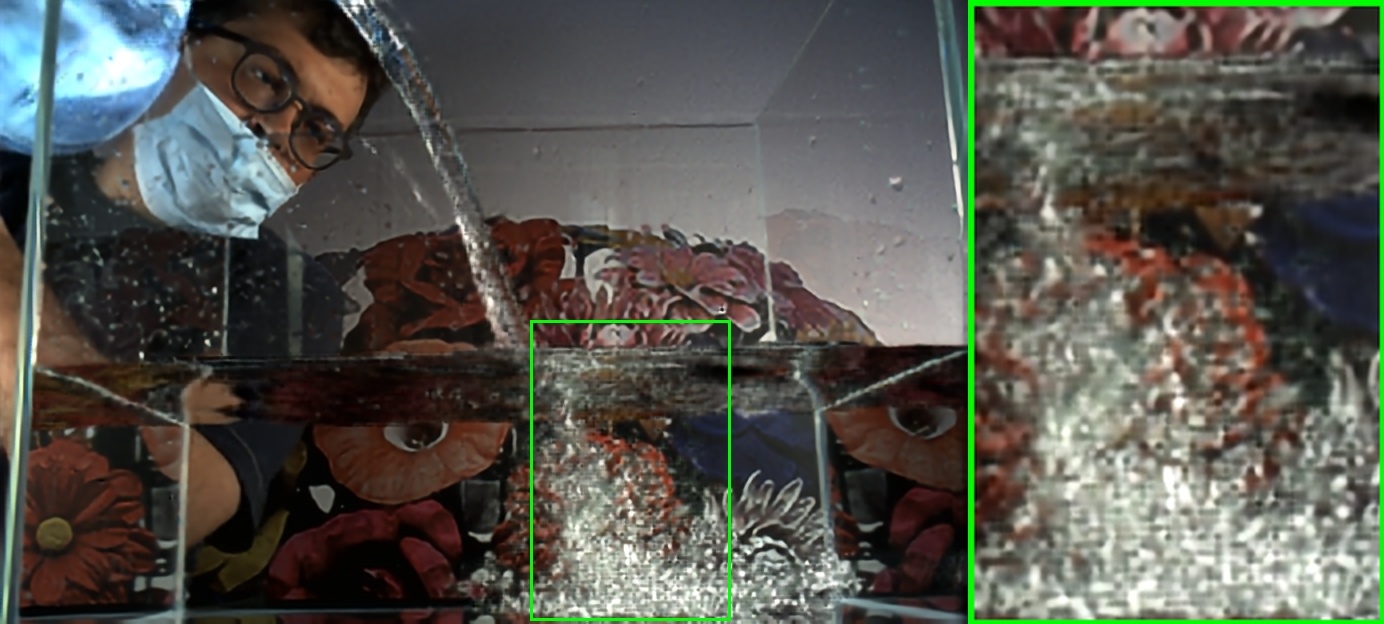}
            \end{tabular}
            \\ \\ \\
            \begin{tabular}{c}
            \rotatebox[origin=c]{90}{\makecell{\(\mathcal{T}=1\)\\}}
            \\ \\ \\ \\ \\ \\ \\ \\ \\ \\
            \end{tabular}
            &
            \begin{tabular}{c@{\hskip 0.005\linewidth}c@{\hskip 0.005\linewidth}c@{\hskip 0.005\linewidth}c}
                \includegraphics[width=0.23\linewidth]{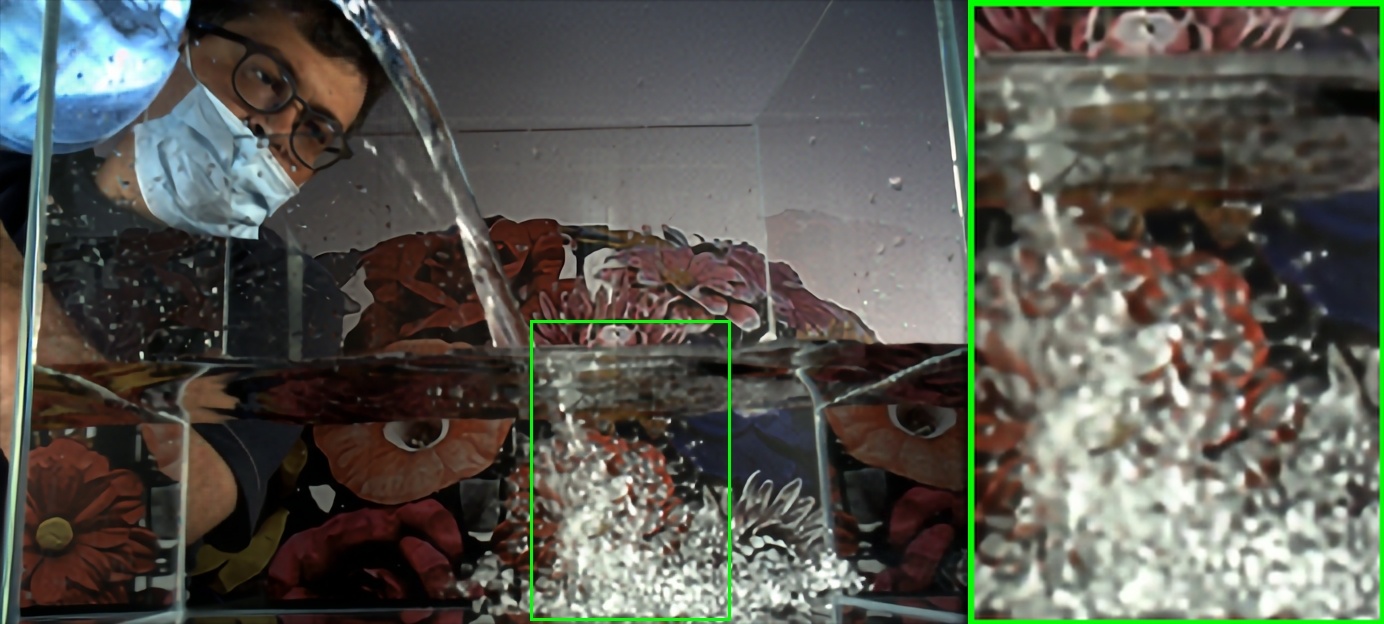}&
                \includegraphics[width=0.23\linewidth]{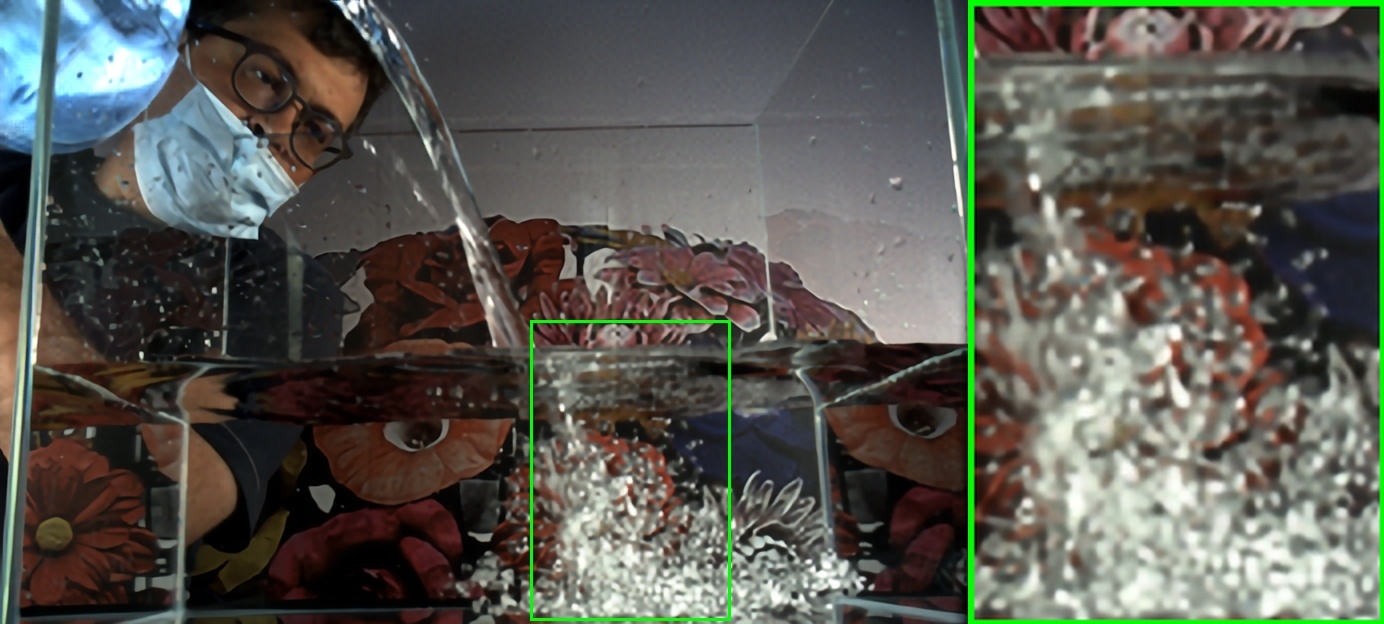}&
                \includegraphics[width=0.23\linewidth]{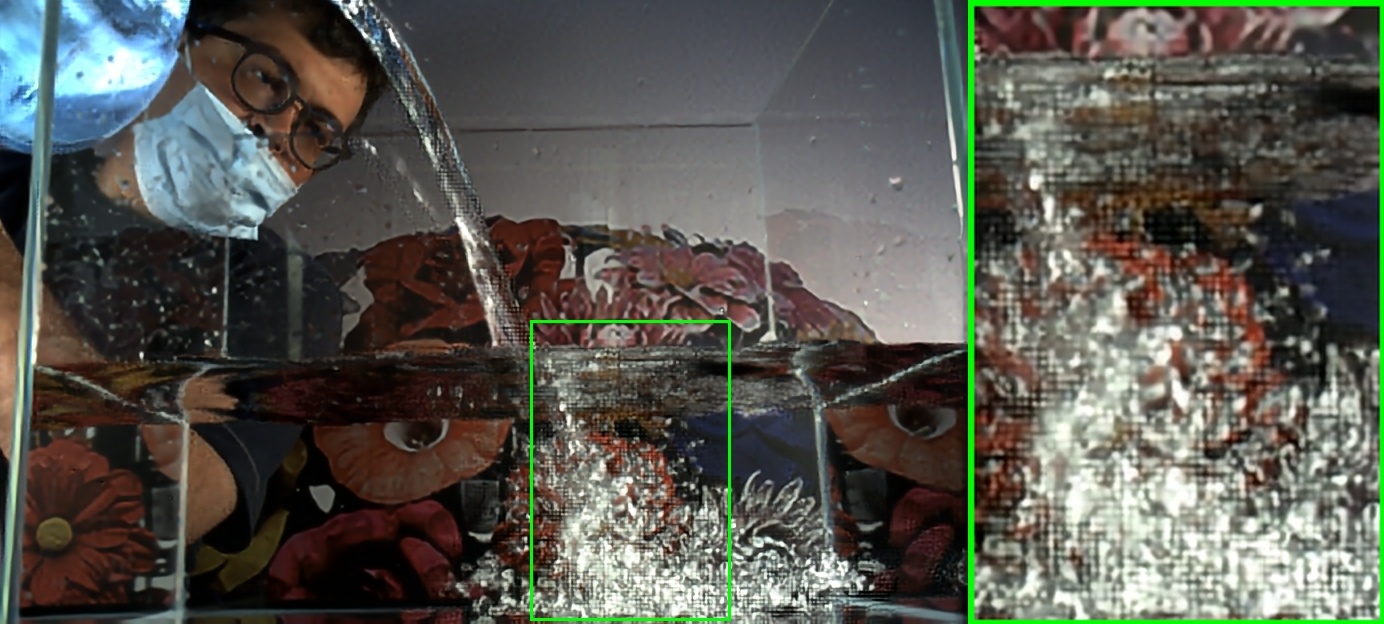}&
                \includegraphics[width=0.23\linewidth]{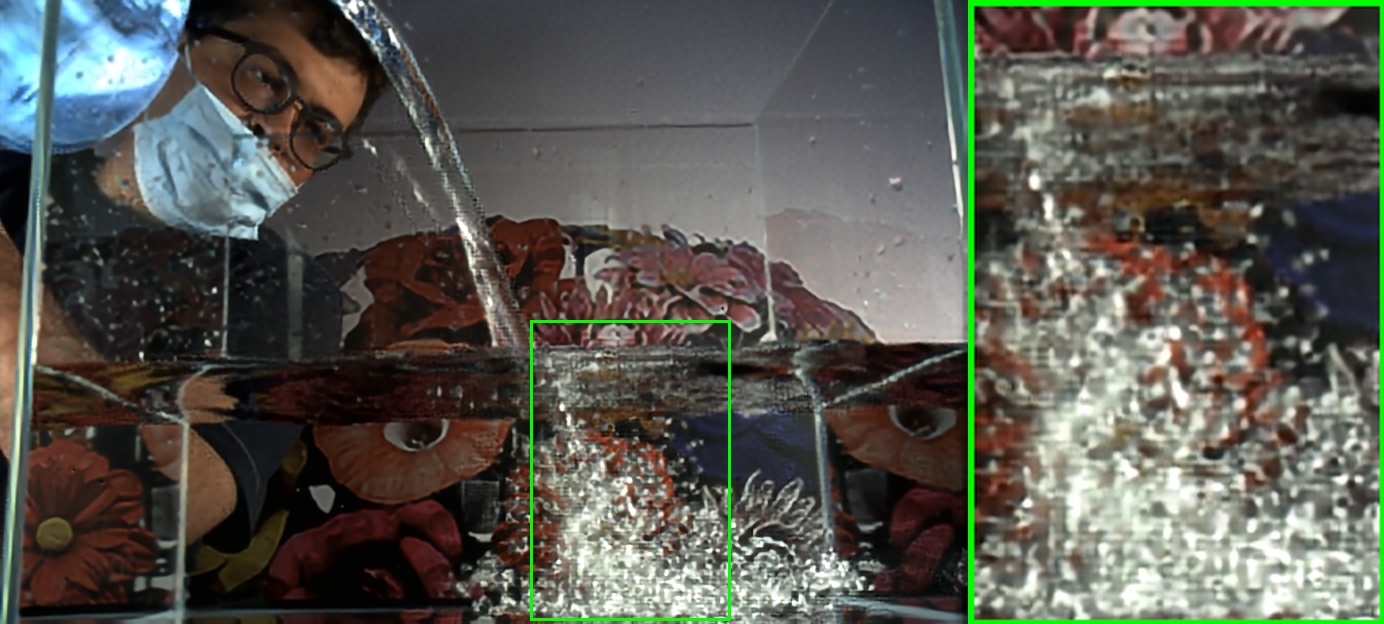}
            \\ \\
            VideoINR~\cite{chen2022videoinr}&
            MoTIF~\cite{chen2023motif}&
            \textbf{EvEnhancer (Ours)}&
            \textbf{EvEnhancer-light (Ours)}
            \end{tabular}
        \end{tabular}
    \caption{
       Qualitative comparison for In-dist. scale (\(t=8,s=4\)) on the BS-ERGB dataset \cite{tulyakov2022time}. Best zoom in for better visualization.
    }
    \label{fig:s_event2}
\end{figure*}
\renewcommand{\arraystretch}{1.}

%% file: fig/exp/exp_suppl_adobe240.tex
\setlength{\tabcolsep}{0.1pt}
\renewcommand{\arraystretch}{0.1}

\begin{figure*}[t]
    \centering\
         \begin{tabular}{cc}
            \rotatebox[origin=c]{90}{\makecell{\(\mathcal{T}=0\)\\}}  &
            \begin{tabular}{c@{\hskip 0.005\linewidth}c@{\hskip 0.005\linewidth}c@{\hskip 0.005\linewidth}c}
                \includegraphics[width=0.23\linewidth]{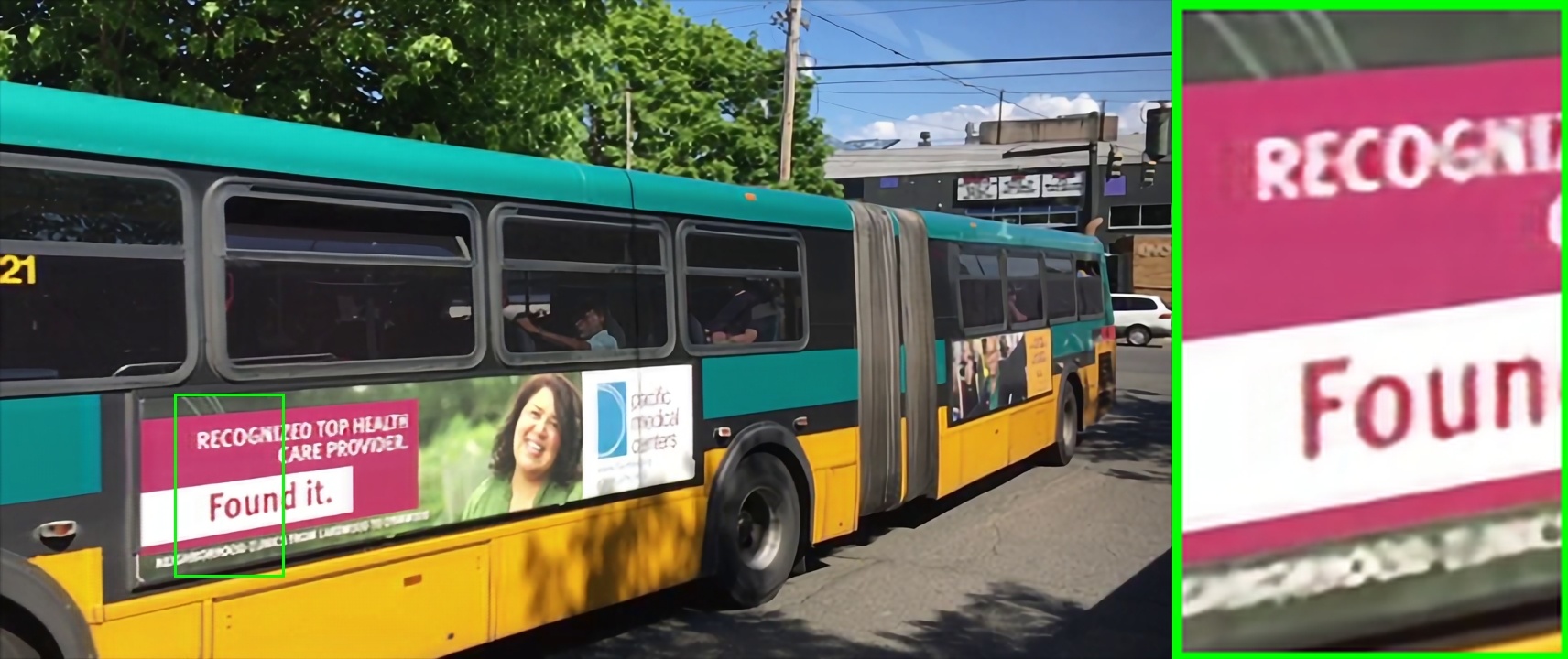}&
                \includegraphics[width=0.23\linewidth]{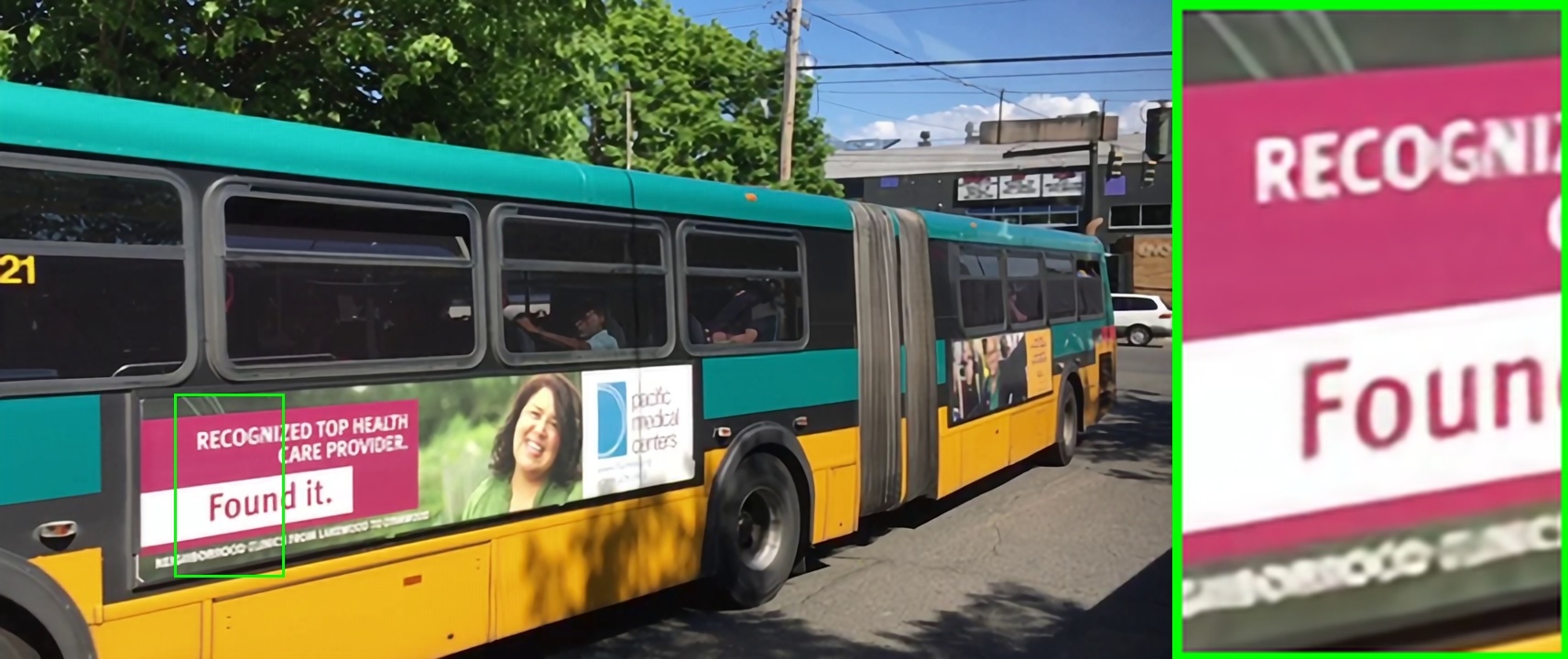}&
                \includegraphics[width=0.23\linewidth]{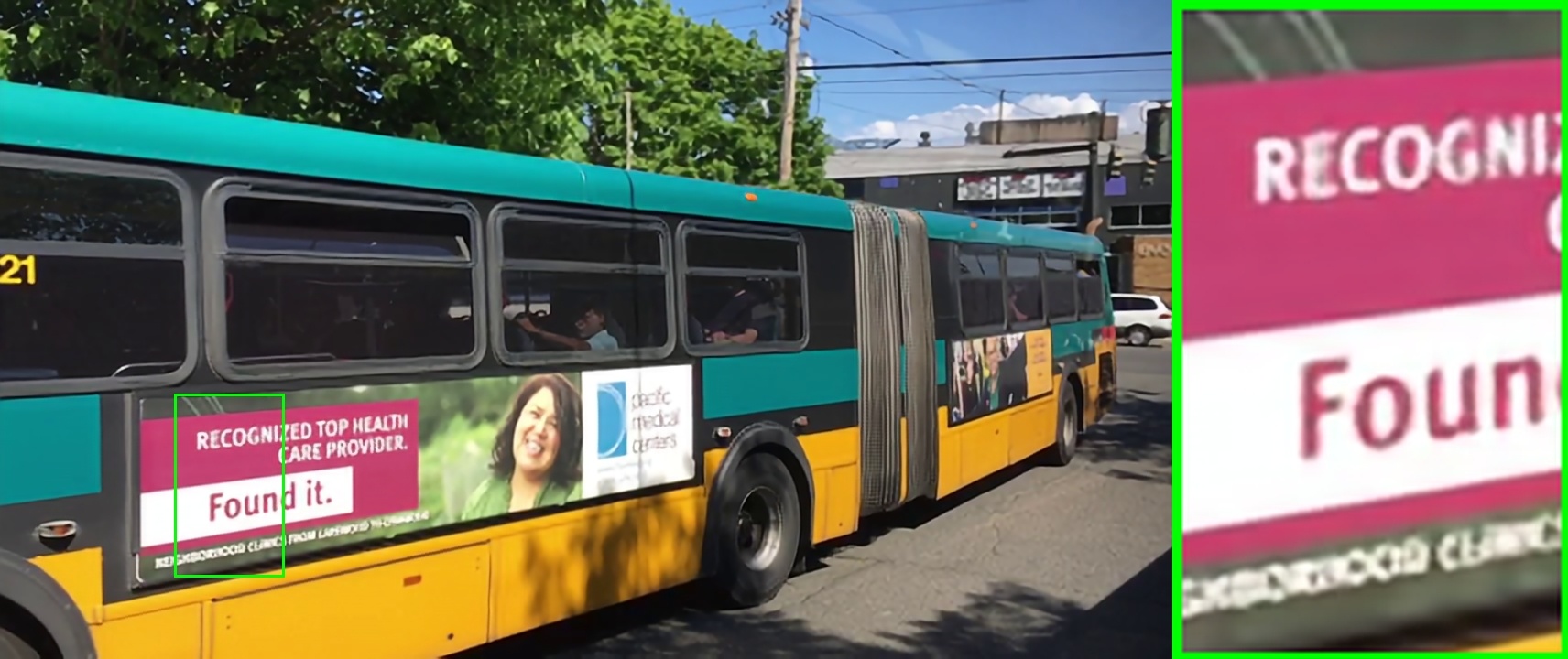}&
                \includegraphics[width=0.23\linewidth]{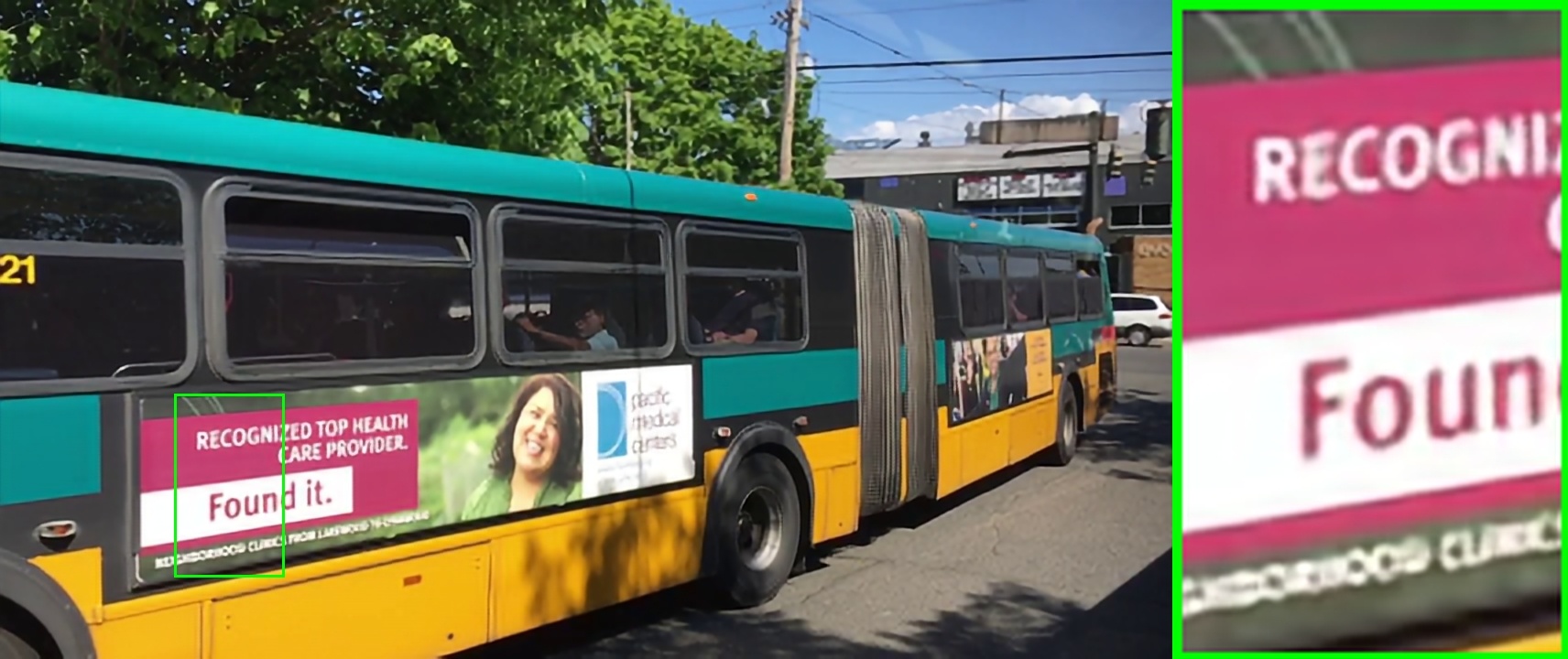}
            \end{tabular}
            \\ \\ \\
            \rotatebox[origin=c]{90}{\makecell{\(\mathcal{T}=0.125\)\\}}  &
            \begin{tabular}{c@{\hskip 0.005\linewidth}c@{\hskip 0.005\linewidth}c@{\hskip 0.005\linewidth}c}
                \includegraphics[width=0.23\linewidth]{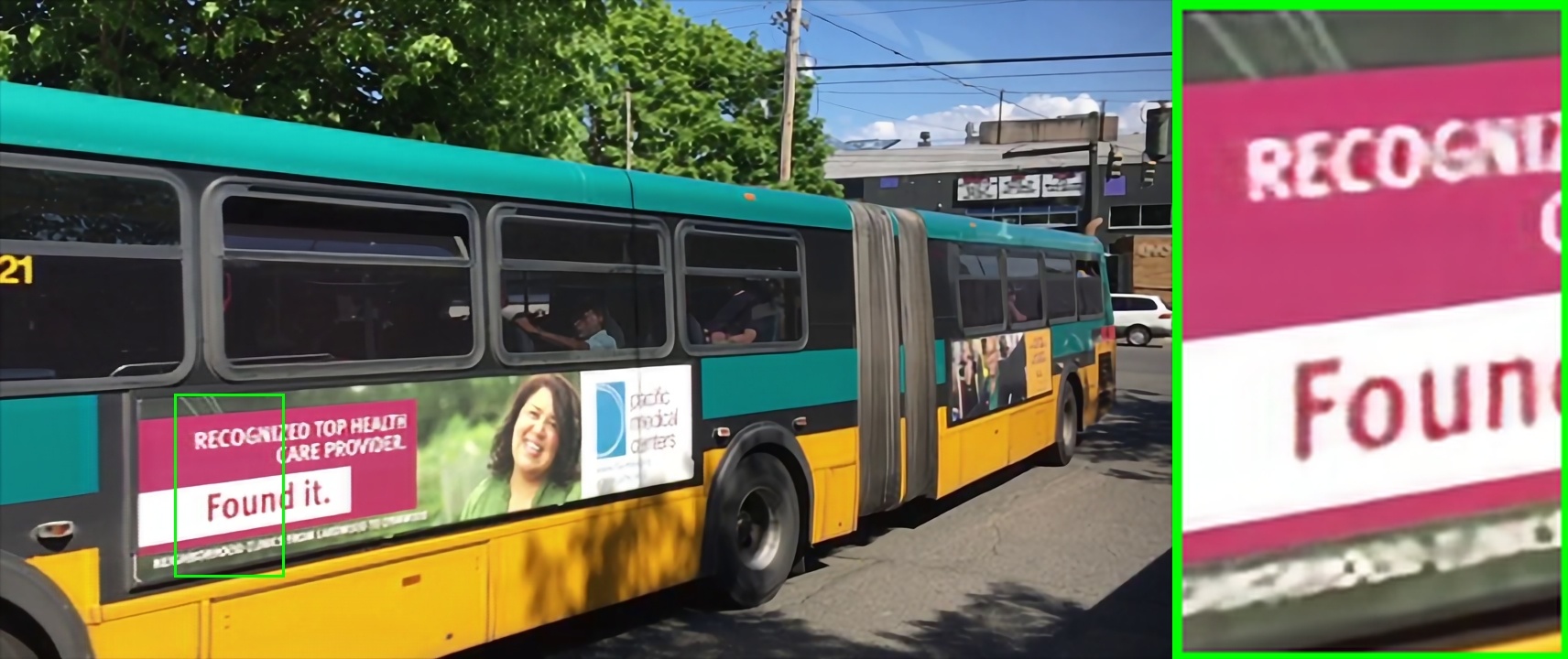}&
                \includegraphics[width=0.23\linewidth]{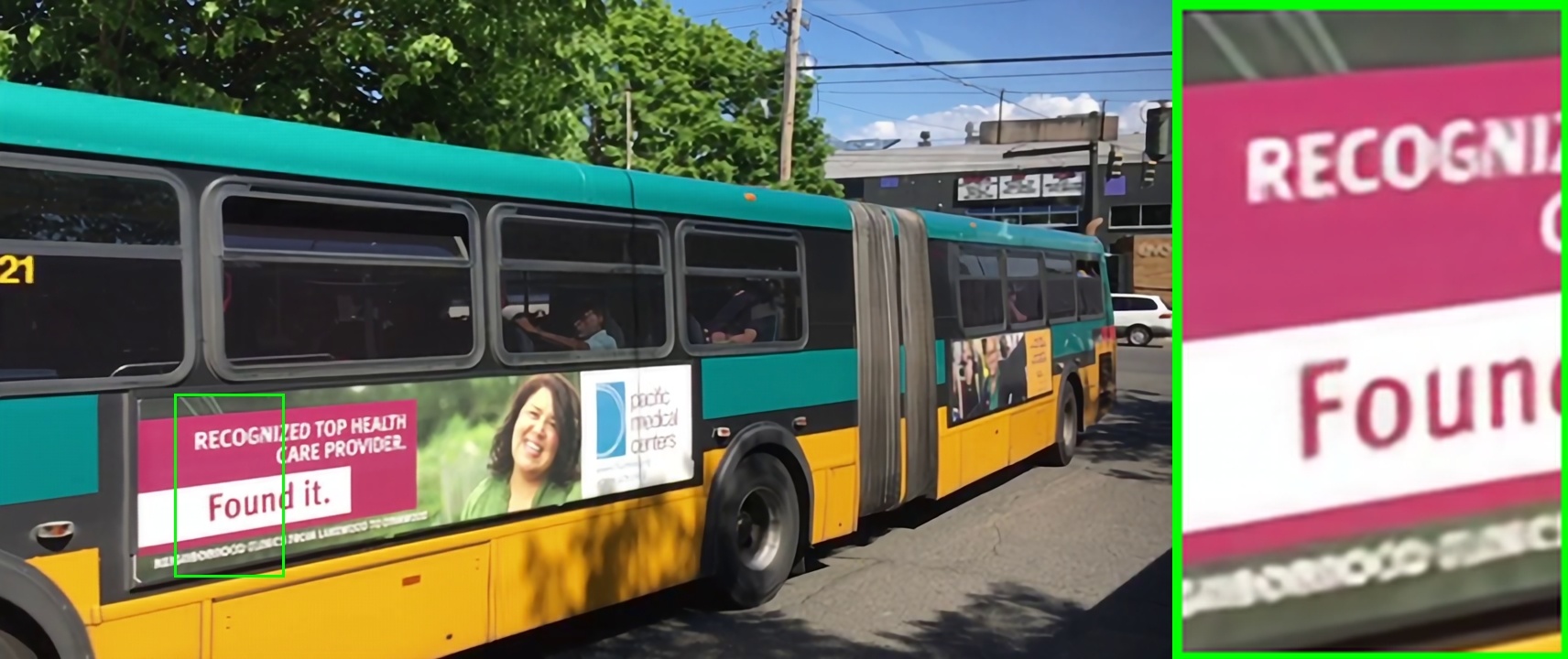}&
                \includegraphics[width=0.23\linewidth]{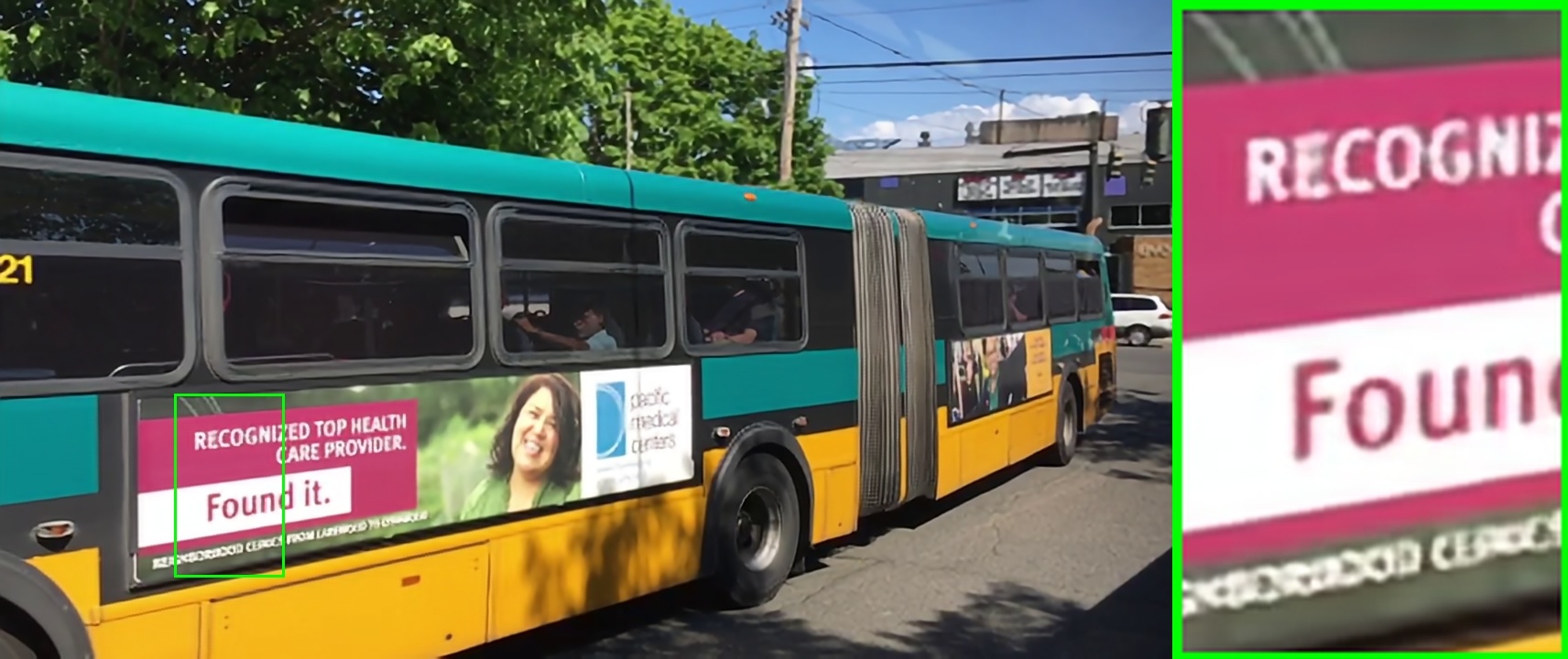}&
                \includegraphics[width=0.23\linewidth]{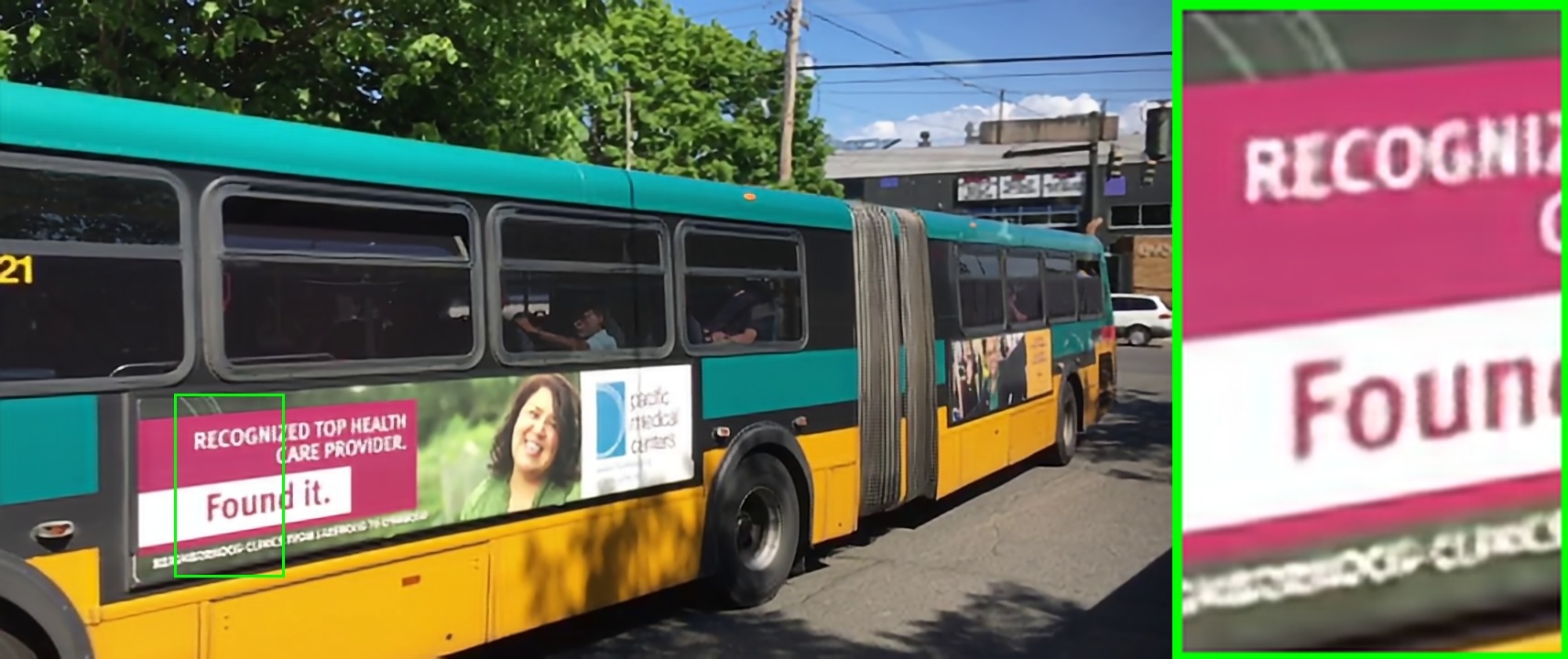}
            \end{tabular}
            \\ \\ \\
            \rotatebox[origin=c]{90}{\makecell{\(\mathcal{T}=0.25\)\\}}  &
            \begin{tabular}{c@{\hskip 0.005\linewidth}c@{\hskip 0.005\linewidth}c@{\hskip 0.005\linewidth}c}
                \includegraphics[width=0.23\linewidth]{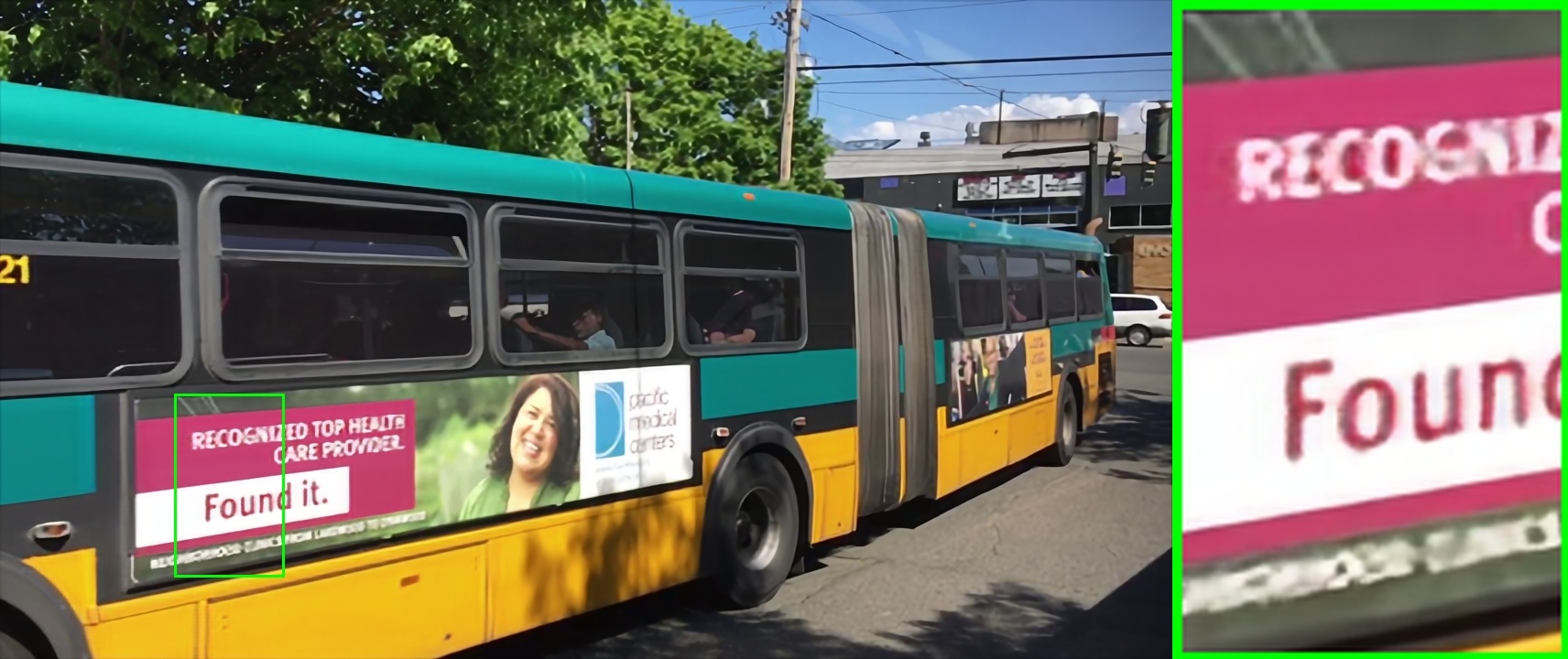}&
                \includegraphics[width=0.23\linewidth]{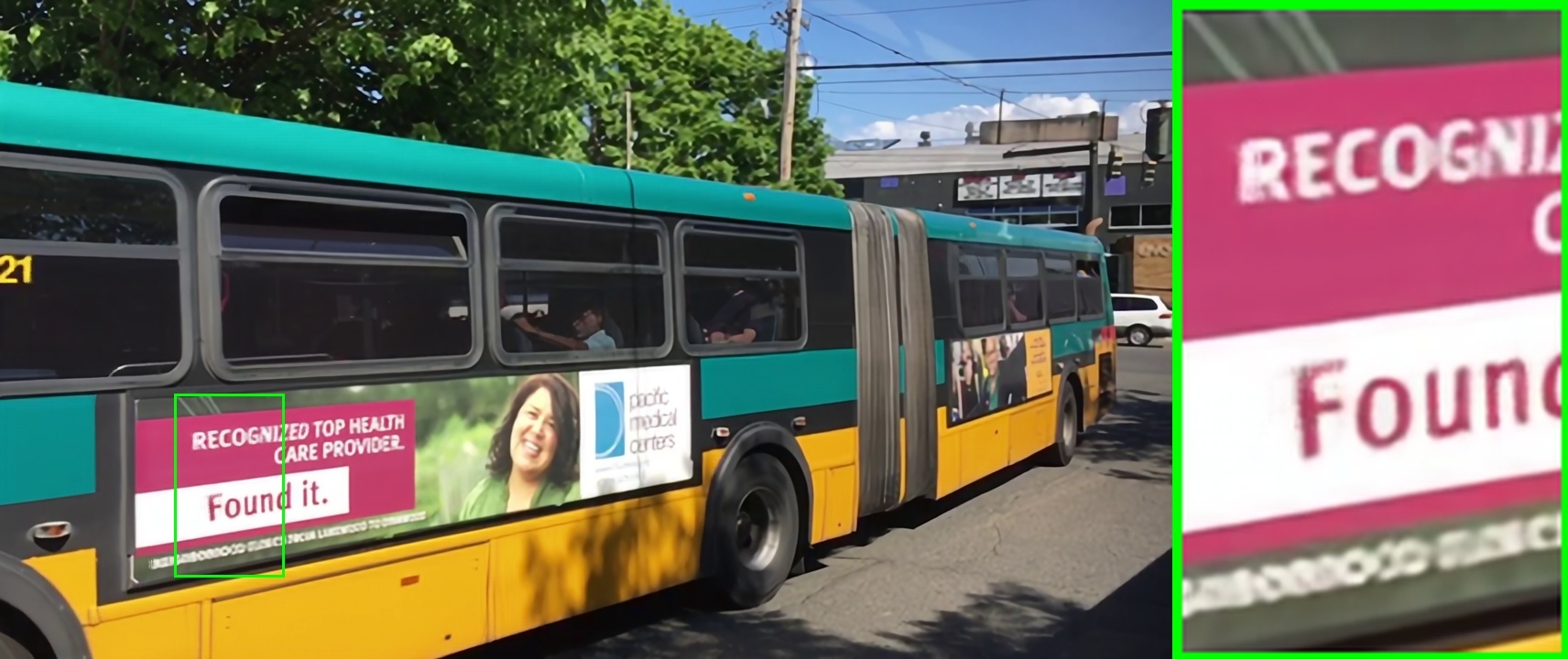}&
                \includegraphics[width=0.23\linewidth]{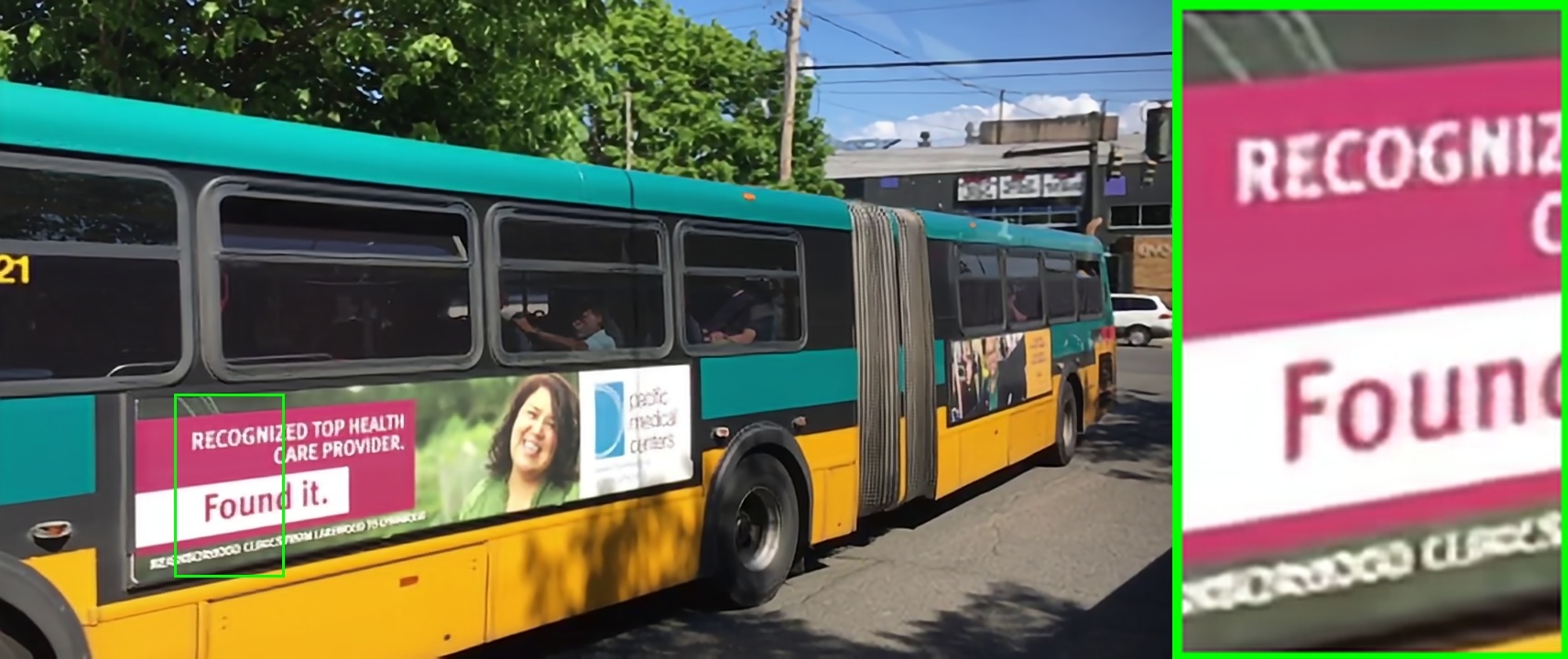}&
                \includegraphics[width=0.23\linewidth]{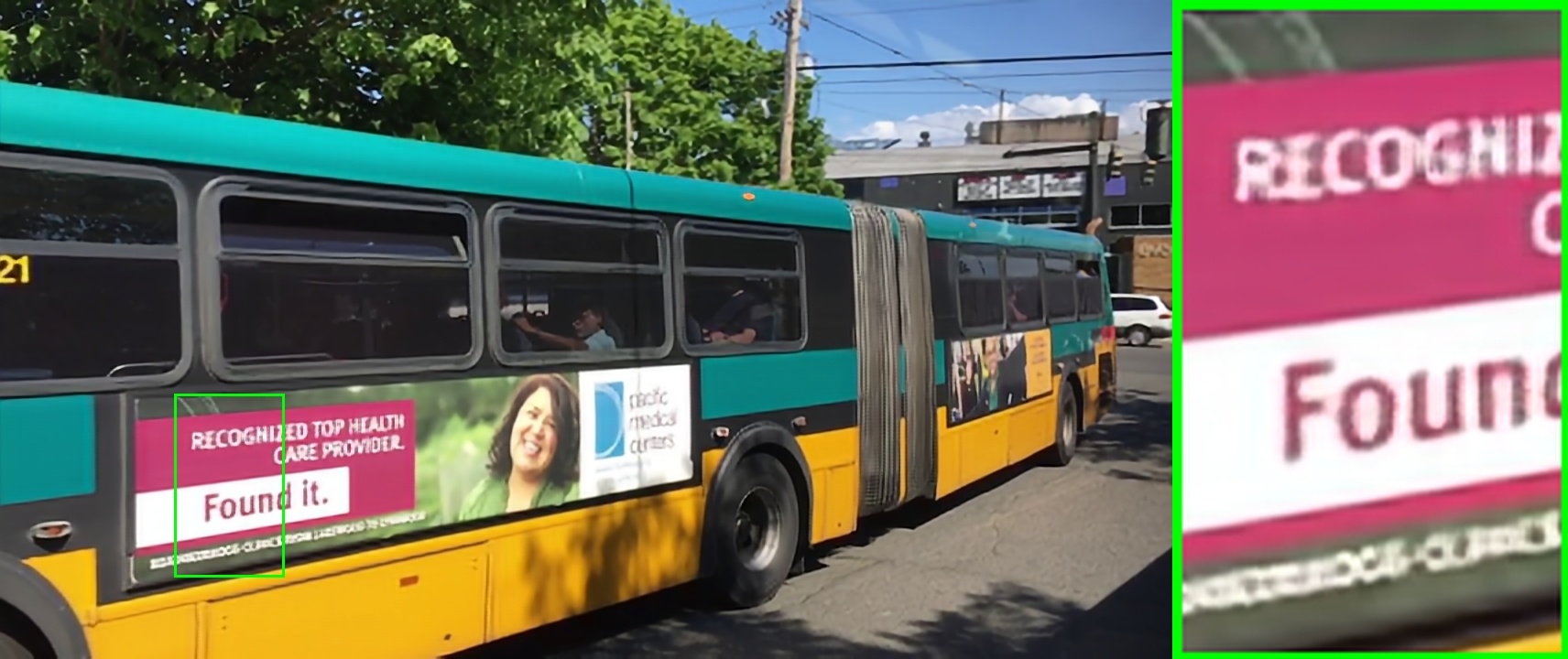}
            \end{tabular}
            \\ \\ \\
            \rotatebox[origin=c]{90}{\makecell{\(\mathcal{T}=0.375\)\\}}  &
            \begin{tabular}{c@{\hskip 0.005\linewidth}c@{\hskip 0.005\linewidth}c@{\hskip 0.005\linewidth}c}
                 \includegraphics[width=0.23\linewidth]{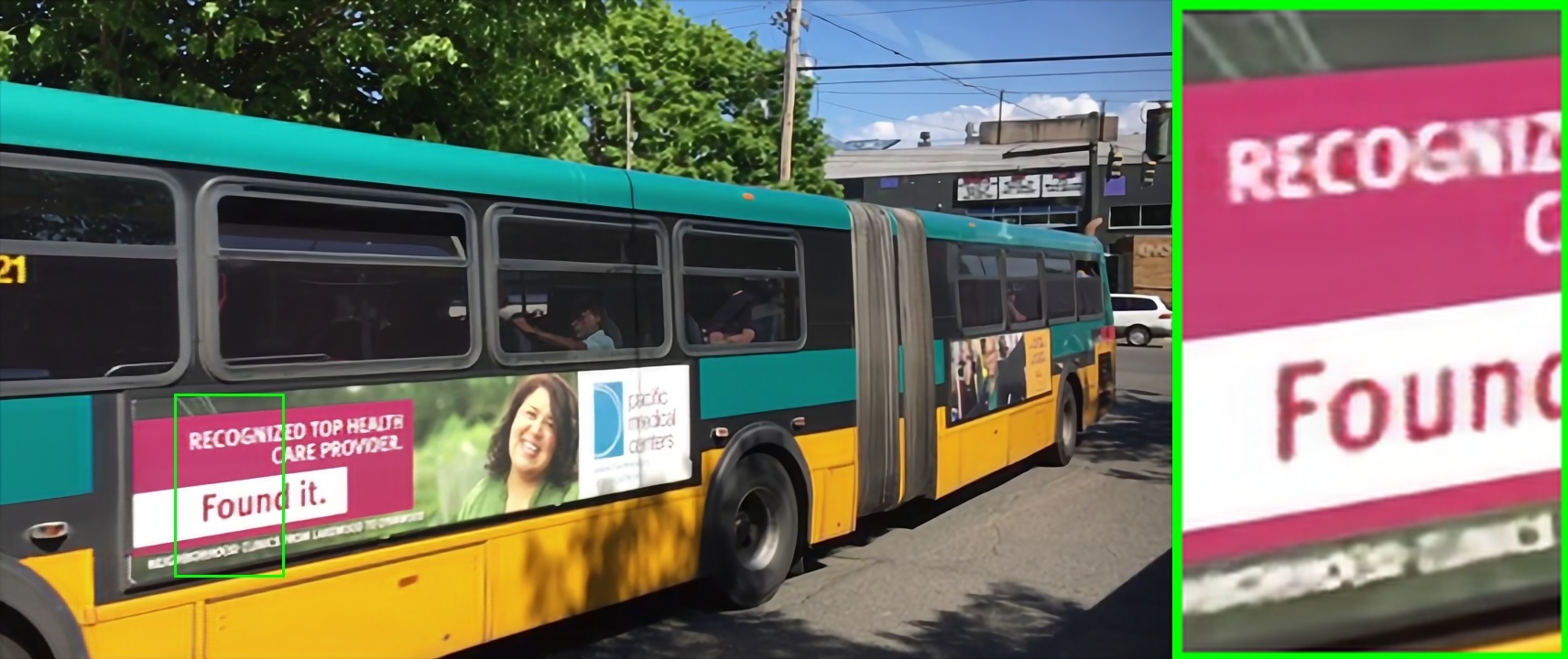}&
                \includegraphics[width=0.23\linewidth]{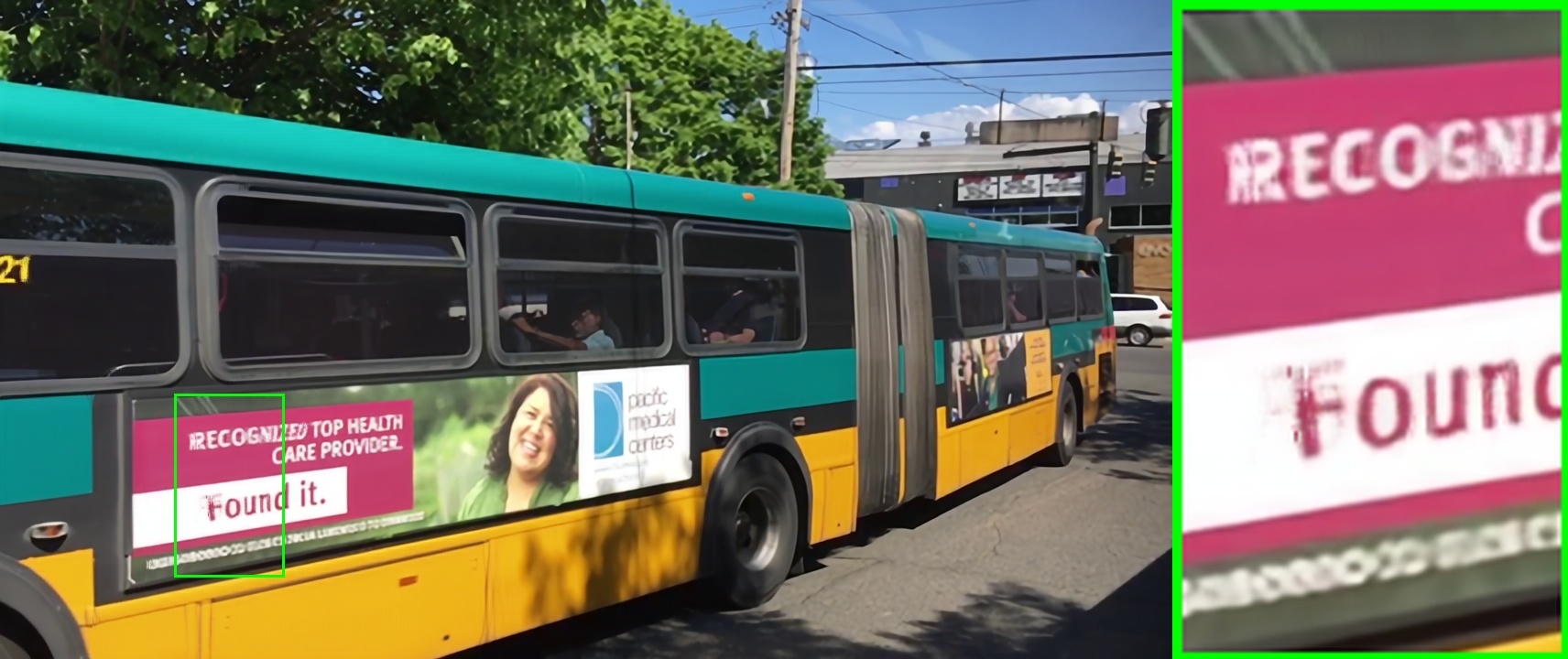}&
                \includegraphics[width=0.23\linewidth]{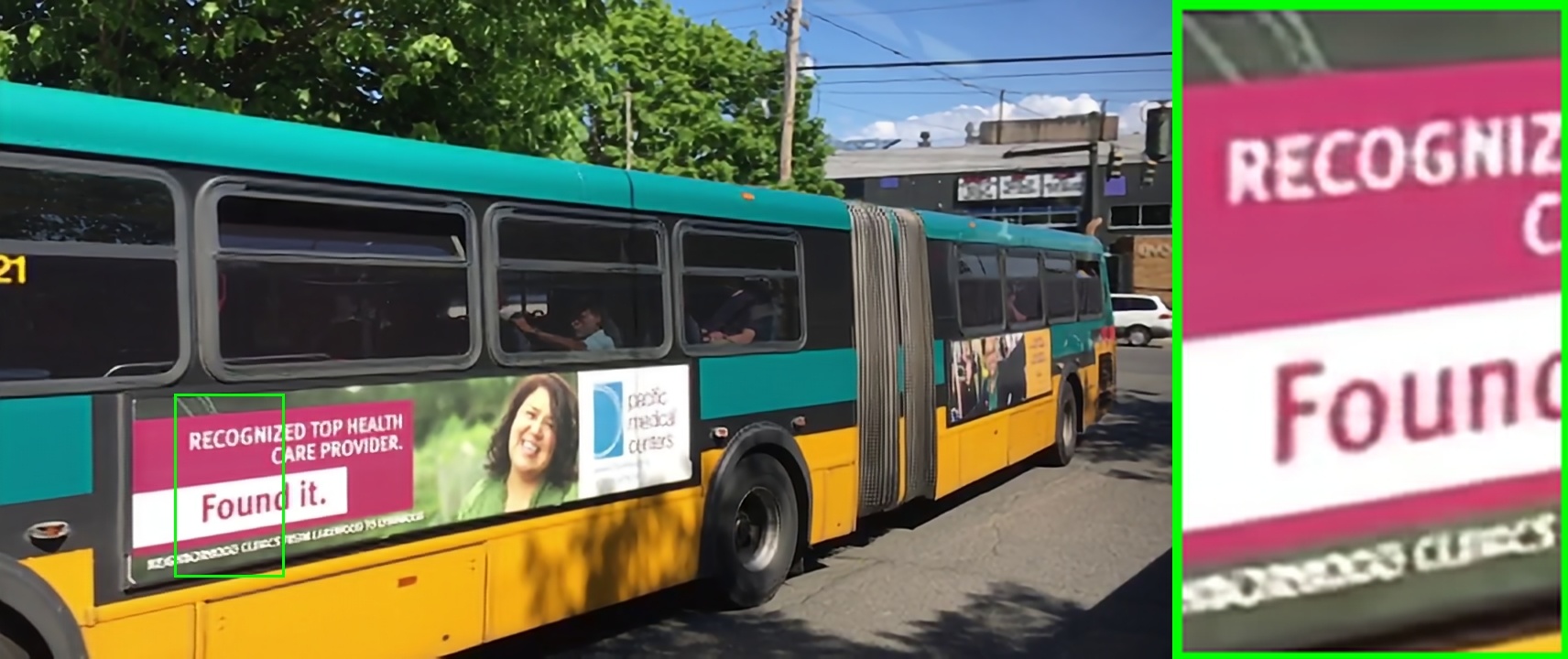}&
                \includegraphics[width=0.23\linewidth]{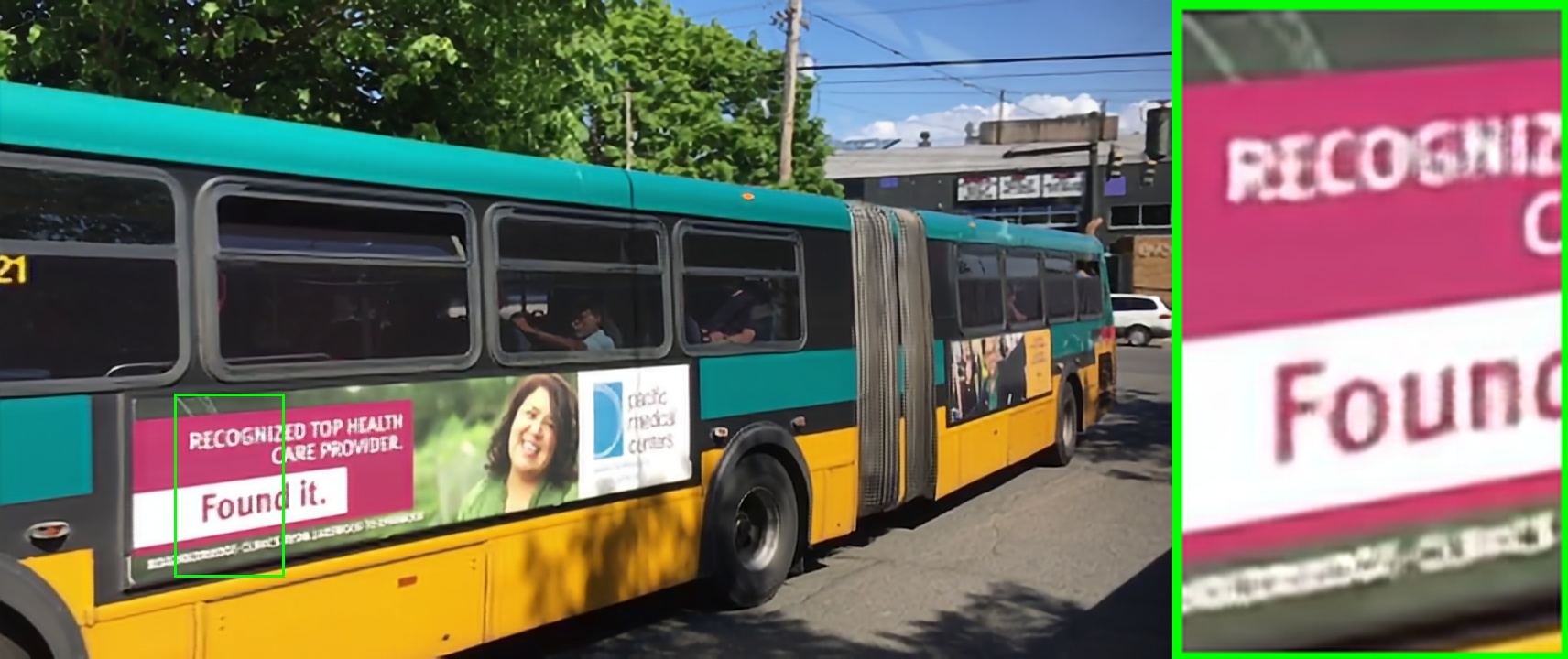}
            \end{tabular}
            \\ \\ \\
            \rotatebox[origin=c]{90}{\makecell{\(\mathcal{T}=0.5\)\\}}  &
            \begin{tabular}{c@{\hskip 0.005\linewidth}c@{\hskip 0.005\linewidth}c@{\hskip 0.005\linewidth}c}
                \includegraphics[width=0.23\linewidth]{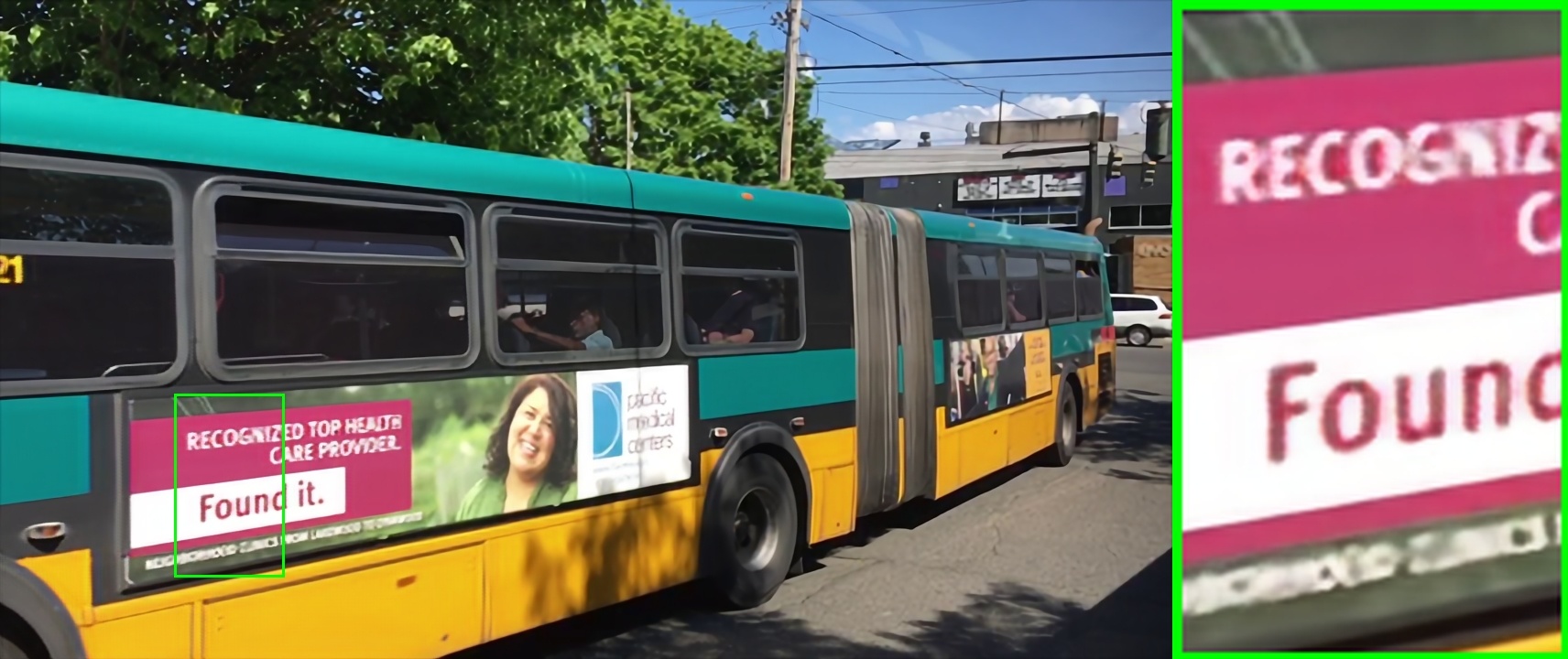}&
                \includegraphics[width=0.23\linewidth]{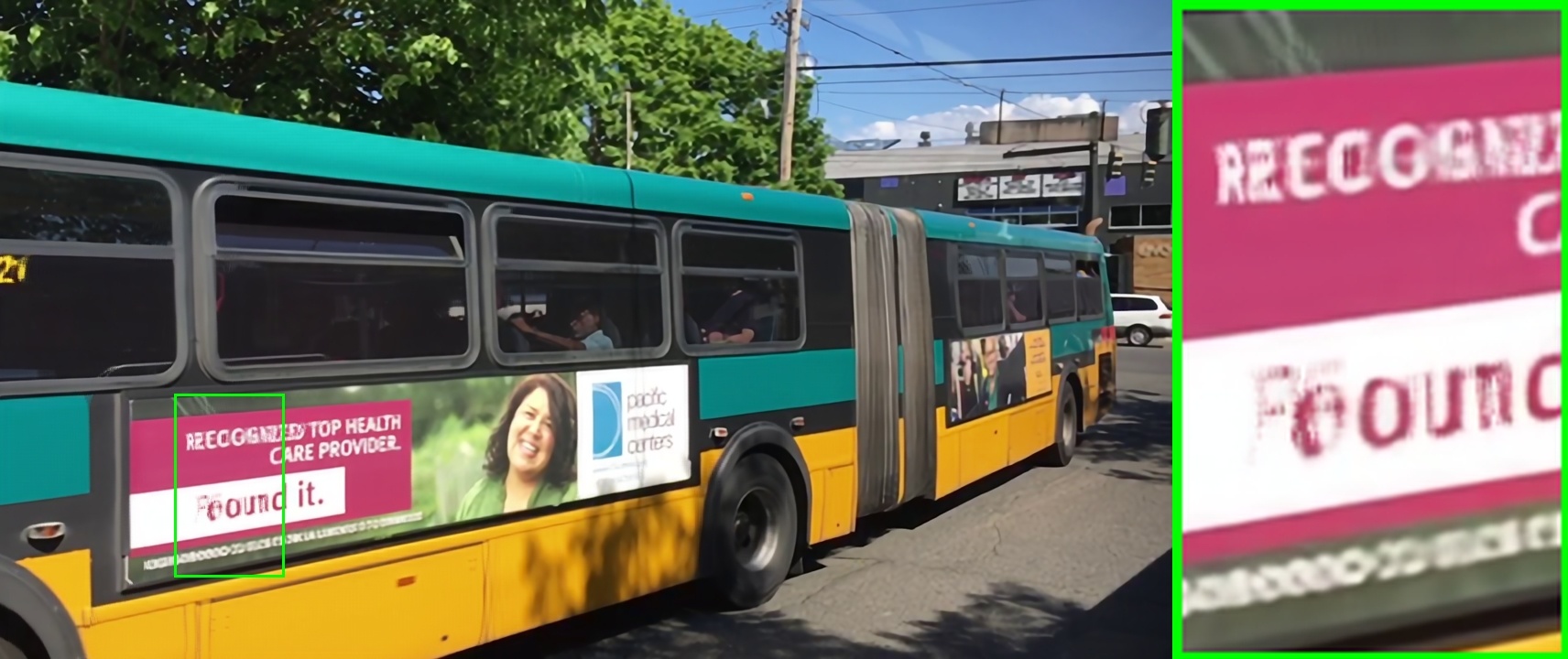}&
                \includegraphics[width=0.23\linewidth]{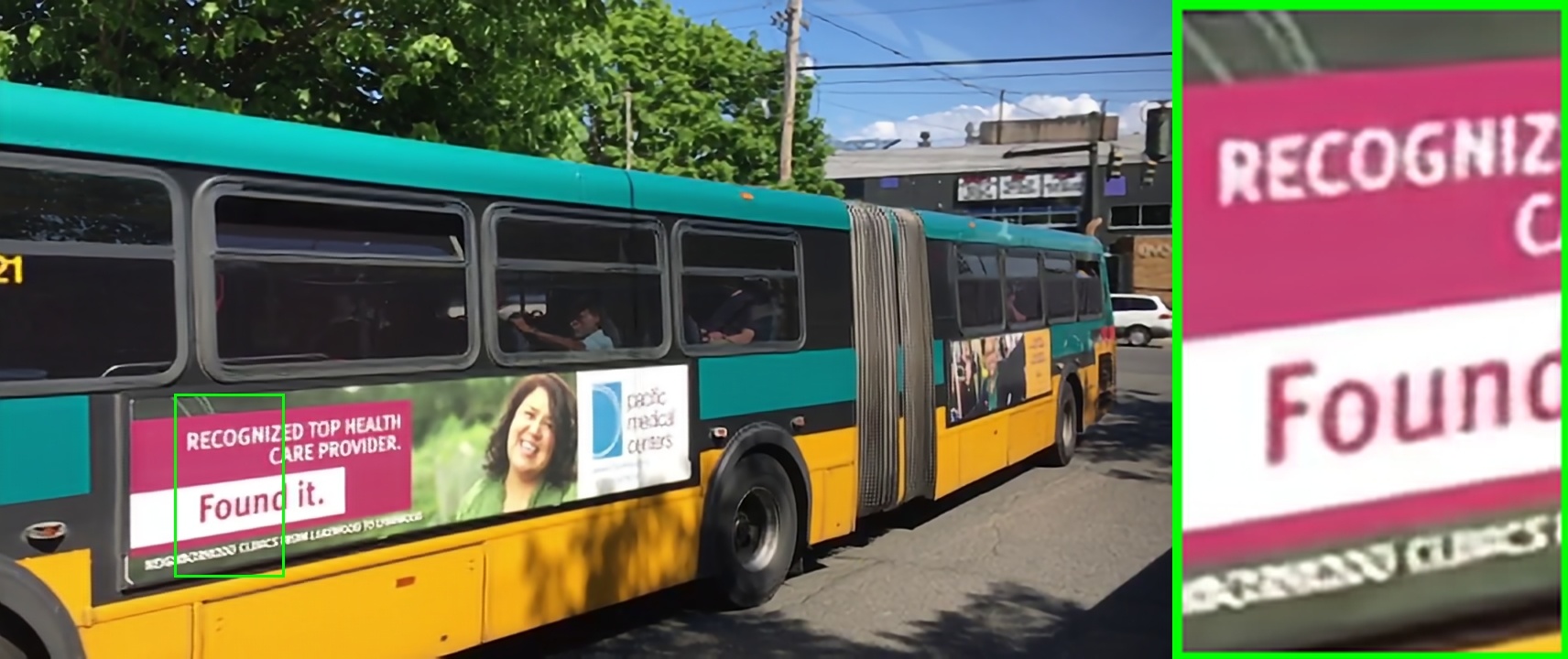}&
                \includegraphics[width=0.23\linewidth]{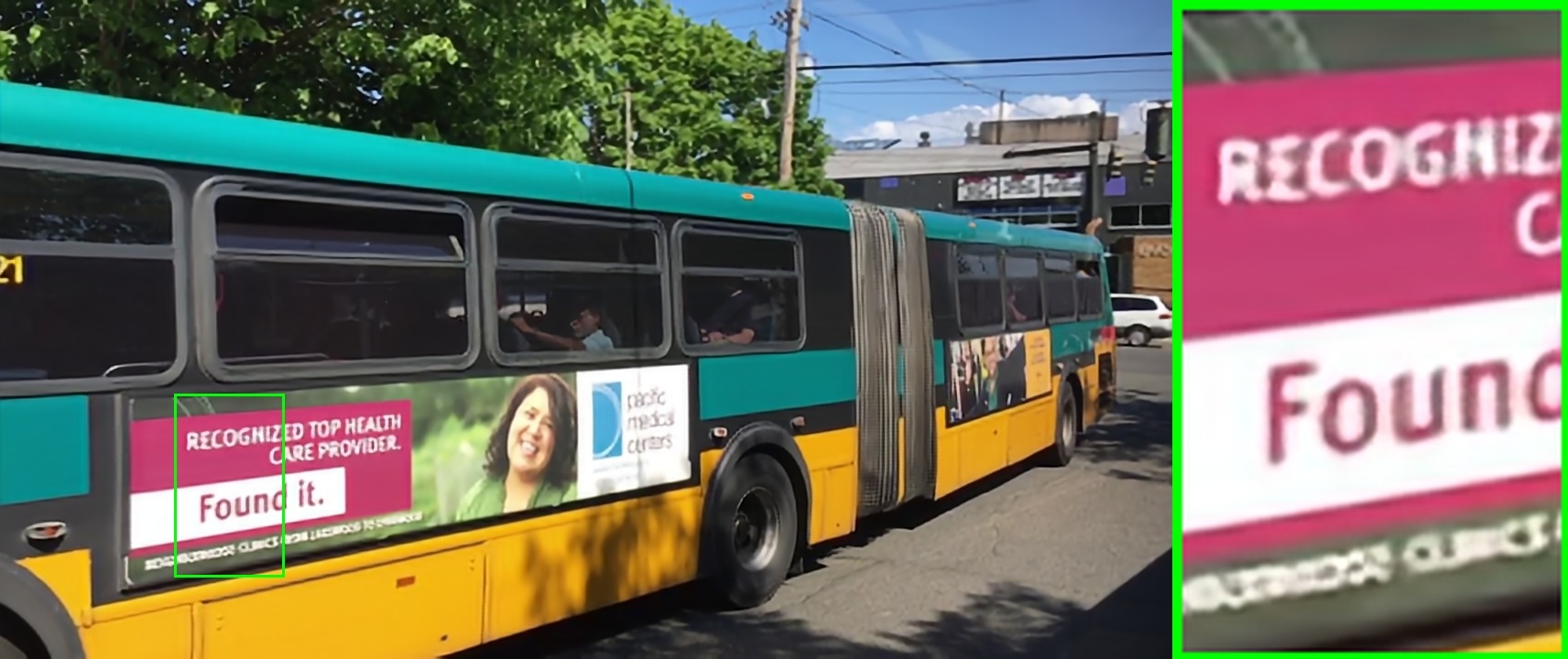}
            \end{tabular}
            \\ \\ \\
            \rotatebox[origin=c]{90}{\makecell{\(\mathcal{T}=0.625\)\\}}  &
            \begin{tabular}{c@{\hskip 0.005\linewidth}c@{\hskip 0.005\linewidth}c@{\hskip 0.005\linewidth}c}
                \includegraphics[width=0.23\linewidth]{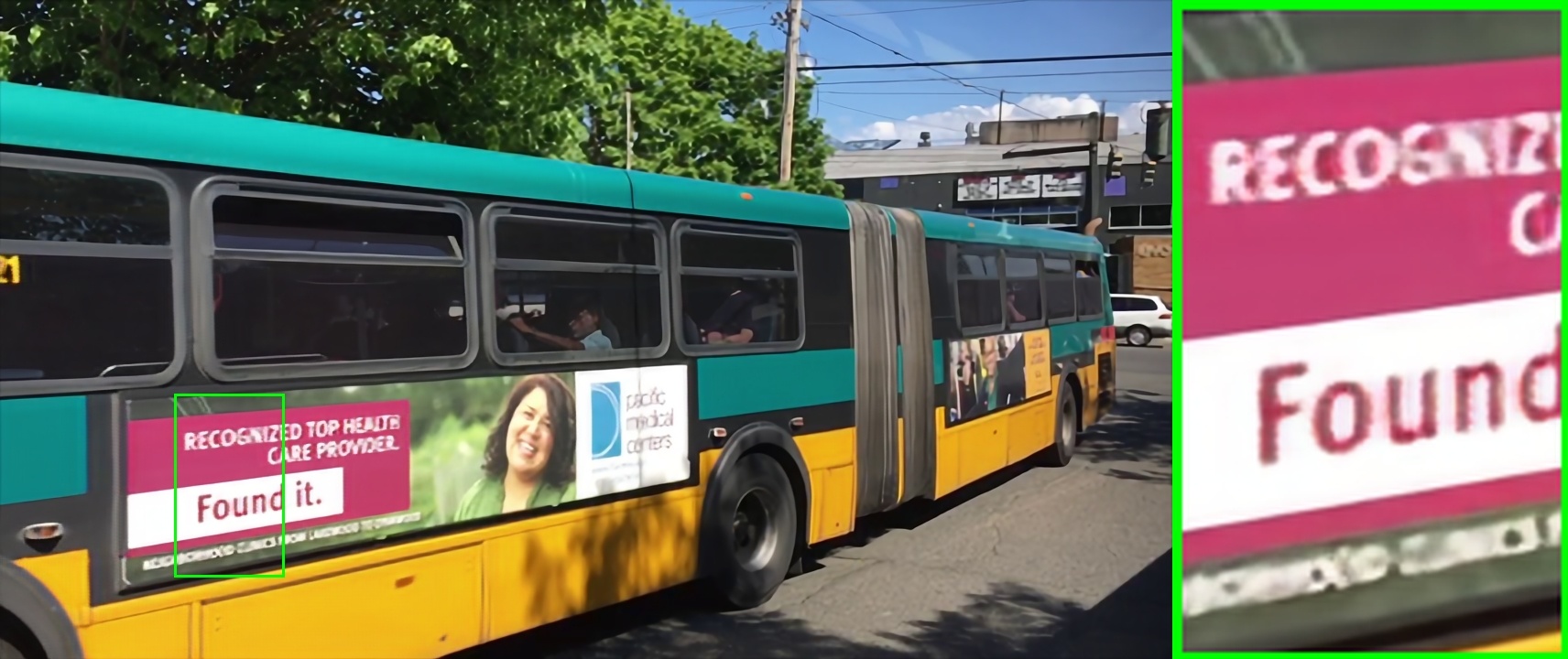}&
                \includegraphics[width=0.23\linewidth]{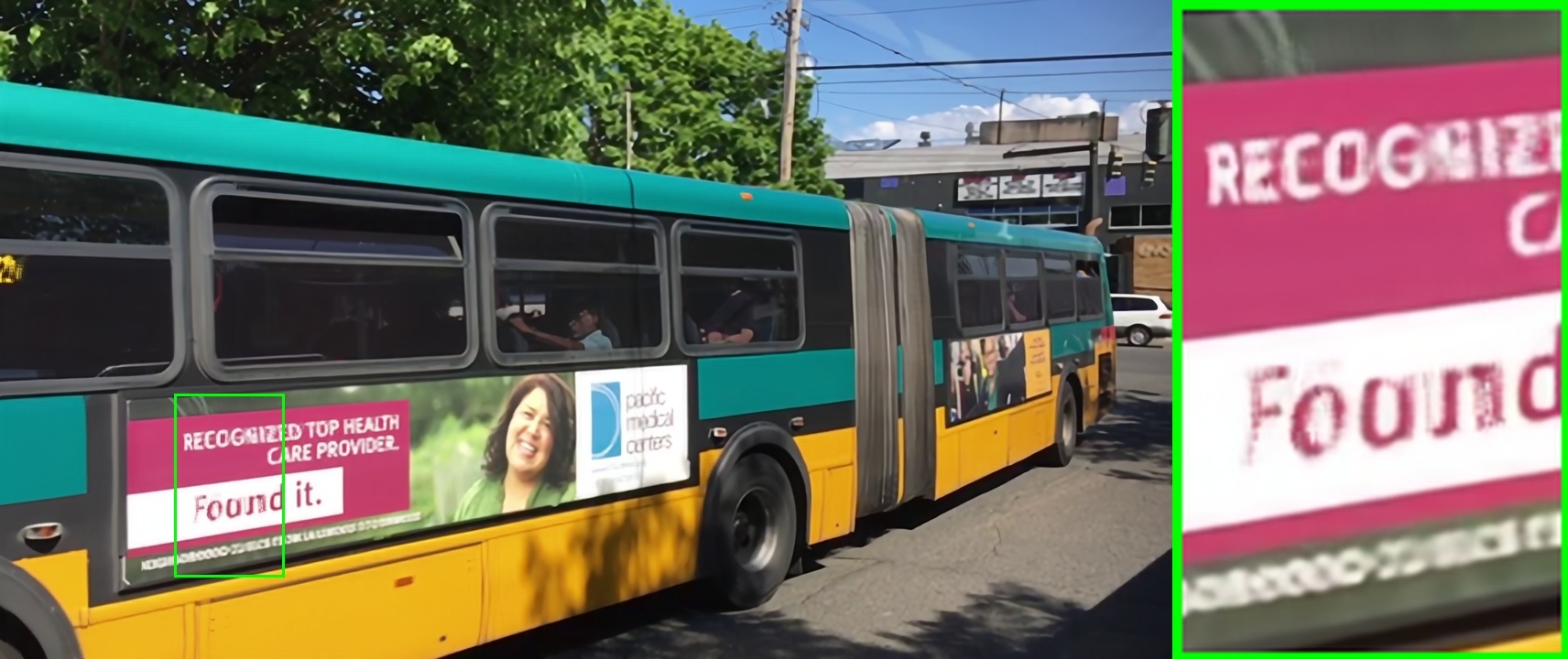}&
                \includegraphics[width=0.23\linewidth]{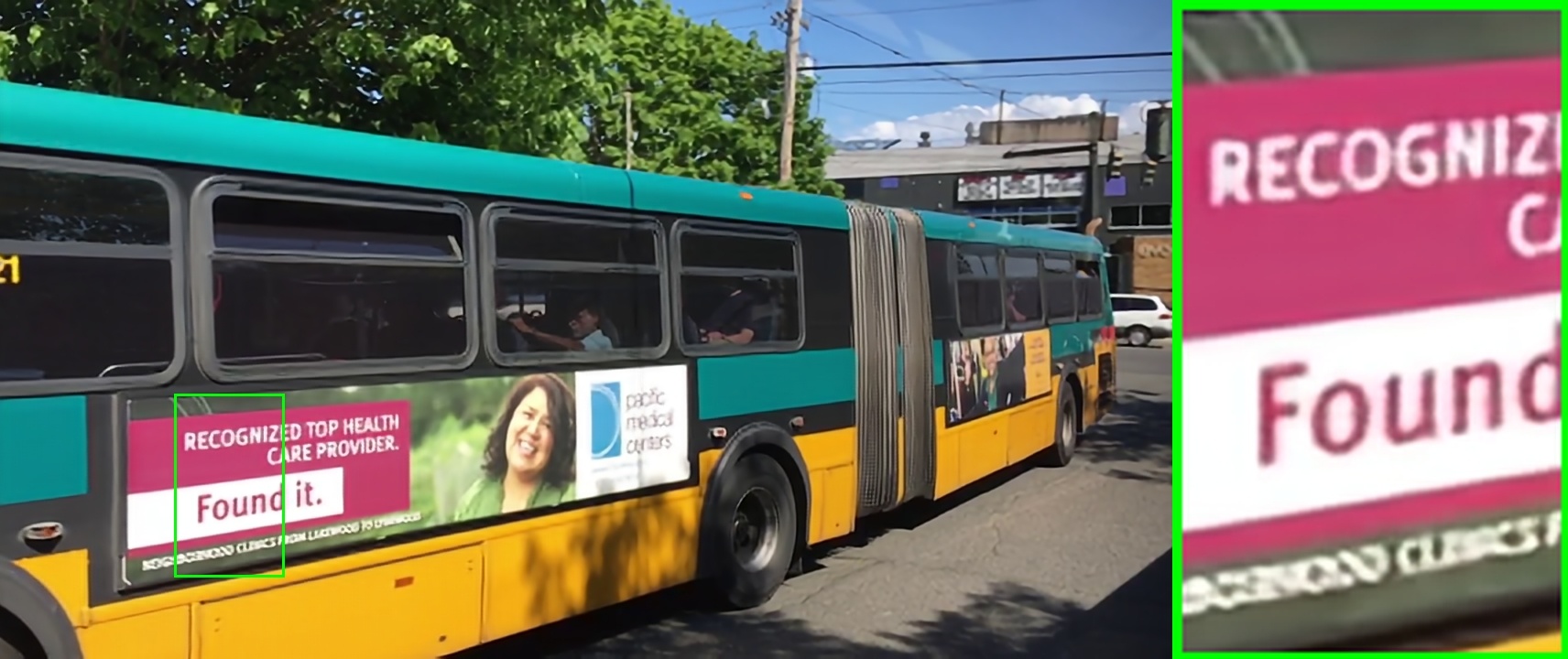}&
                \includegraphics[width=0.23\linewidth]{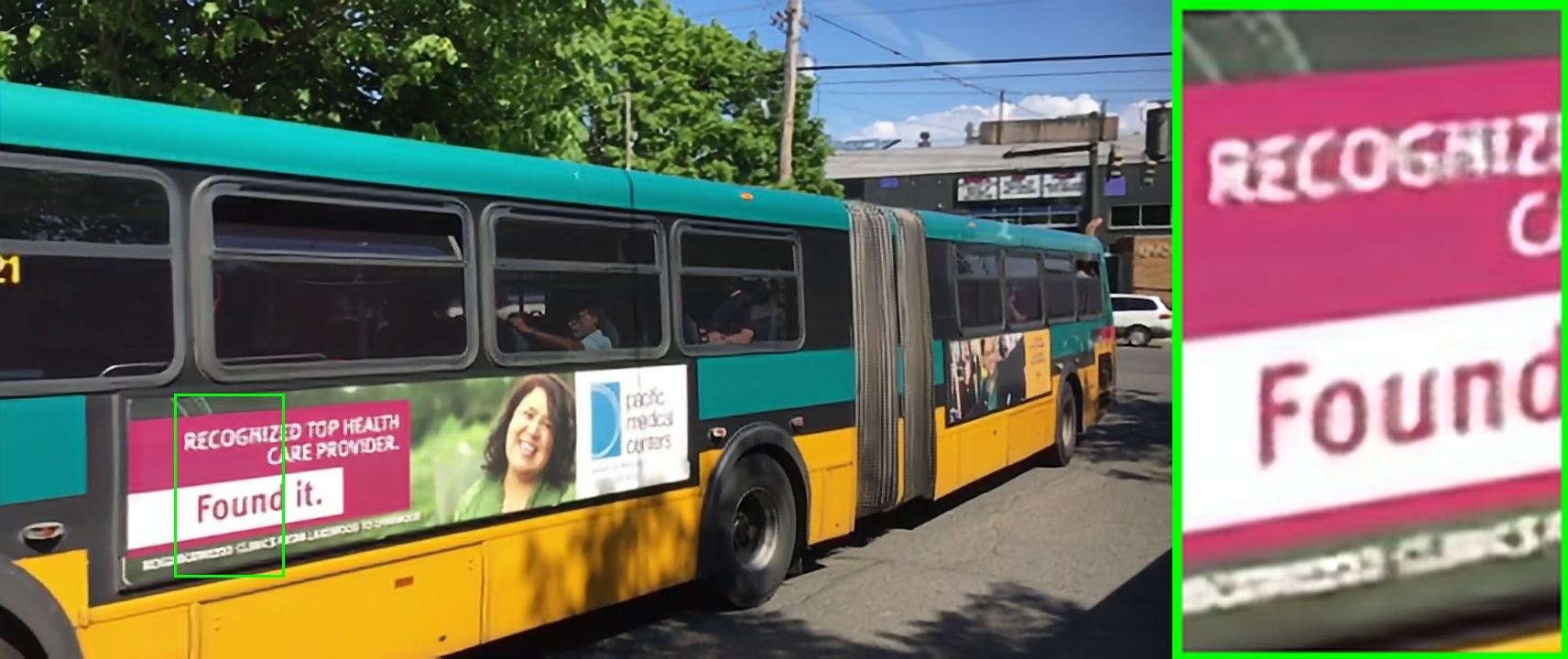}
            \end{tabular}
            \\ \\ \\
            \rotatebox[origin=c]{90}{\makecell{\(\mathcal{T}=0.75\)\\}}  &
            \begin{tabular}{c@{\hskip 0.005\linewidth}c@{\hskip 0.005\linewidth}c@{\hskip 0.005\linewidth}c}
                \includegraphics[width=0.23\linewidth]{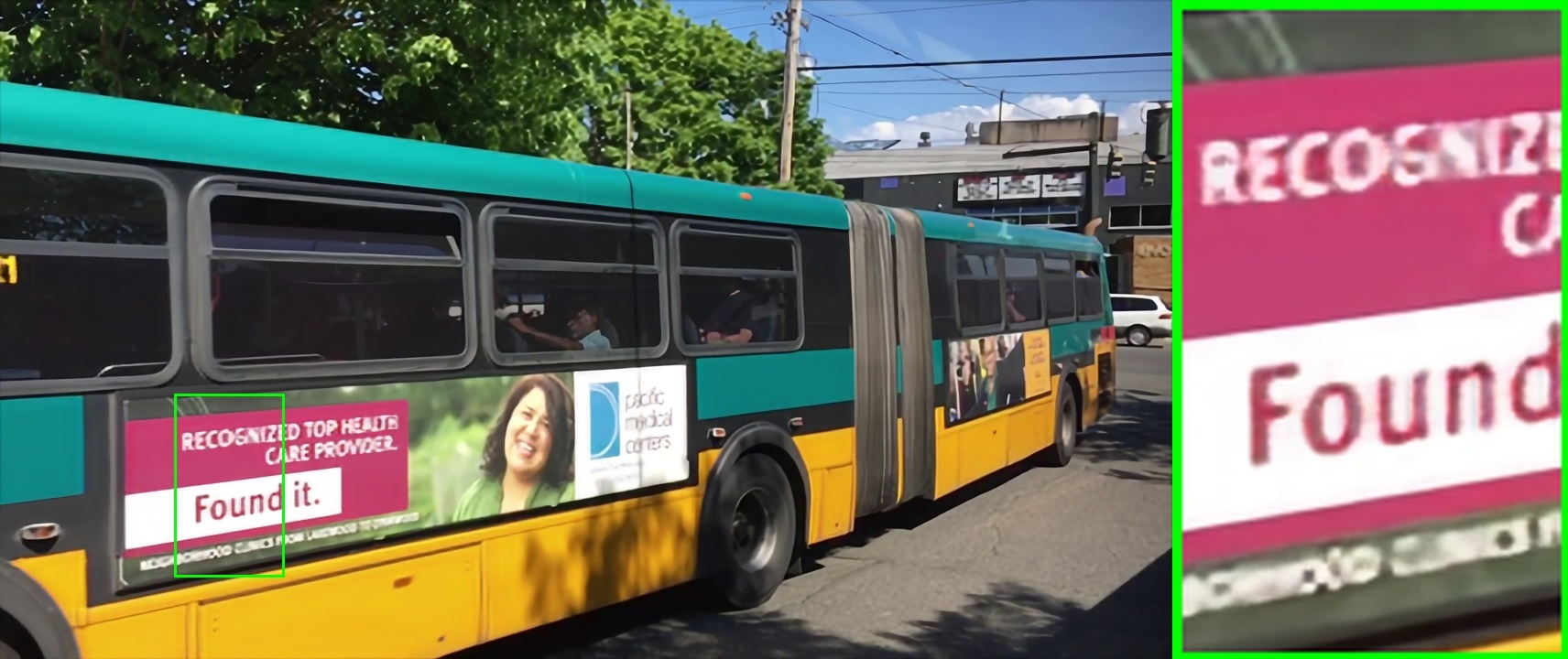}&
                \includegraphics[width=0.23\linewidth]{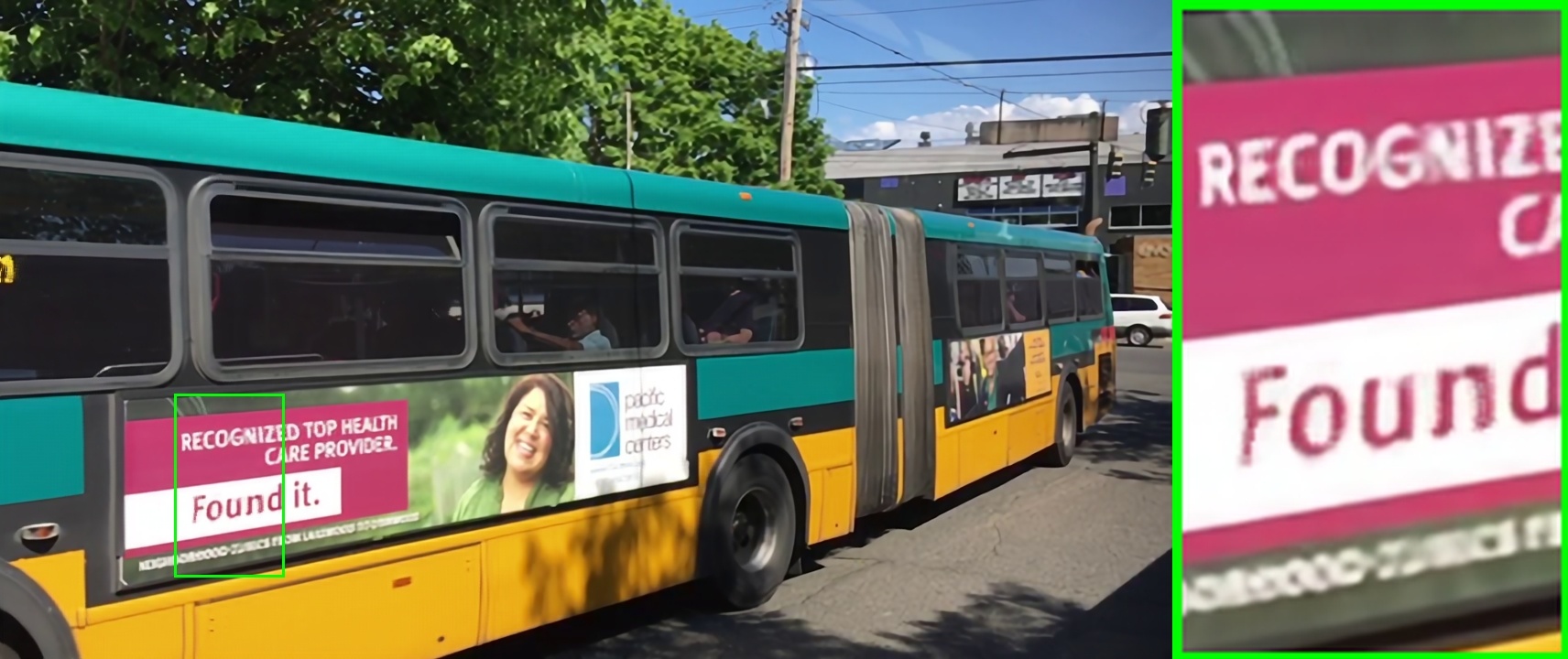}&
                \includegraphics[width=0.23\linewidth]{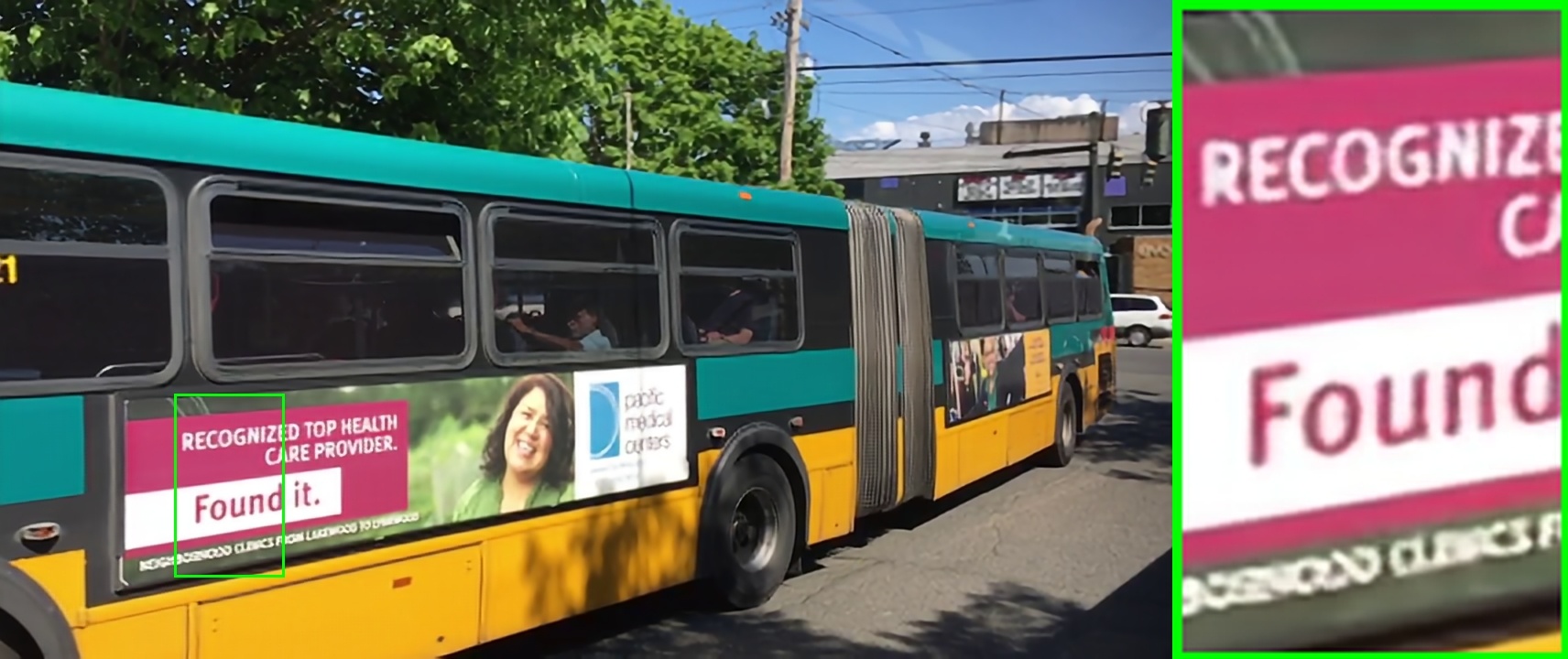}&
                \includegraphics[width=0.23\linewidth]{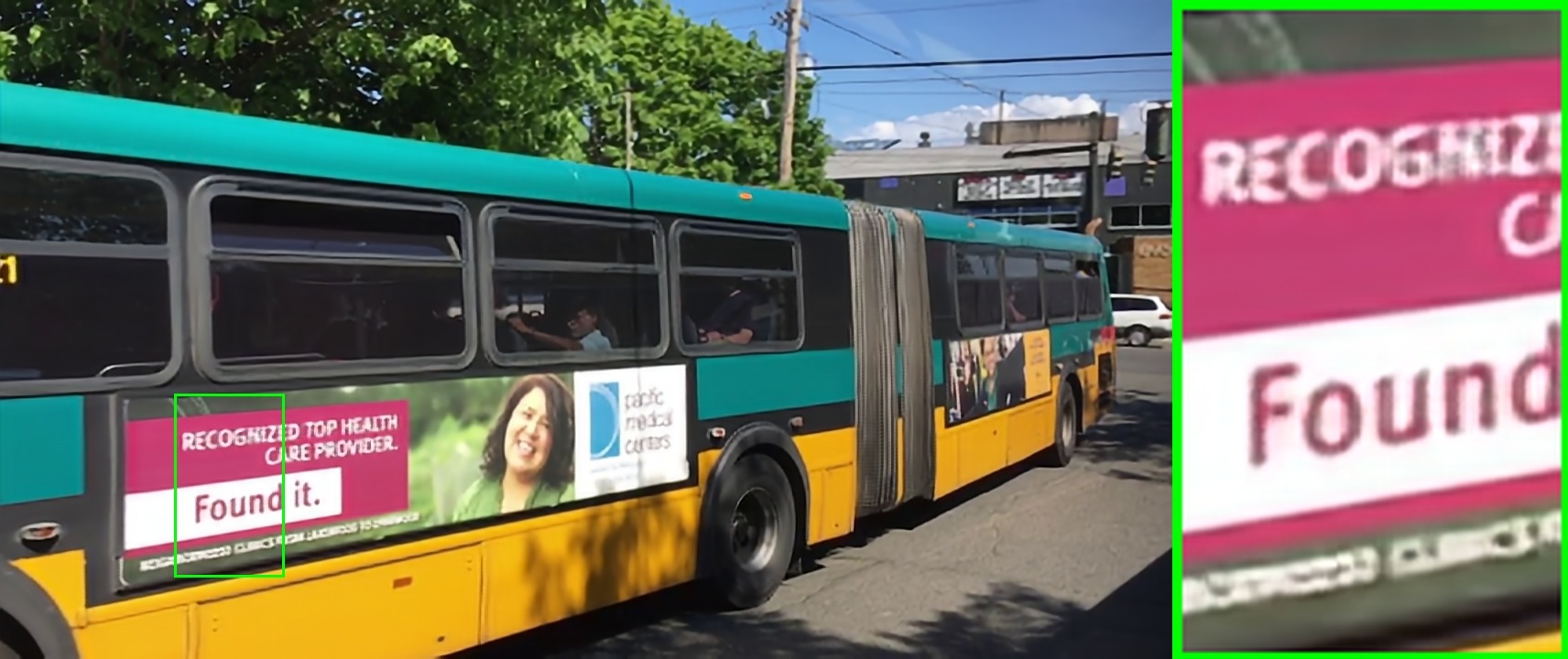}
            \end{tabular}
            \\ \\ \\
            \rotatebox[origin=c]{90}{\makecell{\(\mathcal{T}=0.875\)\\}}  &
            \begin{tabular}{c@{\hskip 0.005\linewidth}c@{\hskip 0.005\linewidth}c@{\hskip 0.005\linewidth}c}
                \includegraphics[width=0.23\linewidth]{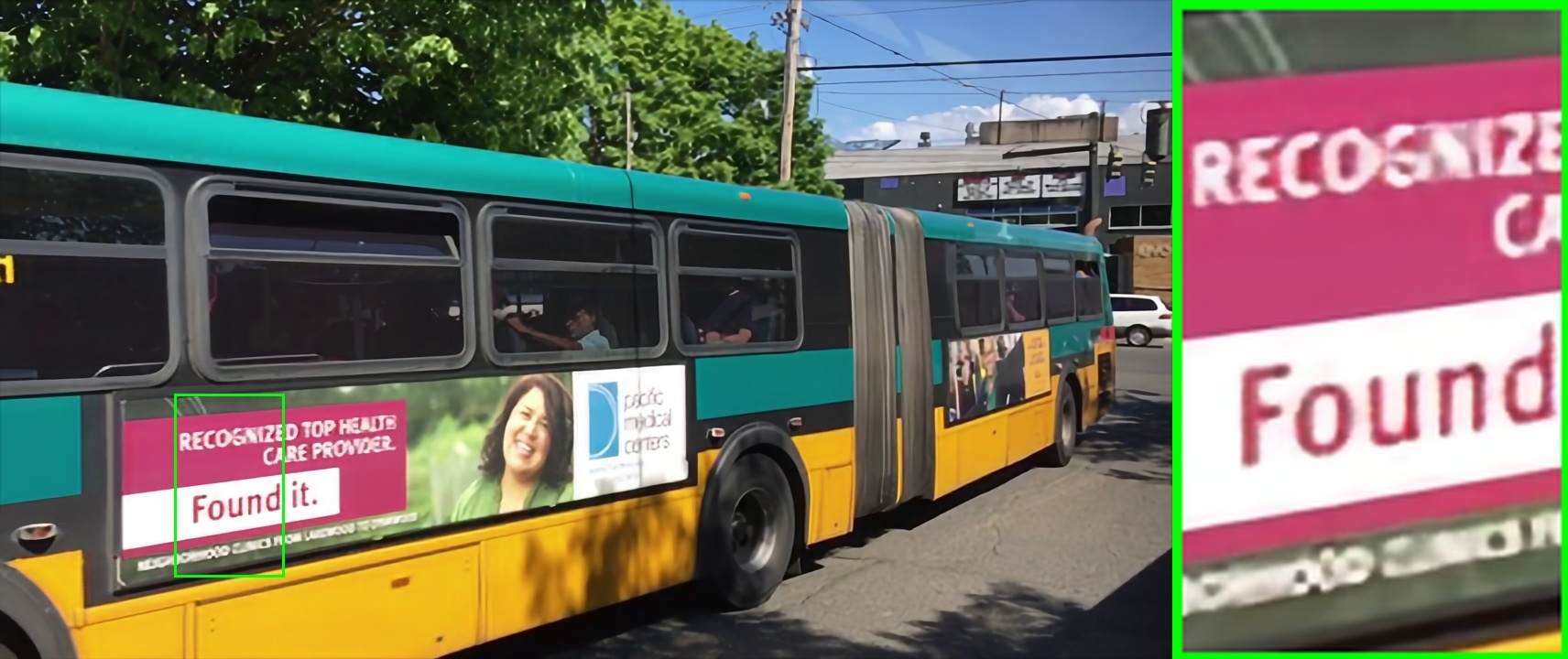}&
                \includegraphics[width=0.23\linewidth]{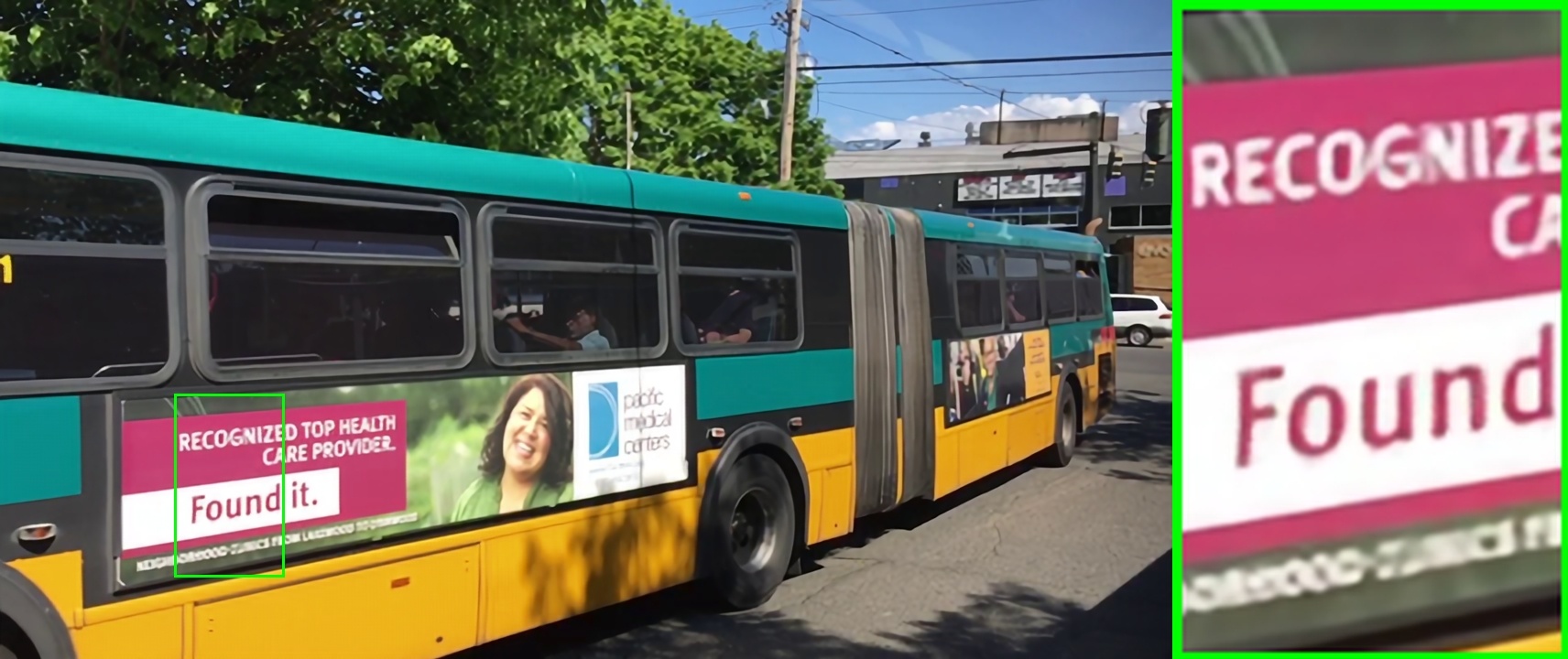}&
                \includegraphics[width=0.23\linewidth]{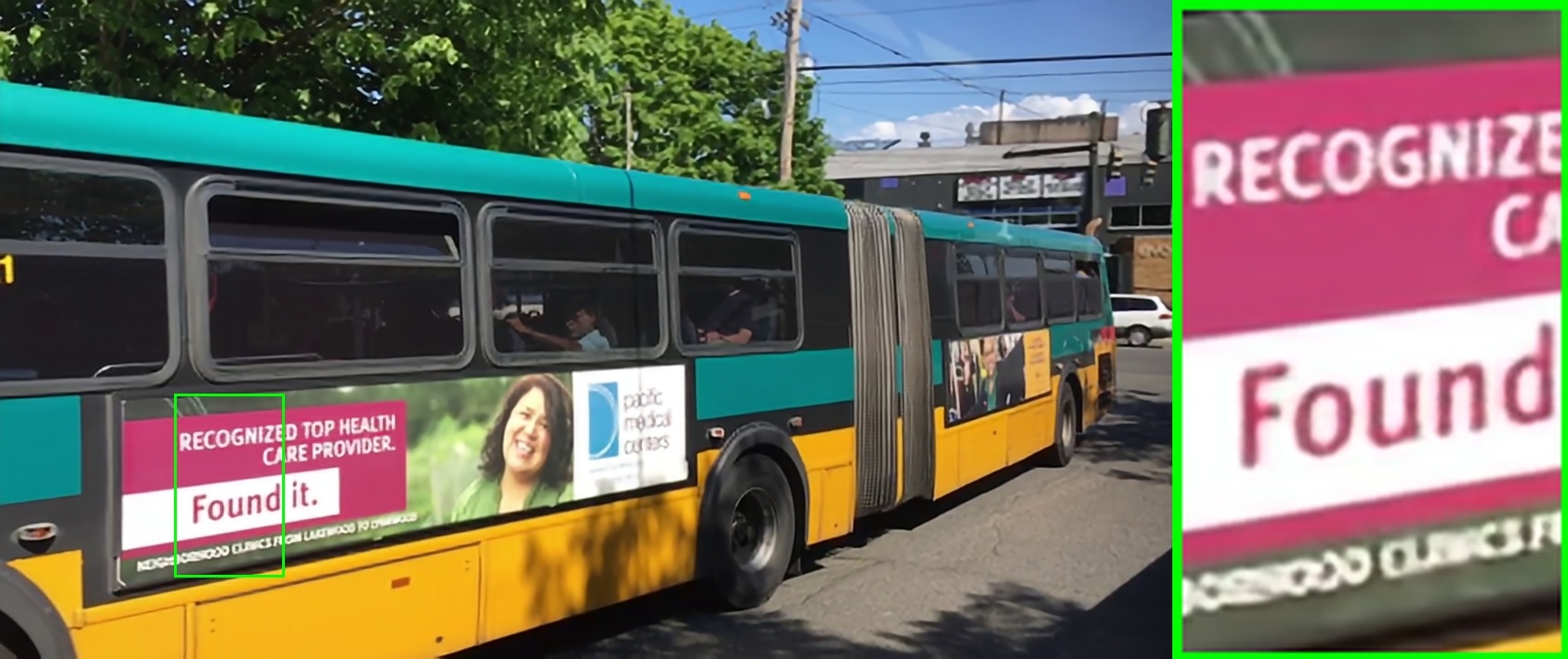}&
                \includegraphics[width=0.23\linewidth]{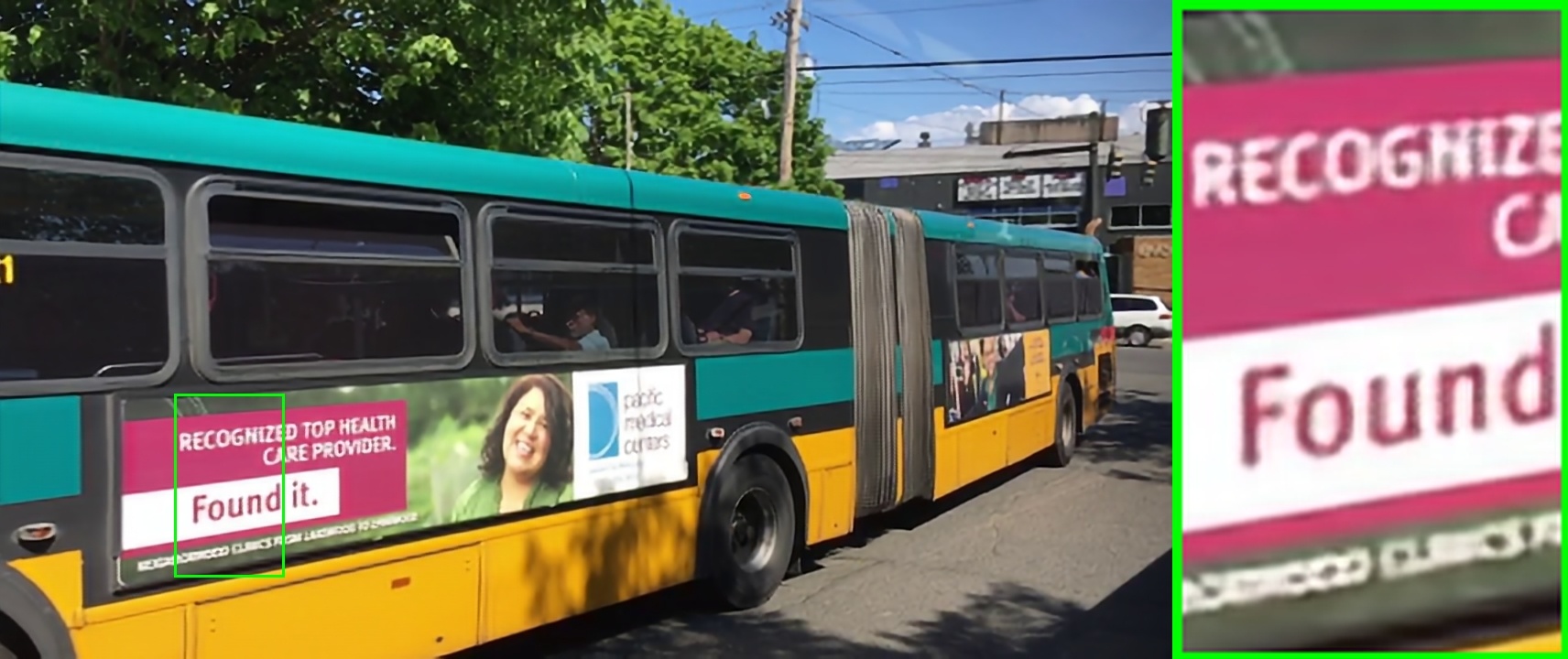}
            \end{tabular}
            \\ \\ \\
            \begin{tabular}{c}
            \rotatebox[origin=c]{90}{\makecell{\(\mathcal{T}=1\)\\}}
            \\ \\ \\ \\ \\ \\ \\ \\ \\ \\
            \end{tabular}
            &
            \begin{tabular}{c@{\hskip 0.005\linewidth}c@{\hskip 0.005\linewidth}c@{\hskip 0.005\linewidth}c}
                \includegraphics[width=0.23\linewidth]{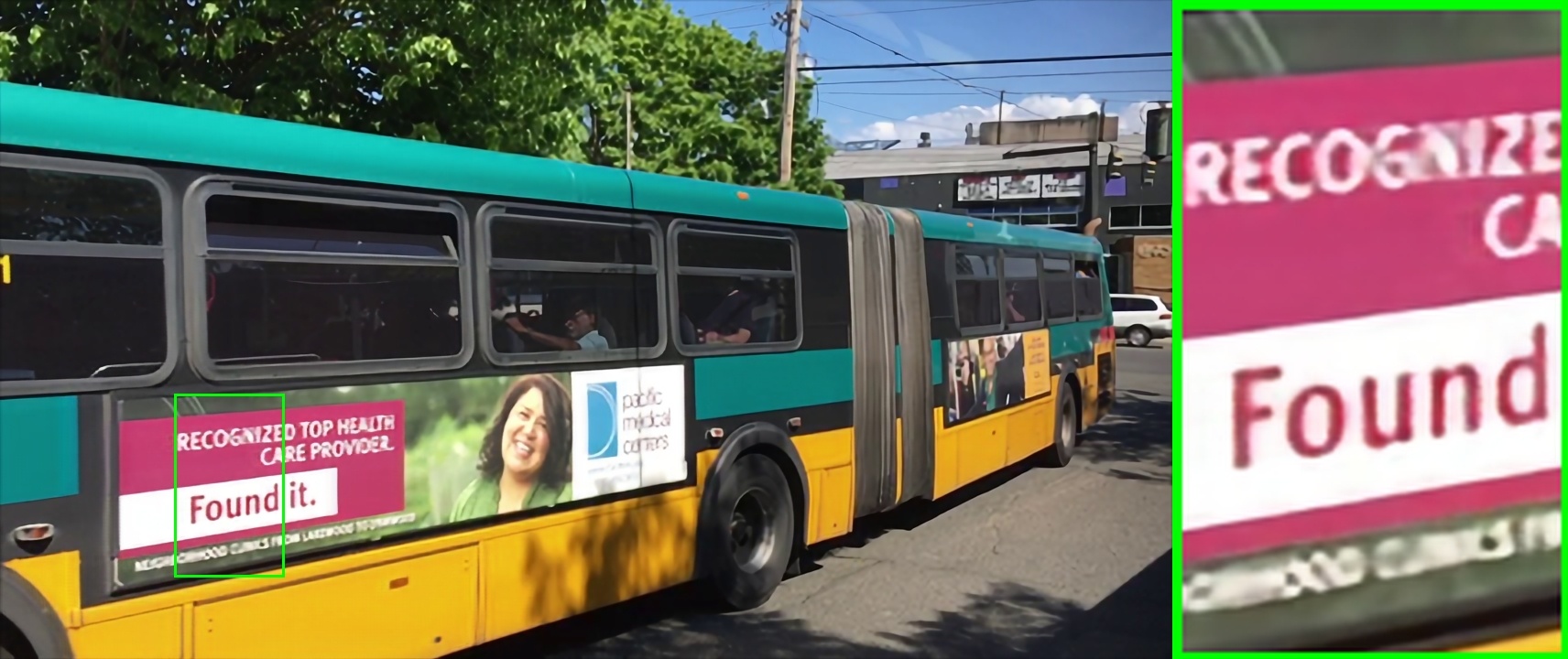}&
                \includegraphics[width=0.23\linewidth]{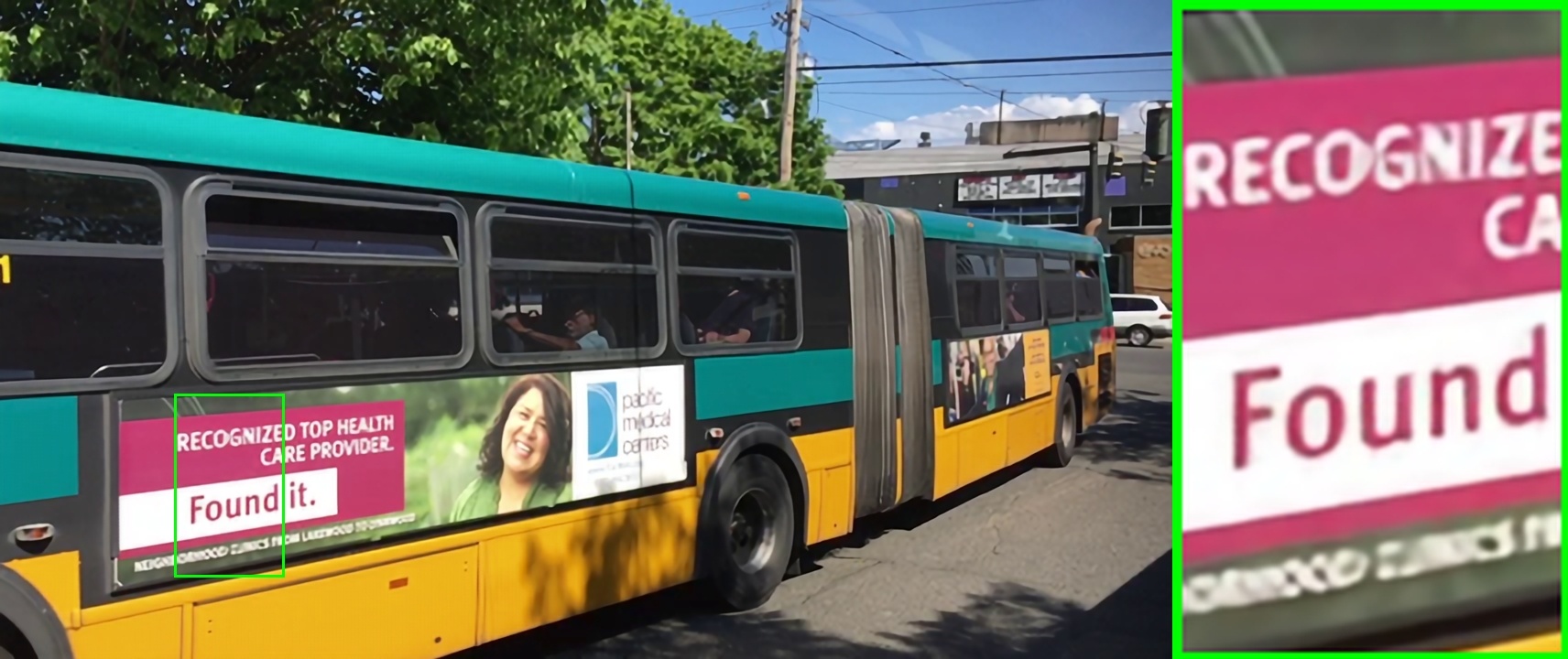}&
                \includegraphics[width=0.23\linewidth]{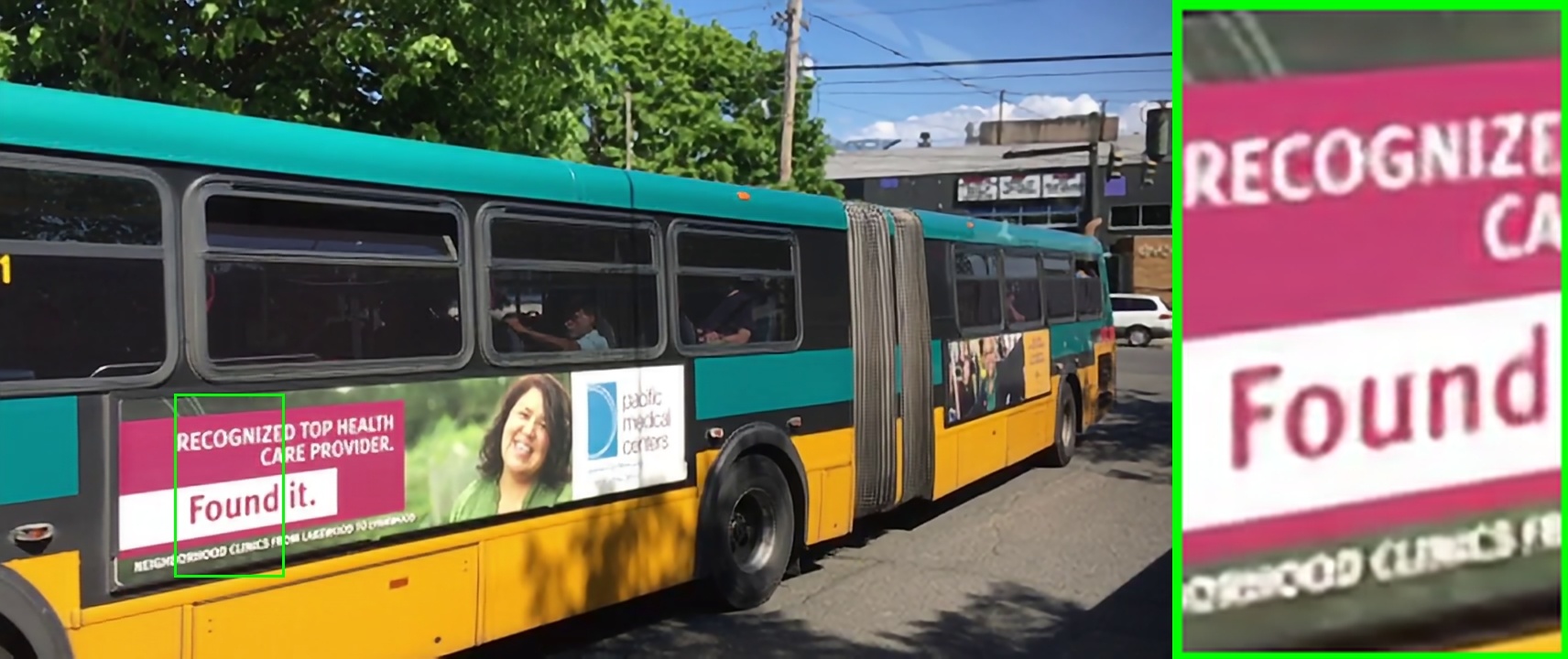}&
                \includegraphics[width=0.23\linewidth]{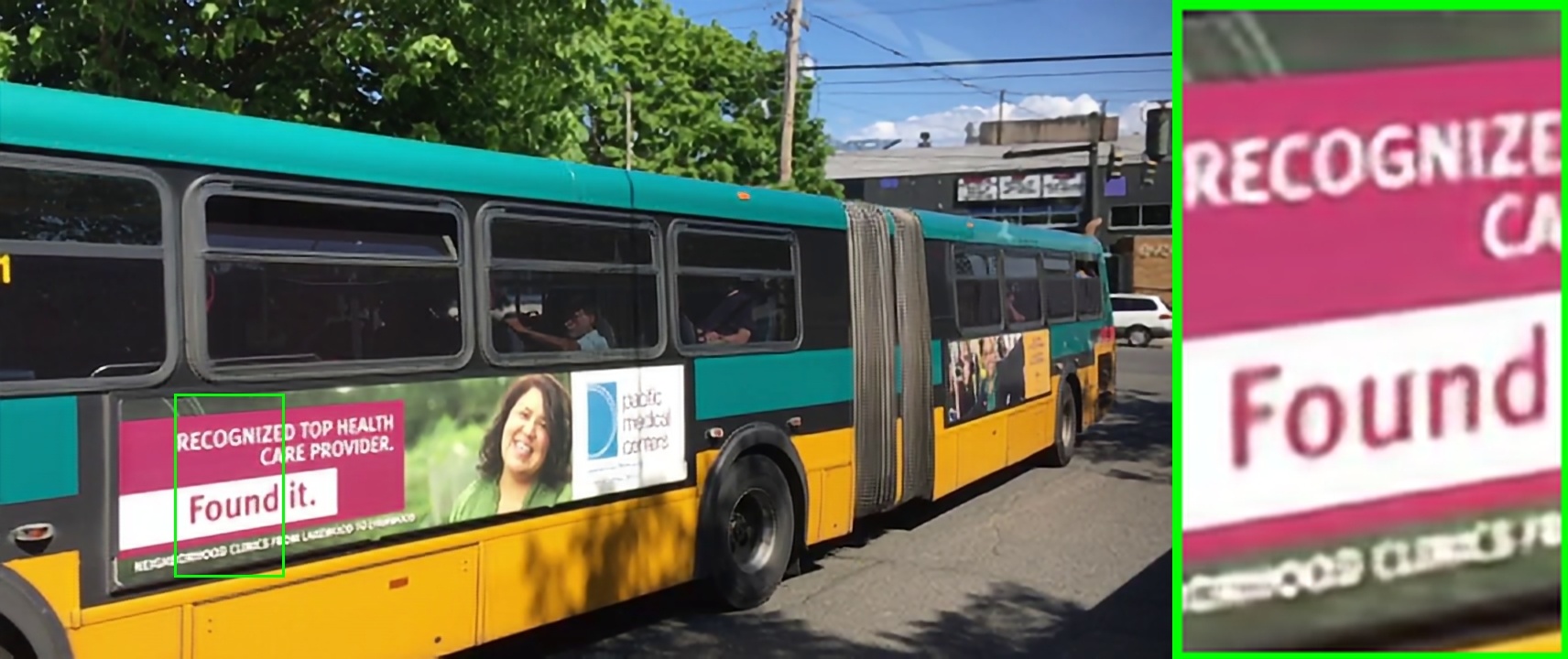}
            \\ \\
            VideoINR~\cite{chen2022videoinr}&
            MoTIF~\cite{chen2023motif}&
            \textbf{EvEnhancer (Ours)}&
            \textbf{EvEnhancer-light (Ours)}
            \end{tabular}
        \end{tabular}
    \caption{
       Qualitative comparison for In-dist. scale (\(t=8,s=4\)) on the Adobe240 dataset \cite{su2017deep}. Best zoom in for better visualization.
    }
    \label{fig:s_adobe}
\end{figure*}
\renewcommand{\arraystretch}{1.}

%% file: fig/exp/exp_suppl_gopro_t12s6.tex
\setlength{\tabcolsep}{0.1pt}
\renewcommand{\arraystretch}{0.1}

\begin{figure*}[ht!]
    \centering\
         \begin{tabular}{cc}
            \rotatebox[origin=c]{90}{\makecell{\footnotesize \(\mathcal{T}=0\)\\}}  &
            \begin{tabular}{c@{\hskip 0.005\linewidth}c@{\hskip 0.005\linewidth}c@{\hskip 0.005\linewidth}c}
                \includegraphics[width=0.23\linewidth]{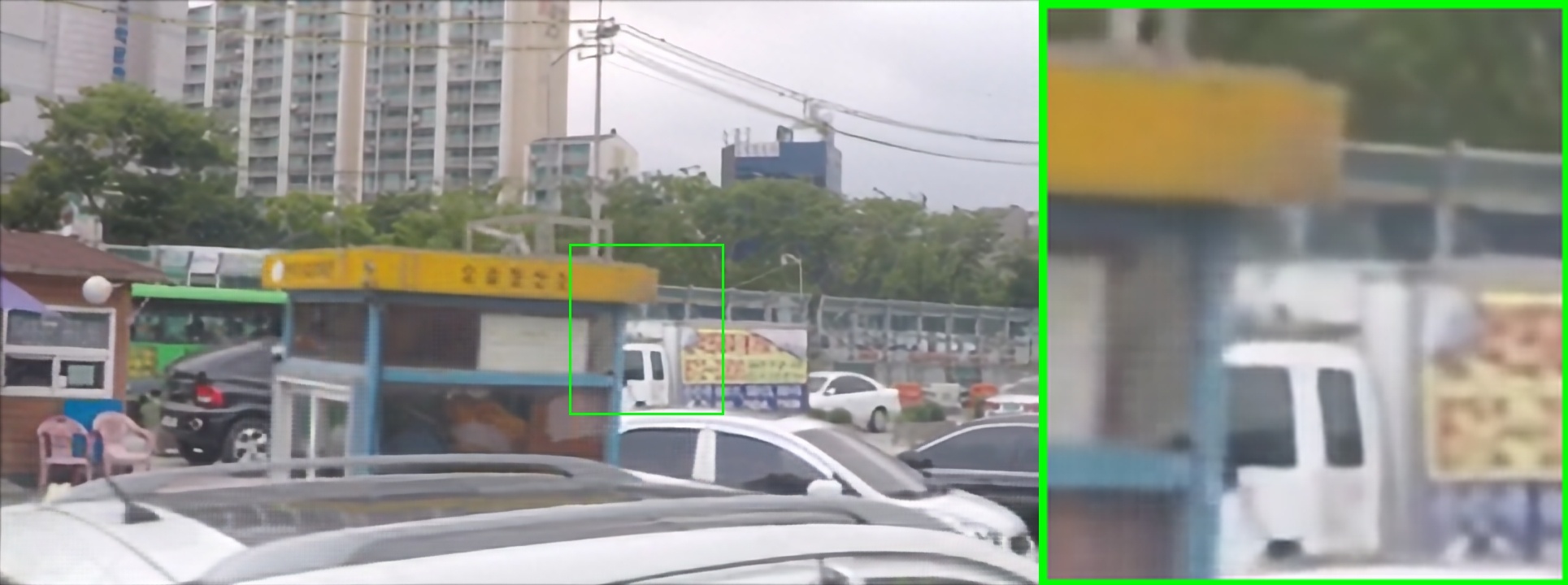}&
                \includegraphics[width=0.23\linewidth]{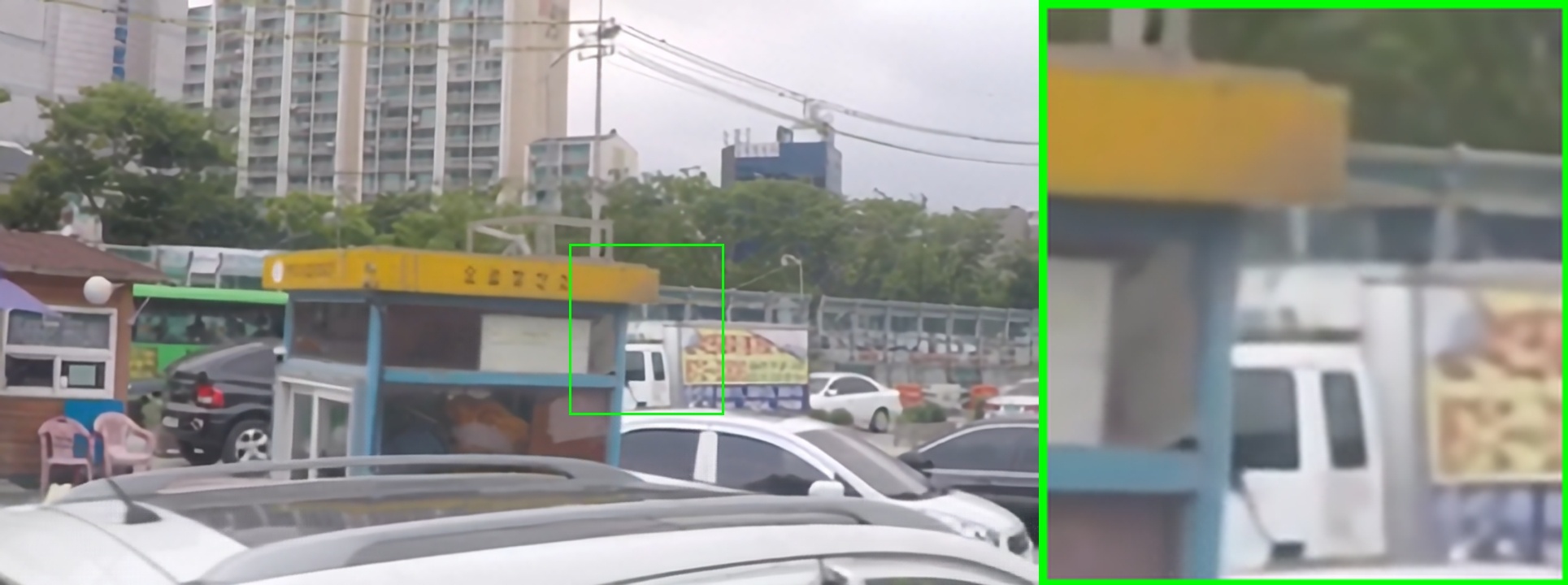}&
                \includegraphics[width=0.23\linewidth]{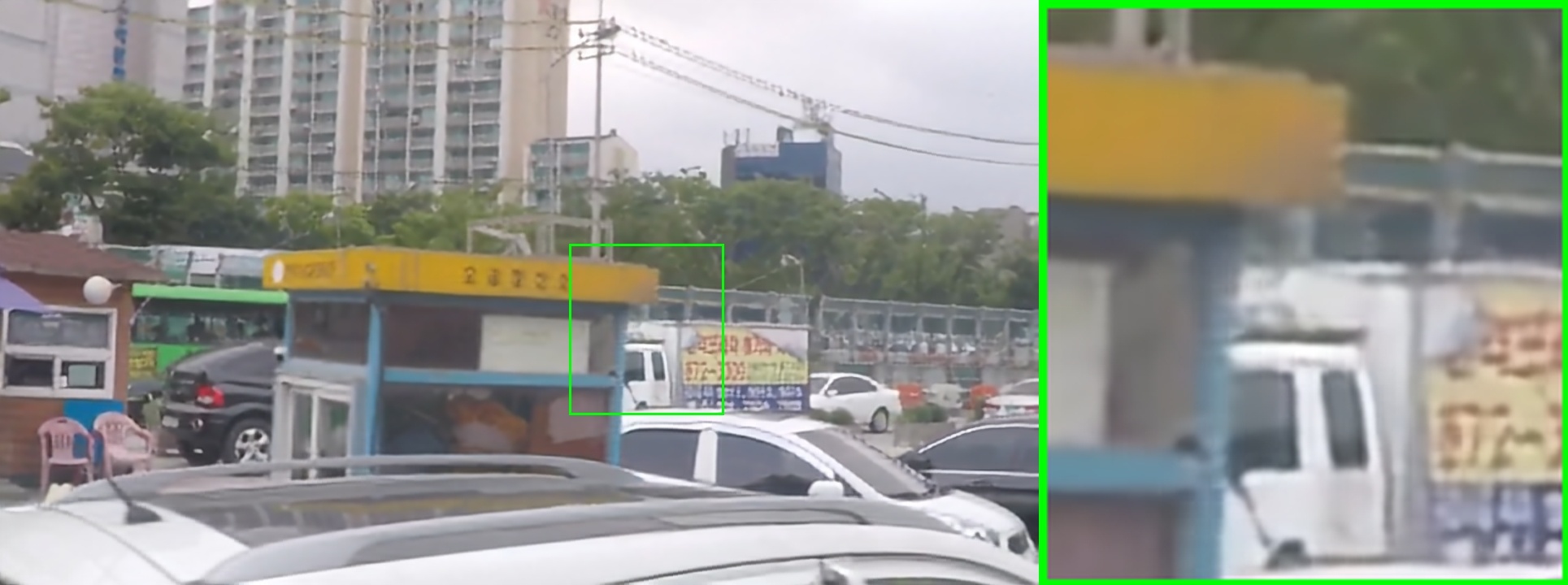}&
                \includegraphics[width=0.23\linewidth]{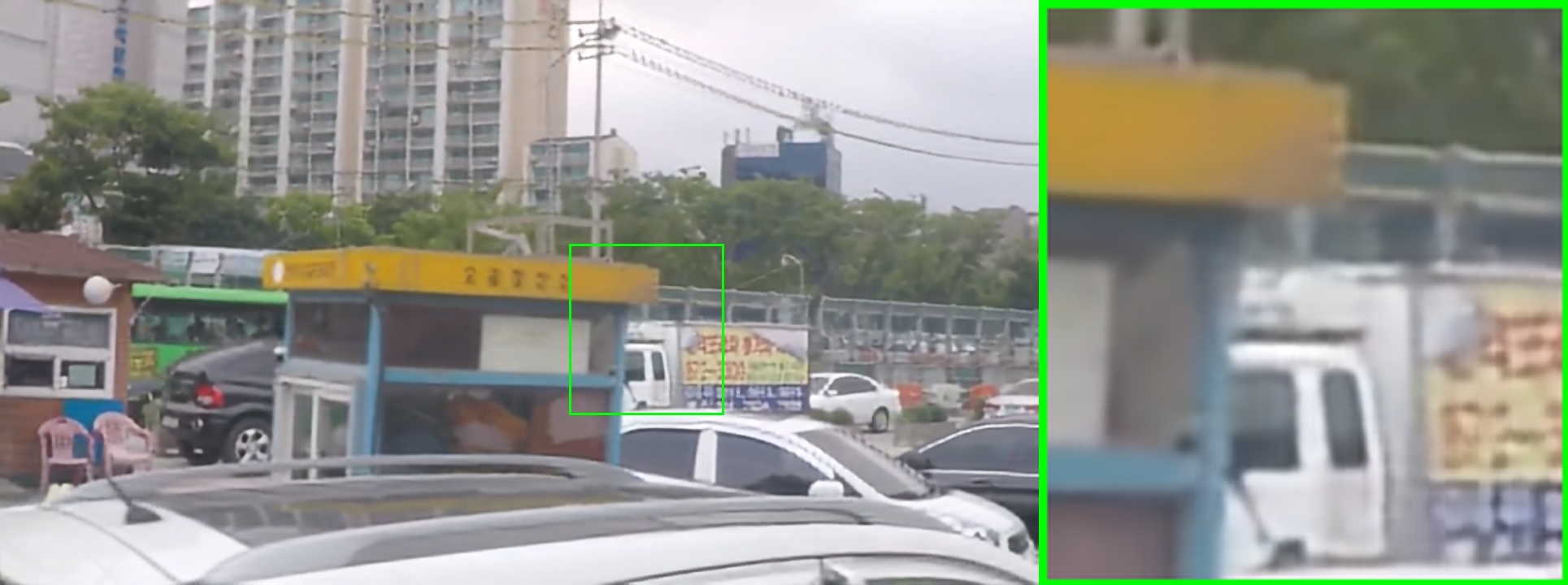}
            \end{tabular}
            \\ \\ \\
            \rotatebox[origin=c]{90}{\makecell{\footnotesize \(\mathcal{T}=0.083\)\\}}  &
            \begin{tabular}{c@{\hskip 0.005\linewidth}c@{\hskip 0.005\linewidth}c@{\hskip 0.005\linewidth}c}
                \includegraphics[width=0.23\linewidth]{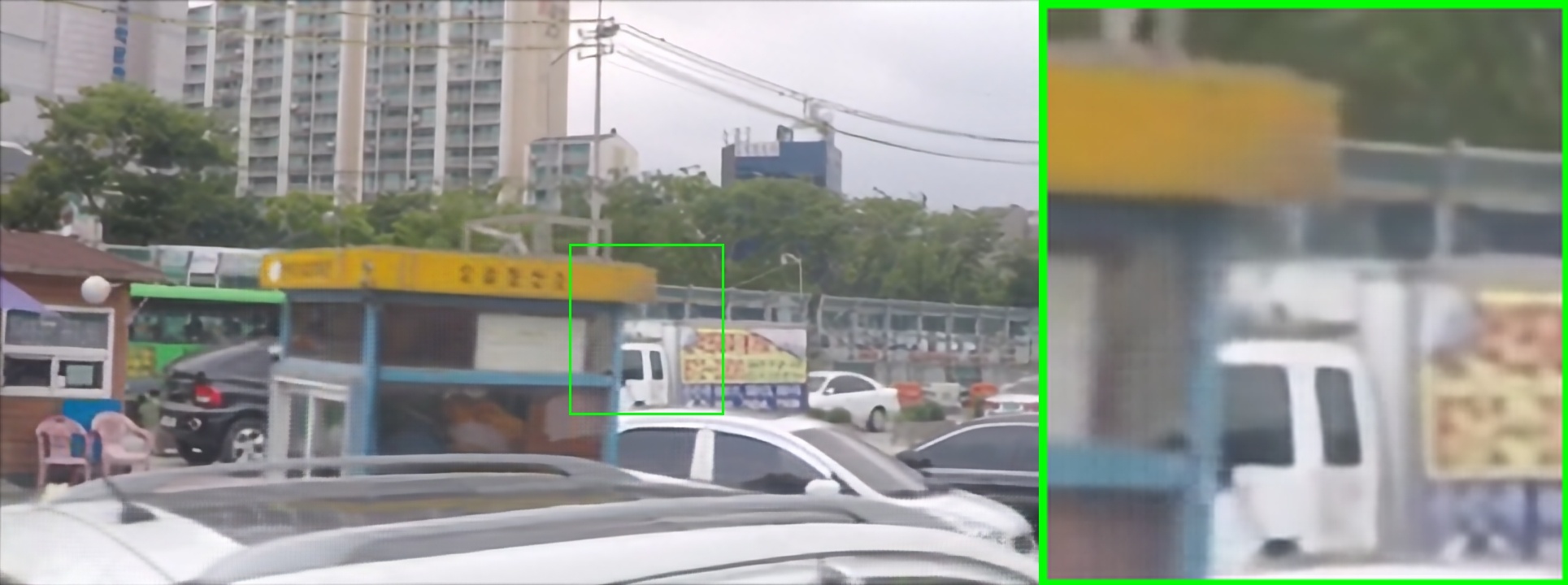}&
                \includegraphics[width=0.23\linewidth]{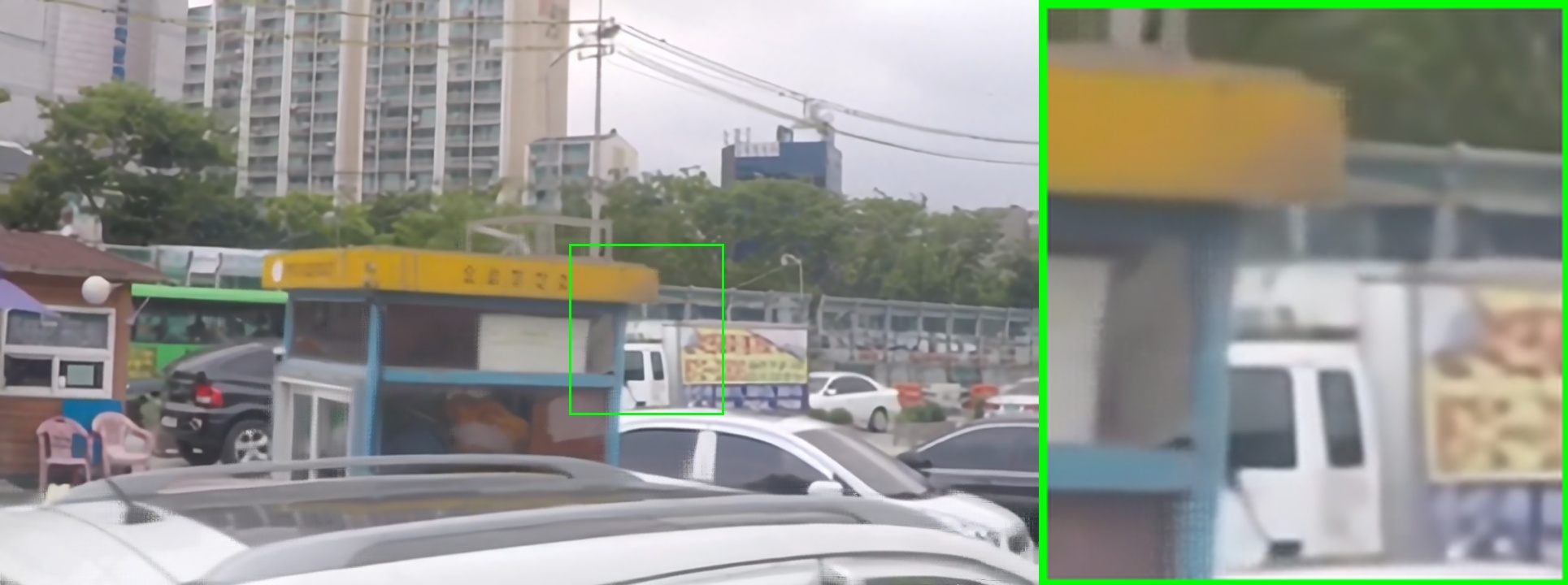}&
                \includegraphics[width=0.23\linewidth]{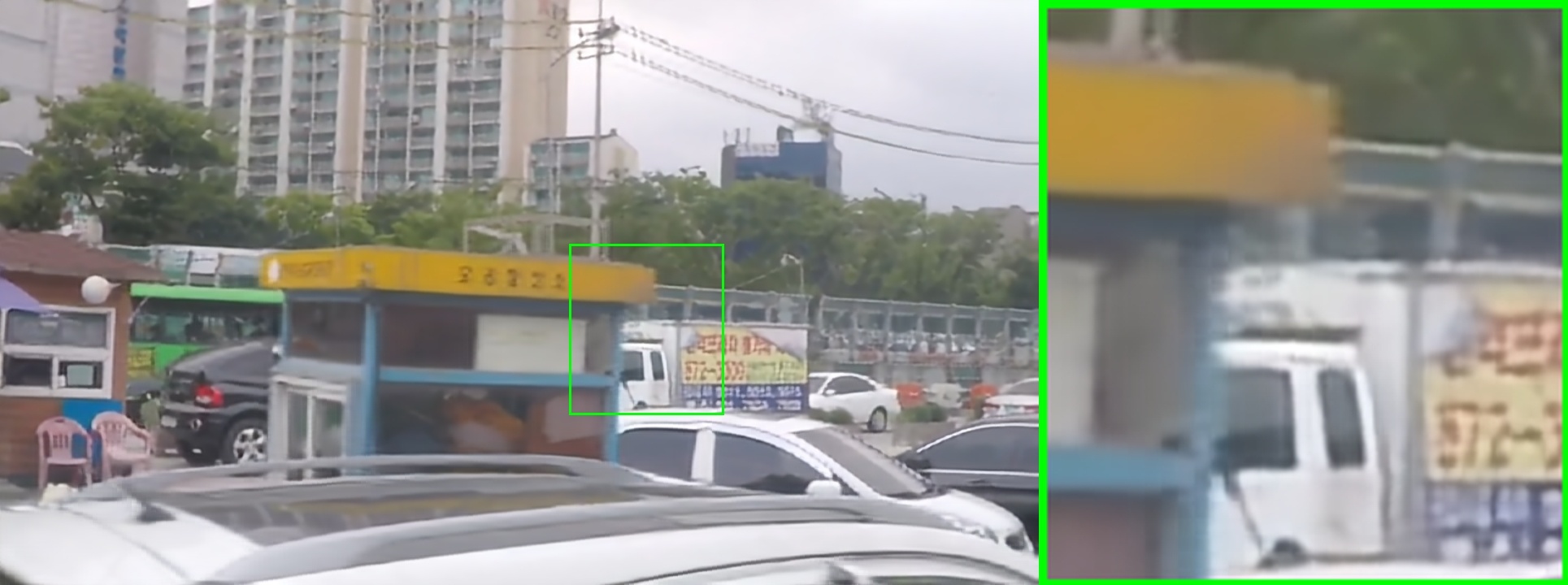}&
                \includegraphics[width=0.23\linewidth]{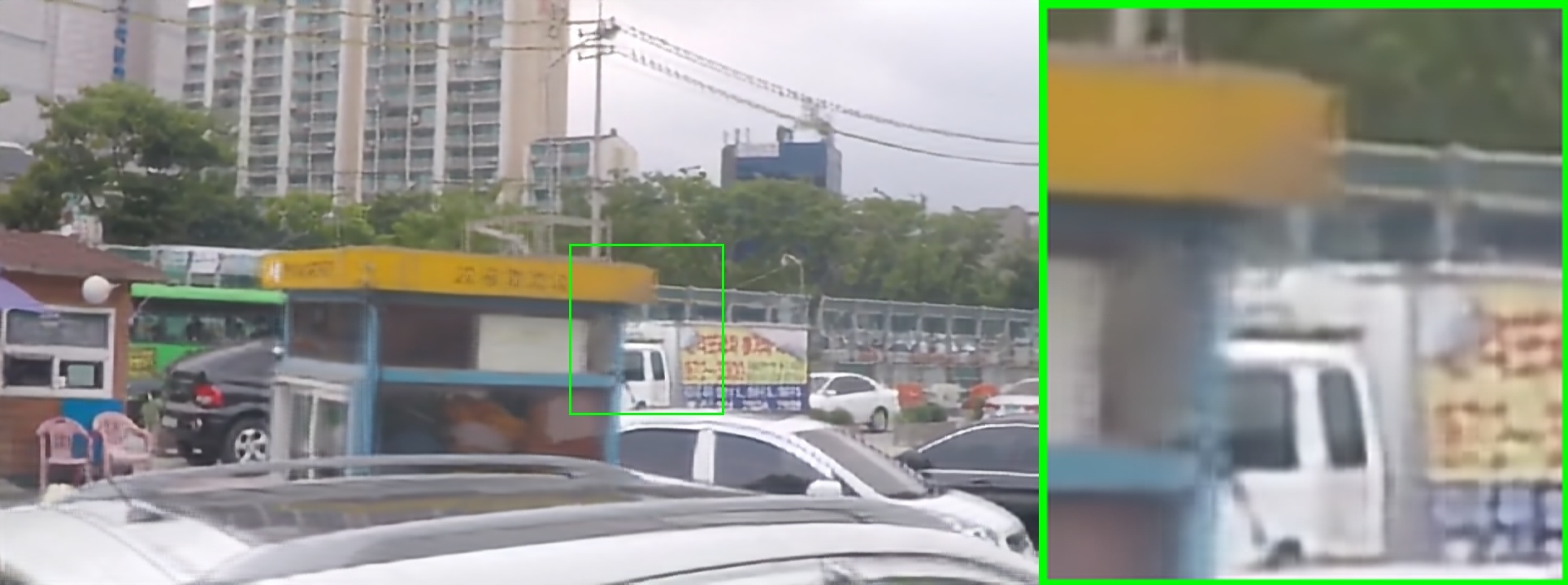}
            \end{tabular}
            \\ \\ \\
            \rotatebox[origin=c]{90}{\makecell{\footnotesize \(\mathcal{T}=0.167\)\\}}  &
            \begin{tabular}{c@{\hskip 0.005\linewidth}c@{\hskip 0.005\linewidth}c@{\hskip 0.005\linewidth}c}
                \includegraphics[width=0.23\linewidth]{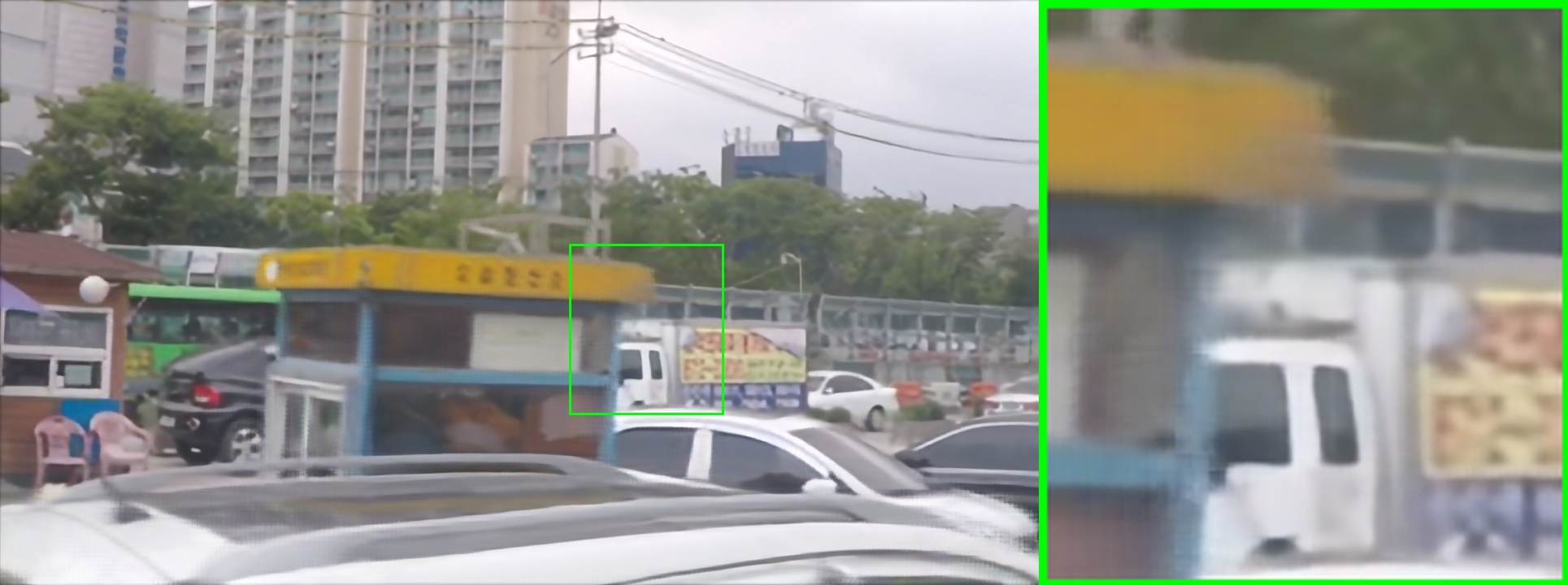}&
                \includegraphics[width=0.23\linewidth]{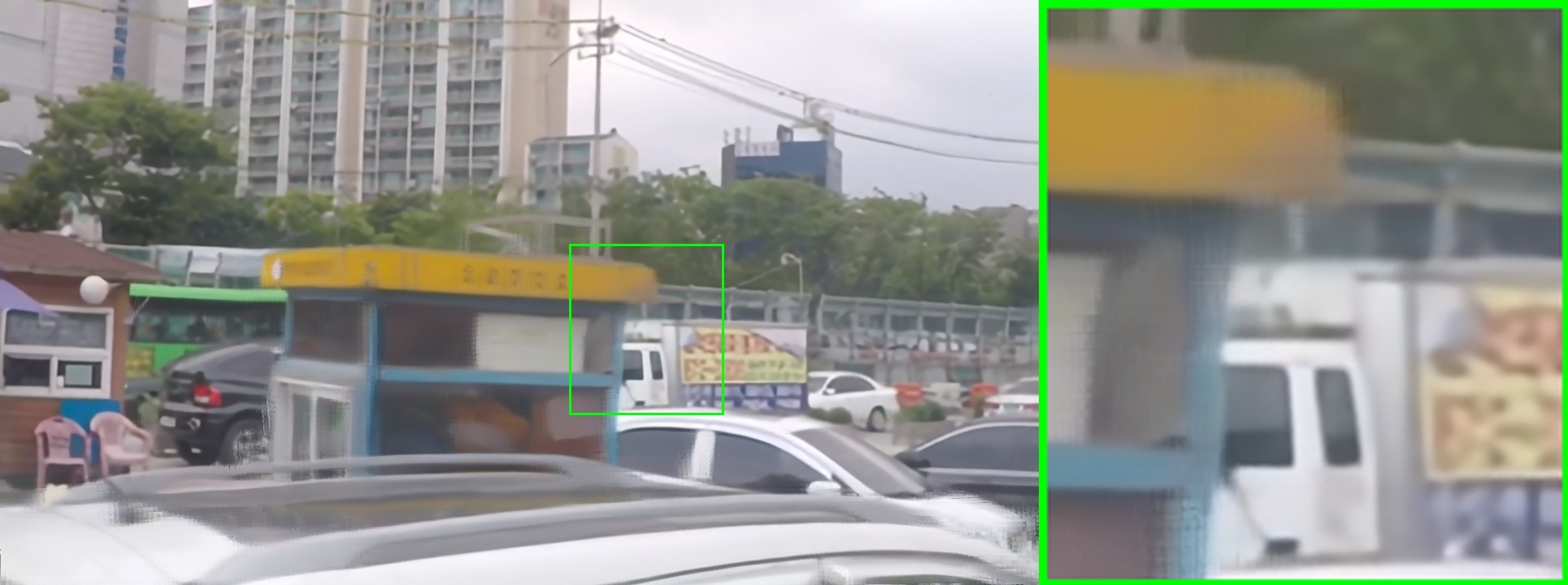}&
                \includegraphics[width=0.23\linewidth]{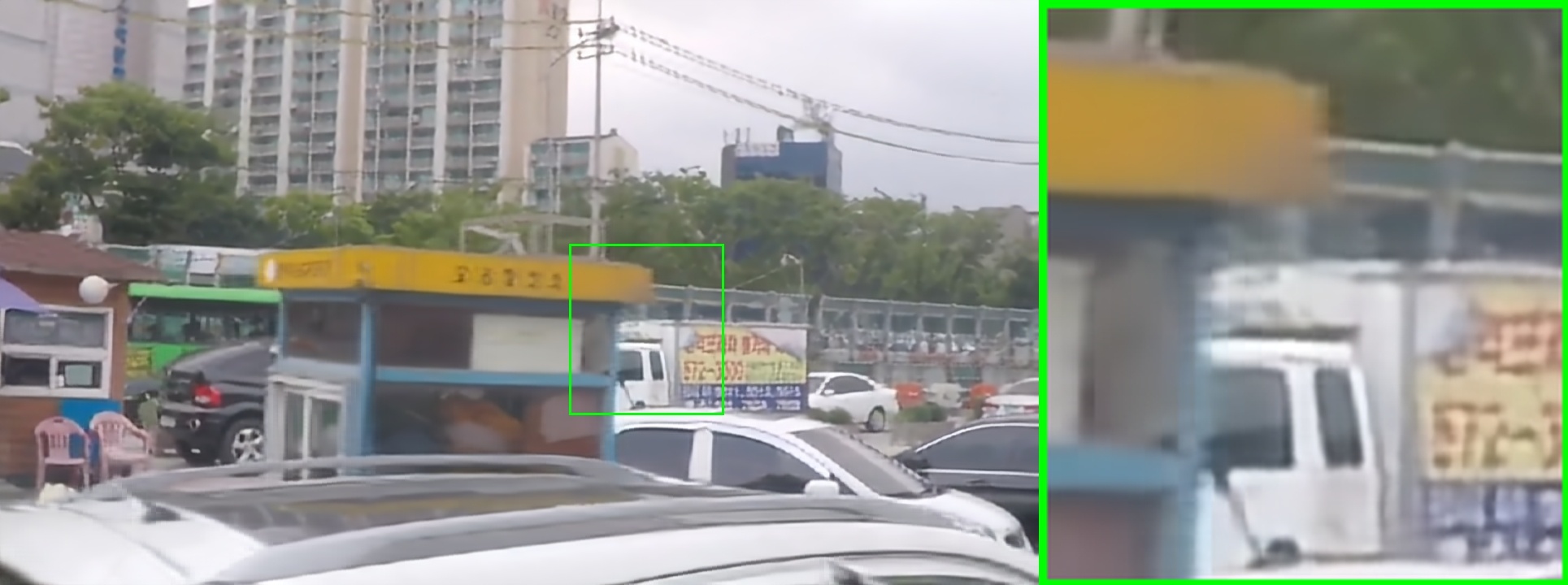}&
                \includegraphics[width=0.23\linewidth]{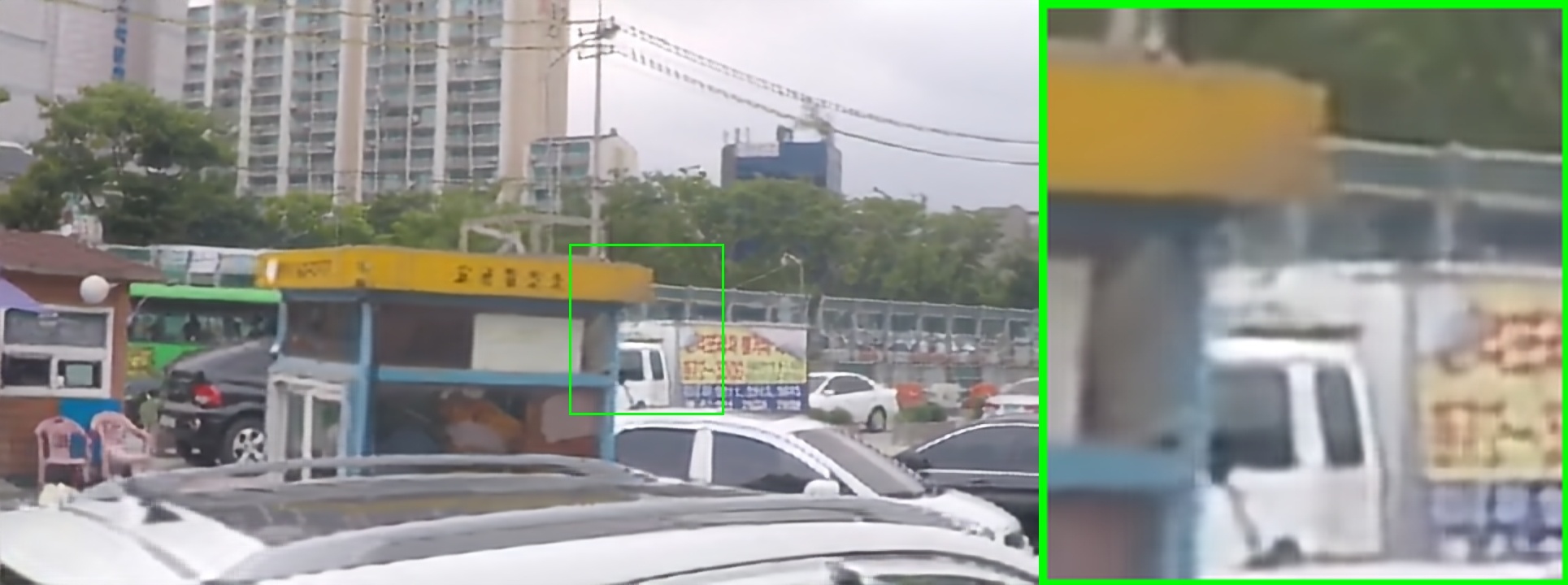}
            \end{tabular}
            \\ \\ \\
            \rotatebox[origin=c]{90}{\makecell{\footnotesize \(\mathcal{T}=0.250\)\\}}  &
            \begin{tabular}{c@{\hskip 0.005\linewidth}c@{\hskip 0.005\linewidth}c@{\hskip 0.005\linewidth}c}
                 \includegraphics[width=0.23\linewidth]{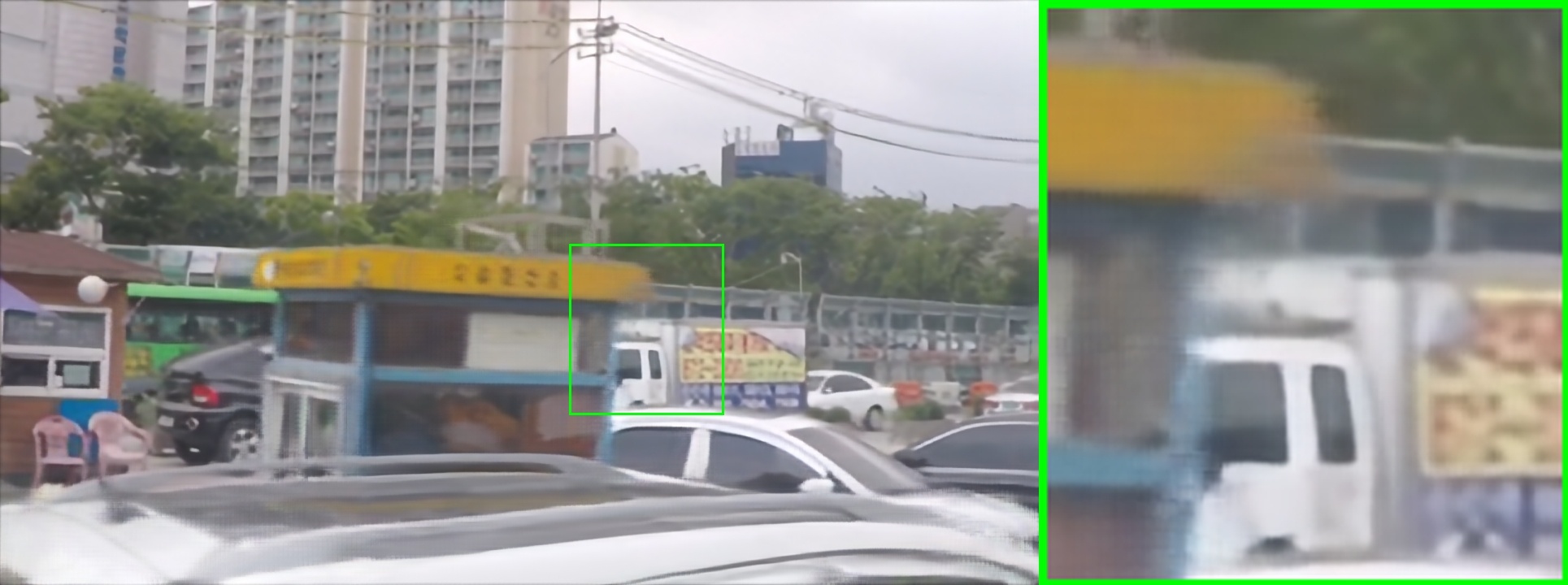}&
                \includegraphics[width=0.23\linewidth]{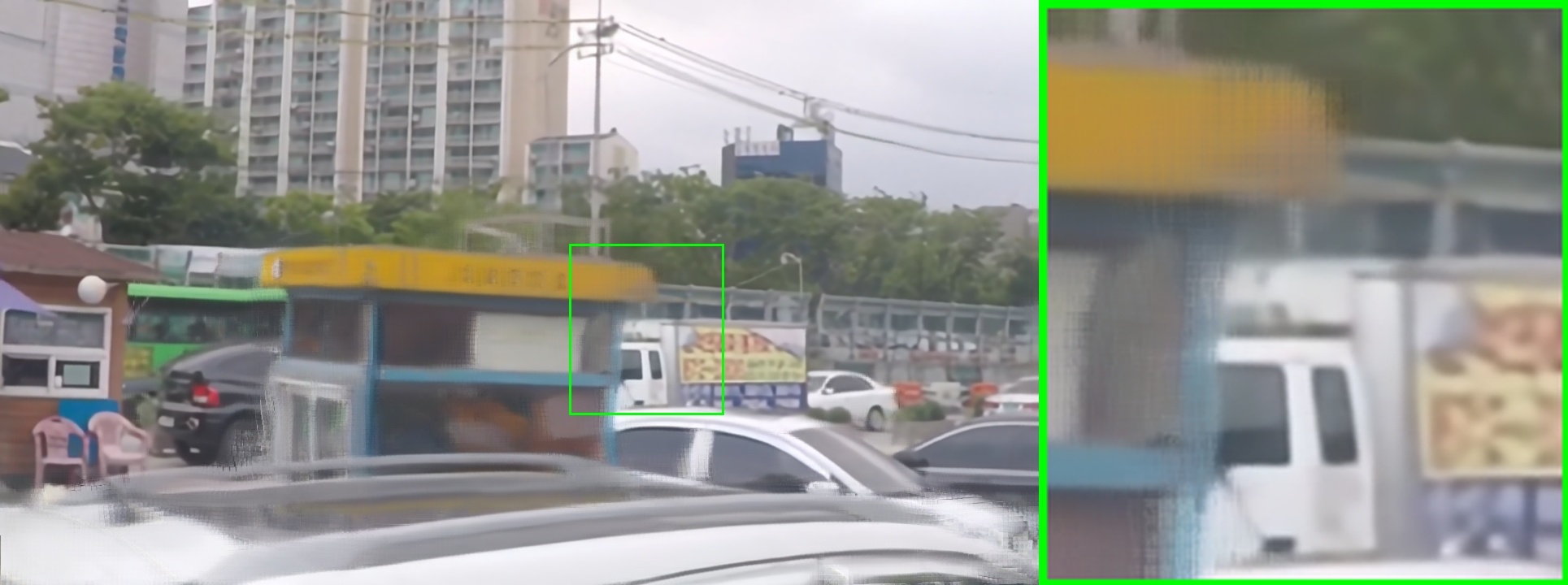}&
                \includegraphics[width=0.23\linewidth]{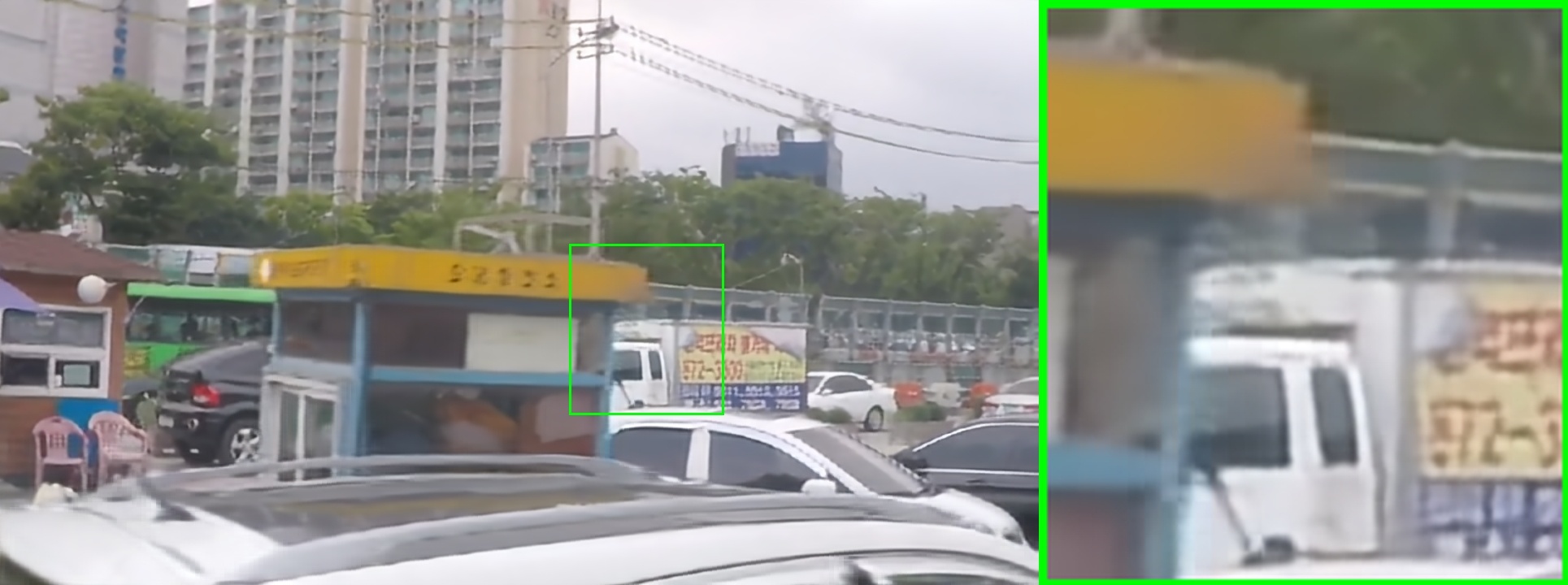}&
                \includegraphics[width=0.23\linewidth]{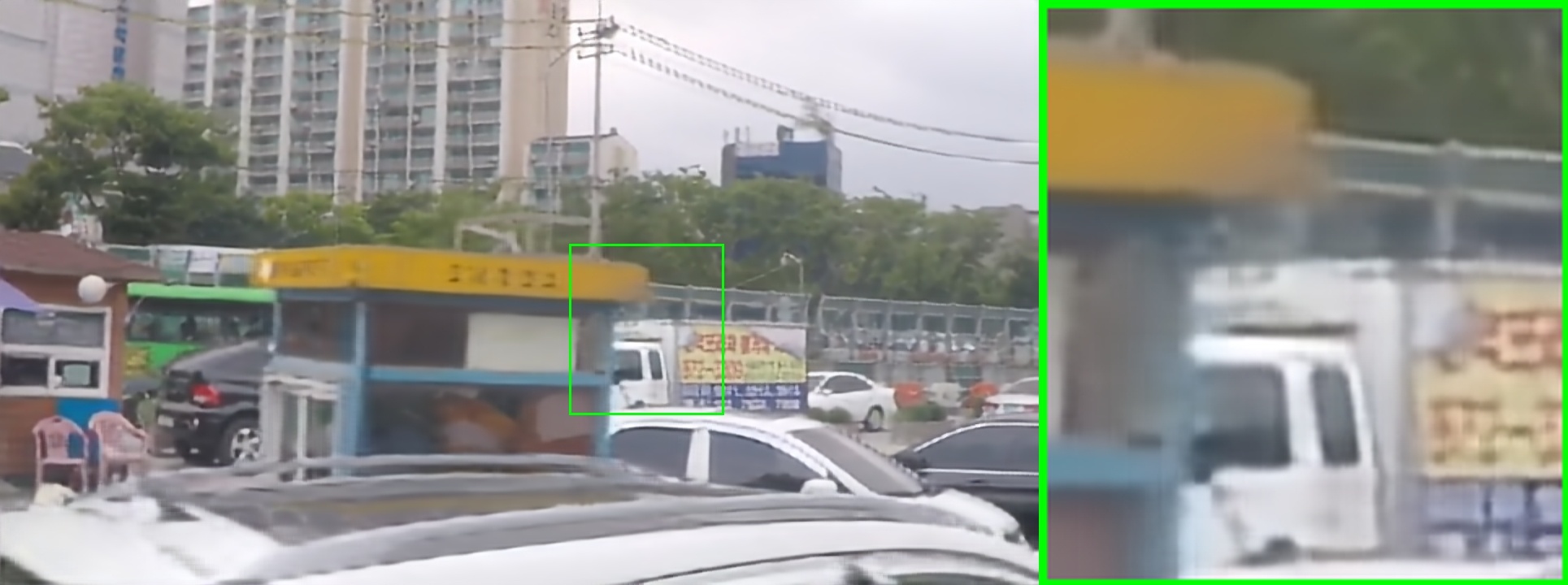}
            \end{tabular}
            \\ \\ \\
            \rotatebox[origin=c]{90}{\makecell{\footnotesize \(\mathcal{T}=0.333\)\\}}  &
            \begin{tabular}{c@{\hskip 0.005\linewidth}c@{\hskip 0.005\linewidth}c@{\hskip 0.005\linewidth}c}
                \includegraphics[width=0.23\linewidth]{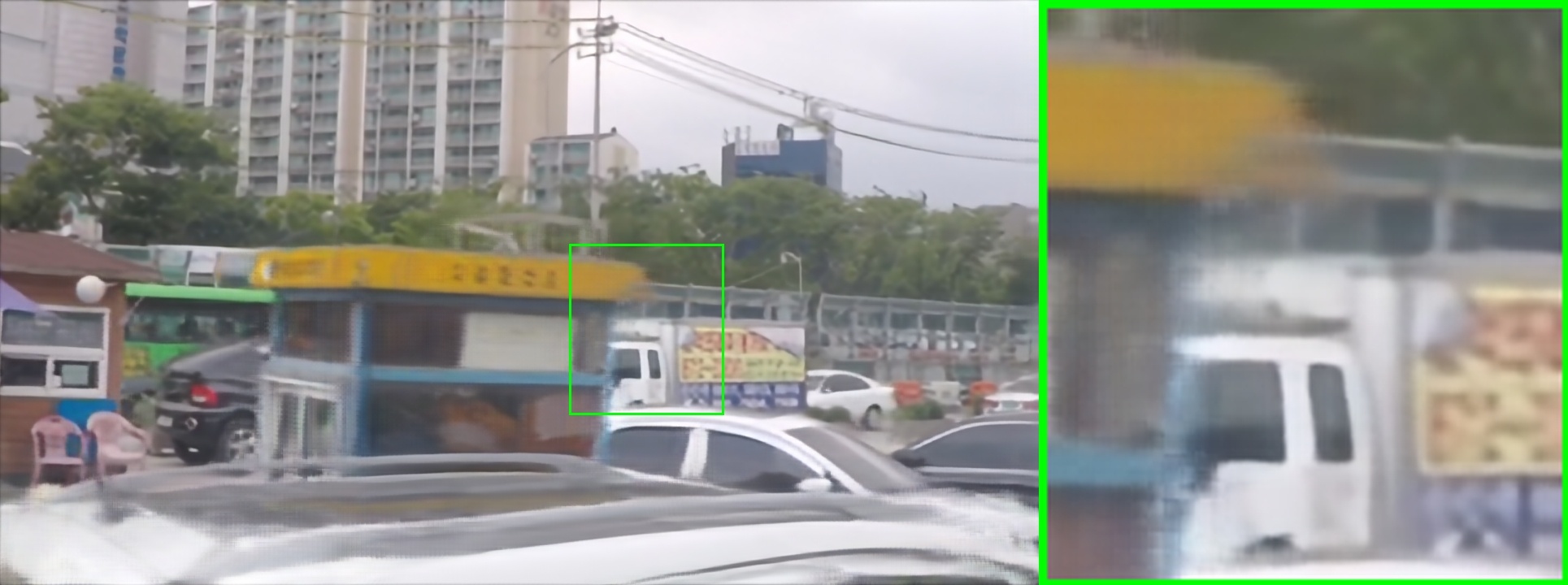}&
                \includegraphics[width=0.23\linewidth]{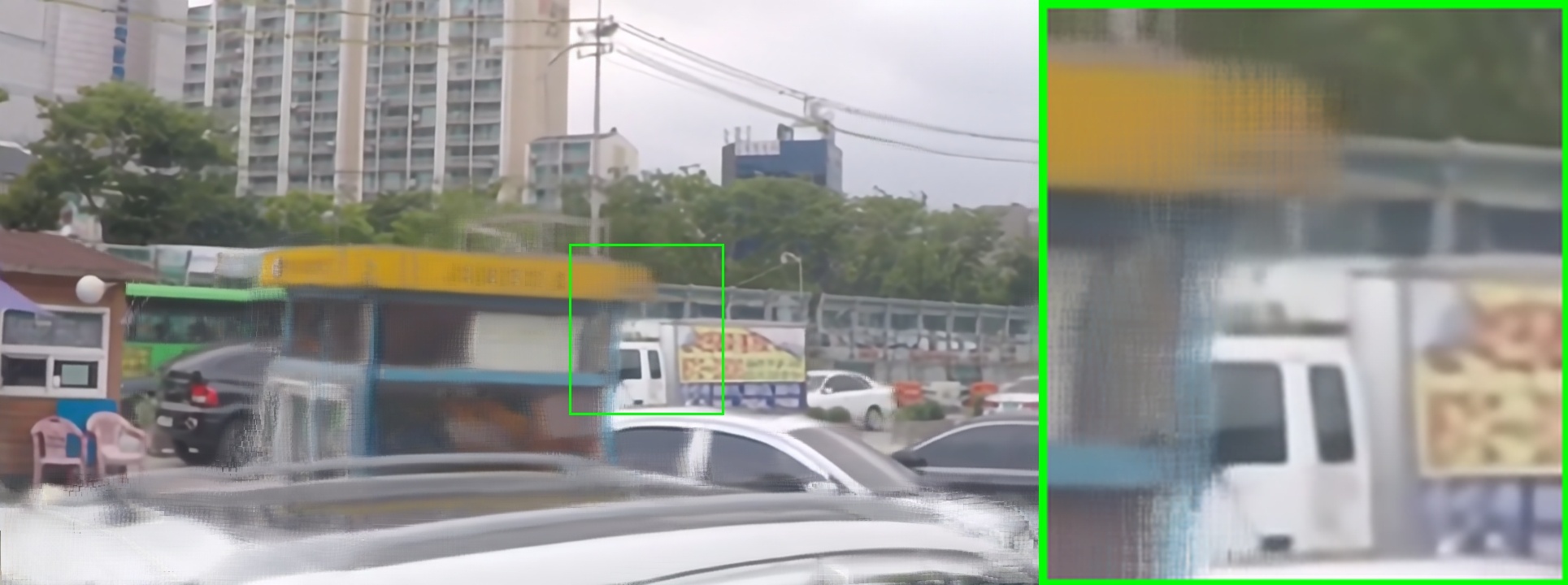}&
                \includegraphics[width=0.23\linewidth]{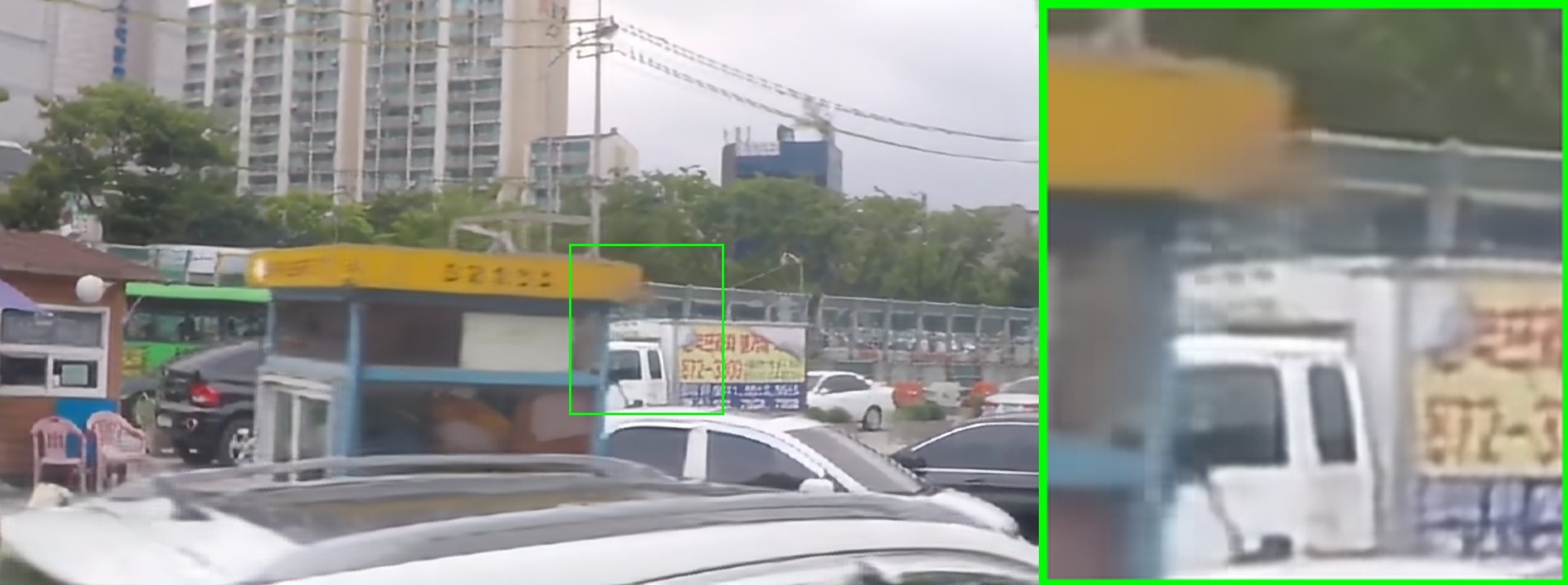}&
                \includegraphics[width=0.23\linewidth]{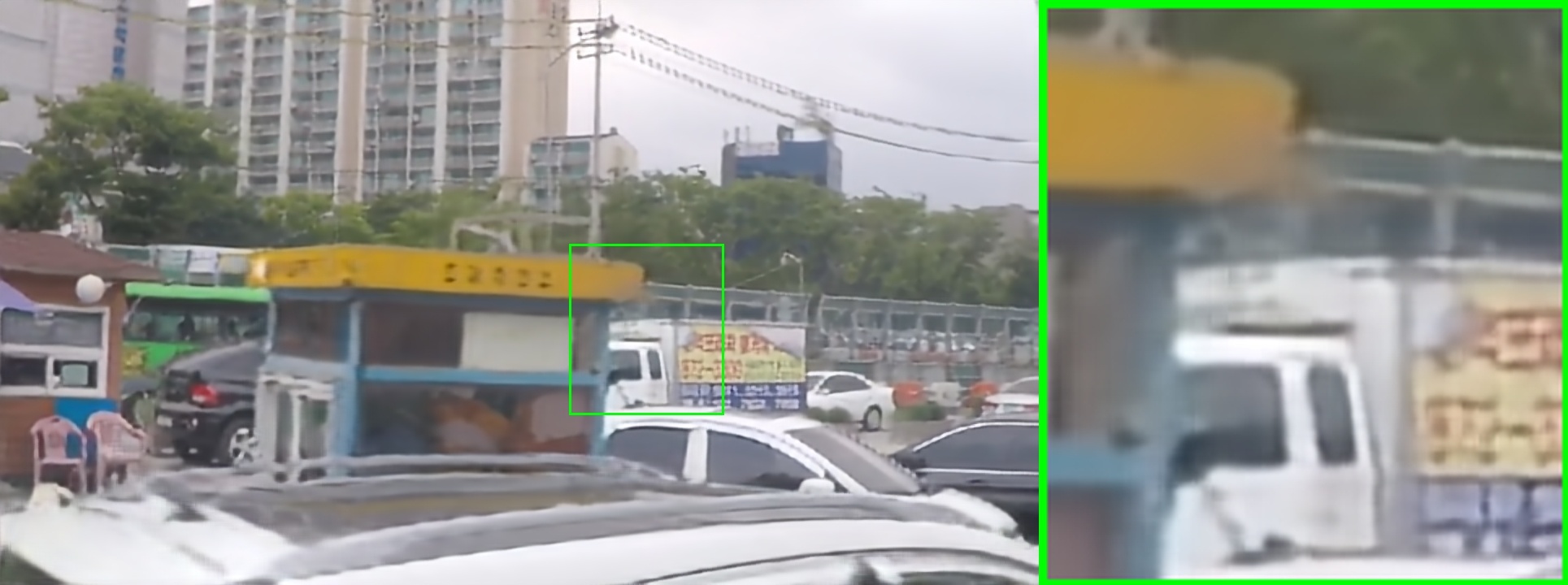}
            \end{tabular}
            \\ \\ \\
            \rotatebox[origin=c]{90}{\makecell{\footnotesize \(\mathcal{T}=0.417\)\\}}  &
            \begin{tabular}{c@{\hskip 0.005\linewidth}c@{\hskip 0.005\linewidth}c@{\hskip 0.005\linewidth}c}
                \includegraphics[width=0.23\linewidth]{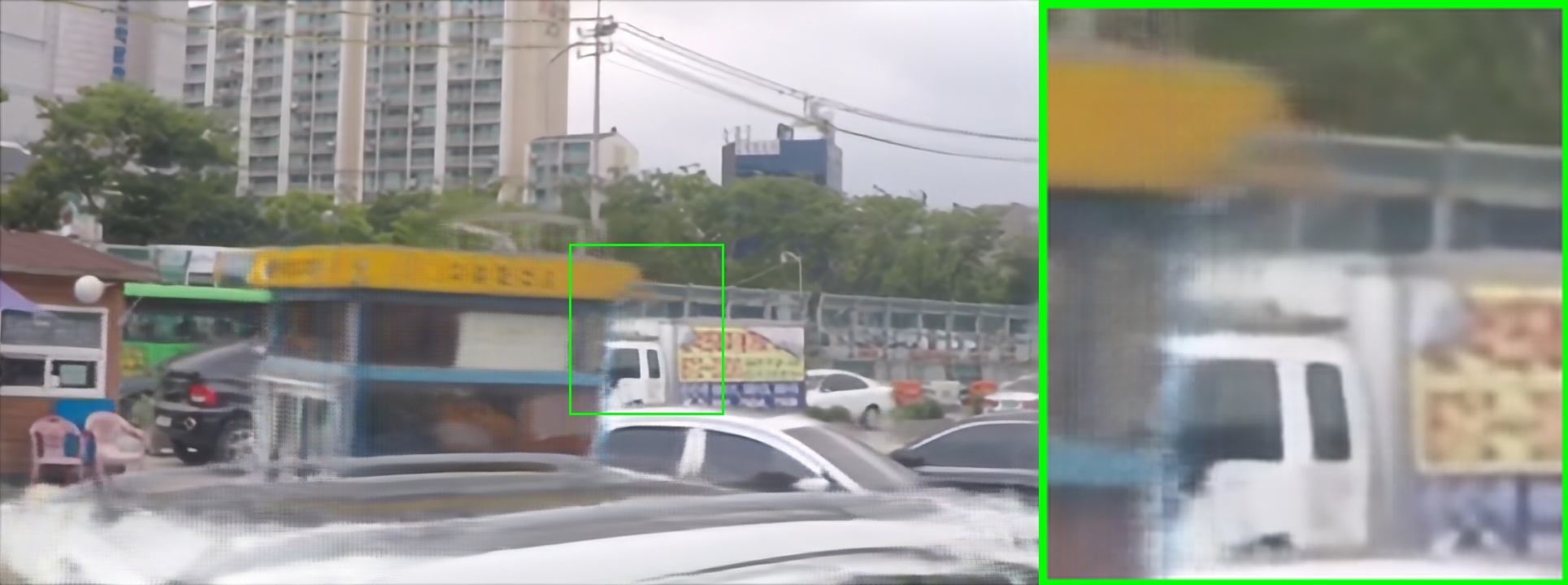}&
                \includegraphics[width=0.23\linewidth]{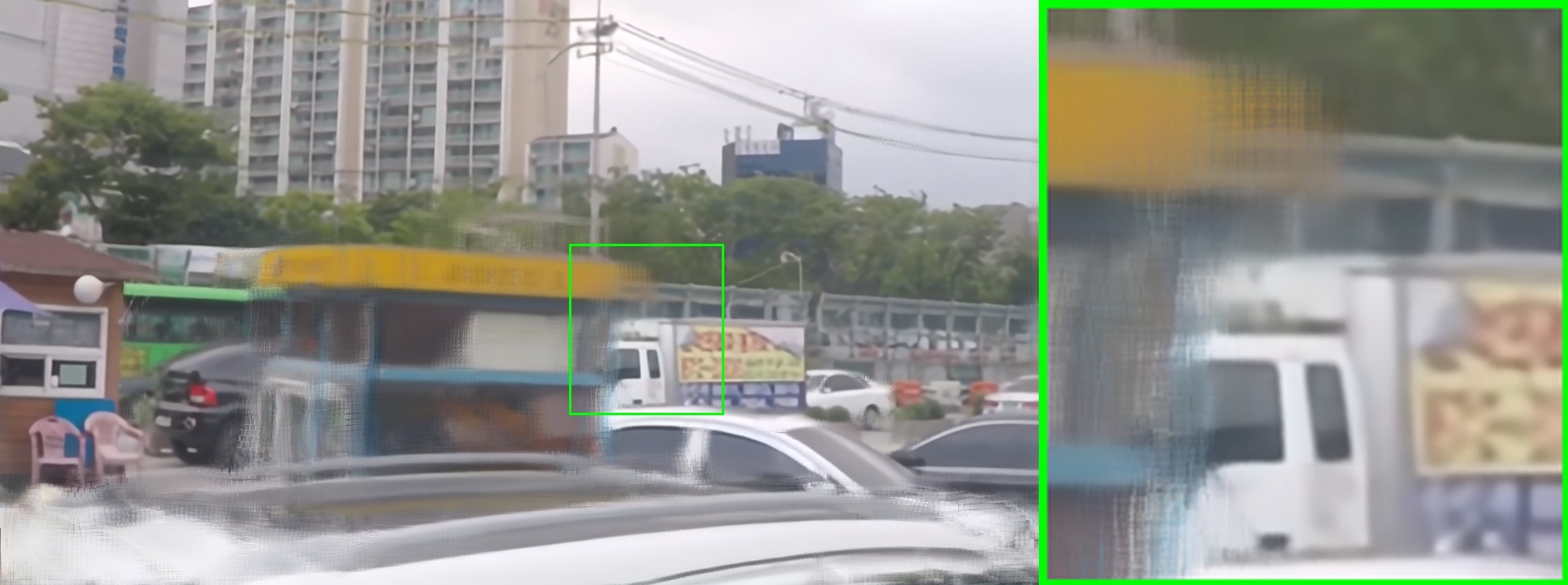}&
                \includegraphics[width=0.23\linewidth]{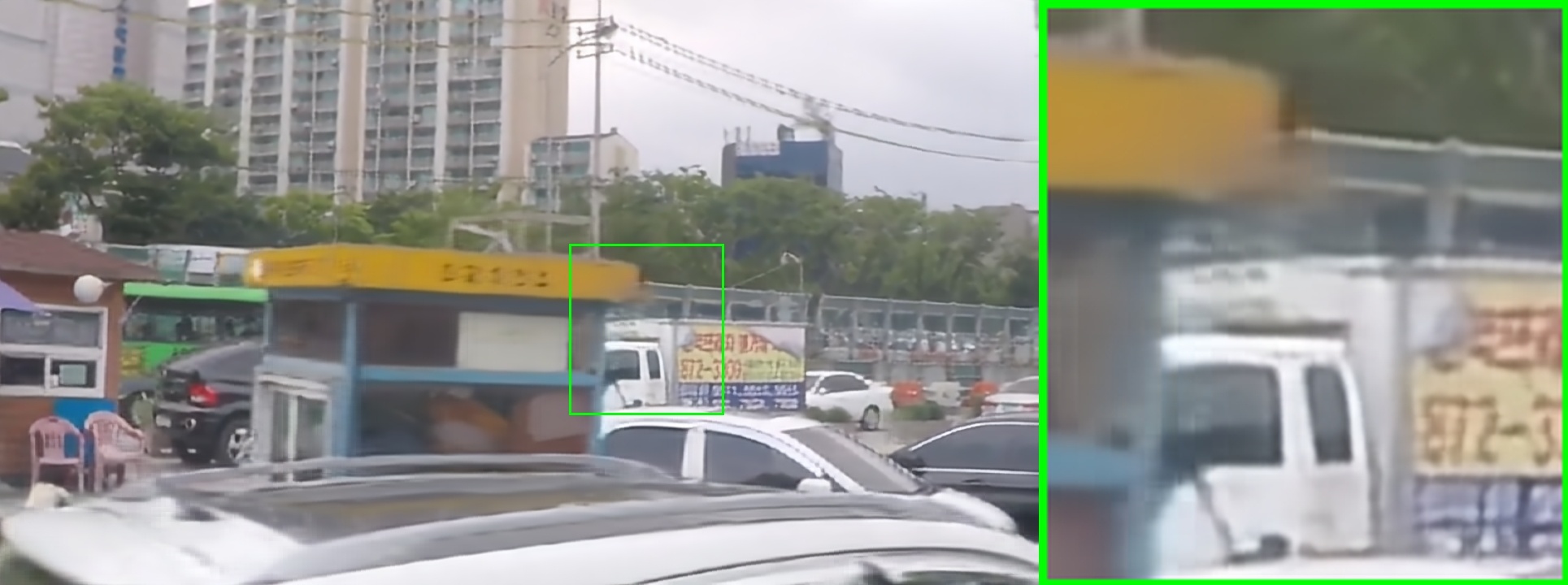}&
                \includegraphics[width=0.23\linewidth]{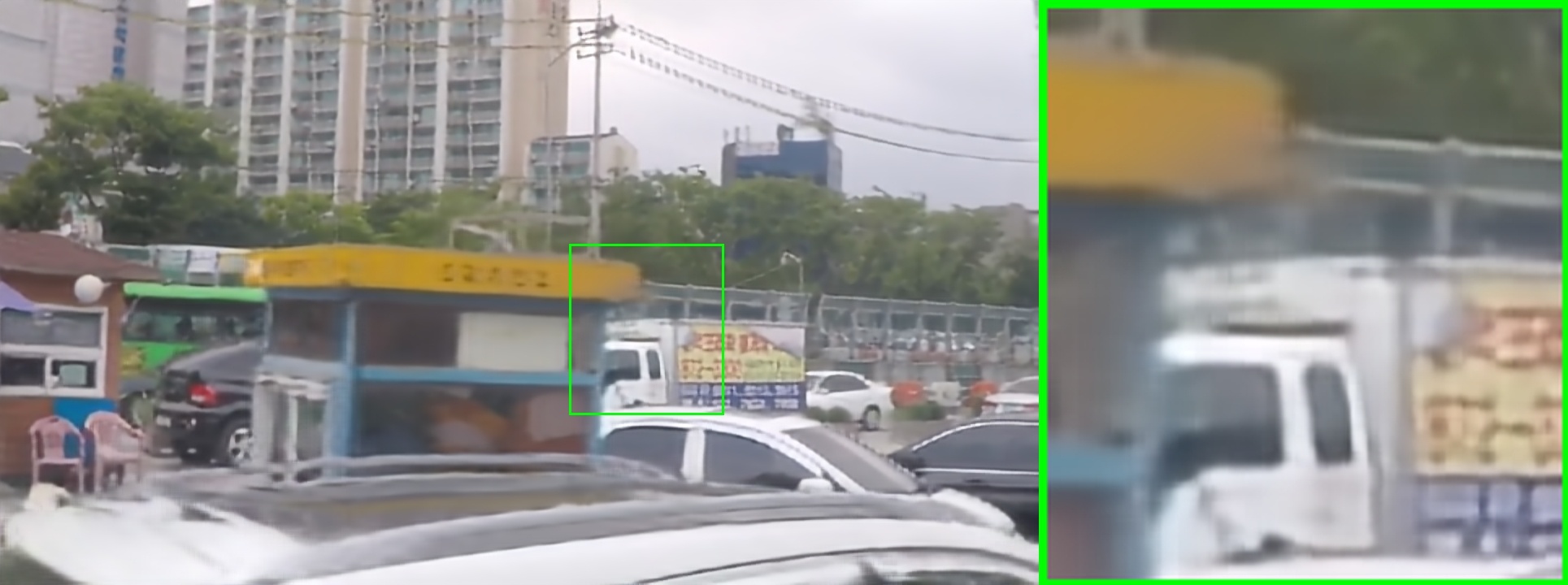}
            \end{tabular}
            \\ \\ \\
            \rotatebox[origin=c]{90}{\makecell{\footnotesize \(\mathcal{T}=0.500\)\\}}  &
            \begin{tabular}{c@{\hskip 0.005\linewidth}c@{\hskip 0.005\linewidth}c@{\hskip 0.005\linewidth}c}
                \includegraphics[width=0.23\linewidth]{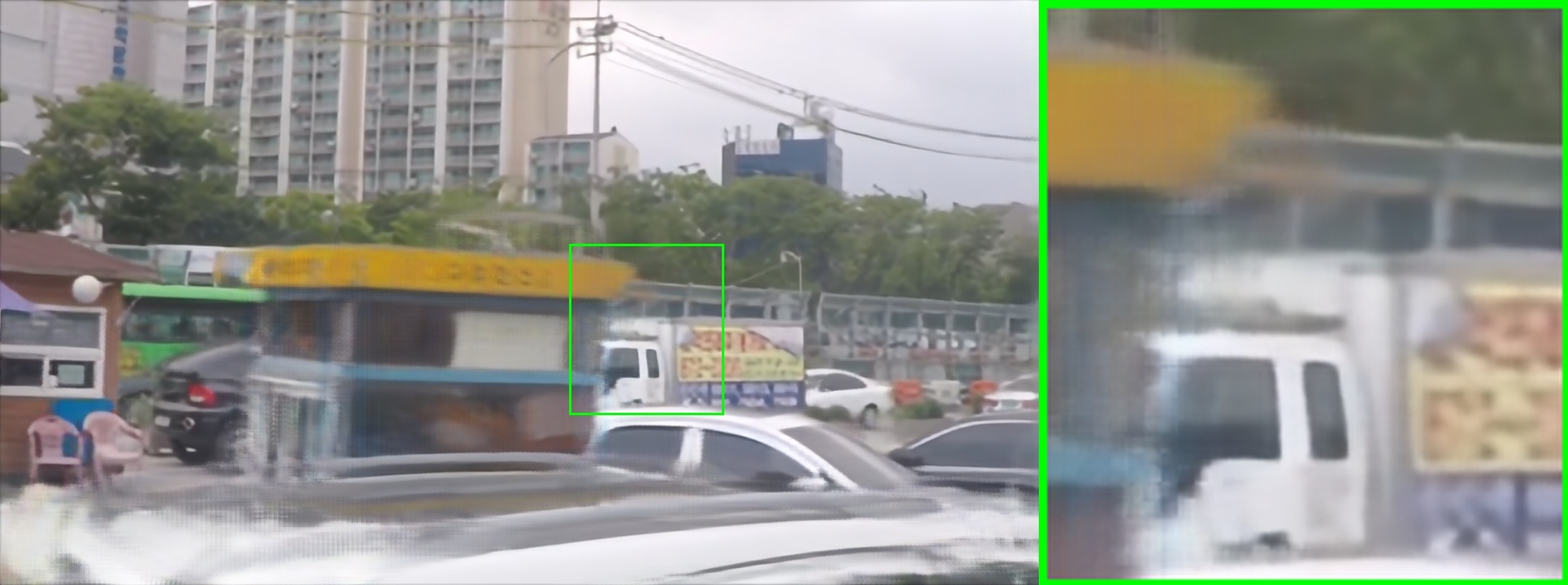}&
                \includegraphics[width=0.23\linewidth]{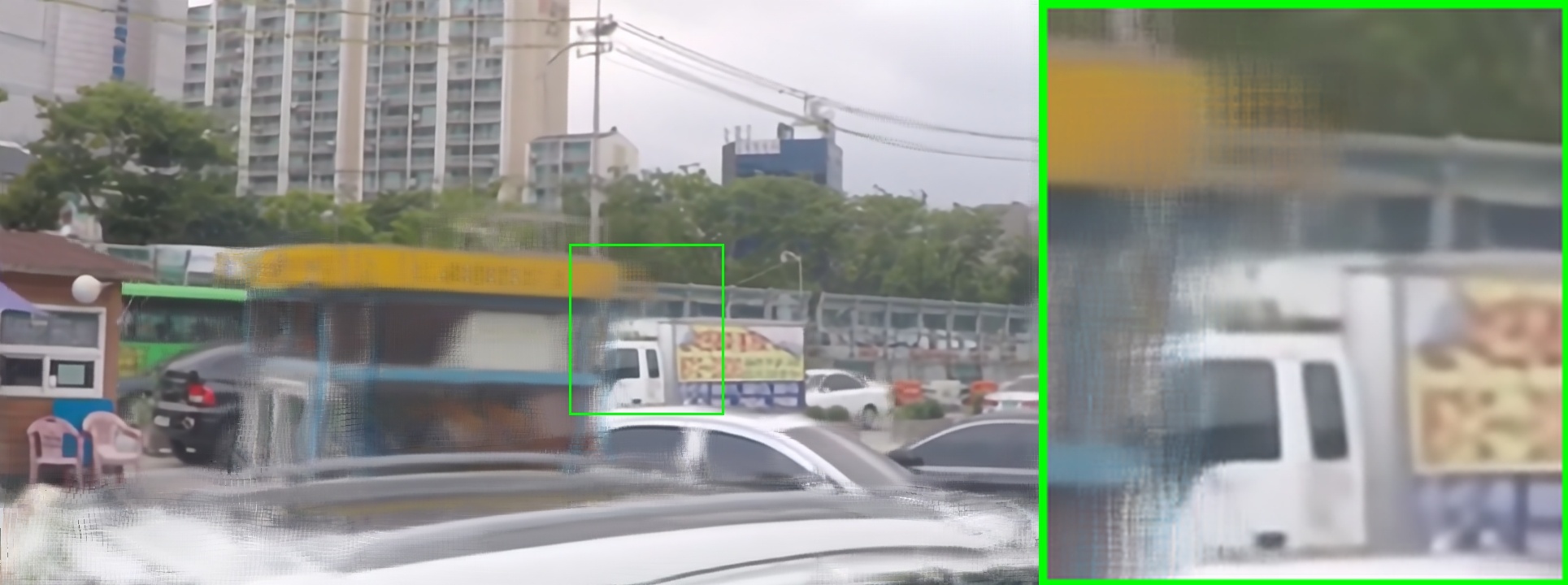}&
                \includegraphics[width=0.23\linewidth]{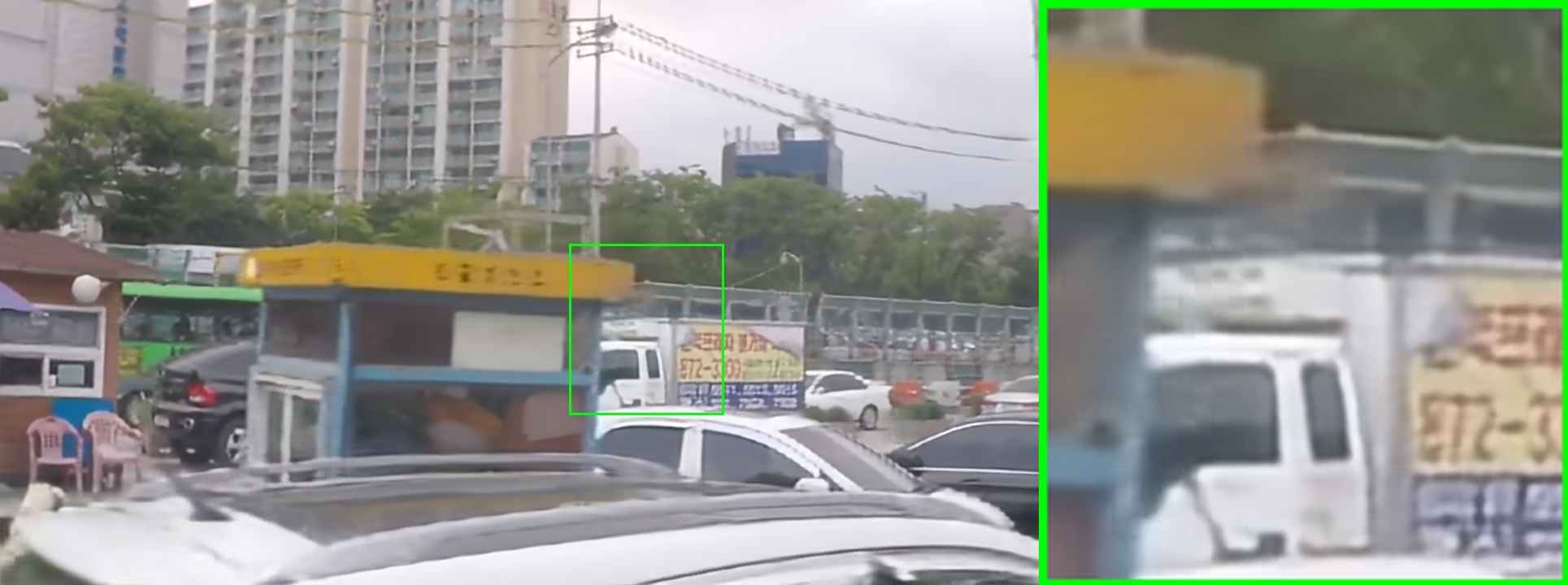}&
                \includegraphics[width=0.23\linewidth]{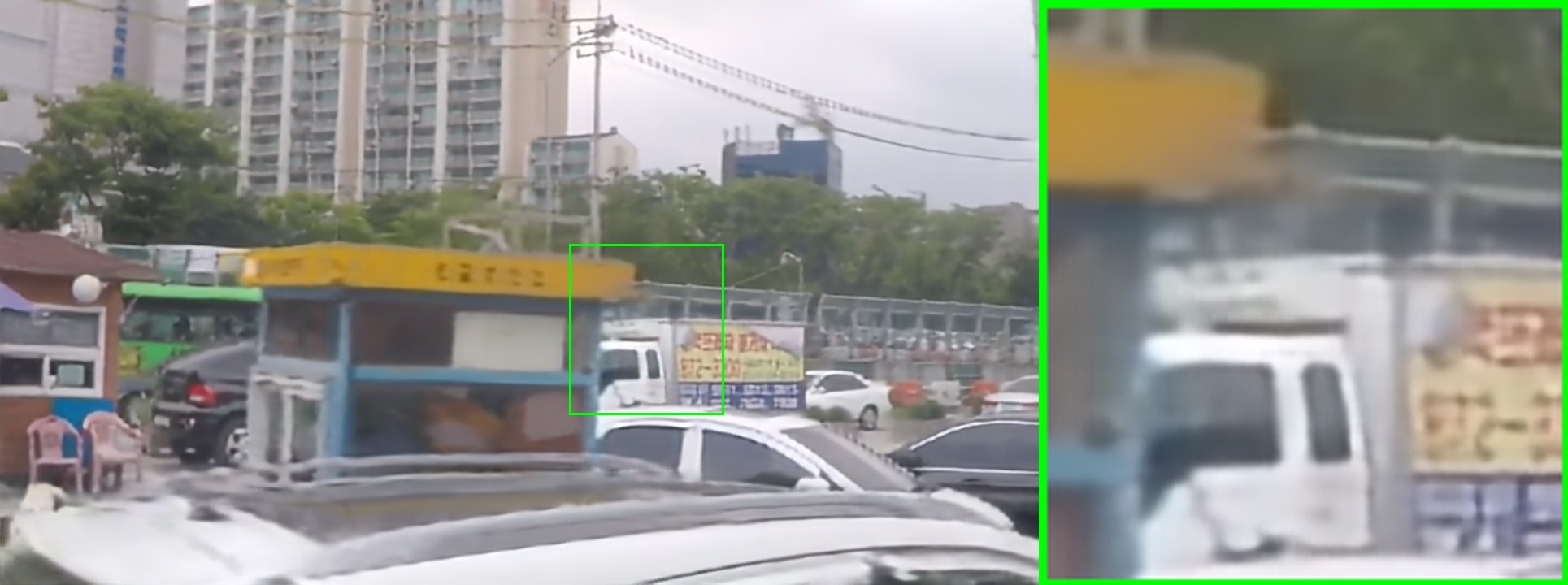}
            \end{tabular}
            \\ \\ \\
            \rotatebox[origin=c]{90}{\makecell{\footnotesize \(\mathcal{T}=0.583\)\\}}  &
            \begin{tabular}{c@{\hskip 0.005\linewidth}c@{\hskip 0.005\linewidth}c@{\hskip 0.005\linewidth}c}
                \includegraphics[width=0.23\linewidth]{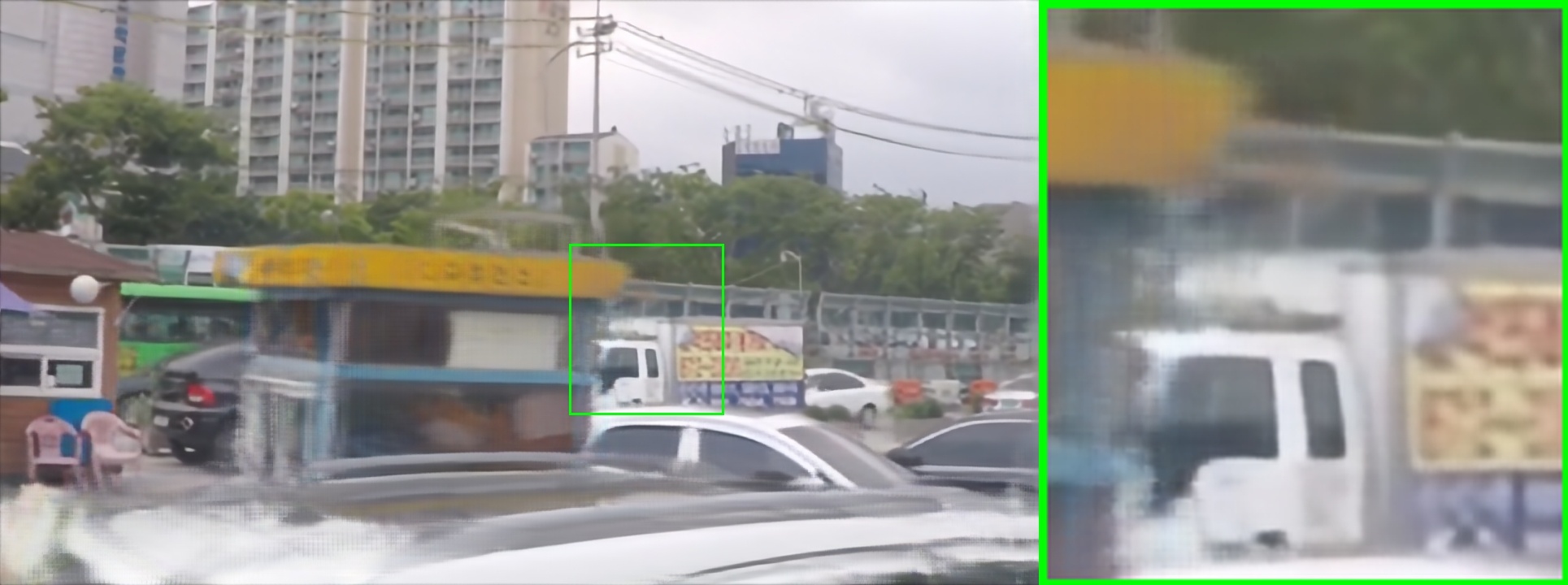}&
                \includegraphics[width=0.23\linewidth]{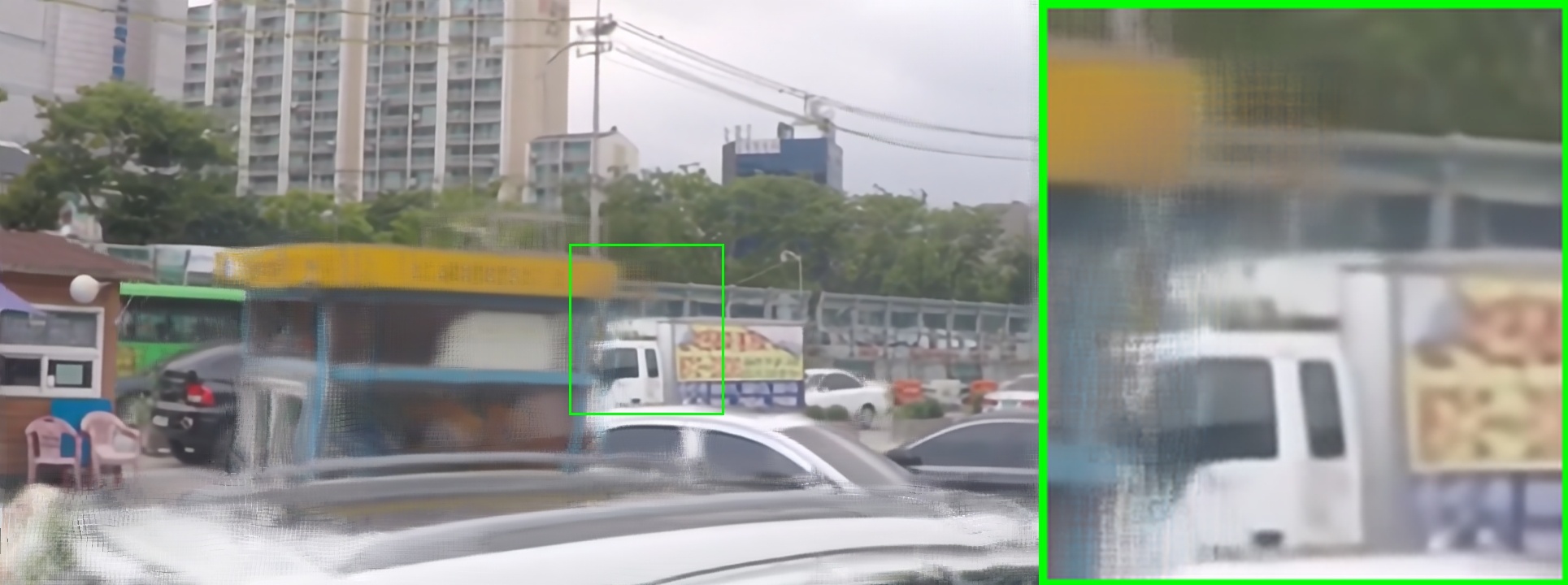}&
                \includegraphics[width=0.23\linewidth]{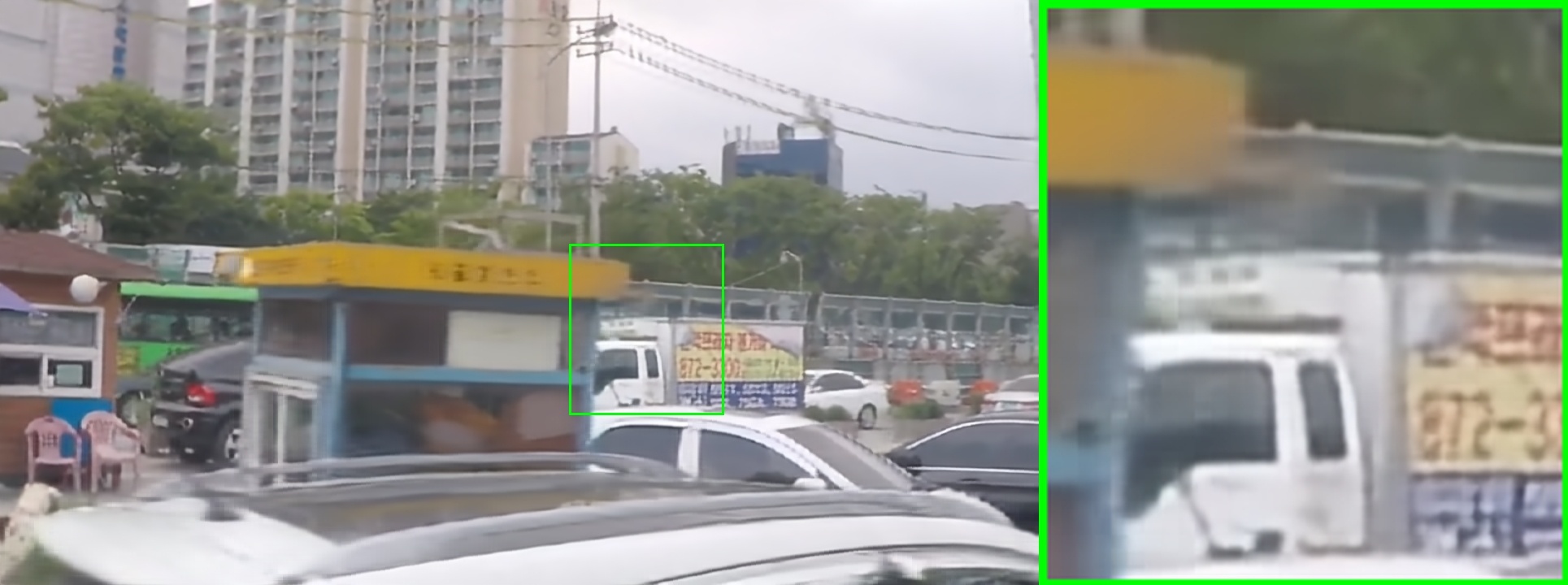}&
                \includegraphics[width=0.23\linewidth]{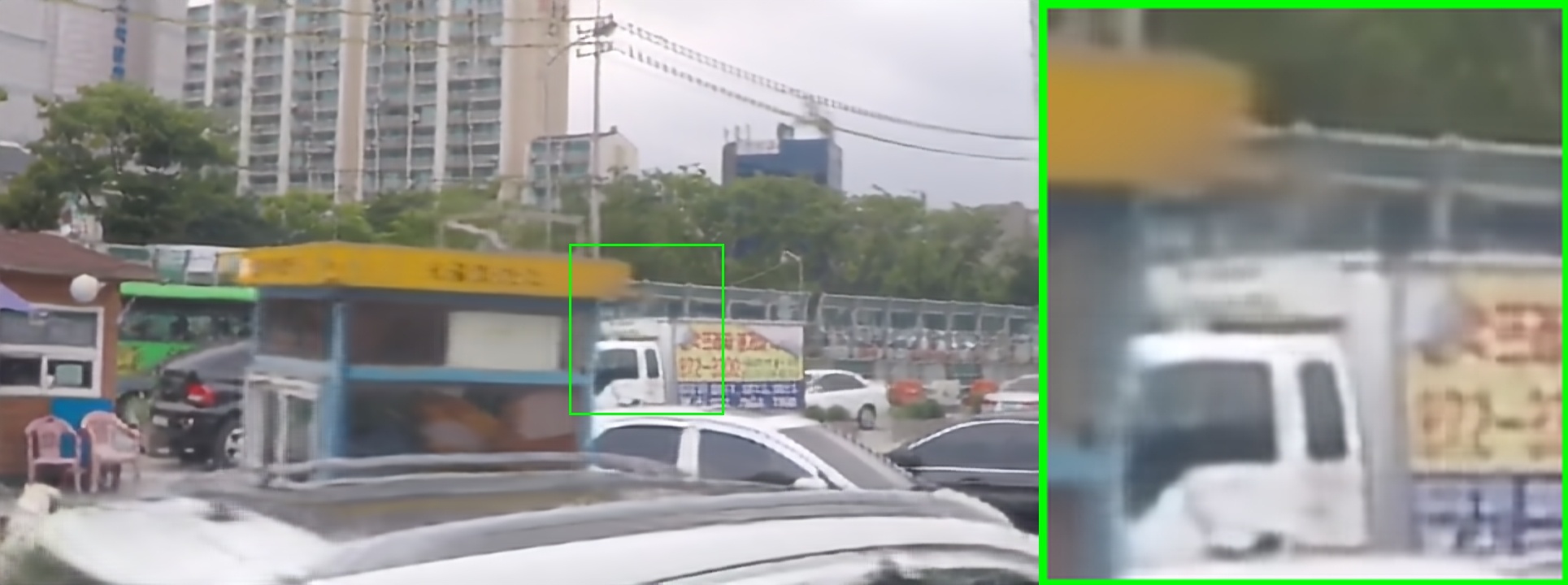}
            \end{tabular}
            \\ \\ \\
            \rotatebox[origin=c]{90}{\makecell{\footnotesize \(\mathcal{T}=0.667\)\\}}  &
            \begin{tabular}{c@{\hskip 0.005\linewidth}c@{\hskip 0.005\linewidth}c@{\hskip 0.005\linewidth}c}
                \includegraphics[width=0.23\linewidth]{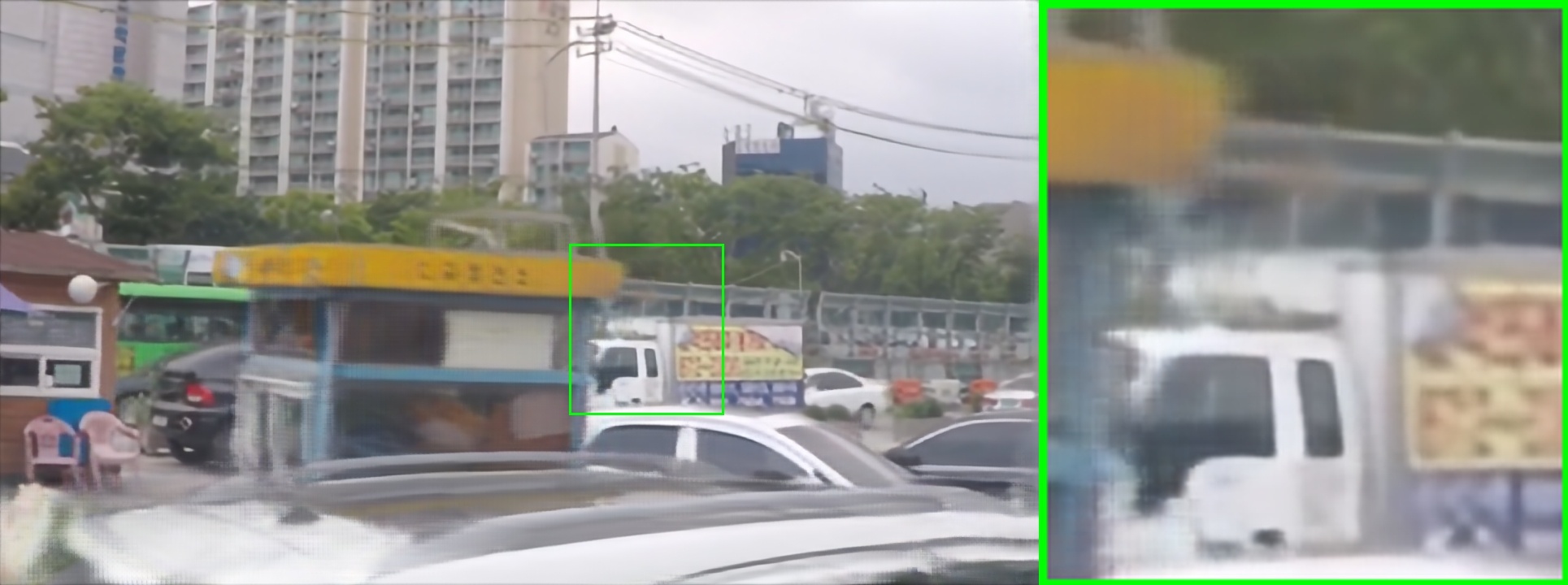}&
                \includegraphics[width=0.23\linewidth]{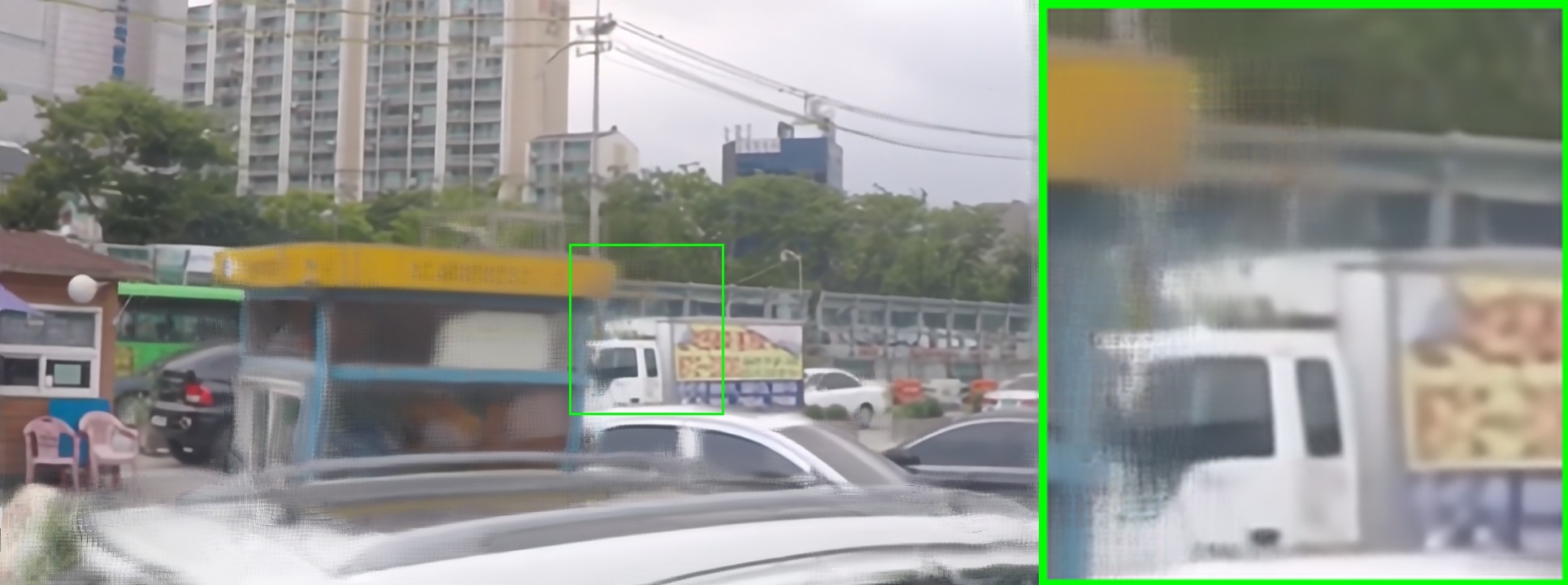}&
                \includegraphics[width=0.23\linewidth]{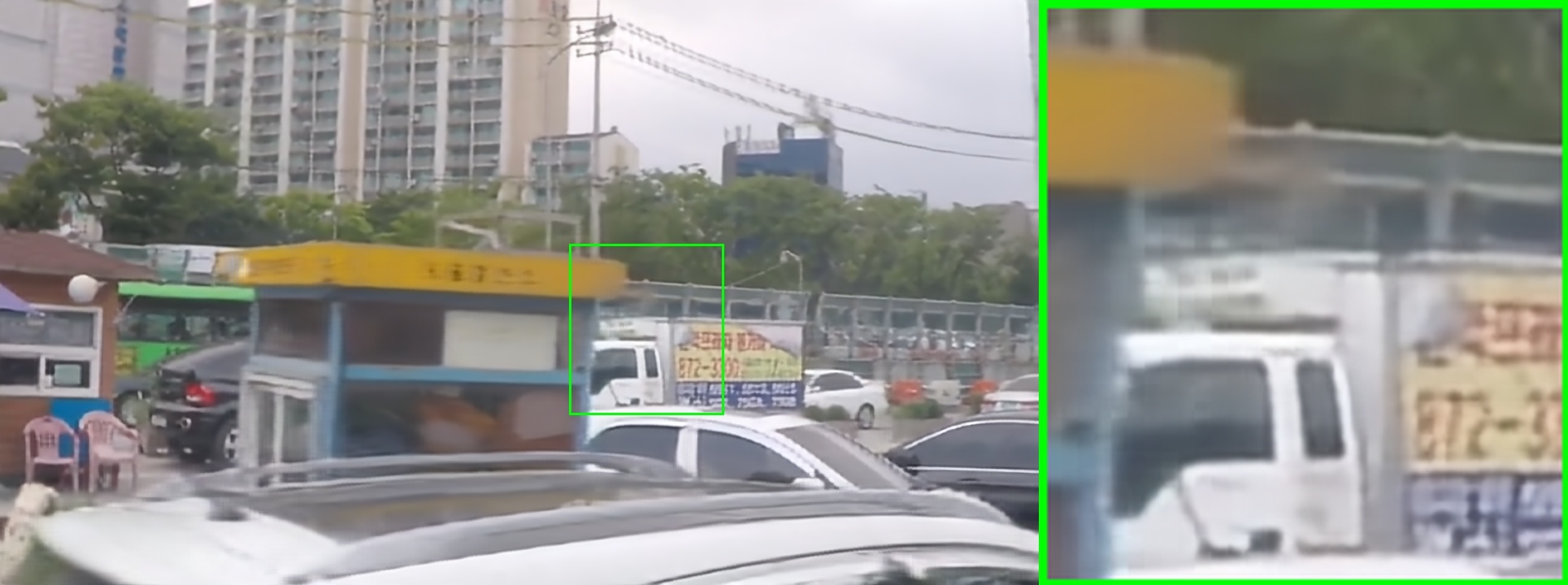}&
                \includegraphics[width=0.23\linewidth]{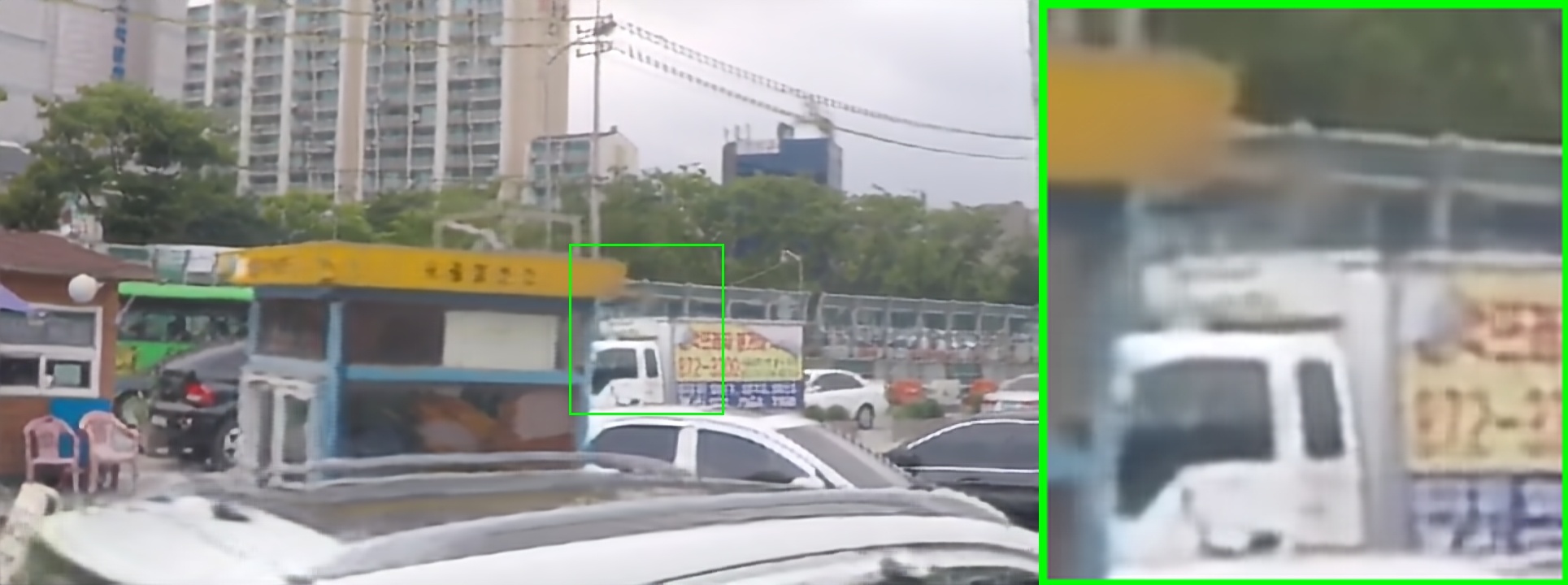}
            \end{tabular}
            \\ \\ \\
            \rotatebox[origin=c]{90}{\makecell{\footnotesize \(\mathcal{T}=0.750\)\\}}  &
            \begin{tabular}{c@{\hskip 0.005\linewidth}c@{\hskip 0.005\linewidth}c@{\hskip 0.005\linewidth}c}
                \includegraphics[width=0.23\linewidth]{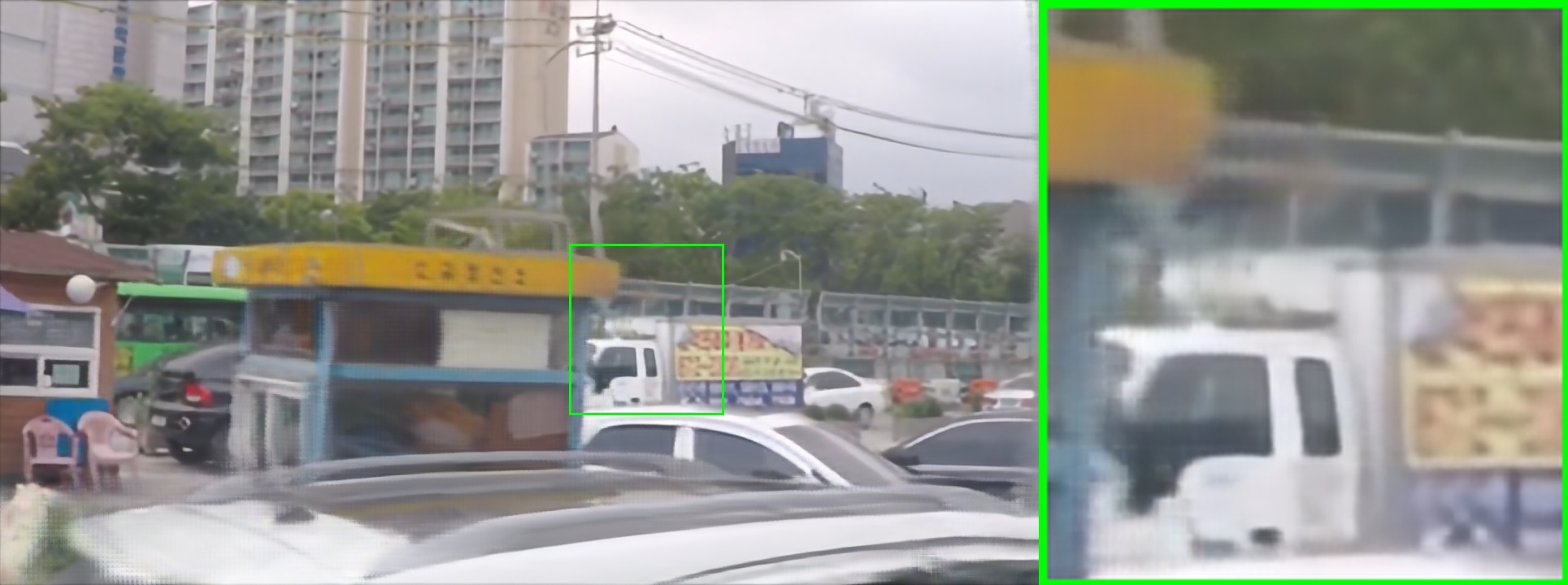}&
                \includegraphics[width=0.23\linewidth]{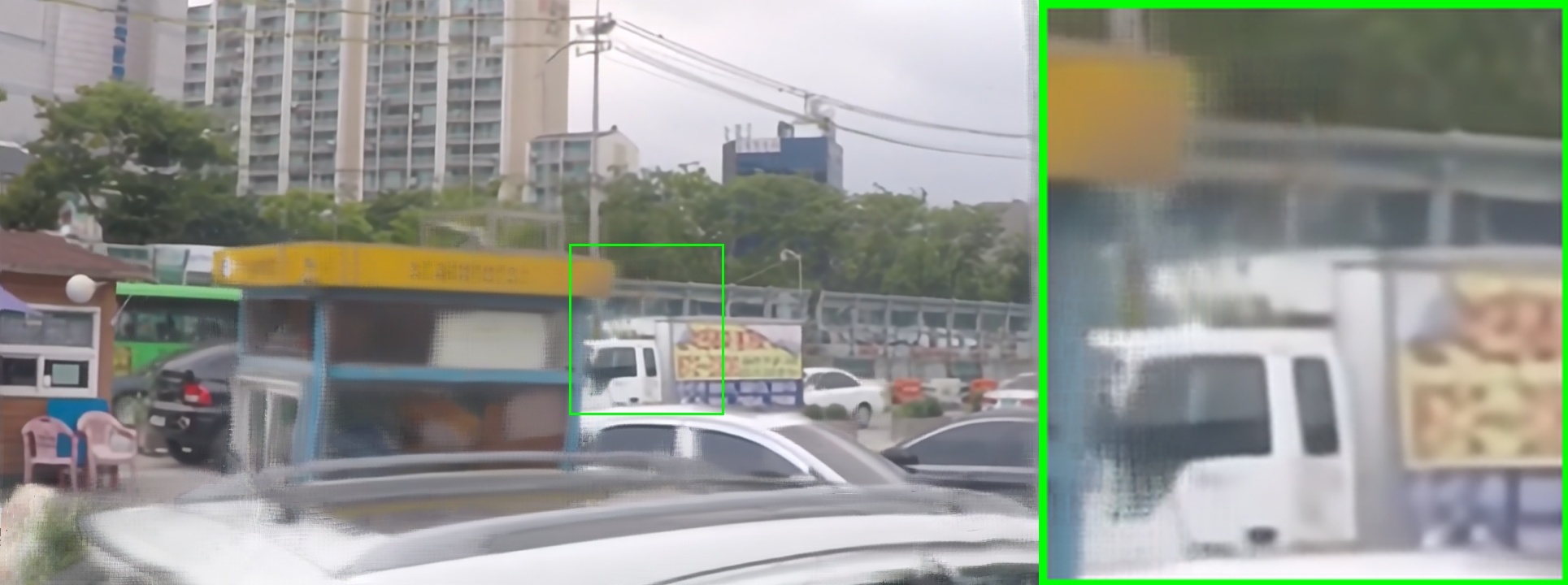}&
                \includegraphics[width=0.23\linewidth]{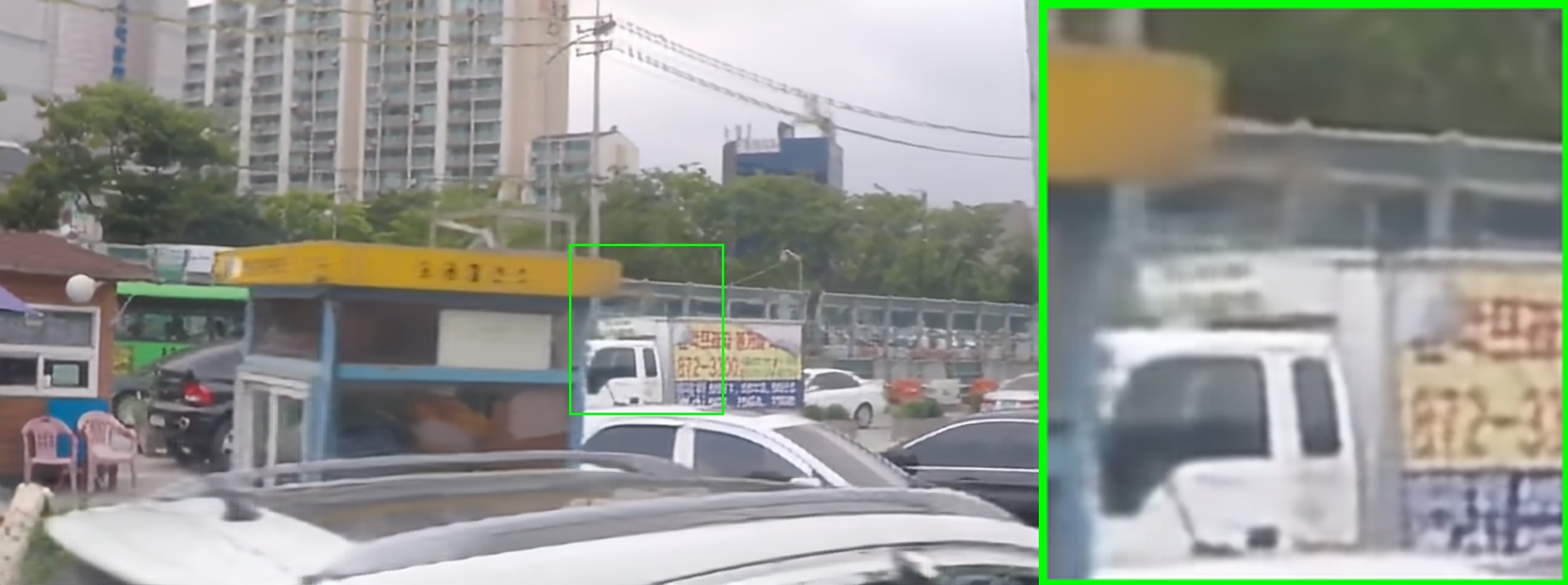}&
                \includegraphics[width=0.23\linewidth]{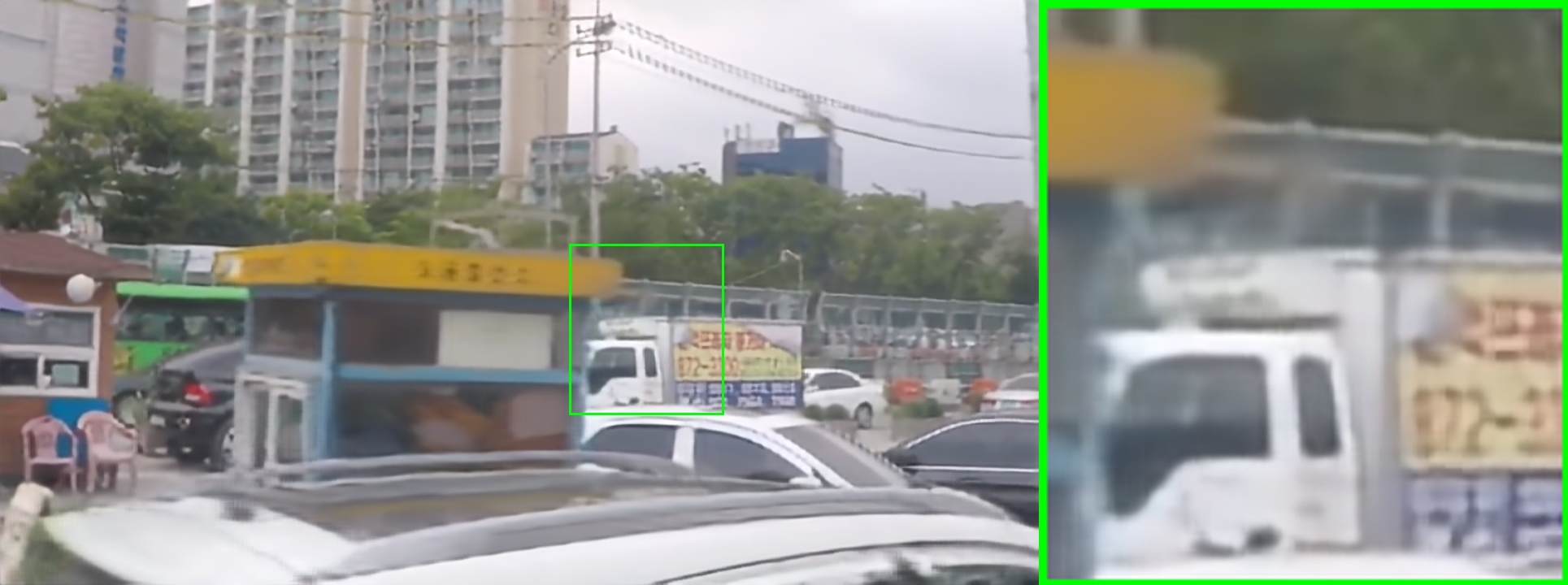}
            \end{tabular}
            \\ \\ \\
            \rotatebox[origin=c]{90}{\makecell{\footnotesize \(\mathcal{T}=0.833\)\\}}  &
            \begin{tabular}{c@{\hskip 0.005\linewidth}c@{\hskip 0.005\linewidth}c@{\hskip 0.005\linewidth}c}
                \includegraphics[width=0.23\linewidth]{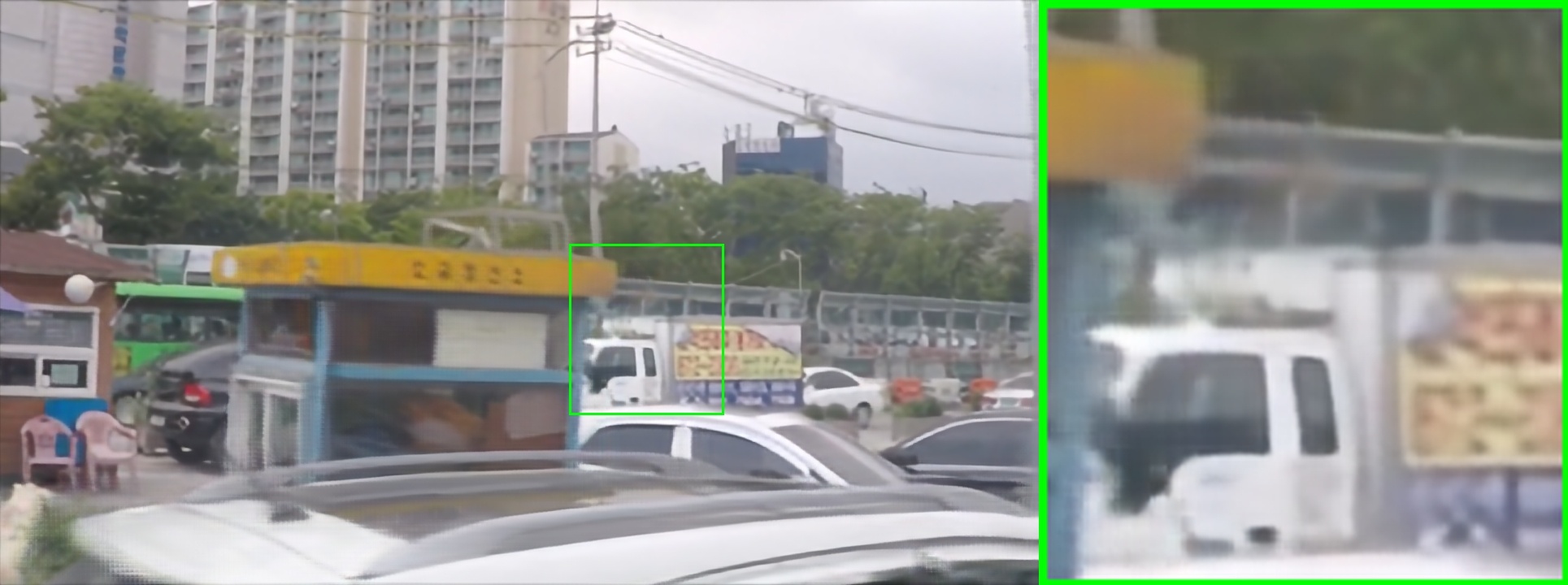}&
                \includegraphics[width=0.23\linewidth]{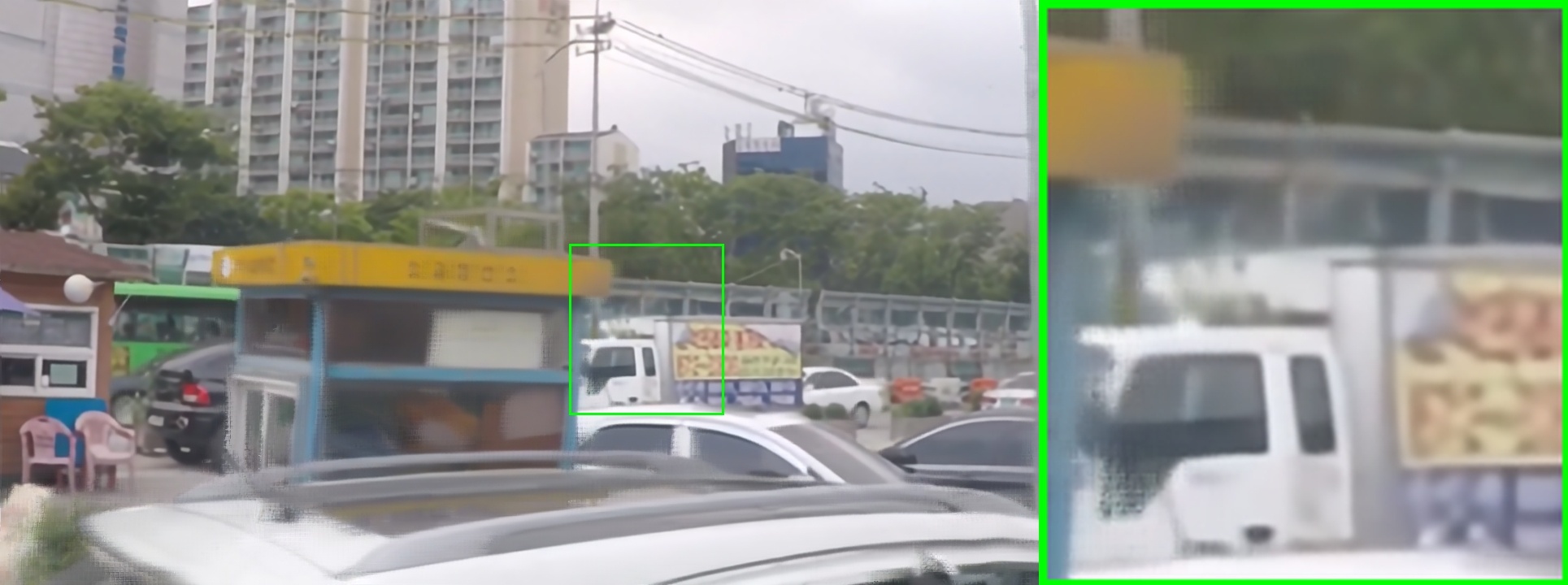}&
                \includegraphics[width=0.23\linewidth]{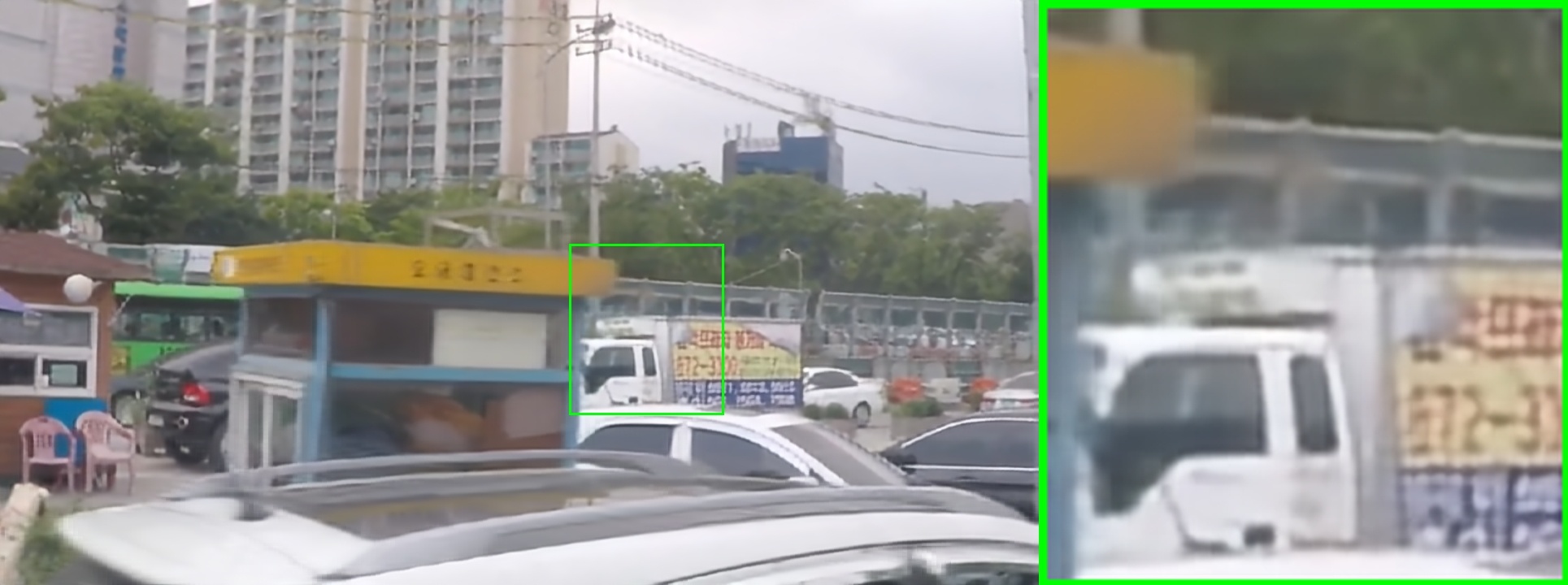}&
                \includegraphics[width=0.23\linewidth]{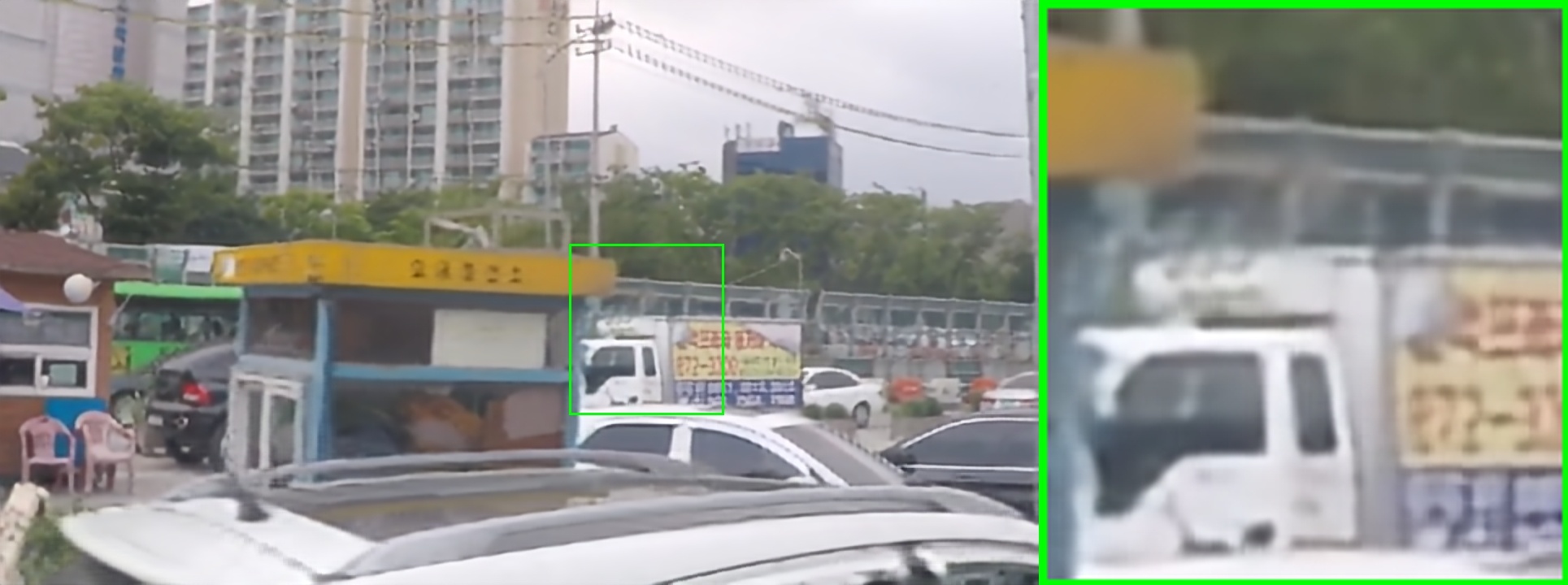}
            \end{tabular}
            \\ \\ \\
            \rotatebox[origin=c]{90}{\makecell{\footnotesize \(\mathcal{T}=0.917\)\\}}  &
            \begin{tabular}{c@{\hskip 0.005\linewidth}c@{\hskip 0.005\linewidth}c@{\hskip 0.005\linewidth}c}
                \includegraphics[width=0.23\linewidth]{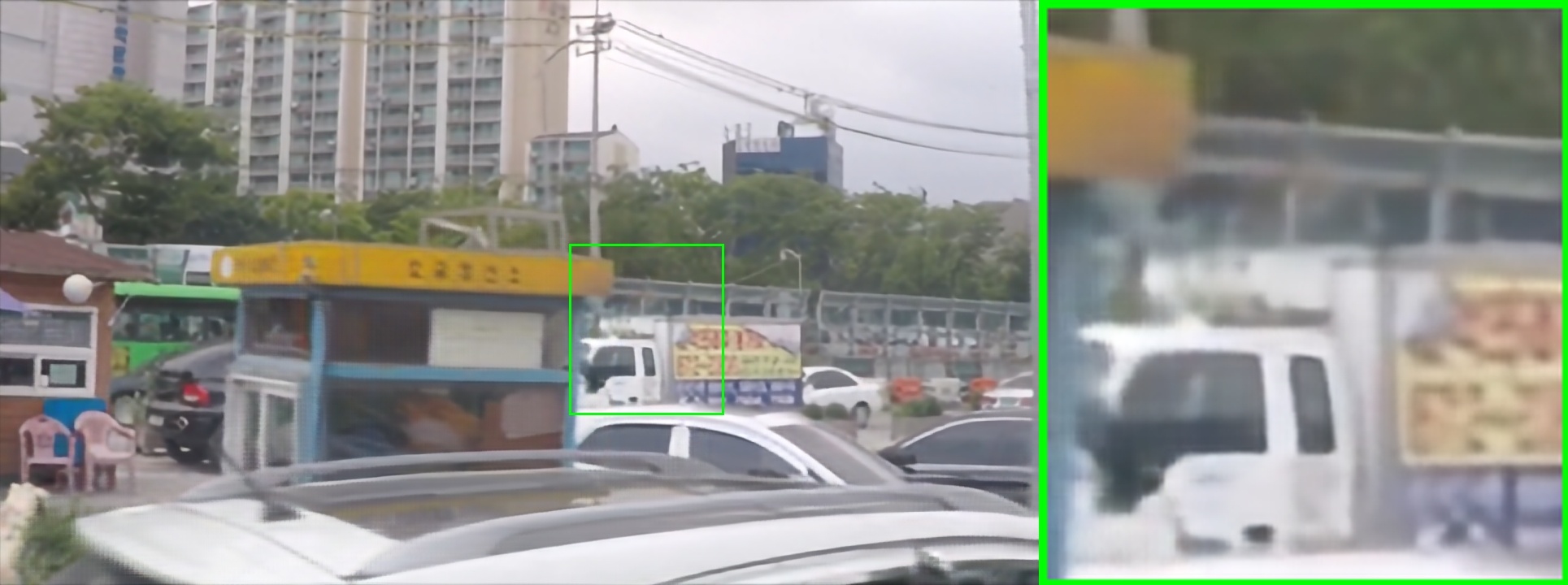}&
                \includegraphics[width=0.23\linewidth]{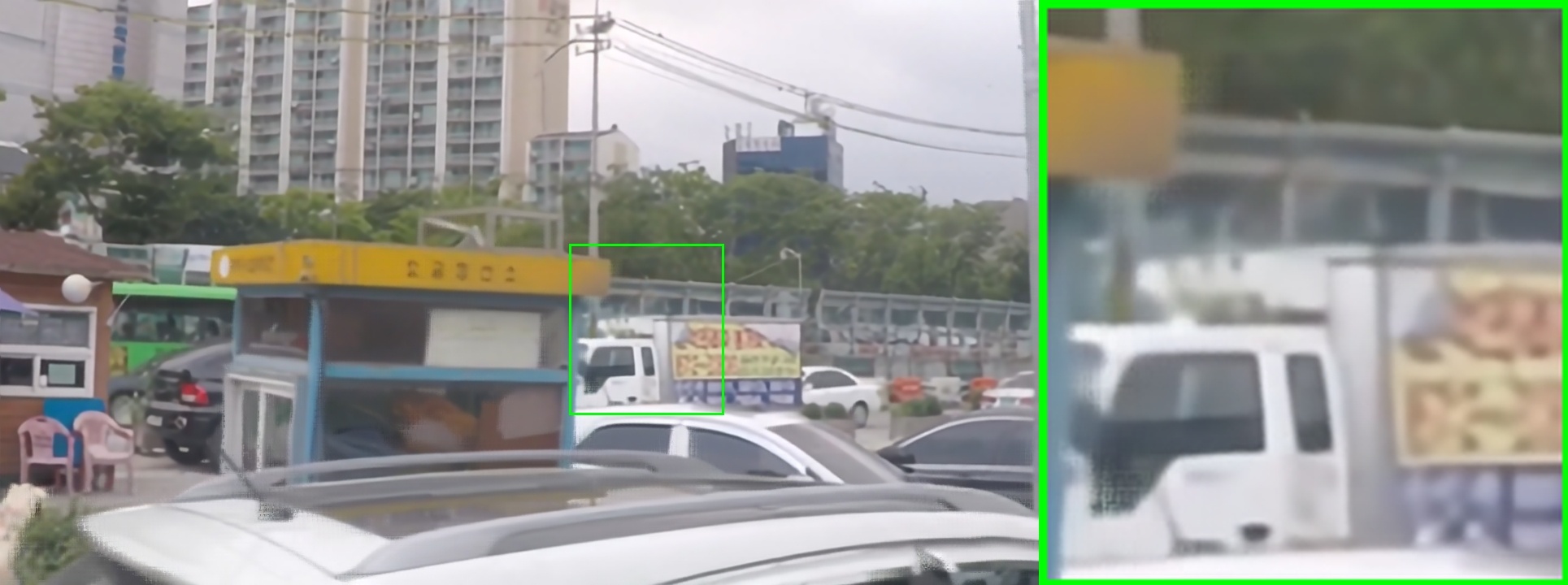}&
                \includegraphics[width=0.23\linewidth]{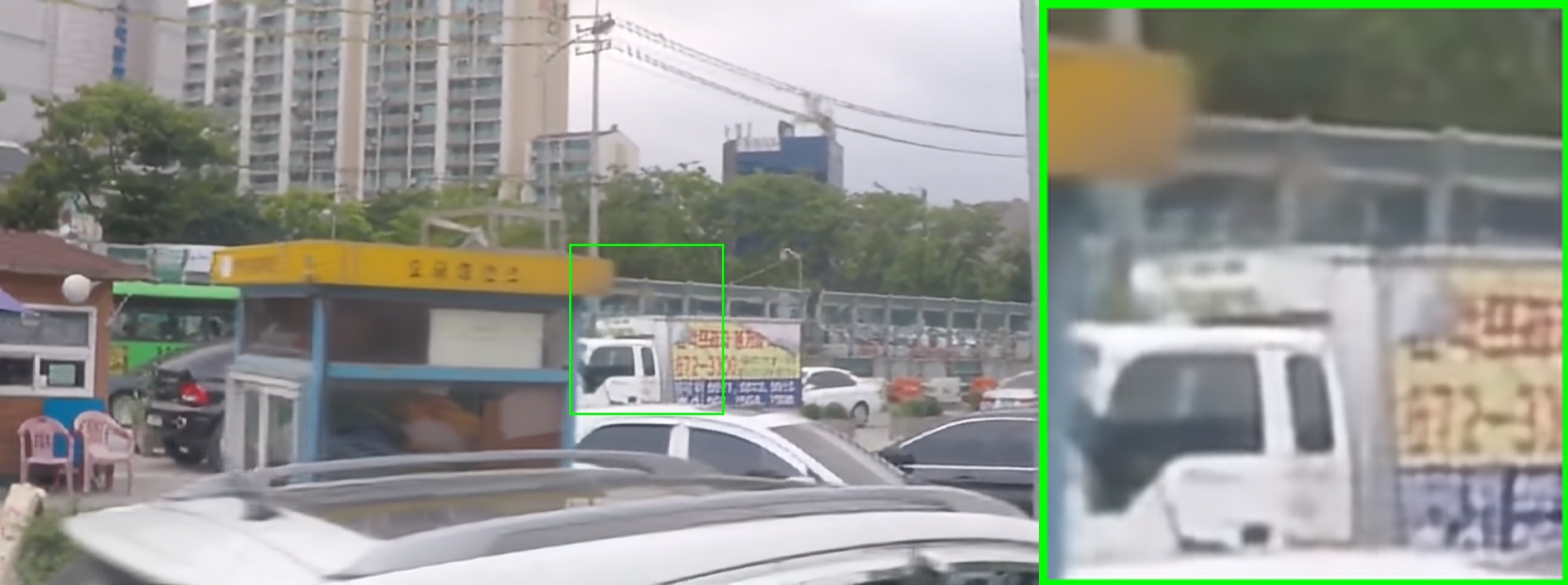}&
                \includegraphics[width=0.23\linewidth]{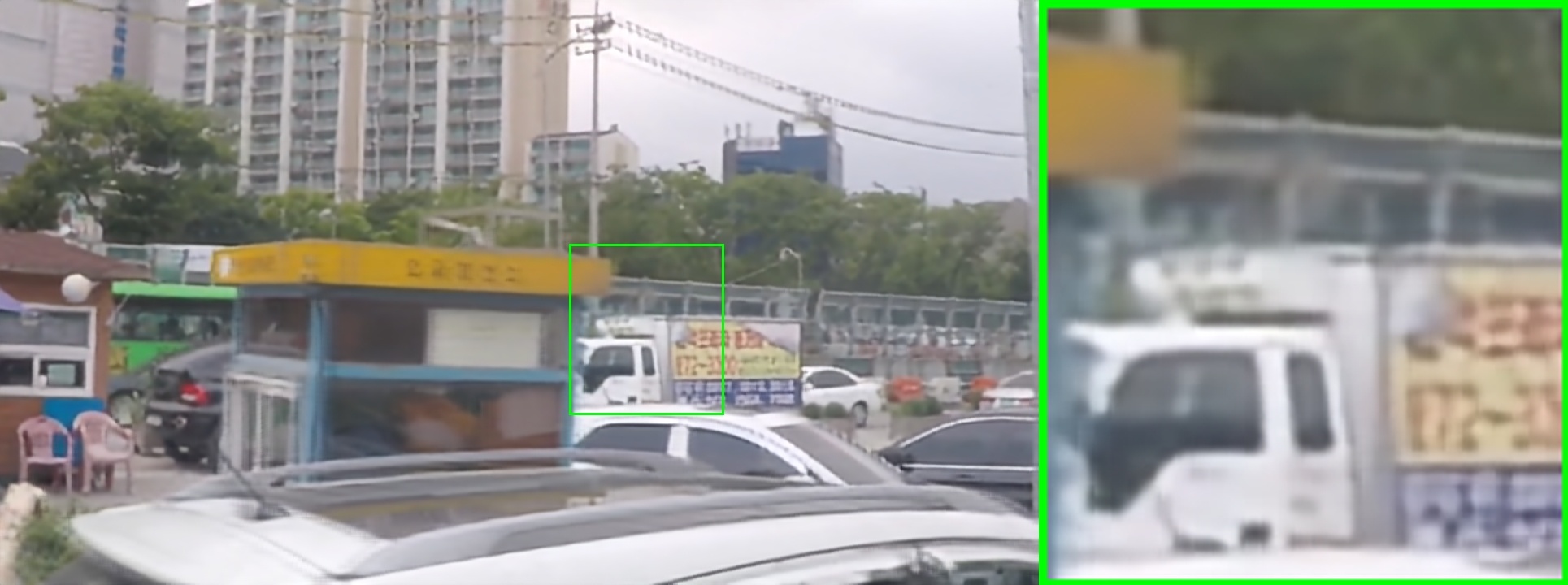}
            \end{tabular}
            \\ \\ \\
            \begin{tabular}{c}
            \rotatebox[origin=c]{90}{\makecell{\footnotesize \(\mathcal{T}=1\)\\}}
            \\ \\ \\ \\ \\ \\ \\ \\ \\ \\
            \end{tabular}
            &
            \begin{tabular}{c@{\hskip 0.005\linewidth}c@{\hskip 0.005\linewidth}c@{\hskip 0.005\linewidth}c}
                \includegraphics[width=0.23\linewidth]{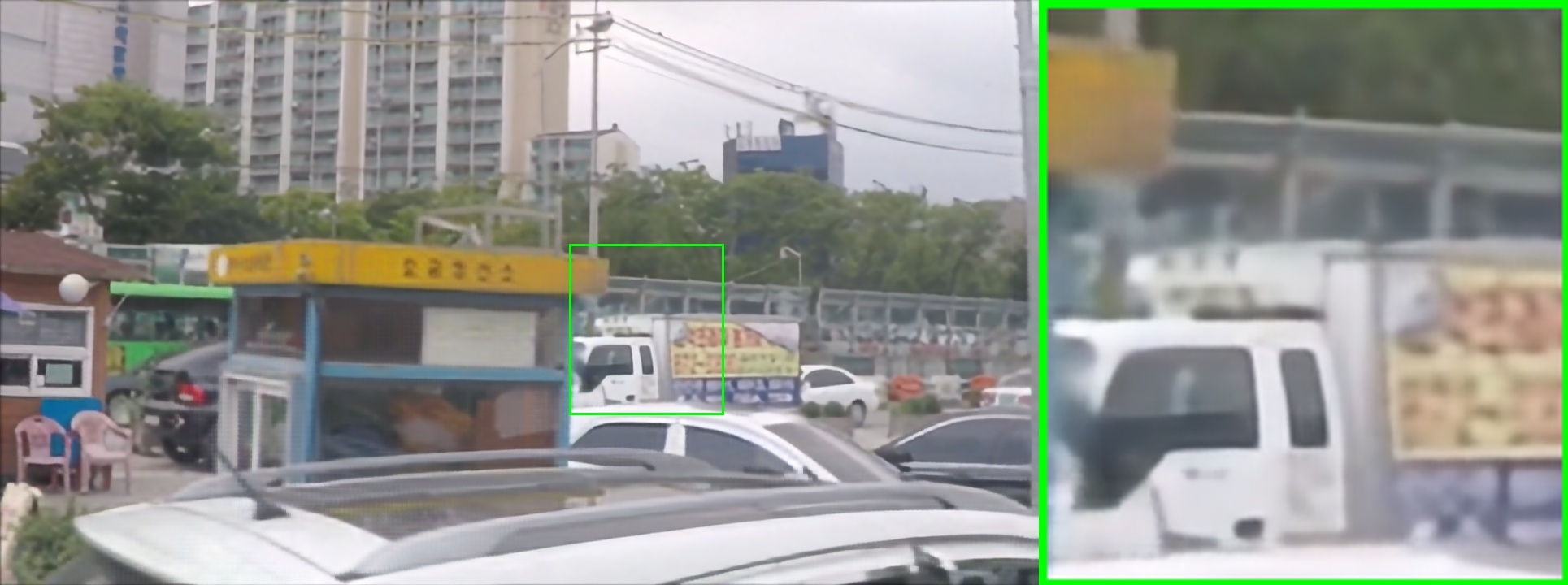}&
                \includegraphics[width=0.23\linewidth]{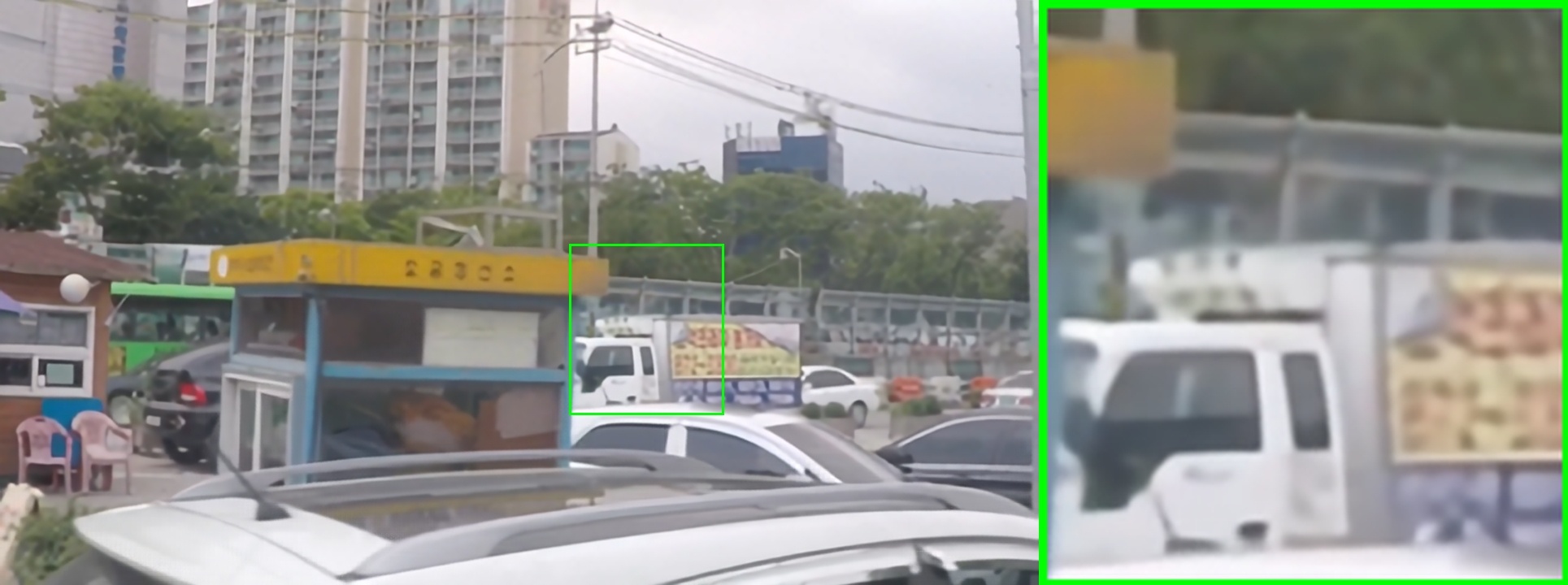}&
                \includegraphics[width=0.23\linewidth]{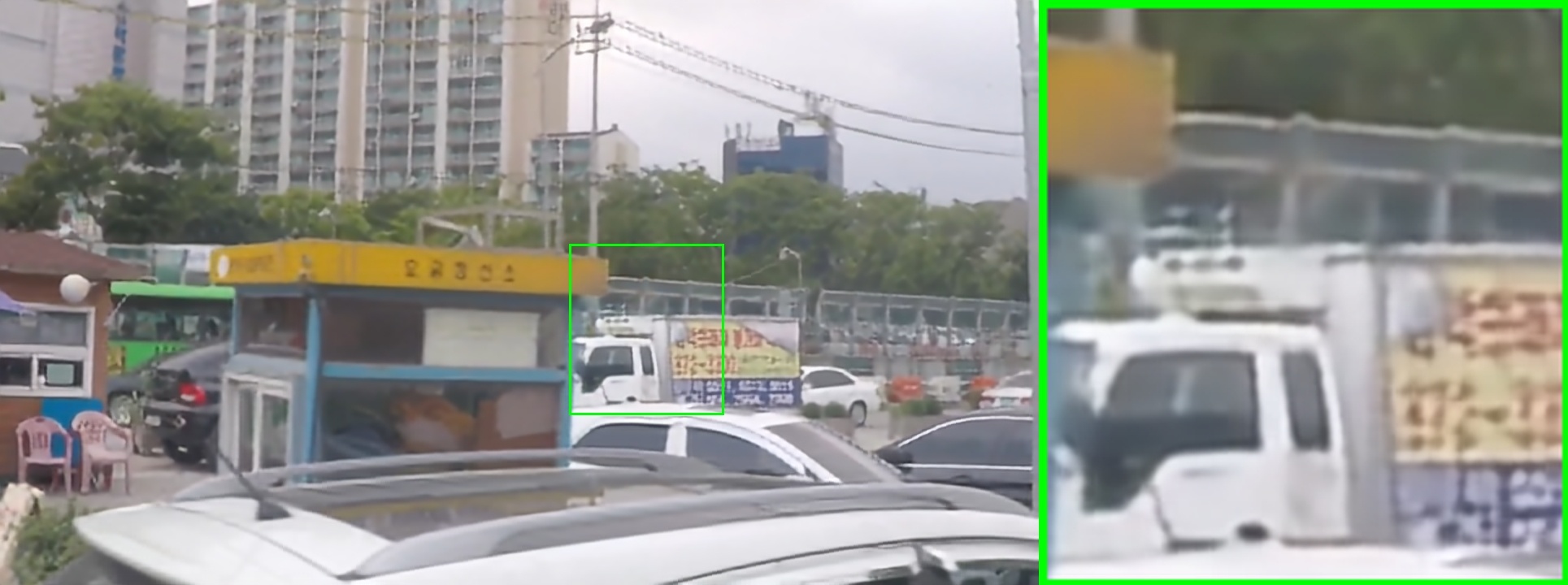}&
                \includegraphics[width=0.23\linewidth]{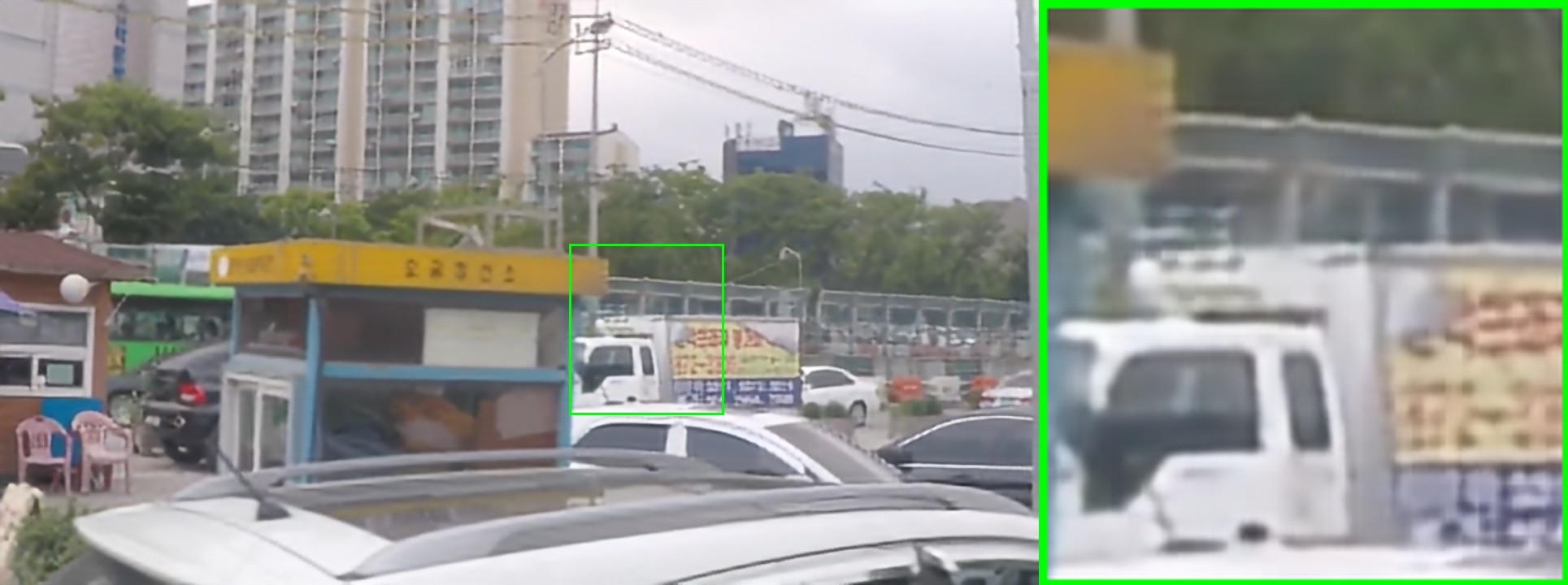}
            \\ \\
            VideoINR~\cite{chen2022videoinr}&
            MoTIF~\cite{chen2023motif}&
            \textbf{EvEnhancer (Ours)}&
            \textbf{EvEnhancer-light (Ours)}
            \end{tabular}
        \end{tabular}
    \caption{
       Qualitative comparison for OOD scale (\(t=12,s=6\)) on the GoPro dataset \cite{nah2017deep}. Best zoom in for better visualization.
    }
    \label{fig:s_gopro_t12s6}
\end{figure*}
\renewcommand{\arraystretch}{1.}

%% file: fig/exp/exp_suppl_gopro_t6sx.tex
\begin{figure*}[t]
    \centering\
        \renewcommand{\arraystretch}{0.5}
         \begin{tabular}{c@{\hskip 0.01\linewidth}c}
            \renewcommand{\arraystretch}{0}
            \begin{tabular}{c@{\hskip 0.005\linewidth}c@{\hskip 0.005\linewidth}c}
                \multicolumn{3}{c}{VideoINR~\cite{chen2022videoinr}}\\
                \includegraphics[width=0.065\linewidth]{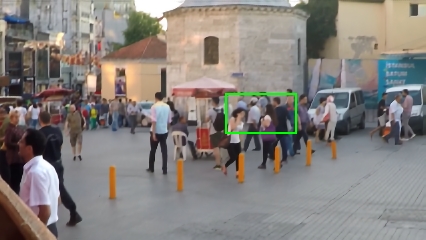}&
                \includegraphics[width=0.13\linewidth]{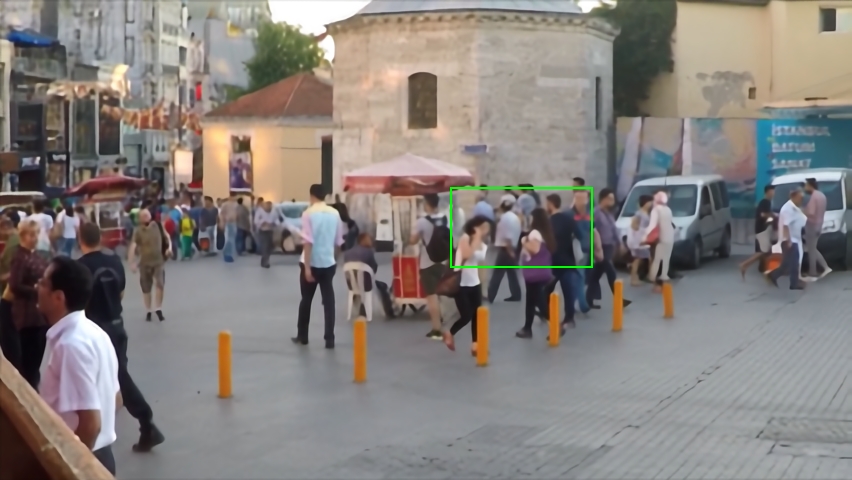}&
                \includegraphics[width=0.26\linewidth]{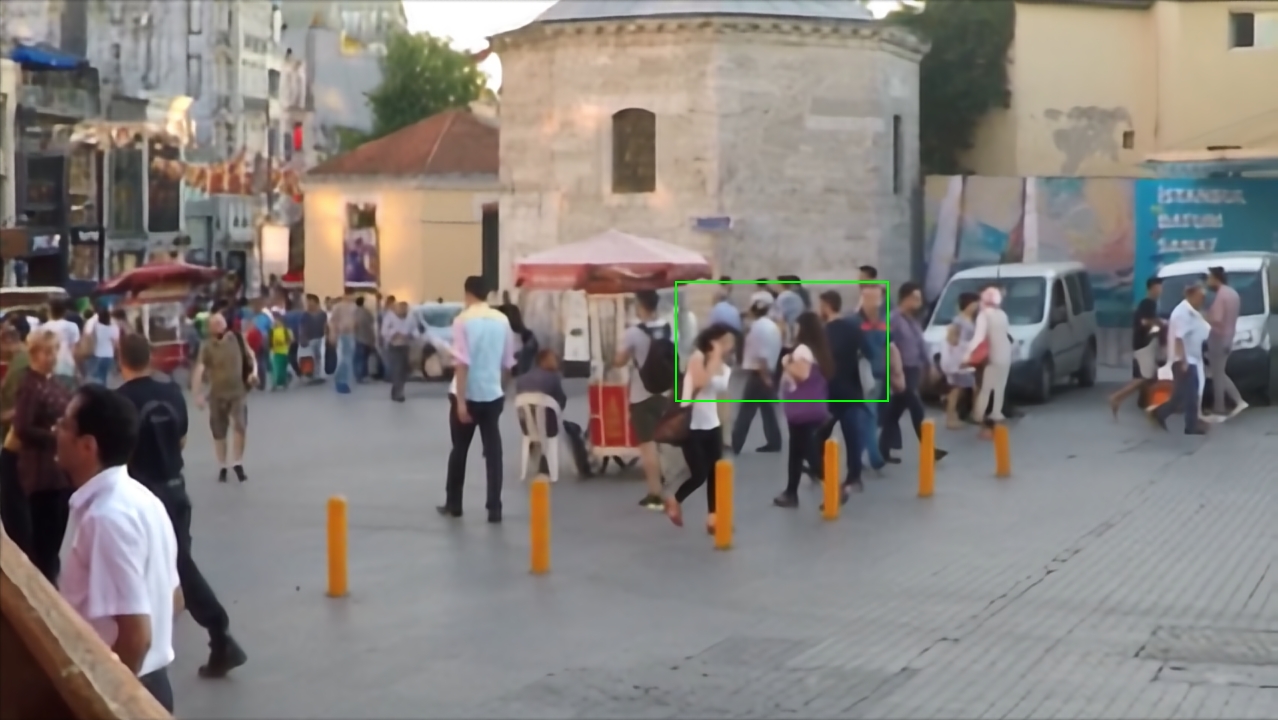} \\
                \includegraphics[width=0.065\linewidth]{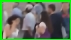}&
                \includegraphics[width=0.13\linewidth]{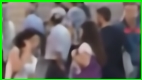}&
                \includegraphics[width=0.26\linewidth]{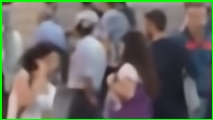}
            \\ \\
            \(s=2\)&
            \(s=4\)&
            \(s=6\)
            \end{tabular}
            &
            \renewcommand{\arraystretch}{0}
            \begin{tabular}{c@{\hskip 0.005\linewidth}c@{\hskip 0.005\linewidth}c}
                \multicolumn{3}{c}{MoTIF~\cite{chen2023motif}}\\
                \includegraphics[width=0.065\linewidth]{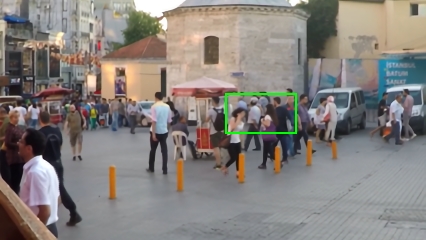}&
                \includegraphics[width=0.13\linewidth]{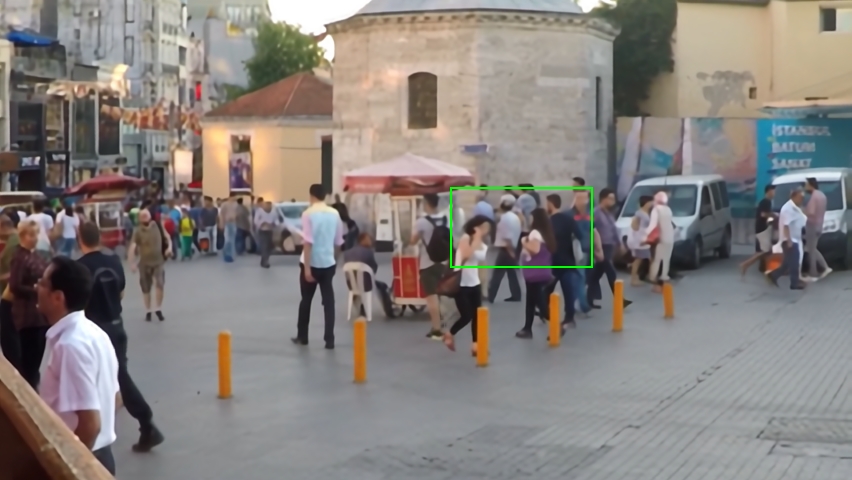}&
                \includegraphics[width=0.26\linewidth]{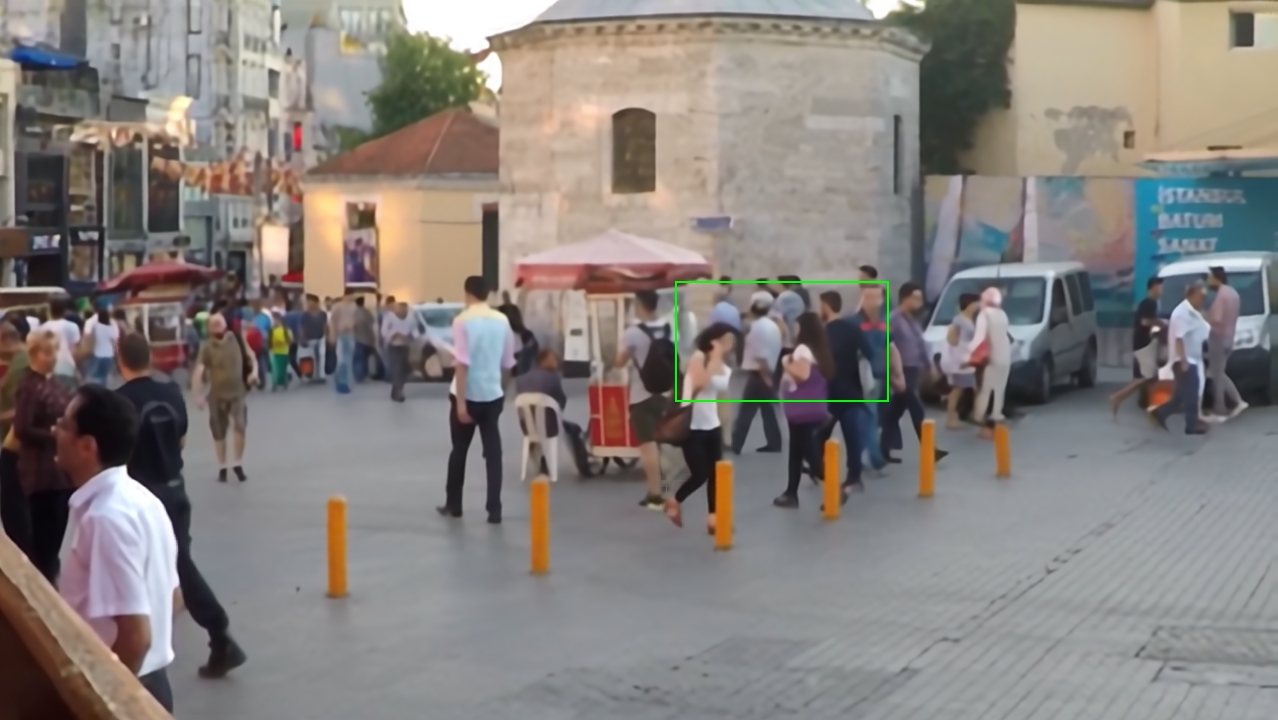} \\
                \includegraphics[width=0.065\linewidth]{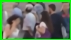}&
                \includegraphics[width=0.13\linewidth]{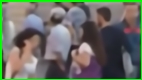}&
                \includegraphics[width=0.26\linewidth]{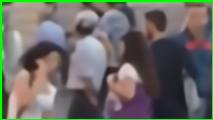}
            \\ \\
            \(s=2\)&
            \(s=4\)&
            \(s=6\)
            \end{tabular}
            \\ \\ \\
            \renewcommand{\arraystretch}{0}
            \begin{tabular}{c@{\hskip 0.005\linewidth}c@{\hskip 0.005\linewidth}c}
                \multicolumn{3}{c}{\textbf{EvEnhancer (Ours)}}\\
                \includegraphics[width=0.065\linewidth]{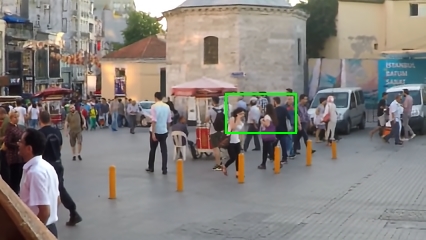}&
                \includegraphics[width=0.13\linewidth]{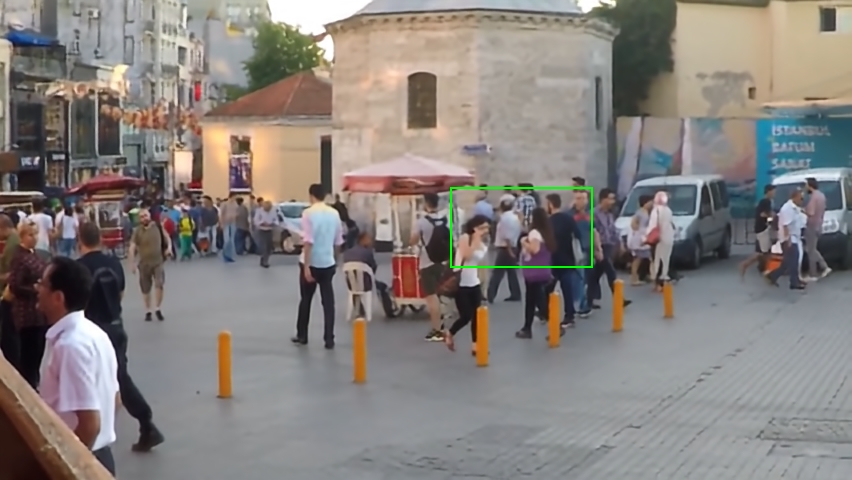}&
                \includegraphics[width=0.26\linewidth]{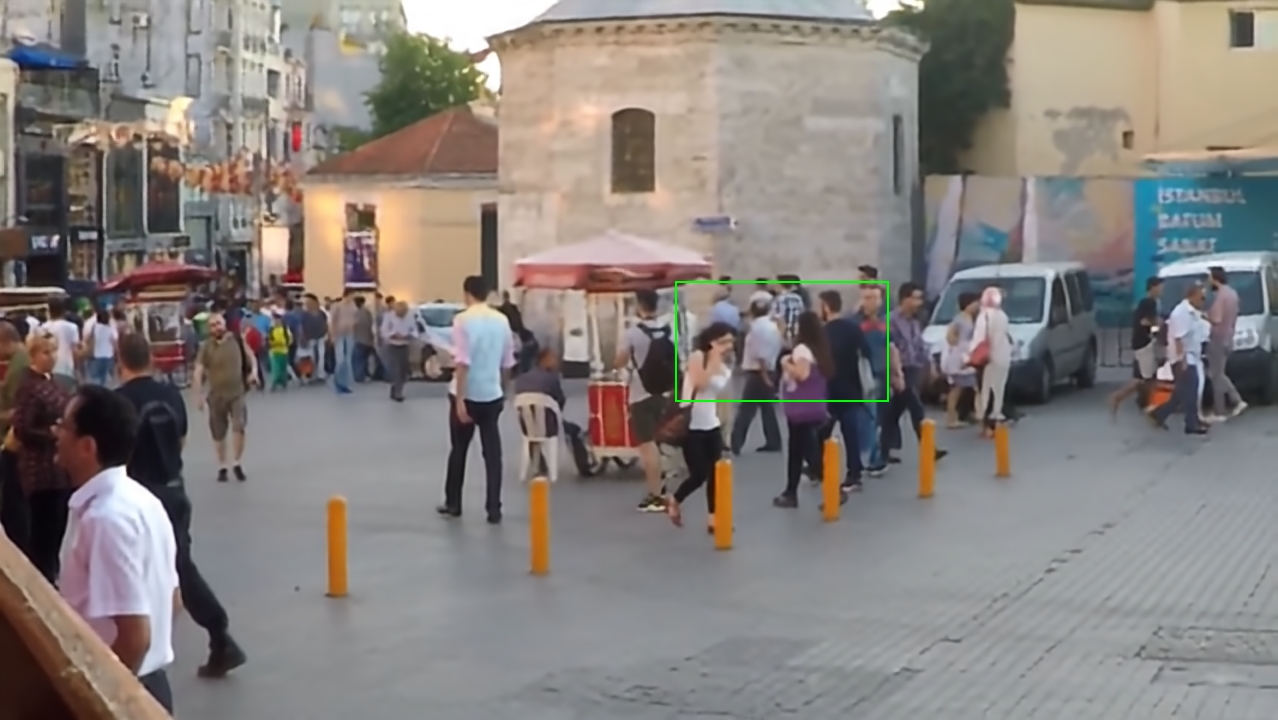} \\
                \includegraphics[width=0.065\linewidth]{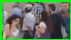}&
                \includegraphics[width=0.13\linewidth]{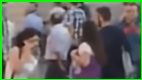}&
                \includegraphics[width=0.26\linewidth]{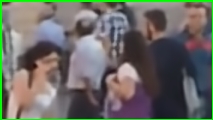}
            \\ \\
            \(s=2\)&
            \(s=4\)&
            \(s=6\)
            \end{tabular}
            &
            \renewcommand{\arraystretch}{0}
            \begin{tabular}{c@{\hskip 0.005\linewidth}c@{\hskip 0.005\linewidth}c}
                \multicolumn{3}{c}{\textbf{EvEnhancer-light (Ours)}}\\
                \includegraphics[width=0.065\linewidth]{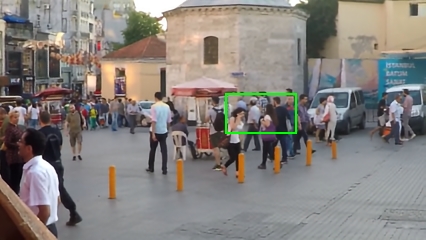}&
                \includegraphics[width=0.13\linewidth]{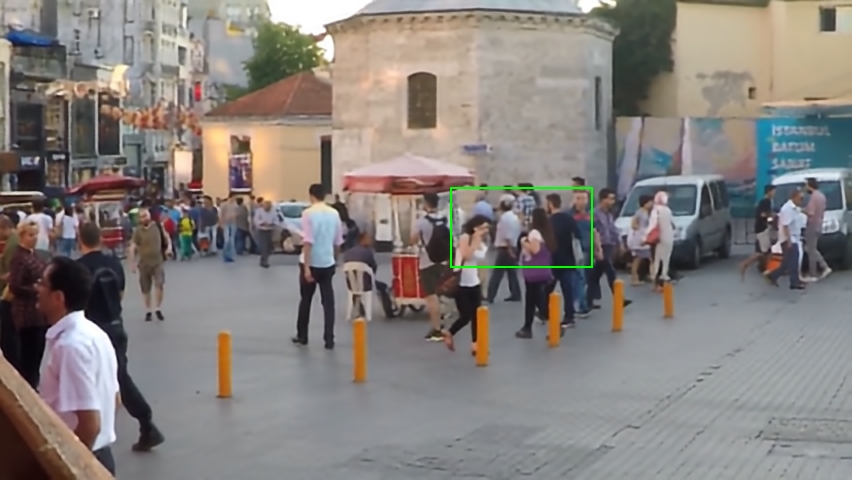}&
                \includegraphics[width=0.26\linewidth]{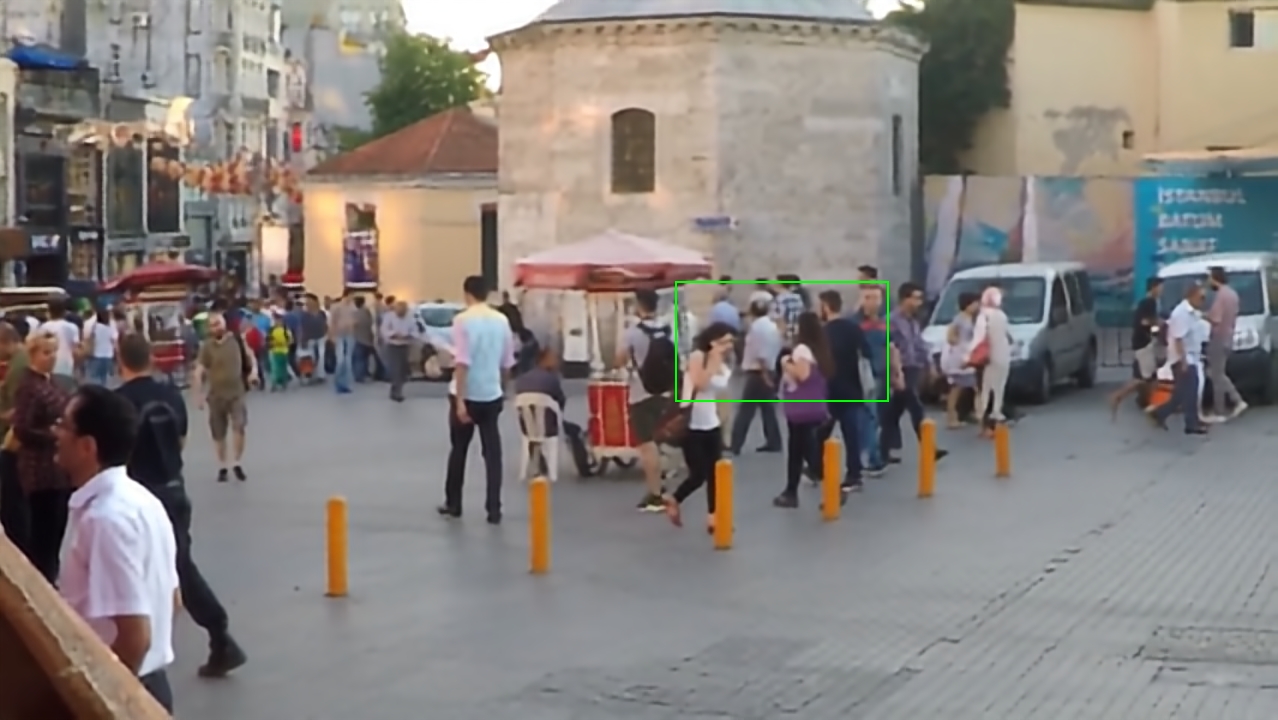} \\
                \includegraphics[width=0.065\linewidth]{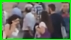}&
                \includegraphics[width=0.13\linewidth]{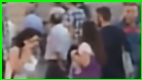}&
                \includegraphics[width=0.26\linewidth]{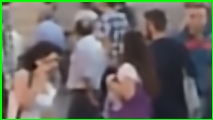}
            \\ \\
            \(s=2\)&
            \(s=4\)&
            \(s=6\)
            \end{tabular}
        \end{tabular}
    \caption{
       Qualitative comparison for different spatial scale (\(s=2,4,6\)), and fixed temporal scale \(t=6\) on the GoPro dataset \cite{nah2017deep}. We display the center frame at \(\mathcal{T}=0.5\). Best zoom in for better visualization.
    }
    \label{fig:s_gopro_t12sx}
\end{figure*}
\renewcommand{\arraystretch}{1.}

%% file: main.bbl
\begin{thebibliography}{94}
\providecommand{\natexlab}[1]{#1}
\providecommand{\url}[1]{\texttt{#1}}
\expandafter\ifx\csname urlstyle\endcsname\relax
  \providecommand{\doi}[1]{doi: #1}\else
  \providecommand{\doi}{doi: \begingroup \urlstyle{rm}\Url}\fi

\bibitem[Bao et~al.(2019)Bao, Lai, Ma, Zhang, Gao, and Yang]{bao2019depth}
Wenbo Bao, Wei-Sheng Lai, Chao Ma, Xiaoyun Zhang, Zhiyong Gao, and Ming-Hsuan Yang.
\newblock Depth-aware video frame interpolation.
\newblock In \emph{Proceedings of the IEEE/CVF conference on computer vision and pattern recognition}, pages 3703--3712, 2019.

\bibitem[Chan et~al.(2021)Chan, Wang, Yu, Dong, and Loy]{chan2021basicvsr}
Kelvin~CK Chan, Xintao Wang, Ke Yu, Chao Dong, and Chen~Change Loy.
\newblock Basicvsr: The search for essential components in video super-resolution and beyond.
\newblock In \emph{Proceedings of the IEEE/CVF conference on computer vision and pattern recognition}, pages 4947--4956, 2021.

\bibitem[Chan et~al.(2022)Chan, Zhou, Xu, and Loy]{chan2022basicvsr++}
Kelvin~CK Chan, Shangchen Zhou, Xiangyu Xu, and Chen~Change Loy.
\newblock Basicvsr++: Improving video super-resolution with enhanced propagation and alignment.
\newblock In \emph{Proceedings of the IEEE/CVF conference on computer vision and pattern recognition}, pages 5972--5981, 2022.

\bibitem[Chen et~al.(2023{\natexlab{a}})Chen, Xu, Hong, Tsai, Kuo, and Lee]{chen2023cascaded}
Hao-Wei Chen, Yu-Syuan Xu, Min-Fong Hong, Yi-Min Tsai, Hsien-Kai Kuo, and Chun-Yi Lee.
\newblock Cascaded local implicit transformer for arbitrary-scale super-resolution.
\newblock In \emph{Proceedings of the IEEE/CVF Conference on Computer Vision and Pattern Recognition}, pages 18257--18267, 2023{\natexlab{a}}.

\bibitem[Chen et~al.(2021)Chen, Liu, and Wang]{chen2021learning}
Yinbo Chen, Sifei Liu, and Xiaolong Wang.
\newblock Learning continuous image representation with local implicit image function.
\newblock In \emph{Proceedings of the IEEE/CVF conference on computer vision and pattern recognition}, pages 8628--8638, 2021.

\bibitem[Chen et~al.(2023{\natexlab{b}})Chen, Chen, Lin, and Peng]{chen2023motif}
Yi-Hsin Chen, Si-Cun Chen, Yen-Yu Lin, and Wen-Hsiao Peng.
\newblock Motif: Learning motion trajectories with local implicit neural functions for continuous space-time video super-resolution.
\newblock In \emph{Proceedings of the IEEE/CVF International Conference on Computer Vision}, pages 23131--23141, 2023{\natexlab{b}}.

\bibitem[Chen et~al.(2022)Chen, Chen, Liu, Xu, Goel, Wang, Shi, and Wang]{chen2022videoinr}
Zeyuan Chen, Yinbo Chen, Jingwen Liu, Xingqian Xu, Vidit Goel, Zhangyang Wang, Humphrey Shi, and Xiaolong Wang.
\newblock Videoinr: Learning video implicit neural representation for continuous space-time super-resolution.
\newblock In \emph{Proceedings of the IEEE/CVF Conference on Computer Vision and Pattern Recognition}, pages 2047--2057, 2022.

\bibitem[Cheng and Chen(2020)]{cheng2020video}
Xianhang Cheng and Zhenzhong Chen.
\newblock Video frame interpolation via deformable separable convolution.
\newblock In \emph{Proceedings of the AAAI Conference on Artificial Intelligence}, pages 10607--10614, 2020.

\bibitem[Chi et~al.(2020)Chi, Mohammadi~Nasiri, Liu, Lu, Tang, and Plataniotis]{chi2020all}
Zhixiang Chi, Rasoul Mohammadi~Nasiri, Zheng Liu, Juwei Lu, Jin Tang, and Konstantinos~N Plataniotis.
\newblock All at once: Temporally adaptive multi-frame interpolation with advanced motion modeling.
\newblock In \emph{Computer Vision--ECCV 2020: 16th European Conference, Glasgow, UK, August 23--28, 2020, Proceedings, Part XXVII 16}, pages 107--123. Springer, 2020.

\bibitem[Dosovitskiy et~al.(2015)Dosovitskiy, Fischer, Ilg, Hausser, Hazirbas, Golkov, Van Der~Smagt, Cremers, and Brox]{dosovitskiy2015flownet}
Alexey Dosovitskiy, Philipp Fischer, Eddy Ilg, Philip Hausser, Caner Hazirbas, Vladimir Golkov, Patrick Van Der~Smagt, Daniel Cremers, and Thomas Brox.
\newblock Flownet: Learning optical flow with convolutional networks.
\newblock In \emph{Proceedings of the IEEE international conference on computer vision}, pages 2758--2766, 2015.

\bibitem[Gehrig et~al.(2020)Gehrig, Gehrig, Hidalgo-Carri{\'o}, and Scaramuzza]{gehrig2020video}
Daniel Gehrig, Mathias Gehrig, Javier Hidalgo-Carri{\'o}, and Davide Scaramuzza.
\newblock Video to events: Recycling video datasets for event cameras.
\newblock In \emph{CVPR}, pages 3586--3595, 2020.

\bibitem[Geng et~al.(2022)Geng, Liang, Ding, and Zharkov]{geng2022rstt}
Zhicheng Geng, Luming Liang, Tianyu Ding, and Ilya Zharkov.
\newblock Rstt: Real-time spatial temporal transformer for space-time video super-resolution.
\newblock In \emph{Proceedings of the IEEE/CVF conference on computer vision and pattern recognition}, pages 17441--17451, 2022.

\bibitem[Han et~al.(2021)Han, Yang, Zhou, Xu, and Shi]{han2021evintsr}
Jin Han, Yixin Yang, Chu Zhou, Chao Xu, and Boxin Shi.
\newblock Evintsr-net: Event guided multiple latent frames reconstruction and super-resolution.
\newblock In \emph{Proceedings of the IEEE/CVF International Conference on Computer Vision}, pages 4882--4891, 2021.

\bibitem[Haris et~al.(2019)Haris, Shakhnarovich, and Ukita]{haris2019recurrent}
Muhammad Haris, Gregory Shakhnarovich, and Norimichi Ukita.
\newblock Recurrent back-projection network for video super-resolution.
\newblock In \emph{Proceedings of the IEEE/CVF conference on computer vision and pattern recognition}, pages 3897--3906, 2019.

\bibitem[Haris et~al.(2020)Haris, Shakhnarovich, and Ukita]{haris2020space}
Muhammad Haris, Greg Shakhnarovich, and Norimichi Ukita.
\newblock Space-time-aware multi-resolution video enhancement.
\newblock In \emph{Proceedings of the IEEE/CVF conference on computer vision and pattern recognition}, pages 2859--2868, 2020.

\bibitem[Hassani et~al.(2023)Hassani, Walton, Li, Li, and Shi]{hassani2023neighborhood}
Ali Hassani, Steven Walton, Jiachen Li, Shen Li, and Humphrey Shi.
\newblock Neighborhood attention transformer.
\newblock In \emph{Proceedings of the IEEE/CVF Conference on Computer Vision and Pattern Recognition}, pages 6185--6194, 2023.

\bibitem[He et~al.(2022)He, You, Qiao, Jia, Zhang, Wang, Lu, Wang, and Liao]{he2022timereplayer}
Weihua He, Kaichao You, Zhendong Qiao, Xu Jia, Ziyang Zhang, Wenhui Wang, Huchuan Lu, Yaoyuan Wang, and Jianxing Liao.
\newblock Timereplayer: Unlocking the potential of event cameras for video interpolation.
\newblock In \emph{Proceedings of the IEEE/CVF Conference on Computer Vision and Pattern Recognition}, pages 17804--17813, 2022.

\bibitem[He and Jin(2024)]{he2024latent}
Zongyao He and Zhi Jin.
\newblock Latent modulated function for computational optimal continuous image representation.
\newblock In \emph{Proceedings of the IEEE/CVF Conference on Computer Vision and Pattern Recognition}, pages 26026--26035, 2024.

\bibitem[Hou et~al.(2022)Hou, Ghildyal, and Liu]{hou2022perceptual}
Qiqi Hou, Abhijay Ghildyal, and Feng Liu.
\newblock A perceptual quality metric for video frame interpolation.
\newblock In \emph{European Conference on Computer Vision}, pages 234--253. Springer, 2022.

\bibitem[Hu et~al.(2022)Hu, Jiang, Liao, Xiao, Jiang, and Wang]{hu2022spatial}
Mengshun Hu, Kui Jiang, Liang Liao, Jing Xiao, Junjun Jiang, and Zheng Wang.
\newblock Spatial-temporal space hand-in-hand: Spatial-temporal video super-resolution via cycle-projected mutual learning.
\newblock In \emph{Proceedings of the IEEE/CVF conference on computer vision and pattern recognition}, pages 3574--3583, 2022.

\bibitem[Hu et~al.(2023{\natexlab{a}})Hu, Jiang, Wang, Bai, and Hu]{hu2023cycmunet+}
Mengshun Hu, Kui Jiang, Zheng Wang, Xiang Bai, and Ruimin Hu.
\newblock Cycmunet+: Cycle-projected mutual learning for spatial-temporal video super-resolution.
\newblock \emph{IEEE Transactions on Pattern Analysis and Machine Intelligence}, 2023{\natexlab{a}}.

\bibitem[Hu et~al.(2023{\natexlab{b}})Hu, Niklaus, Zhang, Sclaroff, and Saenko]{hu2023video}
Ping Hu, Simon Niklaus, Lu Zhang, Stan Sclaroff, and Kate Saenko.
\newblock Video frame interpolation with many-to-many splatting and spatial selective refinement.
\newblock \emph{IEEE Transactions on Pattern Analysis and Machine Intelligence}, 2023{\natexlab{b}}.

\bibitem[Huang et~al.(2024{\natexlab{a}})Huang, Li, Chu, Liu, and Lu]{huang2024arbitrary}
Cong Huang, Jiahao Li, Lei Chu, Dong Liu, and Yan Lu.
\newblock Arbitrary-scale video super-resolution guided by dynamic context.
\newblock In \emph{Proceedings of the AAAI Conference on Artificial Intelligence}, pages 2294--2302, 2024{\natexlab{a}}.

\bibitem[Huang et~al.(2017)Huang, Wang, and Wang]{huang2017video}
Yan Huang, Wei Wang, and Liang Wang.
\newblock Video super-resolution via bidirectional recurrent convolutional networks.
\newblock \emph{IEEE transactions on pattern analysis and machine intelligence}, 40\penalty0 (4):\penalty0 1015--1028, 2017.

\bibitem[Huang et~al.(2022)Huang, Zhang, Heng, Shi, and Zhou]{huang2022real}
Zhewei Huang, Tianyuan Zhang, Wen Heng, Boxin Shi, and Shuchang Zhou.
\newblock Real-time intermediate flow estimation for video frame interpolation.
\newblock In \emph{European Conference on Computer Vision}, pages 624--642. Springer, 2022.

\bibitem[Huang et~al.(2024{\natexlab{b}})Huang, Huang, Hu, Hu, Xu, and Zhou]{huang2024scale}
Zhewei Huang, Ailin Huang, Xiaotao Hu, Chen Hu, Jun Xu, and Shuchang Zhou.
\newblock Scale-adaptive feature aggregation for efficient space-time video super-resolution.
\newblock In \emph{Proceedings of the IEEE/CVF Winter Conference on Applications of Computer Vision}, pages 4228--4239, 2024{\natexlab{b}}.

\bibitem[Ilg et~al.(2017)Ilg, Mayer, Saikia, Keuper, Dosovitskiy, and Brox]{ilg2017flownet}
Eddy Ilg, Nikolaus Mayer, Tonmoy Saikia, Margret Keuper, Alexey Dosovitskiy, and Thomas Brox.
\newblock Flownet 2.0: Evolution of optical flow estimation with deep networks.
\newblock In \emph{Proceedings of the IEEE conference on computer vision and pattern recognition}, pages 2462--2470, 2017.

\bibitem[Isobe et~al.(2020{\natexlab{a}})Isobe, Li, Jia, Yuan, Slabaugh, Xu, Li, Wang, and Tian]{isobe2020video}
Takashi Isobe, Songjiang Li, Xu Jia, Shanxin Yuan, Gregory Slabaugh, Chunjing Xu, Ya-Li Li, Shengjin Wang, and Qi Tian.
\newblock Video super-resolution with temporal group attention.
\newblock In \emph{Proceedings of the IEEE/CVF conference on computer vision and pattern recognition}, pages 8008--8017, 2020{\natexlab{a}}.

\bibitem[Isobe et~al.(2020{\natexlab{b}})Isobe, Zhu, Jia, and Wang]{isobe2020revisiting}
Takashi Isobe, Fang Zhu, Xu Jia, and Shengjin Wang.
\newblock Revisiting temporal modeling for video super-resolution.
\newblock \emph{arXiv preprint arXiv:2008.05765}, 2020{\natexlab{b}}.

\bibitem[Jiang et~al.(2018)Jiang, Sun, Jampani, Yang, Learned-Miller, and Kautz]{jiang2018super}
Huaizu Jiang, Deqing Sun, Varun Jampani, Ming-Hsuan Yang, Erik Learned-Miller, and Jan Kautz.
\newblock Super slomo: High quality estimation of multiple intermediate frames for video interpolation.
\newblock In \emph{Proceedings of the IEEE conference on computer vision and pattern recognition}, pages 9000--9008, 2018.

\bibitem[Jing et~al.(2021)Jing, Yang, Wang, Song, and Tao]{jing2021turning}
Yongcheng Jing, Yiding Yang, Xinchao Wang, Mingli Song, and Dacheng Tao.
\newblock Turning frequency to resolution: Video super-resolution via event cameras.
\newblock In \emph{Proceedings of the IEEE/CVF Conference on Computer Vision and Pattern Recognition}, pages 7772--7781, 2021.

\bibitem[Kai et~al.(2024)Kai, Lu, Zhang, and Sun]{kai2024evtexture}
Dachun Kai, Jiayao Lu, Yueyi Zhang, and Xiaoyan Sun.
\newblock {E}v{T}exture: {E}vent-driven {T}exture {E}nhancement for {V}ideo {S}uper-{R}esolution.
\newblock In \emph{Proceedings of the 41st International Conference on Machine Learning}, pages 22817--22839. PMLR, 2024.

\bibitem[Kalluri et~al.(2023)Kalluri, Pathak, Chandraker, and Tran]{kalluri2023flavr}
Tarun Kalluri, Deepak Pathak, Manmohan Chandraker, and Du Tran.
\newblock Flavr: Flow-agnostic video representations for fast frame interpolation.
\newblock In \emph{Proceedings of the IEEE/CVF winter conference on applications of computer vision}, pages 2071--2082, 2023.

\bibitem[Kim et~al.(2020)Kim, Oh, and Kim]{kim2020fisr}
Soo~Ye Kim, Jihyong Oh, and Munchurl Kim.
\newblock Fisr: Deep joint frame interpolation and super-resolution with a multi-scale temporal loss.
\newblock In \emph{Proceedings of the AAAI Conference on Artificial Intelligence}, pages 11278--11286, 2020.

\bibitem[Kim et~al.(2023)Kim, Chae, Jang, and Yoon]{kim2023event}
Taewoo Kim, Yujeong Chae, Hyun-Kurl Jang, and Kuk-Jin Yoon.
\newblock Event-based video frame interpolation with cross-modal asymmetric bidirectional motion fields.
\newblock In \emph{Proceedings of the IEEE/CVF Conference on Computer Vision and Pattern Recognition}, pages 18032--18042, 2023.

\bibitem[Kingma and Ba(2014)]{kingma2014adam}
Diederik~P Kingma and Jimmy Ba.
\newblock Adam: A method for stochastic optimization.
\newblock \emph{arXiv preprint arXiv:1412.6980}, 2014.

\bibitem[Lai et~al.(2017)Lai, Huang, Ahuja, and Yang]{lai2017deep}
Wei-Sheng Lai, Jia-Bin Huang, Narendra Ahuja, and Ming-Hsuan Yang.
\newblock Deep laplacian pyramid networks for fast and accurate super-resolution.
\newblock In \emph{Proceedings of the IEEE conference on computer vision and pattern recognition}, pages 624--632, 2017.

\bibitem[Lee et~al.(2020)Lee, Kim, Chung, Pak, Ban, and Lee]{lee2020adacof}
Hyeongmin Lee, Taeoh Kim, Tae-young Chung, Daehyun Pak, Yuseok Ban, and Sangyoun Lee.
\newblock Adacof: Adaptive collaboration of flows for video frame interpolation.
\newblock In \emph{Proceedings of the IEEE/CVF conference on computer vision and pattern recognition}, pages 5316--5325, 2020.

\bibitem[Lee and Jin(2022)]{lee2022local}
Jaewon Lee and Kyong~Hwan Jin.
\newblock Local texture estimator for implicit representation function.
\newblock In \emph{Proceedings of the IEEE/CVF conference on computer vision and pattern recognition}, pages 1929--1938, 2022.

\bibitem[Li et~al.(2020{\natexlab{a}})Li, Bai, and Zhao]{li2020learning}
Feng Li, Huihui Bai, and Yao Zhao.
\newblock Learning a deep dual attention network for video super-resolution.
\newblock \emph{IEEE transactions on image processing}, 29:\penalty0 4474--4488, 2020{\natexlab{a}}.

\bibitem[Li et~al.(2024{\natexlab{a}})Li, Wu, Li, Bai, Cong, and Zhao]{li2024enhanced}
Feng Li, Yixuan Wu, Anqi Li, Huihui Bai, Runmin Cong, and Yao Zhao.
\newblock Enhanced video super-resolution network towards compressed data.
\newblock \emph{ACM Transactions on Multimedia Computing, Communications and Applications}, 20\penalty0 (7):\penalty0 1--21, 2024{\natexlab{a}}.

\bibitem[Li et~al.(2019)Li, He, Du, Zhang, Xu, and Tao]{li2019fast}
Sheng Li, Fengxiang He, Bo Du, Lefei Zhang, Yonghao Xu, and Dacheng Tao.
\newblock Fast spatio-temporal residual network for video super-resolution.
\newblock In \emph{Proceedings of the IEEE/CVF Conference on Computer Vision and Pattern Recognition}, pages 10522--10531, 2019.

\bibitem[Li et~al.(2020{\natexlab{b}})Li, Tao, Guo, Qi, Lu, and Jia]{li2020mucan}
Wenbo Li, Xin Tao, Taian Guo, Lu Qi, Jiangbo Lu, and Jiaya Jia.
\newblock Mucan: Multi-correspondence aggregation network for video super-resolution.
\newblock In \emph{Computer Vision--ECCV 2020: 16th European Conference, Glasgow, UK, August 23--28, 2020, Proceedings, Part X 16}, pages 335--351. Springer, 2020{\natexlab{b}}.

\bibitem[Li et~al.(2024{\natexlab{b}})Li, Liu, Shang, Liu, Wan, and Feng]{li2024savsr}
Zekun Li, Hongying Liu, Fanhua Shang, Yuanyuan Liu, Liang Wan, and Wei Feng.
\newblock Savsr: Arbitrary-scale video super-resolution via a learned scale-adaptive network.
\newblock In \emph{Proceedings of the AAAI Conference on Artificial Intelligence}, pages 3288--3296, 2024{\natexlab{b}}.

\bibitem[Liang et~al.(2022)Liang, Fan, Xiang, Ranjan, Ilg, Green, Cao, Zhang, Timofte, and Gool]{liang2022recurrent}
Jingyun Liang, Yuchen Fan, Xiaoyu Xiang, Rakesh Ranjan, Eddy Ilg, Simon Green, Jiezhang Cao, Kai Zhang, Radu Timofte, and Luc~V Gool.
\newblock Recurrent video restoration transformer with guided deformable attention.
\newblock \emph{Advances in Neural Information Processing Systems}, 35:\penalty0 378--393, 2022.

\bibitem[Lin et~al.(2020)Lin, Zhang, Pan, Jiang, Zou, Wang, Chen, and Ren]{lin2020learning}
Songnan Lin, Jiawei Zhang, Jinshan Pan, Zhe Jiang, Dongqing Zou, Yongtian Wang, Jing Chen, and Jimmy Ren.
\newblock Learning event-driven video deblurring and interpolation.
\newblock In \emph{Computer Vision--ECCV 2020: 16th European Conference, Glasgow, UK, August 23--28, 2020, Proceedings, Part VIII 16}, pages 695--710. Springer, 2020.

\bibitem[Liu et~al.(2022)Liu, Yang, Fu, and Qian]{liu2022learning}
Chengxu Liu, Huan Yang, Jianlong Fu, and Xueming Qian.
\newblock Learning trajectory-aware transformer for video super-resolution.
\newblock In \emph{Proceedings of the IEEE/CVF conference on computer vision and pattern recognition}, pages 5687--5696, 2022.

\bibitem[Liu et~al.(2024)Liu, Zhang, Zhao, and Wang]{liu2024sparse}
Chunxu Liu, Guozhen Zhang, Rui Zhao, and Limin Wang.
\newblock Sparse global matching for video frame interpolation with large motion.
\newblock In \emph{Proceedings of the IEEE/CVF Conference on Computer Vision and Pattern Recognition}, pages 19125--19134, 2024.

\bibitem[Lu et~al.(2022)Lu, Wu, Lin, Lu, and Jia]{lu2022video}
Liying Lu, Ruizheng Wu, Huaijia Lin, Jiangbo Lu, and Jiaya Jia.
\newblock Video frame interpolation with transformer.
\newblock In \emph{Proceedings of the IEEE/CVF Conference on Computer Vision and Pattern Recognition}, pages 3532--3542, 2022.

\bibitem[Lu et~al.(2023)Lu, Wang, Liu, Wang, and Wang]{lu2023learning}
Yunfan Lu, Zipeng Wang, Minjie Liu, Hongjian Wang, and Lin Wang.
\newblock Learning spatial-temporal implicit neural representations for event-guided video super-resolution.
\newblock In \emph{Proceedings of the IEEE/CVF Conference on Computer Vision and Pattern Recognition}, pages 1557--1567, 2023.

\bibitem[Lu et~al.(2024)Lu, Wang, Wang, and Xiong]{lu2024hr}
Yunfan Lu, Zipeng Wang, Yusheng Wang, and Hui Xiong.
\newblock Hr-inr: Continuous space-time video super-resolution via event camera.
\newblock \emph{arXiv preprint arXiv:2405.13389}, 2024.

\bibitem[Mildenhall et~al.(2021)Mildenhall, Srinivasan, Tancik, Barron, Ramamoorthi, and Ng]{mildenhall2021nerf}
Ben Mildenhall, Pratul~P Srinivasan, Matthew Tancik, Jonathan~T Barron, Ravi Ramamoorthi, and Ren Ng.
\newblock Nerf: Representing scenes as neural radiance fields for view synthesis.
\newblock \emph{Communications of the ACM}, 65\penalty0 (1):\penalty0 99--106, 2021.

\bibitem[Nah et~al.(2017)Nah, Hyun~Kim, and Mu~Lee]{nah2017deep}
Seungjun Nah, Tae Hyun~Kim, and Kyoung Mu~Lee.
\newblock Deep multi-scale convolutional neural network for dynamic scene deblurring.
\newblock In \emph{Proceedings of the IEEE conference on computer vision and pattern recognition}, pages 3883--3891, 2017.

\bibitem[Niklaus and Liu(2020)]{niklaus2020softmax}
Simon Niklaus and Feng Liu.
\newblock Softmax splatting for video frame interpolation.
\newblock In \emph{Proceedings of the IEEE/CVF conference on computer vision and pattern recognition}, pages 5437--5446, 2020.

\bibitem[Niklaus et~al.(2017{\natexlab{a}})Niklaus, Mai, and Liu]{niklaus2017video1}
Simon Niklaus, Long Mai, and Feng Liu.
\newblock Video frame interpolation via adaptive separable convolution.
\newblock In \emph{Proceedings of the IEEE international conference on computer vision}, pages 261--270, 2017{\natexlab{a}}.

\bibitem[Niklaus et~al.(2017{\natexlab{b}})Niklaus, Mai, and Liu]{niklaus2017video2}
Simon Niklaus, Long Mai, and Feng Liu.
\newblock Video frame interpolation via adaptive convolution.
\newblock In \emph{Proceedings of the IEEE conference on computer vision and pattern recognition}, pages 670--679, 2017{\natexlab{b}}.

\bibitem[Niklaus et~al.(2023)Niklaus, Hu, and Chen]{niklaus2023splatting}
Simon Niklaus, Ping Hu, and Jiawen Chen.
\newblock Splatting-based synthesis for video frame interpolation.
\newblock In \emph{Proceedings of the IEEE/CVF Winter Conference on Applications of Computer Vision}, pages 713--723, 2023.

\bibitem[Park et~al.(2021)Park, Lee, and Kim]{park2021asymmetric}
Junheum Park, Chul Lee, and Chang-Su Kim.
\newblock Asymmetric bilateral motion estimation for video frame interpolation.
\newblock In \emph{Proceedings of the IEEE/CVF International Conference on Computer Vision}, pages 14539--14548, 2021.

\bibitem[Qiu et~al.(2023)Qiu, Yang, Fu, Liu, Xu, and Fu]{qiu2023learning}
Zhongwei Qiu, Huan Yang, Jianlong Fu, Daochang Liu, Chang Xu, and Dongmei Fu.
\newblock Learning degradation-robust spatiotemporal frequency-transformer for video super-resolution.
\newblock \emph{IEEE Transactions on Pattern Analysis and Machine Intelligence}, 2023.

\bibitem[Rahimi and Tekalp(2023)]{rahimi2023spatio}
Nasrin Rahimi and A~Murat Tekalp.
\newblock Spatio-temporal perception-distortion trade-off in learned video sr.
\newblock In \emph{2023 IEEE International Conference on Image Processing (ICIP)}, pages 1400--1404. IEEE, 2023.

\bibitem[Reda et~al.(2019)Reda, Sun, Dundar, Shoeybi, Liu, Shih, Tao, Kautz, and Catanzaro]{reda2019unsupervised}
Fitsum~A Reda, Deqing Sun, Aysegul Dundar, Mohammad Shoeybi, Guilin Liu, Kevin~J Shih, Andrew Tao, Jan Kautz, and Bryan Catanzaro.
\newblock Unsupervised video interpolation using cycle consistency.
\newblock In \emph{Proceedings of the IEEE/CVF international conference on computer Vision}, pages 892--900, 2019.

\bibitem[Shahar et~al.(2011)Shahar, Faktor, and Irani]{shahar2011space}
Oded Shahar, Alon Faktor, and Michal Irani.
\newblock \emph{Space-time super-resolution from a single video}.
\newblock IEEE, 2011.

\bibitem[Shang et~al.(2025)Shang, Ren, Zhang, Fang, Zuo, and Ma]{shang2025arbitrary}
Wei Shang, Dongwei Ren, Wanying Zhang, Yuming Fang, Wangmeng Zuo, and Kede Ma.
\newblock Arbitrary-scale video super-resolution with structural and textural priors.
\newblock In \emph{European Conference on Computer Vision}, pages 73--90. Springer, 2025.

\bibitem[Shechtman et~al.(2005)Shechtman, Caspi, and Irani]{shechtman2005space}
Eli Shechtman, Yaron Caspi, and Michal Irani.
\newblock Space-time super-resolution.
\newblock \emph{IEEE Transactions on Pattern Analysis and Machine Intelligence}, 27\penalty0 (4):\penalty0 531--545, 2005.

\bibitem[Shi et~al.(2022)Shi, Xu, Liu, Chen, and Yang]{shi2022video}
Zhihao Shi, Xiangyu Xu, Xiaohong Liu, Jun Chen, and Ming-Hsuan Yang.
\newblock Video frame interpolation transformer.
\newblock In \emph{Proceedings of the IEEE/CVF Conference on Computer Vision and Pattern Recognition}, pages 17482--17491, 2022.

\bibitem[Su et~al.(2017)Su, Delbracio, Wang, Sapiro, Heidrich, and Wang]{su2017deep}
Shuochen Su, Mauricio Delbracio, Jue Wang, Guillermo Sapiro, Wolfgang Heidrich, and Oliver Wang.
\newblock Deep video deblurring for hand-held cameras.
\newblock In \emph{Proceedings of the IEEE conference on computer vision and pattern recognition}, pages 1279--1288, 2017.

\bibitem[Sun et~al.(2018)Sun, Yang, Liu, and Kautz]{sun2018pwc}
Deqing Sun, Xiaodong Yang, Ming-Yu Liu, and Jan Kautz.
\newblock Pwc-net: Cnns for optical flow using pyramid, warping, and cost volume.
\newblock In \emph{Proceedings of the IEEE conference on computer vision and pattern recognition}, pages 8934--8943, 2018.

\bibitem[Sun et~al.(2023)Sun, Sakaridis, Liang, Sun, Cao, Zhang, Jiang, Wang, and Van~Gool]{sun2023event}
Lei Sun, Christos Sakaridis, Jingyun Liang, Peng Sun, Jiezhang Cao, Kai Zhang, Qi Jiang, Kaiwei Wang, and Luc Van~Gool.
\newblock Event-based frame interpolation with ad-hoc deblurring.
\newblock In \emph{Proceedings of the IEEE/CVF Conference on Computer Vision and Pattern Recognition}, pages 18043--18052, 2023.

\bibitem[Tian et~al.(2020)Tian, Zhang, Fu, and Xu]{tian2020tdan}
Yapeng Tian, Yulun Zhang, Yun Fu, and Chenliang Xu.
\newblock Tdan: Temporally-deformable alignment network for video super-resolution.
\newblock In \emph{Proceedings of the IEEE/CVF conference on computer vision and pattern recognition}, pages 3360--3369, 2020.

\bibitem[Tulyakov et~al.(2021)Tulyakov, Gehrig, Georgoulis, Erbach, Gehrig, Li, and Scaramuzza]{tulyakov2021time}
Stepan Tulyakov, Daniel Gehrig, Stamatios Georgoulis, Julius Erbach, Mathias Gehrig, Yuanyou Li, and Davide Scaramuzza.
\newblock Time lens: Event-based video frame interpolation.
\newblock In \emph{Proceedings of the IEEE/CVF conference on computer vision and pattern recognition}, pages 16155--16164, 2021.

\bibitem[Tulyakov et~al.(2022)Tulyakov, Bochicchio, Gehrig, Georgoulis, Li, and Scaramuzza]{tulyakov2022time}
Stepan Tulyakov, Alfredo Bochicchio, Daniel Gehrig, Stamatios Georgoulis, Yuanyou Li, and Davide Scaramuzza.
\newblock Time lens++: Event-based frame interpolation with parametric non-linear flow and multi-scale fusion.
\newblock In \emph{Proceedings of the IEEE/CVF Conference on Computer Vision and Pattern Recognition}, pages 17755--17764, 2022.

\bibitem[Wang et~al.(2022)Wang, Yang, Liao, and Zhou]{wang2022bi}
Hai Wang, Wenming Yang, Qingmin Liao, and Jie Zhou.
\newblock Bi-rstu: Bidirectional recurrent upsampling network for space-time video super-resolution.
\newblock \emph{IEEE Transactions on Multimedia}, 2022.

\bibitem[Wang et~al.(2023{\natexlab{a}})Wang, Xiang, Tian, Yang, and Liao]{wang2023stdan}
Hai Wang, Xiaoyu Xiang, Yapeng Tian, Wenming Yang, and Qingmin Liao.
\newblock Stdan: deformable attention network for space-time video super-resolution.
\newblock \emph{IEEE Transactions on Neural Networks and Learning Systems}, 2023{\natexlab{a}}.

\bibitem[Wang et~al.(2019{\natexlab{a}})Wang, Guo, Lin, Deng, and An]{wang2019learning}
Longguang Wang, Yulan Guo, Zaiping Lin, Xinpu Deng, and Wei An.
\newblock Learning for video super-resolution through hr optical flow estimation.
\newblock In \emph{Computer Vision--ACCV 2018: 14th Asian Conference on Computer Vision, Perth, Australia, December 2--6, 2018, Revised Selected Papers, Part I 14}, pages 514--529. Springer, 2019{\natexlab{a}}.

\bibitem[Wang et~al.(2020)Wang, Li, Zhu, Tian, and Shan]{wang2020dual}
Li Wang, Dong Li, Yousong Zhu, Lu Tian, and Yi Shan.
\newblock Dual super-resolution learning for semantic segmentation.
\newblock In \emph{Proceedings of the IEEE/CVF conference on computer vision and pattern recognition}, pages 3774--3783, 2020.

\bibitem[Wang et~al.(2018)Wang, Girshick, Gupta, and He]{wang2018non}
Xiaolong Wang, Ross Girshick, Abhinav Gupta, and Kaiming He.
\newblock Non-local neural networks.
\newblock In \emph{Proceedings of the IEEE conference on computer vision and pattern recognition}, pages 7794--7803, 2018.

\bibitem[Wang et~al.(2019{\natexlab{b}})Wang, Chan, Yu, Dong, and Change~Loy]{wang2019edvr}
Xintao Wang, Kelvin~CK Chan, Ke Yu, Chao Dong, and Chen Change~Loy.
\newblock Edvr: Video restoration with enhanced deformable convolutional networks.
\newblock In \emph{Proceedings of the IEEE/CVF conference on computer vision and pattern recognition workshops}, pages 0--0, 2019{\natexlab{b}}.

\bibitem[Wang et~al.(2023{\natexlab{b}})Wang, Isobe, Jia, Tao, Lu, and Tai]{wang2023compression}
Yingwei Wang, Takashi Isobe, Xu Jia, Xin Tao, Huchuan Lu, and Yu-Wing Tai.
\newblock Compression-aware video super-resolution.
\newblock In \emph{Proceedings of the IEEE/CVF Conference on Computer Vision and Pattern Recognition}, pages 2012--2021, 2023{\natexlab{b}}.

\bibitem[Wu et~al.(2024)Wu, Tao, Li, Wang, Liu, and Zheng]{wu2024perception}
Guangyang Wu, Xin Tao, Changlin Li, Wenyi Wang, Xiaohong Liu, and Qingqing Zheng.
\newblock Perception-oriented video frame interpolation via asymmetric blending.
\newblock In \emph{Proceedings of the IEEE/CVF Conference on Computer Vision and Pattern Recognition}, pages 2753--2762, 2024.

\bibitem[Wu et~al.(2022)Wu, You, He, Yang, Tian, Wang, Zhang, and Liao]{wu2022video}
Song Wu, Kaichao You, Weihua He, Chen Yang, Yang Tian, Yaoyuan Wang, Ziyang Zhang, and Jianxing Liao.
\newblock Video interpolation by event-driven anisotropic adjustment of optical flow.
\newblock In \emph{European Conference on Computer Vision}, pages 267--283. Springer, 2022.

\bibitem[Xiang et~al.(2020)Xiang, Tian, Zhang, Fu, Allebach, and Xu]{xiang2020zooming}
Xiaoyu Xiang, Yapeng Tian, Yulun Zhang, Yun Fu, Jan~P Allebach, and Chenliang Xu.
\newblock Zooming slow-mo: Fast and accurate one-stage space-time video super-resolution.
\newblock In \emph{Proceedings of the IEEE/CVF conference on computer vision and pattern recognition}, pages 3370--3379, 2020.

\bibitem[Xu et~al.(2024)Xu, Liu, Yao, Lin, and Zhao]{xu2024ibvc}
Chenming Xu, Meiqin Liu, Chao Yao, Weisi Lin, and Yao Zhao.
\newblock Ibvc: Interpolation-driven b-frame video compression.
\newblock \emph{Pattern Recognition}, 153:\penalty0 110465, 2024.

\bibitem[Xu et~al.(2021)Xu, Xu, Li, Wang, Sun, and Cheng]{xu2021temporal}
Gang Xu, Jun Xu, Zhen Li, Liang Wang, Xing Sun, and Ming-Ming Cheng.
\newblock Temporal modulation network for controllable space-time video super-resolution.
\newblock In \emph{Proceedings of the IEEE/CVF conference on computer vision and pattern recognition}, pages 6388--6397, 2021.

\bibitem[Xu et~al.(2019)Xu, Siyao, Sun, Yin, and Yang]{xu2019quadratic}
Xiangyu Xu, Li Siyao, Wenxiu Sun, Qian Yin, and Ming-Hsuan Yang.
\newblock Quadratic video interpolation.
\newblock \emph{Advances in Neural Information Processing Systems}, 32, 2019.

\bibitem[Yang et~al.(2021)Yang, Shen, Yue, and Li]{yang2021implicit}
Jingyu Yang, Sheng Shen, Huanjing Yue, and Kun Li.
\newblock Implicit transformer network for screen content image continuous super-resolution.
\newblock \emph{Advances in Neural Information Processing Systems}, 34:\penalty0 13304--13315, 2021.

\bibitem[Yi et~al.(2019)Yi, Wang, Jiang, Jiang, and Ma]{yi2019progressive}
Peng Yi, Zhongyuan Wang, Kui Jiang, Junjun Jiang, and Jiayi Ma.
\newblock Progressive fusion video super-resolution network via exploiting non-local spatio-temporal correlations.
\newblock In \emph{Proceedings of the IEEE/CVF international conference on computer vision}, pages 3106--3115, 2019.

\bibitem[Yoo et~al.(2023)Yoo, Lee, and Jung]{yoo2023video}
Jun-Sang Yoo, Hongjae Lee, and Seung-Won Jung.
\newblock Video object segmentation-aware video frame interpolation.
\newblock In \emph{Proceedings of the IEEE/CVF International Conference on Computer Vision}, pages 12322--12333, 2023.

\bibitem[Yu et~al.(2022)Yu, Liu, Bo, and Mei]{yu2022memory}
Jiyang Yu, Jingen Liu, Liefeng Bo, and Tao Mei.
\newblock Memory-augmented non-local attention for video super-resolution.
\newblock In \emph{Proceedings of the IEEE/CVF conference on computer vision and pattern recognition}, pages 17834--17843, 2022.

\bibitem[Yu et~al.(2021)Yu, Zhang, Liu, Zou, Chen, Liu, and Ren]{yu2021training}
Zhiyang Yu, Yu Zhang, Deyuan Liu, Dongqing Zou, Xijun Chen, Yebin Liu, and Jimmy~S Ren.
\newblock Training weakly supervised video frame interpolation with events.
\newblock In \emph{Proceedings of the IEEE/CVF International Conference on Computer Vision}, pages 14589--14598, 2021.

\bibitem[Zhang et~al.(2023)Zhang, Zhu, Wang, Chen, Wu, and Wang]{zhang2023extracting}
Guozhen Zhang, Yuhan Zhu, Haonan Wang, Youxin Chen, Gangshan Wu, and Limin Wang.
\newblock Extracting motion and appearance via inter-frame attention for efficient video frame interpolation.
\newblock In \emph{Proceedings of the IEEE/CVF Conference on Computer Vision and Pattern Recognition}, pages 5682--5692, 2023.

\bibitem[Zhang et~al.(2019)Zhang, Liu, and Xiong]{zhang2019two}
Haochen Zhang, Dong Liu, and Zhiwei Xiong.
\newblock Two-stream action recognition-oriented video super-resolution.
\newblock In \emph{Proceedings of the IEEE/CVF international conference on computer vision}, pages 8799--8808, 2019.

\bibitem[Zhou et~al.(2022)Zhou, Li, Lu, Han, and Lu]{zhou2022revisiting}
Kun Zhou, Wenbo Li, Liying Lu, Xiaoguang Han, and Jiangbo Lu.
\newblock Revisiting temporal alignment for video restoration.
\newblock In \emph{Proceedings of the IEEE/CVF conference on computer vision and pattern recognition}, pages 6053--6062, 2022.

\bibitem[Zhou et~al.(2024)Zhou, Zhang, Zhao, Wang, Li, and Gu]{zhou2024video}
Xingyu Zhou, Leheng Zhang, Xiaorui Zhao, Keze Wang, Leida Li, and Shuhang Gu.
\newblock Video super-resolution transformer with masked inter\&intra-frame attention.
\newblock In \emph{Proceedings of the IEEE/CVF Conference on Computer Vision and Pattern Recognition}, pages 25399--25408, 2024.

\bibitem[Zhu et~al.(2019)Zhu, Hu, Lin, and Dai]{zhu2019deformable}
Xizhou Zhu, Han Hu, Stephen Lin, and Jifeng Dai.
\newblock Deformable convnets v2: More deformable, better results.
\newblock In \emph{Proceedings of the IEEE/CVF conference on computer vision and pattern recognition}, pages 9308--9316, 2019.

\end{thebibliography}
